# Non-collinear Magnetoelectronics

Arne Brataas

*Department of Physics, Norwegian University of Science and Technology, N-7491 Trondheim, Norway*

Gerrit E. W. Bauer

*Kavli Institute of NanoScience, Delft University of Technology, 2628 CJ Delft, The Netherlands*

Paul J. Kelly

*Faculty of Science and Technology and MESA+ Research Institute, University of Twente, P.O. Box 217, 7500 AE Enschede, The Netherlands*

## Abstract

The electron transport properties of hybrid ferromagnetic|normal metal structures such as multilayers and spin valves depend on the relative orientation of the magnetization direction of the ferromagnetic elements. Whereas the contrast in the resistance for parallel and antiparallel magnetizations, the so-called Giant Magnetoresistance, is relatively well understood for quite some time, a coherent picture for non-collinear magnetoelectronic circuits and devices has evolved only recently. We review here such a theory for electron charge and spin transport with general magnetization directions that is based on the semiclassical concept of a vector *spin accumulation.* In conjunction with first-principles calculations of scattering matrices many phenomena, e.g. the current-induced spin-transfer torque, can be understood and predicted quantitatively for different material combinations.

**Contents**

## I. INTRODUCTION

The term magnetoelectronics has to a large extent been synonymous with the giant magnetoresistance (GMR) in ferromagnetic multilayers and tunnel junctions [1, 2], *i.e.* the modulation of the electron transport by the magnetic-field-induced configuration changes of the magnetization profile. Much of the interest in magnetoelectronic phenomena is motivated by its technological potential. The dependence of the electrical resistance of ferromagnetic/normal metal spin valves and multilayers on applied magnetic fields has been employed in read heads for mass data storage devices. Magnetic random access memories (MRAMs) are based on the related effect of tunnelling magnetoresistance (TMR) between two ferromagnets separated by a tunnelling barrier. MRAMs have the advantage to be non-volatile, which means that no applied voltage is necessary to maintain a given memory state and are therefore serious contenders of flash memories in applications like reprogrammable logics and processors. These and other applications are reviewed in Refs. [3–5].

Basic research in magnetoelectronics is moving rapidly to smaller structures and novel materials. The unifying concept is that of spin-accumulation, *i.e.* the non-equilibrium magnetization which is injected into a non-magnetic material by a ferromagnetic contact by an applied voltage, which has been pioneered by Johnson and Silsbee [6, 7]. The smallest lateral structures that can be fabricated by advanced lithography are of the order of 100 nm [8, 9], which is of the same order of magnitude as for state of the art semiconductor structures. Basic research focuses on new physical phenomena and functionalities of ferromagnets, with topics like "spin transistors" [10–15], ferromagnetic single electron transistors [16], hybrid superconductor-ferromagnet structures [17–19], spin-injection into semiconductors [20, 21], molecules and carbon nanotubes [22], the fabrication of sophisticated magnetoelectronic structures at mesoscopic length scales [23] that can be used to analyze spin precession in diffuse metals [9, 24], and field-induced magnetism in semiconductors [25]. An important breakthrough in magnetoelectronics is the prediction [26–28] and subsequent observation of spin-current induced magnetization reversal in layered structures fabricated into pillars with diameters of about 50 nanometers [8, 29].

In order to keep this review manageable, we chose to *not* discuss in depth several interesting aspects of mesoscopic and nanoscale magnetoelectronics. From the outset we exclude many topics which are fit under a common umbrella of "spintronics", like macroscopic quantum coherence of magnetism, single-electron spin manipulation in semiconductor quantum dots [30], or topics related to the use of spin in quantum information processing [31]. The competition between superconducting and ferromagnetic order parameters in small structures has been the topic of many recent studies that are not covered here.[1] Instead we address an axiomatic *ab initio* theory of the DC transport properties of metallic magnetoelectronic circuits and devices as a function of the magnetization direction of the ferromagnetic elements and the applied potentials. The emphasis is on our own work and interests, but we try to give sufficient cross references to put it into perspective. The present theory provides a comprehensive recipe to understand and compute the spin-current induced magnetization or spin transfer torque [26, 27] and is discussed in some detail. However, the study of time dependent phenomena, such as the dynamics of the magnetization reversal, the spin pumping, enhanced damping of the magnetization dynamics, and dynamic cross talk in multilayers

[1] Andreev scattering at normal or ferromagnetic point contacts with superconductors can be treated by the instruments discussed in the following chapters [19].

[32], is subject of a separate review paper [33].

The theoretical formalism most appropriate for much of magnetoelectronics to date is a semiclassical circuit theory [12, 34], in which the key parameters are expressed in terms of the scattering matrix of the current limiting elements. The latter are accessible to first-principles calculation, even in disordered systems [35, 36]. It is our aim to provide a complete exposure of the first-principles circuit theory in this review. Much of the physics to be discussed here is relevant for the GMR in the current perpendicular to the plane (CPP) configuration as reviewed in [37–39]. Two recent reviews in the form of a monograph on the experimental and computational aspects of the giant magnetoresistance [40] and an anthology, which additionally addresses the tunneling magnetoresistance [41], are complementary to the present one. Semiconductor spintronics has been reviewed in [42, 43]. An in many aspects different point of view on transport in layered metallic systems is expressed in [44].

### A. Spin current and spin accumulation

A ferromagnet [45] is characterized by a phase transition at a critical (Curie) temperature $T_c$ at which the spin-rotational symmetry is broken by a collective ordering of the electron spins creating a macroscopic magnetic moment. Ferromagnetism is driven by the strong exchange interaction based on the Coulomb interaction and the Pauli principle, corresponding to very high $T_c$'s (*e.g.* 1400 K for cobalt). The dipole-dipole interaction in larger samples of ferromagnets usually causes the ferromagnet to be divided into domains of coherent magnetization, which minimize the energy of the macroscopic magnetic field outside the sample. The domains are separated by domain walls in which the order parameter rotates between two bulk values. The domain walls are a source of electron scattering that vanishes when all domains are reoriented by an external magnetic field. The domain wall magnetoresistance (DMR) shares analogies with the GMR and has attracted some attention (for a review see [46]). It is often a bulk effect, but domain walls can be also trapped by constrictions [47, 48]. The current induced motion of domain walls in small wires has recently attracted a lot of attention [49–52], but we had to abandon this topic for the present review.

In metallic ferromagnets, the differences between electronic bands and scattering cross-sections of impurities for majority and minority spins at the Fermi energy cause spin-dependent mobilities. In the presence of applied electric fields and not too strong spin-flip scattering processes, a two-channel resistor model is applicable, according to which currents of two different species flow in parallel. The difference between spin-up and spin-down electric currents is called a *spin-current.* It is a tensor, with a direction of flow and a spin-polarization parallel to the equilibrium magnetization vector. An imbalance between the electrochemical potentials is called *spin-accumulation*, which is a vector, parallel again to the magnetization. Spin accumulation is a non-equilibrium phenomenon, but its lifetime is usually much longer than all other relaxation time scales. Spin-flip scattering by spin-orbit interaction and magnetic impurities and disorder destroys a non-equilibrium spin-accumulation. Its importance depends strongly on material and material purity. Here and in most theoretical approaches to magnetoelectronics spin-flip scattering is treated phenomenologically in terms of the spin-flip diffusion length, i.e. the length scale over which an injected spin accumulation loses its polarization, that is typically $\ell_{\mathrm{sd}}^F \sim 5$ nm (Permalloy, Py) $-$ 50 nm (Co). In the bulk of metallic ferromagnets the spin accumulation vanishes beyond a skin depth of $\ell_{\mathrm{sd}}^F$ although spin currents persist.

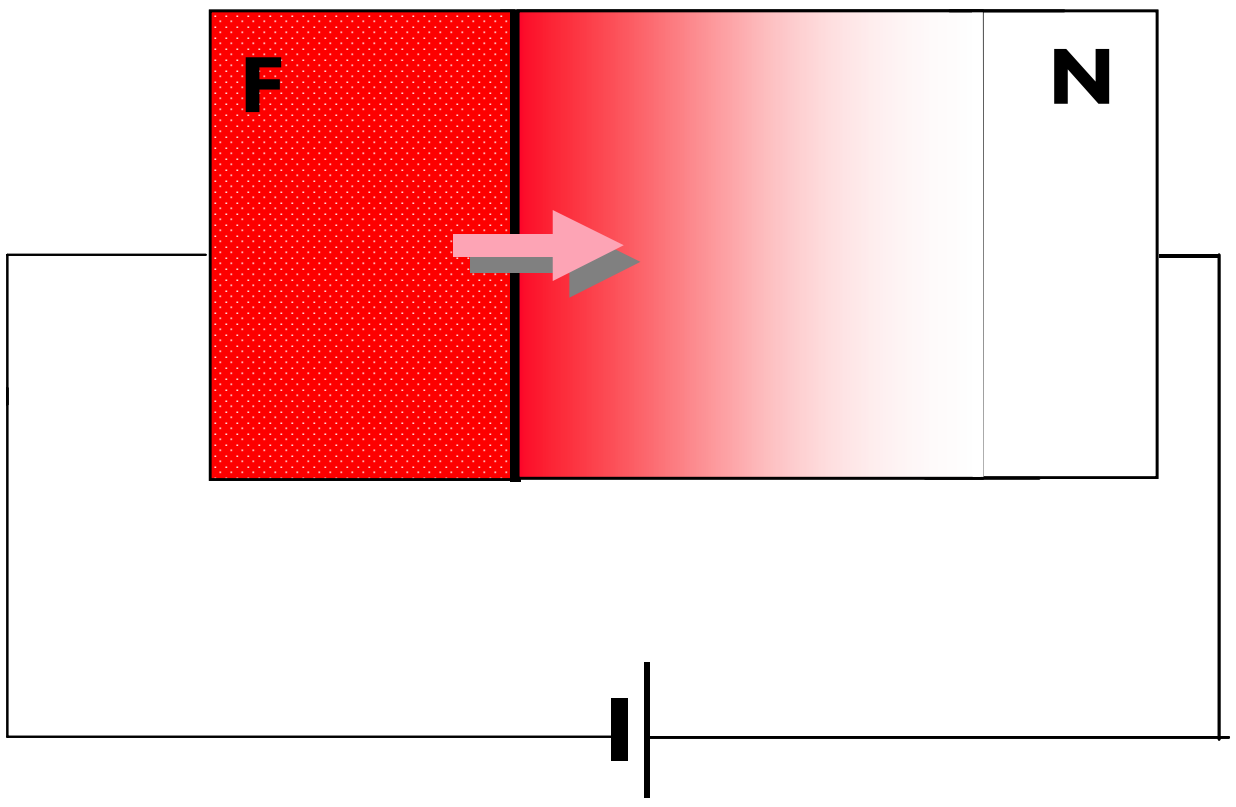

FIG. 1: The non-equilibrium magnetization of the spin accumulation injected from a ferromagnet (F) into a normal metal (N), which decays over a length scale given by the spin-flip diffusion length $\ell_{\mathrm{sd}}^N$.The spin accumulation in the ferromagnet is not shown, being small compared to the equilibrium magnetization and more localized to the interface since usually $\ell_{\mathrm{sd}}^F \ll \ell_{\mathrm{sd}}^N$.

The quantum mechanics of solids explains why cobalt is a ferromagnet, but copper is not. Nevertheless, much of magnetoelectronics is based on the notion that also copper can be magnetized. This magic is done by applying a voltage over a the ferromagnetic (F) | normal metal (N) contact (Fig. 1). Via the ferromagnet a spin-polarized current is then injected into the normal metal [53–55]. The result is a spin accumulation at the interface that extends into the ferromagnet by the spin-flip diffusion length $\ell_{\mathrm{sd}}^F$ introduced above. The non-magnetic metal is effectively magnetized over a decay length corresponding to the spin-flip diffusion length $\ell_{\mathrm{sd}}^N$ that can be very large compared to the typical $\ell_{\mathrm{sd}}^F$, *e.g.* about 1 $\mu$m in copper [23], well above the smallest structures created by microelectronic fabrication technology. The spin accumulation is a vector that in F|N bilayers (see Fig. 1) is collinear to the ferromagnetic magnetization, *i.e.* parallel or antiparallel, de-

pending on the spin-dependent interface and bulk conductances. The direction of the spin accumulation may precess around an applied magnetic field as a function of position. In non-collinear (*i.e.* neither parallel nor antiparallel) spin valves, schematically F($\uparrow$)|N|F($\nearrow$), or other devices with two or more ferromagnetic contacts whose magnetizations are not parallel, the injected spin currents are also non-collinear, and the resulting spin accumulation can point in arbitrary directions, depending on the details of the device materials and magnetic configuration within the spin-coherent region defined by the spin-flip diffusion lengths. The manipulation of electronic properties via the long range spin-coherence carried by the spin accumulation [12] is a main challenge of modern magnetoelectronics.

Typical magnetoelectronic structures are made from ferromagnetic metals like iron, cobalt or the magnetically soft permalloy (Py), a Ni/Fe alloy. The normal metals are typically Al, Cu or Cr, where the spin-density wave in the latter is usually disregarded in studies of transport. These metals cannot be grown as perfectly as strongly bonded tetrahedral semiconductors; moreover, the Fermi wavelength is of the order of the interatomic distances. These systems are said to be "dirty", meaning that size quantization effects on the transport properties may be disregarded [56]. In this limit the physics is most adequately described by semiclassical theories on the level of Boltzmann or diffusion equations. The spin accumulation is then just the difference in the local chemical potentials for up and down spin [6]. Valet and Fert [57] analyzed the giant magnetoresistance of magnetic multilayers in the perpendicular configuration. They used a spin-polarized linear Boltzmann equation to derive a diffusion equation including spin-flip scattering, that for vanishing spin-flip scattering reduces to the two-channel series resistor model [58]. The total current can then be interpreted as two parallel spin-up and spin-down electron currents, which are in turn limited by resistors in series that represent interfaces and bulk scattering. However, the discontinuities in the electronic structure at interfaces occur on an atomic scale and cannot be treated semiclassically. Quantum mechanical calculations have shown that interface scattering is very significant [59, 60] and often dominates the device properties. Regions in which electron scattering has to be treated phase-coherently can be incorporated into semiclassical theories in the form of boundary conditions for the distribution functions on both sides. In Refs. [61, 62] it is shown how this can be carried out on the level of the diffusion equation. Combined with first-principles calculations of the interface scattering matrix, this allows *parameter-free* predictions of spin and charge transport in collinear magnetoelectronic devices [61, 62].

When magnetization vectors and spin-accumulations are not collinear with the spin-quantization ($z-$)axis, the two-channel resistor model cannot be used anymore. The concept of up and down spin states must be replaced by a representation in terms of $2\times2$ matrices in Pauli spin space with non-diagonal terms that reflect the spin-coherence, analogous to the anomalous Green functions in superconductivity that reflect the electron-hole coherence in the superconducting state. The spin accumulation in normal metals can be manipulated easily via the magnetization direction of the ferromagnetic source and drain contacts or an applied magnetic field. The latter causes the spin accumulation to precess around the field direction vector [6, 63]. The associated dephasing by elastic impurity scattering is also called "Hanle effect", which has been recently remeasured [24].

A phase difference in the superconducting order parameter is equivalent to a supercurrent. Analogously, gradients in the magnetic order parameter induce persistent spin currents. The ground state is equivalent to a configuration in which these *equilibrium* spin currents vanish. These arguments can be used to explain the celebrated non-local exchange coupling in magnetic F|N|F spin valves and multilayers [64, 65]. Here we are mainly interested in the *non-equilibrium* charge and spin currents that flow under the influence of externally applied voltages. The spin currents are tensors with a direction and a polarization. When the magnetization directions in the systems are not collinear, the polarization (magnetization) directions of the non-equilibrium accumulations and currents are not parallel or antiparallel with the magnetizations. This gives rise to interesting physics like the spin transfer effect in spin-valves [26, 27] (see below).

The magnetization configurations may vary in time when subject to sufficiently strong non-collinear magnetic fields or electric currents. The time scale of the magnetization motion set by the Larmor frequency is usually much longer than the electron dwell times. Even during the process of magnetization reversal, magnetic devices may usually be treated in an adiabatic approximation, *i.e.* charge and spin currents are governed by the instantaneous magnetic configurations. The magnetizations dynamics are then governed by the parametric torques due to spin currents and magnetic fields.

### B. Magnetoelectronic circuits and devices

Transport in hybrid metallic systems in the presence of long-range correlations in an order parameter can be described by a generalization of Kirchhoff's theory of electronic circuits when the electronic phase is sufficiently scrambled in parts of the system, the "nodes". This approach has been pioneered by Nazarov [66, 67] for electronic networks with superconducting elements, and adapted to magnetoelectronic circuits in [12]. Circuit theory can been applied, for instance, to devices such as perpendicular spin valves, the Johnson spin transistor [11] or the 4-terminal dot by Zaffalon and van Wees [9]. It can be interpreted as a generalization of the two-channel series resistor model [57, 58, 61] to multi-terminal and non-collinear situations. A formalism for disordered non-collinear magnetoelectronic systems based on Random Matrix Theory has been developed by Waintal et al. [56] with emphasis on the spin-transfer torque. Its philosophy is completely different from the Green-function based circuit theory, but both turn out to be equivalent in limiting cases [34].

Magnetoelectronic circuit theory can be derived from a given Stoner Hamiltonian in terms of the Keldysh non-equilibrium Green function formalism in spin space [68]. It comes down to a finite-element formulation of the diffusion equation with quantum mechanical boundary conditions between distribution functions on both sides of a resistor. The initial step is an analysis of the circuit or device topology by dividing it into reservoirs, resistors and nodes that can be real or fictitious. The expressions are importantly simplified by assuming that the electron distributions in the nodes are isotropic. This implies the presence of sufficient disorder (or chaotic scattering). Ferromagnetic transition metals have high critical temperature and exchange splittings. The ferromagnetic coherence length is therefore assumed much smaller than the mean free paths [69], which simplifies the formalism [70]. The spin and charge currents through a *contact* connecting two neighboring ferromagnetic and normal metal nodes can then be calculated as a function of the distribution matrices on the adjacent nodes and the $2 \times 2$ conductance tensor composed of the spin-dependent conductances $G^{\uparrow}$ and $G^{\downarrow}$

$$G^{s} = \frac{e^2}{h}\left[M - \sum_{nm} |r_s^{nm}|^2\right] = \frac{e^2}{h}\sum_{nm} |t_s^{nm}|^2 , \tag{1}$$

and the *mixing conductance*

$$G^{s,-s} = \frac{e^2}{h}\left[M - \sum_{nm} r_s^{nm}(r_{-s}^{nm})^*\right] , \tag{2}$$

where $r_s^{nm}$, $t_s^{nm}$ are the reflection and transmission coefficients in a spin-diagonal reference frame, *i.e.* the elements of the *scattering matrix,* and $M$ the number of modes on the normal metal side of the contact. The expressions for the spin-up and down conductances are the Landauer-Büttiker formula in a two-spin-channel model [37, 71]. Experimentally, these parameters have been obtained by extensive measurements on multilayers in the so-called CPP (current perpendicular to the plane) configuration [58]. The complex interface spin-mixing conductances play important roles when the magnetizations are non-collinear as explained in the next subsection.

A requirement for the validity of the circuit theory are nodes with characteristic lengths smaller than the spin-flip diffusion length. When this criterion is not fulfilled, the diffusion equation has to be solved, with boundary conditions governed by the above conductance parameters [63]. The assumption of isotropy can be relaxed to include a drift term, leading to the conclusion that the diagonal [61, 62] and the mixing conductances [34] contain spurious Sharvin resistances. These corrections are essential to make quantitative comparison between *ab initio* calculations with experiments possible.

### C. Spin-transfer torque

A spin accumulation with polarization normal to the magnetization direction cannot penetrate the ferromagnet, but is instead absorbed at the interface, thereby transferring angular momentum to the ferromagnetic order parameter. A large enough torque overcomes the magnetic anisotropy and damping to switch the direction of the magnetization [26, 27]. This spin-transfer phenomenon is believed to be an interesting alternative to the conventional switching in magnetic random access memories since the necessary power scales favorably when the memory elements become smaller.

The spin transfer can be understood in analogy with the Andreev scattering at normal|superconducting interfaces [72]. This is illustrated by Fig. 2 for the simple case that the interface is transparent only to the majority spin. In the coordinate systems of the ferromagnetic magnetization the incoming up-spin electron is not a pure state but a coherent linear combination of "right" and "left" spin states. A simple angular momentum balance of incoming and scattering states shows that the transverse angular momentum of magnitude $\hbar/2$ is seemingly lost, but has been absorbed by the ferromagnet. We can imagine a second elementary scattering process in which a spin-down hole hits the interface from the left below. This causes an identical spin transfer of $\hbar/2$, whereas the transmitted and reflected states cancel the charge and longitudinal momentum current. Both process combined represent a spin-flip reflection at the interface with a spin transfer of $\hbar$. It is equivalent to a spin current polarized transverse to the magnetization that is completely absorbed at the interface. In order to sustain such a transverse spin current, a spin accumulation should be present in N in which the spin-up state is occupied, and the spin down state empty. The scattering process transfers spin from the spin accumulation to the ferromagnetic magnetization, thus reducing the spin accumulation. The *spin-transfer torque* in this simple model is proportional to the spin accumulation times the number of modes in the normal metal. In the presence of conventional scattering processes it is governed by the spin-mixing conductance (2) instead. An analogy with Andreev scattering at a normal metal|superconducting interface is recognized by interpreting the ferromagnet as a condensate of angular momentum, just as a superconductor is a condensate of charge.

A complete theory of the spin-accumulation induced magnetization torque requires a quantitative treatment of the interface scattering but also a description of the whole device that allows computation of the spin accumulation in the normal metal. The magnetoelectronic circuit theory mentioned above [12] has all ingredients available, although the authors initially did not apply it to this phenomenon. The first microscopic treatment explicitly addressing the spin torque in diffuse systems is therefore the random matrix theory in Ref. 56. It has subsequently been demonstrated that the theories are completely equivalent for not too transparent

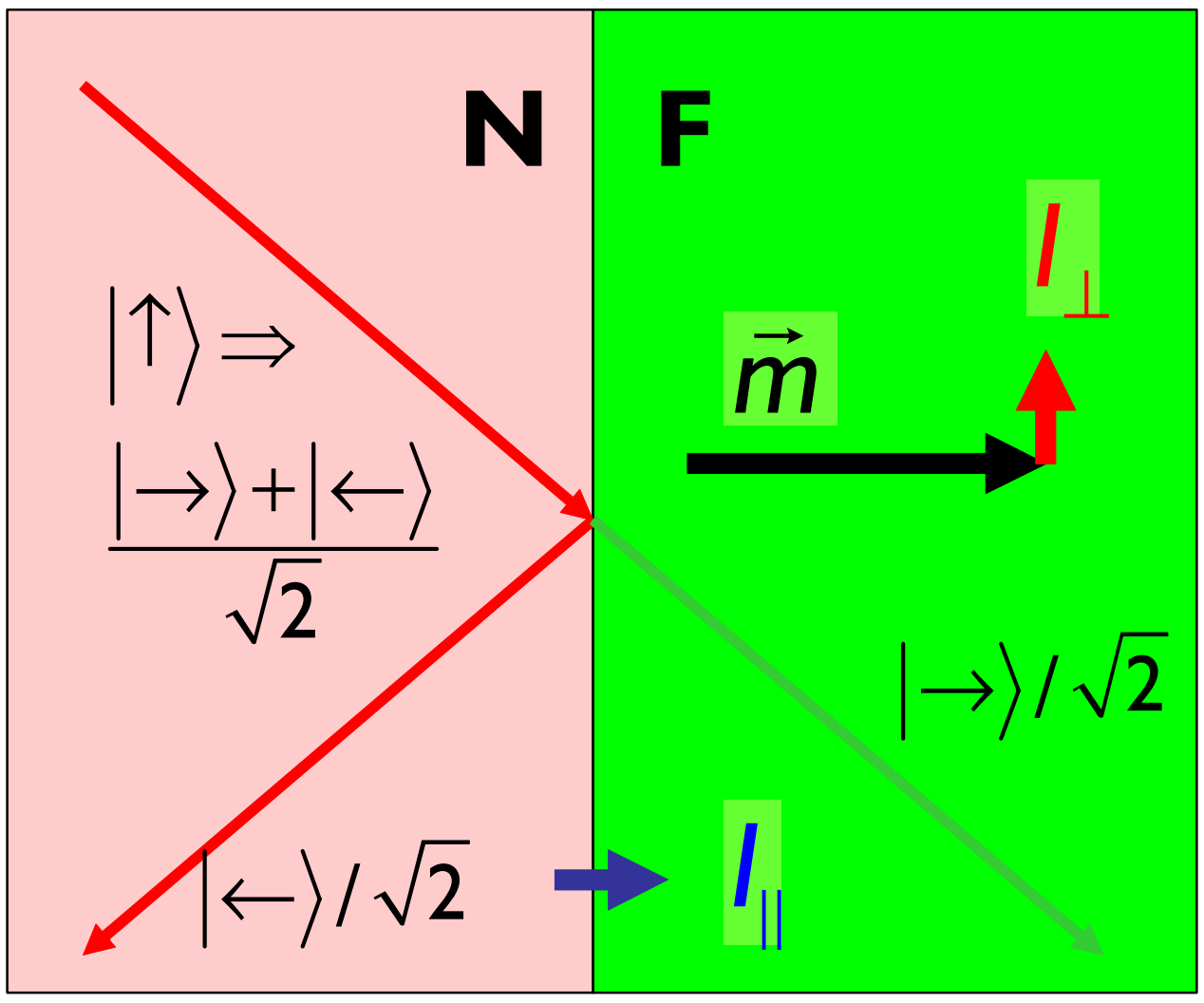

FIG. 2: Illustration of the magnetization torque exerted by a spin current. The magnetization $\vec{m}$ of the ferromagnet is normal to an incoming electron with spin up that can be written as a linear combination of right and left spin states. Assuming that the interface is transparent to only the majority spin, the parallel spin $I_{||}$ and charge currents are conserved, whereas the transverse spin current $I_\perp$ is absorbed and acts as a torque on the magnetization.

interfaces [34].

The spin-mixing conductance at an N|F interface, Eq. (98) can be interpreted as a measure of the angular momentum transfer from the spin accumulation in the normal metal to the ferromagnetic order parameter. By reducing the spin-accumulation in N the spin-transfer torque increases the electrical conductance. The angular magnetoresistance of spin valves is therefore a sensitive measure of the spin-mixing conductance [34, 70]. When sufficiently large, spin-transfer torques cause current-induced magnetization dynamics and reversal [26–29]. The spin-mixing conductance also governs the additional damping of the magnetization dynamics by metallic buffer layers [32, 73]

### D. Ab initio theories

First-principles calculations of transport have a rather short history. Earlier theories of transport in magnetic multilayers relied on the not very realistic model of phase-coherent superlattices with translational periodicity. In one school of thought the Boltzmann equation is expressed in terms of the superlattice (sub)band structure [60, 74, 75]. This approach is formally valid when the broadening due to defects is smaller than the miniband energy splitting, which is not the case in the structures fabricated to date. Another approach is based on the total neglect of any defect scattering in the system, which addresses ballistic point contacts [59, 76]. Both approaches help to understand magnetotransport but we refer to previous reviews for this discussion [37, 40]. In the present context the calculations of the transport properties of single specular and disordered interfaces are relevant as the parameters in the magnetoelectronic circuit theory. The non-local exchange coupling in multilayers can be expressed in terms of the reflection and transmission coefficients of (specular) interfaces and were initially computed for this purpose [65, 77]. However, as noted above, they also govern transport properties, for which they were calculated first in Refs. [61, 78] for interfaces and in [47] for domain walls. Recent first-principle calculations of transport properties also include (interface) disorder [35, 36, 79, 80].

Spin injection into materials other than high-density metals is a topic of considerable interest. Successful spin injection into semiconductors would make it possible to integrate the functionality of magnetoelectronics with the ubiquity of semiconductor electronics. The difficulty of injecting spins into semiconductors with high-density ferromagnets was pointed out in [81]. The problem is the impedance mismatch of highly conducting metals on the one hand and semiconductors with relatively low electron density and conductance on the other. First-principles calculations reveal that a perfect interface such as Fe|InAs can be very spin selective [82, 83]

### E. Overview

This review is organized as follows. In Section II we familiarize the reader with magnetoelectronic circuit theory, a useful guiding principles for most of the physics. In Section III basic transport theory is recapitulated and its extension to non-collinear magnetic systems is explained. We rely on the Stoner model of band magnetism or spin-density functional theory with local single-particle exchange (correlation) potentials. For such a Hamiltonian we discuss elements of the scattering theory of transport. Transport in diffuse systems can be understood from first-principles by two formalisms, Green function theory and random matrix theory. In the presence of sufficient phase randomization a quasi-classical regime is reached in which the quantum-nonlocality is averaged out. It turns out that both formalism are equivalent to the diffusion equation in which the distribution functions are matched at interfaces by quantum mechanical scattering matrices. A general theory of transport in magnetic hybrid structures, devices and circuits is described in Section IV. The circuit theory of magnetoelectronics is derived on the basis of the quasi-classical kinetic equations discussed in the previous chapter. We discuss its generalization to high contact transparencies as relevant for intermetallic interfaces, proving also the equivalence with Random Matrix Theory. Section V is devoted to a discussion of transport by first principles calculations, with special emphasis on interfaces. Section VI is a synthesis of the previous chapters in which general theory is applied to various structures and devices such as non-collinear spin valves, including the spin-accumulation induced magnetization torque in nanopillars, and three or more terminal devices, all with a normal metal central island and variable magnetization directions of the ferromagnetic elements. Whereas the mantra of all previous sections has been that quantum interference effects often can and should be disregarded, it might be worthwhile to remain vigilant towards a possible breakdown of semiclassical approximations, that might give rise to novel physics like the magnetoelectronic spin echo discussed in Chapter VII. An outlook on the field is given in Section VIII. Some technical aspects are deferred to the Appendices.

## II. UNDERSTANDING MAGNETOELECTRONICS

The basis of our present understanding of electronic circuits was founded about 150 years ago by Gustav R. Kirchhoff.[2] He showed how the transport properties of arbitrarily complicated circuits can be understood in terms of the current-voltage relation across single resistive (and subsequently also capacitive or inductive) elements. Central to the present review is a generalization of Kirchhoff's ideas to electronic circuits incorporating ferromagnetic elements [12], that has been inspired by Nazarov's theory for circuits with superconducting terminals [66, 84]. Magnetoelectronic circuit theory is a versatile tool to obtain qualitative and quantitative information about charge and spin-transport that is simple enough to be operated by experimentalists and non-specialists. In order to stimulate a broader acceptance we illustrate in this part in quite some detail how magnetoelectronic circuit theory can be used to gain insights into the magnetization dependent charge and spin currents in two terminal devices. For a practical guide to the use of circuit theory for simple devices we refer to Appendix B.

Let us first recall the treatment of conventional circuits. On a basic level, a topology consisting of nodes that are connected by resistances $R$, or equivalently conductances $G = 1/R$, capacitances $C$, inductances $L$ and current/voltage sources, are sufficient to determine the electrodynamics of conventional passive circuits. We restrict ourselves here to the steady state with DC voltages in the nodes and constant currents through the resistances. Consider a simple element in an electronic circuit consisting of a single resistor sandwiched between two nodes with potentials $V_1$ and $V_2$ as shown in Fig. 3. Ohm's law states that the current

[2] See http://www-gap.dcs.st-and.ac.uk/~history/index.html for biographic details and references.

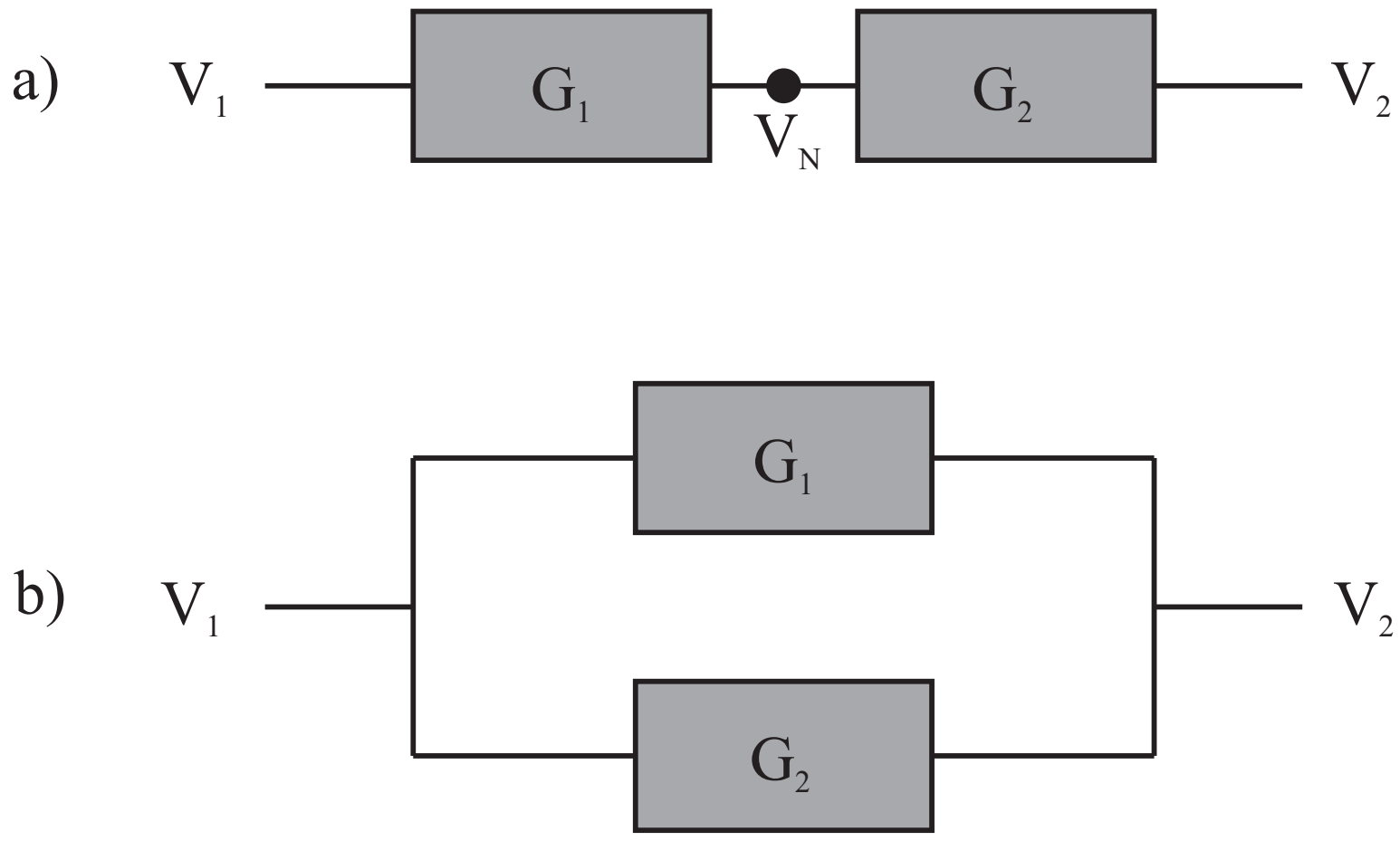

FIG. 4: Electronic circuit consisting of two resistances, a) in series and b) in parallel.

through a basic resistive element is proportional to the voltage difference across the element:

$$I = G\,(V_2 - V_1)\,.$$

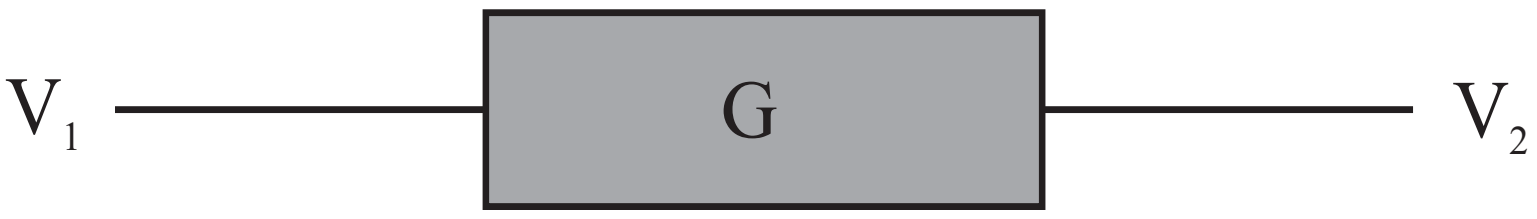

FIG. 3: A basis element of a circuit, a conductance $G$, coupled between two potentials $V_1$ and $V_2$.

Two outer nodes at potentials $V_1$ and $V_2$ can be connected by two resistors with conductances $G_1$ and $G_2$, either in series as in Fig. 4 a) or in parallel as in Fig 4 b). In the series connection we can calculate the, as yet unknown, voltage $V_N$ by making use of *Kirchhoff's 1st Law* of charge conservation, *i.e.* the sum all currents into a node must vanish.

$$\sum_\alpha I_\alpha = 0. \tag{3}$$

For the central node of Fig 4 a) this reads

$$G_1\,(V_1 - V_N) + G_2\,(V_2 - V_N) = 0.$$

from which

$$V_N = \frac{G_1 V_1 + G_2 V_2}{G_1 + G_2}$$

and the current is found to be

$$I = \frac{G_1 G_2}{G_1 + G_2} (V_2 - V_1) = \frac{V_2 - V_1}{R_1 + R_2}.$$

The parallel circuit in Fig. 4 b) in its steady state obeys *Kirchhoff's 2nd law* that the sum of all voltage differences in any closed loop in the circuit vanishes, since they would otherwise be quickly screened by circulating currents. The total current driven by the voltage difference is the sum of the currents passing through the conductances $G_1$ and $G_2$:

$$I = (G_1 + G_2)(V_2 - V_1).$$

Kirchhoff's laws and conventional circuit theory were initially developed and applied to macroscopic circuits which could be analyzed in terms of distinguishable elements. *e.g.* a single resistor is equivalent to two resistors with half resistance in series. This is often not possible in the limit of very small devices in which electrons propagate ballistically and/or when their wave character starts to play a role. For example, the resistance of a thin ballistic wire as depicted in Fig. 5 a) does not depend on its length, in defiance of Ohm's Law. In this case the resistance is purely geometrical, most electrons will be reflected at the boundaries, giving rise to the so-called Sharvin point contact resistance. When the constriction becomes wider the Sharvin resistance becomes smaller and finally negligible compared to the conventional (Ohmic) resistance that is caused by disorder scattering in the bulk. In the intermediate case, the total resistance is well approximated by summing the Sharvin and the Ohmic resistors. Nevertheless, down to the nanoscopic regime circuit theory turns out quite robust since disorder or chaotic scattering is ubiquitous in all but the most dedicated devices. Especially in elemental metal structures in which the Fermi wave lengths are of the order of the atomic spacings, electrons are very sensitive to any kind of disorder and, with few exceptions, transport is diffuse . Even without disorder, circuit theory can often be applied as illustrated in Fig. 5 b) Even though the two constrictions and the island are ballistic, a local potential can be associated to the central node when the electrons reside sufficiently long on the island. This is the case when the classical scattering is chaotic. When there is additional disorder, diffuse scattering conditions are created also for regular geometries like layered thin films. In those cases, voltage drops create currents proportional to a resistance and circuit theory applies.[3] The circuit analogue of the physical system depicted in Fig. 5 b) is therefore Fig. 4 a) such that the total resistance is simply the sum of the resistances of the isolated constrictions. When this approximation does not hold due to residual quantum interference effects, the system has to be represented by a single resistor that is governed by entire phase-coherent volume. Landauer's scattering theory allows us to compute transport in terms of the transmission and reflection coefficients of the resistive elements starting from the Schrödinger equation.

[3] Strictly speaking, this statement is correct only when the contacts are not in the quantum point contact limit, see [56, 72].

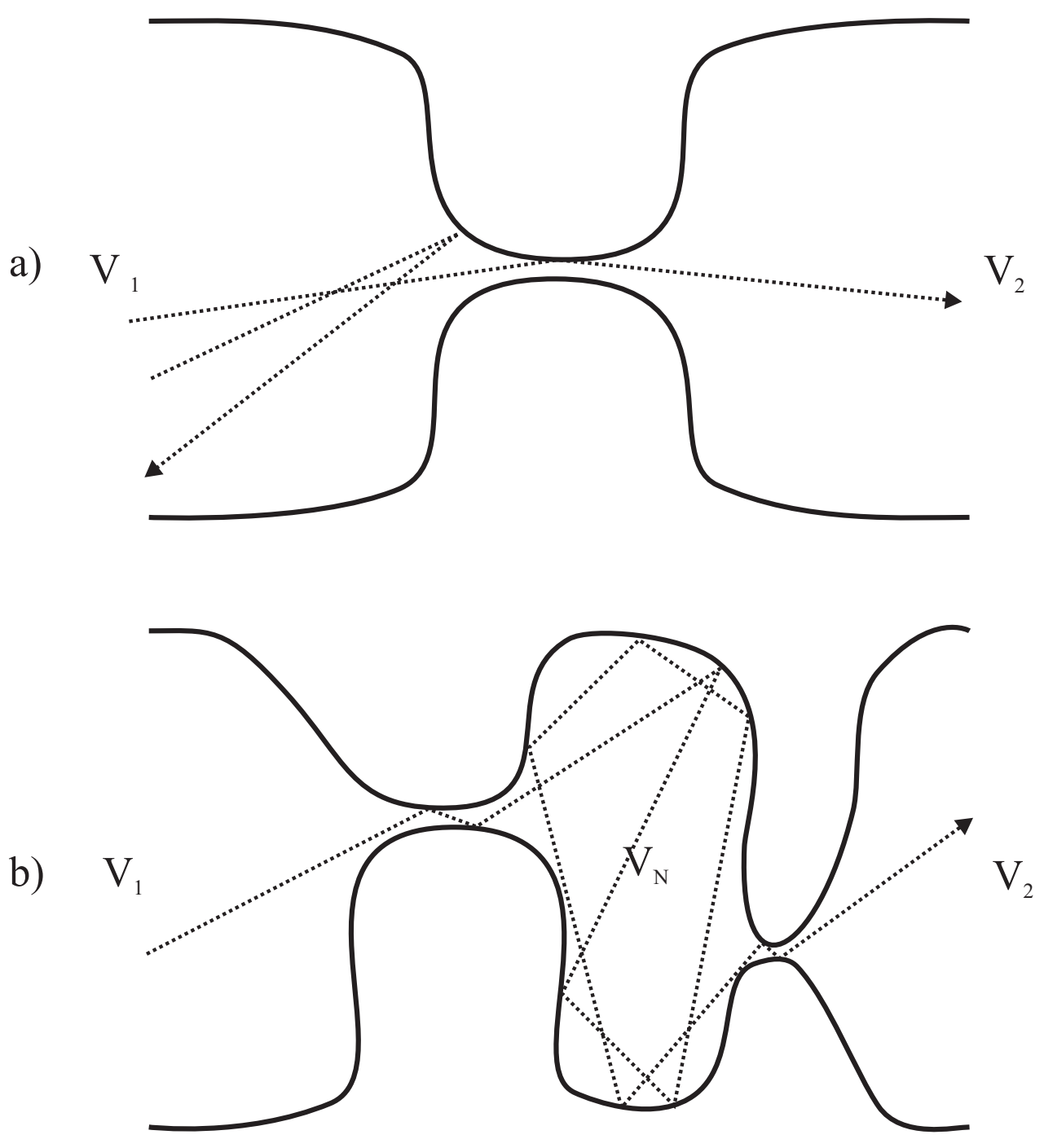

FIG. 5: a) A schematic picture of a narrow constriction/wire that limits the current in a metal connecting two reservoirs. When the wire is shorter than the mean free path, transport is ballistic and the resistance does not scale with inversely with the wire length. Two ballistic trajectories are shown, one where the electron is reflected close to the constriction and one where the electron is transmitted through the wire. b) Two constrictions in series connecting two reservoirs. When the electrons reside sufficiently long on the central island they may be described semiclassically in terms of a local potential $V_N$ . A possible ballistic trajectory for an electron transmitted throughout the structure is shown. The transport limiting elements are point contacts here, but may as wel be tunneling barriers, interfaces or diffuse wires.

In general, the properties of a given device or circuit can be calculated by first prudently separating it into *reservoirs*, *nodes,* and *resistors,* where the latter are the current-limiting elements. The nodes are supposed to have a resistance that is negligibly small and (as in the example of one resistor split into two), may be fictitious. In a multilayer, for example, it is convenient to insert nodes at both sides of an interface, treating the latter as separate resistive element.

Reservoirs represent the "battery poles" that are large thermodynamic baths at thermal equilibrium with a constant bias applied, irrespective of the currents that flow in or out. More precisely, the driving forces for the currents are not the voltage differences, but the electrochemical potential differences.

Circuit theory can be derived and justified formally from the Schrödinger equation by Green function methods. When the electronic potential does not vary much on the scale of a scattering mean free path, the quantum non-locality can be integrated out and the electrons are described by semiclassical distributions functions, that specify position and momentum simultaneously. Disregarding ballistic effects in a node is allowed when its distribution function is isotropic in momentum space, so there is no preferred direction of the electrons except for the drift induced by a chemical potential gradient. This regime is called diffuse transport. Note that these conditions have to be fulfilled only in the nodes. The resistors may display in principle pronounced quantum or ballistic effects, that are most conveniently treated by the scattering theory of transport in terms of the reflection and transmission coefficients.

Ferromagnets have a symmetry-broken ground state, very much like superconductors. A ferromagnet may be interpreted as a condensate of spin angular momentum, just as a superconductor is a condensate of Cooper pairs, *i.e.* charge. Microscopically, the number of up-spins in a given quantization direction parallel to the magnetization differ from that of the down-spin electrons. The large difference in up and down spins cause the familiar macroscopic magnetic field. In metallic ferromagnets like Ni, Co, Fe and their alloys Fermi surfaces for both spins are remarkably different. Also other physical properties become spin-dependent, in particular the electron mobilities. In the absence of strong spin-flip scattering, this directly leads to the so-called *two-channel resistor model* for the ferromagnet.

Let us consider a constriction such as a point contact in a ferromagnetic material, as in Fig. 6 a). The bulk ferromagnets to the left and right can then be described as reservoirs with a given potential difference that drives a current through the constriction. Due to the different electronic structures the transmissivity is different for the two spin states, leading to two different conductances that can be treated in parallel. These conductances can be computed microscopically, *e.g.* by the Landauer formula, as is discussed extensively in later Sections.

When the magnetizations point into the same direction everywhere, the ferromagnetic resistor can be described by two conductances, $G_\uparrow$ for spins aligned parallel with the magnetization and $G_\downarrow$ for spins antiparallel to the magnetization. This "two-current model" is represented by the parallel circuit shown in Fig. 6 b). The current carried by spins aligned to the magnetization is $I_\uparrow = G_\uparrow (V_2 - V_1)$ and the current carried by spins anti-aligned to the magnetization is $I_\downarrow = G_\downarrow (V_2 - V_1)$. The total (charge) current through the constrictions is $I_c = I_\uparrow + I_\downarrow$:

$$I_c = G (V_2 - V_1) .$$

FIG. 6: a) A ferromagnetic metal coupling two particle reservoirs with potentials $V_1$ and $V_2$. b) Circuit model of transport through a single ferromagnetic layer. The conductance of spin-up electrons is $G_\uparrow$ and the conductance of spin-down electrons is $G_\downarrow$.

where the total conductance is

$$G = G_\uparrow + G_\downarrow\,. \tag{4}$$

The spin-polarization of the current or simply *spin-current*:

$$I_s = I_\uparrow - I_\downarrow \equiv PI,$$

where

$$P = \frac{G_\uparrow - G_\downarrow}{G_\uparrow + G_\downarrow} \tag{5}$$

is the polarization of the ferromagnetic resistive element. The precise value of $P$ is governed by the bulk properties of the ferromagnet as well as its interface to the normal metal.

A simple but non-trivial hybrid device is the *spin valve* consisting of two ferromagnetic (F) elements connected by a piece or layers of normal (N) metal. When the magnetizations are collinear, *e.g.* parallel or anti-parallel, the ferromagnetic elements act as two spin-dependent resistors in parallel. Normal metals often have a much higher mobility than ferromagnets, so for convenience we disregard here the normal metal resistance as well, but note that scattering due to bulk impurities can similarly be treated by adding their resistances in series. Since then the potential drop in the spacer is small, we may treat it like a node. For a parallel configuration this leads to the circuit diagram shown in Fig. 7. The potentials, $V_{N\uparrow}$ and $V_{N\downarrow}$ in the normal metal node can be computed now easily using Kirchhoff's laws. The average potential $V_C = \left(V_{N\uparrow} + V_{N\downarrow}\right)/2$ is the conventional voltage felt by the net charge (accumulation) in the node. A new ingredient is the *spin accumulation* in the normal metal $V_S = V_{N\uparrow} - V_{N\downarrow}$ that can be easily calculated by Kirchhoff's rule for the circuit in Fig. 7:

$$V_S = \frac{2G_1G_2\left(P_2 - P_1\right)\left(V_2 - V_1\right)}{\left[G_1\left(1-P_1\right) + G_2\left(1-P_2\right)\right]\left[G_1\left(1+P_1\right) + G_2\left(1+P_2\right)\right]}, \tag{6}$$

where the subscripts refer to the ferromagnets 1 and 2. The spin-accumulation can have either sign, is proportional to the bias voltage $V_2 - V_1$ and vanishes when the two ferromagnetic resistive elements have the same direction and polarizations $P_2 = P_1 \equiv P$. In the same limit, the spin-current $I_s = I_\uparrow - I_\downarrow$

$$\frac{I_S}{V_2 - V_1} = \frac{G_1G_2\left[G_1\left(1-P_1^2\right)P_2 + G_2\left(1-P_2^2\right)P_1\right]}{\left[G_1\left(1-P_1\right) + G_2\left(1-P_2\right)\right]\left[G_1\left(1+P_1\right) + G_2\left(1+P_2\right)\right]} \tag{7}$$

only vanishes with $P$:

$$I_S^{\rm p} \overset{\substack{G_{1/2} \to G \\ P_{1/2} \to P}}{=} \frac{GP}{2}. \tag{8}$$

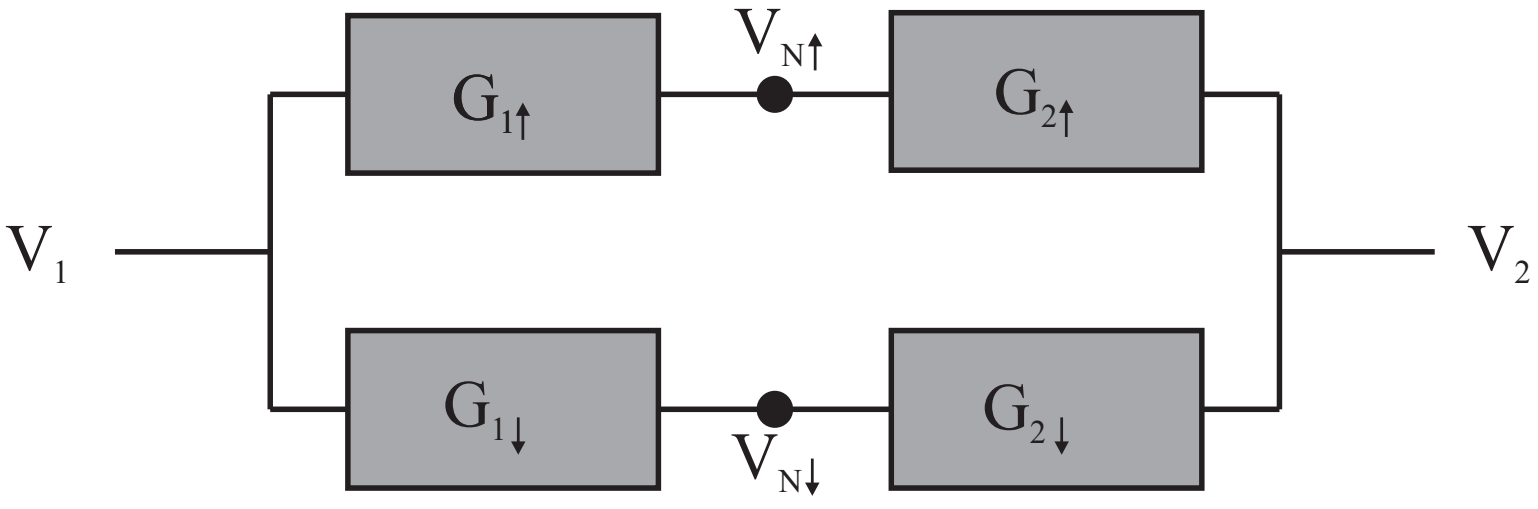

FIG. 7: Two ferromagnets in series.

The transport properties of our spin valve change dramatically when we let the magnetizations point in opposite directions, *i.e.* when the signs of $P_1$ and $P_2$ differ. The total charge through the device is the sum of the up and down spin currents:

$$\begin{aligned} \frac{I_C}{V_2 - V_1} &= \left[\frac{G_{1\uparrow}G_{2\uparrow}}{G_{1\uparrow} + G_{2\uparrow}} + \frac{G_{1\downarrow}G_{2\downarrow}}{G_{1\downarrow} + G_{2\downarrow}}\right] \\ &= \frac{G_1G_2\left[G_2\left(1-P_2^2\right) + G_1\left(1-P_1^2\right)\right]}{\left[G_1\left(1-P_1\right) + G_2\left(1-P_2\right)\right]\left[G_1\left(1+P_1\right) + G_2\left(1+P_2\right)\right]} \end{aligned} \tag{9}$$

For the antiparallel symmetric spin valve:

$$\frac{I_C^{\mathrm{ap}}}{V_2 - V_1} \overset{\substack{G_{1/2} \to G \\ P_1, -P_2 \to P}}{=} \frac{G}{2}\left(1 - P^2\right), \tag{10}$$

which should be compared with the parallel symmetric spin valve:

$$\frac{I_C^{\mathrm{p}}}{V_2 - V_1} \overset{\substack{G_{1/2} \to G \\ P_{1/2} \to P}}{=} \frac{G}{2}. \tag{11}$$

The antiparallel configuration has a higher resistance because the applied voltage is used in part for the kinetic energy cost associated to the accumulation of spins. The relative difference in the currents is called magnetoresistance (MR) ratio (often called "giant", GMR), since a magnetic field can reorient an antiparallel to a parallel configuration by an externally applied magnetic field. For the symmetric spin valve:

$$\mathrm{MR} \equiv \frac{I_C^{\mathrm{p}} - I_C^{\mathrm{ap}}}{I_C^{\mathrm{p}}} = P^2. \tag{12}$$

The magnetoresistance is proportional to the square of the polarization, reflecting the physical mechanism that a spin-polarized current first has to be injected and later detected during transport. The simple circuit theory is usually referred to as *two-channel series resistor model* and its parameters have been determined accurately by fitting extensive series of experiments for the most common material combinations and also by first-principles calculations.

In the picture above we neglected the limited life time of the spin angular momentum of non-equilibrium carriers. A spin can be flipped by spin-orbit interactions and magnetic impurities, and, in ferromagnets, by magnon scattering In circuit theory spin-flip scattering is represented by resistors that connect the up and down spin channels, dissipating the spin accumulation. When occurring in the normal metal, it weakens the contrast between P and AP configurations. In a ferromagnet the distance by which a spin diffuses in a ferromagnet before being flipped, the spin-flip diffusion length, determines the magnetically active region beyond which the bulk ferromagnet does not contribute to the polarization $P$ and the resistance contrast anymore.

Everything up to now is well-known lore for the magnetoelectronic community for almost two decades, since it is nothing but the generally accepted physics of the giant magnetoresistance in the two-channel resistor model. The physics that arises when the magnetization directions of the ferromagnets are non-collinear is the main topic of this review. In spin valves we find a non-trivial dependence of the resistance on angle that is closely related to the spin-current induced magnetization (or spin-transfer) torque pioneered by Slonczewski [26, 85] and Berger

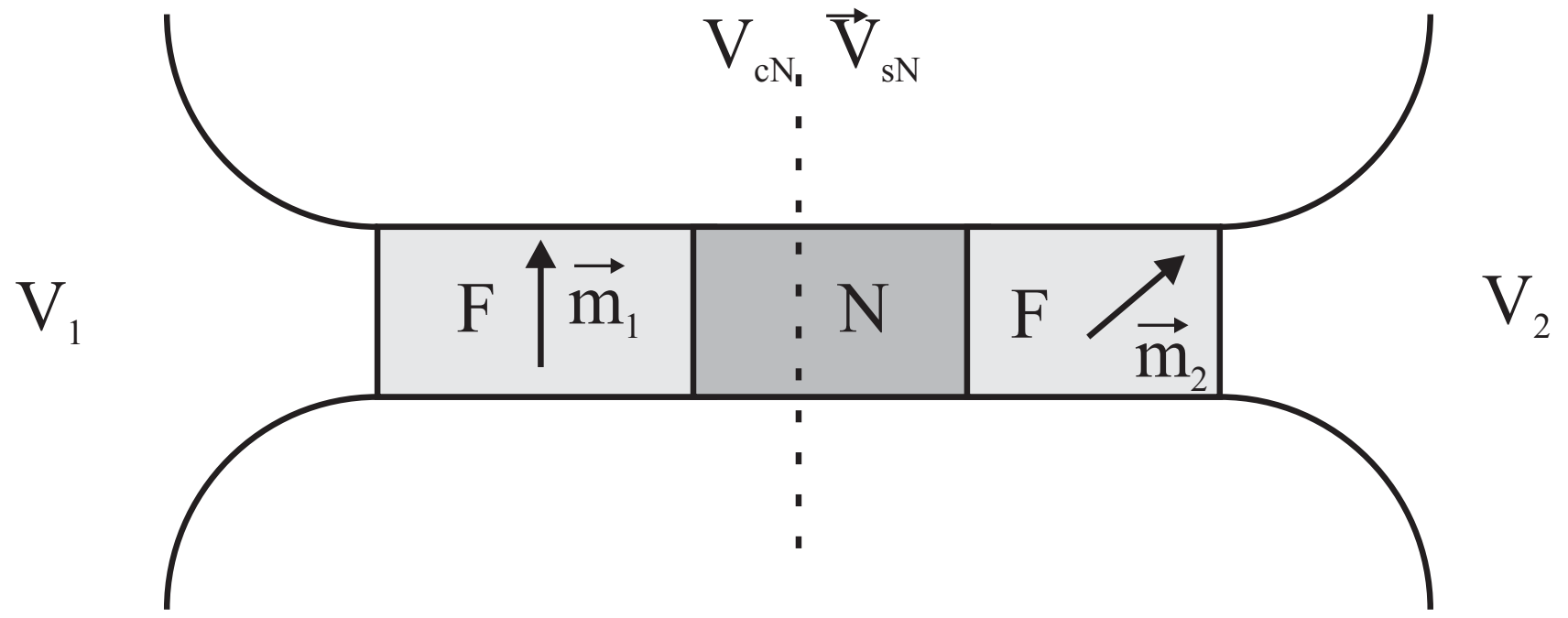

FIG. 8: This is a sketch of a two-terminal spin valve device with non-collinear magnetizations. Electrons flow between the reservoirs biased at potentials $V_1$ and $V_2$. Two monodomain ferromagnets are part of the circuit with magnetization directions $\vec{m}_1$ and $\vec{m}_2$. The transport properties are computed by introducing the charge potential $V_{C,N}$ and the spin potential $\vec{V}_{S,N}$ in the middle normal metal.

[27]. The magnetization dynamics are also strongly modified; these are discussed elsewhere [33]. When the magnetizations are not collinear, the vectorial nature of the spin-accumulation must be fully taken into account, because spin currents are being supplied from ferromagnets with different spin directions. The direction of the spin accumulation vector $\vec{V}_S$ depends on the entire spin-coherent region of the device. Obviously, it is no longer possible to describe a resistor by two effective conductances. Furthermore, interface and bulk resistances cannot be lumped together like resistors in series, but require a radical extension of the conventional circuit theory.

The magnetoelectronic circuit theory for the general non-collinear case requires a generalization of the charge conservation law that allows bookkeeping of the spin angular momentum: The rate of change of the spin accumulation vector in a given node must be equal to the total sum of incoming spin currents. In the absence of spin relaxation and in a steady state this means that the sum of all spin currents must vanish, just like the sum of all charge currents into a node vanishes in Eq. (3). The four parameters describing spin and charge accumulations in a given node can be conveniently lumped together into a $2 \times 2$ matrix in Pauli spin space spanned by the three Pauli spin matrices and the unit matrix. It follows naturally that the parameters of spin-up and down conductances that govern longitudinal spin and charge transport have to be extended to $2 \times 2$ unitary matrices as well. The non-diagonal elements are the so-called mixing conductances that govern the transverse spin currents at N|F interfaces.

Let us illustrate the abstract notions by the non-collinear spin valves sketched in Fig. 8. Electrons can flow from the left reservoir with a potential $V_1$ to the right reservoir with a potential $V_2$ via two ferromagnets with magnetization directions $\vec{m}_1$ and $\vec{m}_2$ with normal metal spacers N. The transport properties are computed by introducing the charge potential $V_{C,\text{N}}$ and the spin-potential $\vec{V}_{S,\text{N}}$ in the middle normal metal that is treated as a node. The dotted line in the figure indicates where we choose to introduce the charge electrochemical potential $V_{C,\text{N}}$ and the spin-accumulation potential $\vec{V}_{S,\text{N}}$. We assume now, for clarity, that the normal metal resistance is much smaller than the resistance of the ferromagnetic elements and disregard spin flip. Generalizations are straightforward and will be discussed in subsequent chapter. The potentials in the normal metal node in are then constant and it does not matter where the dotted line cuts through the normal metal node. The number of net spins in the central node is $s = \mathcal{D}e\left|\vec{V}_{S,\text{N}}\right|$, where $\mathcal{D}$ is the number density of states in the normal metal spacer and $e$ is the electron charge.

A spin current of general polarization that hits the first ferromagnetic metal from the normal metal is in general not collinear to the ferromagnets magnetization direction. Such a current can be decomposed into three polarization components, collinear to the magnetization (longitudinal), $\vec{V}_{S,\text{N}}$, or perpendicular to it (transverse), either in-plane with magnetization and spin accumulation vectors, $\vec{V}_{S,\text{N}} \times \vec{m}_1$, or normal to this plane, $\vec{m}_1 \times \left(\vec{V}_{S,\text{N}} \times \vec{m}_1\right)$. The charge current, $I_{C1}$, from the normal metal into the first ferromagnet can be computed analogous to the two-circuit model for spin-up and spin-down components taking into account the spin accumulation in the normal metal:

$$I_{C1} = G_{1\uparrow}\left(V_{C,\text{N}} + \vec{V}_{S,\text{N}} \cdot \vec{m}_1 - V_1\right) + G_{1\downarrow}\left(V_{C,\text{N}} - \vec{V}_{S,\text{N}} \cdot \vec{m}_1 - V_1\right),$$

where $\pm\vec{V}_{S,\text{N}} \cdot \vec{m}_1$ projects the up and down components of the spin accumulation $\vec{V}_{S,\text{N}}$ on the spin quantization axis of the left ferromagnet and the conductances $G_{1\uparrow}$ and $G_{1\downarrow}$ are the spin-dependent conductances of the left part of the device. The spin current $\vec{I}_{S1}$ from the normal metal into the first ferromagnet consists of a collinear (longitudinal) $\vec{I}_{S1\|}$ and perpendicular (transverse) $\vec{I}_{S1\perp}$ component relative to the magnetization $\vec{m}_1$. The first component

$$\vec{I}_{s1\|} = \vec{m}_1\left[G_{1\uparrow}\left(V_{C,\text{N}} + \vec{V}_{S,\text{N}} \cdot \vec{m}_1 - V_1\right) - G_{1\downarrow}\left(V_{C,\text{N}} - \vec{V}_{S,\text{N}} \cdot \vec{m}_1 - V_1\right)\right].$$

is easy to understand, involving only the conventional conductances $G_{1\uparrow}$ and $G_{1\downarrow}$ for spins aligned parallel and antiparallel to the magnetization direction, quite analogous to the spin current for collinear systems.

In order to understand the transverse spin current it is essential to realize that a spin state not collinear to the magnetization is not an eigenstate (majority or minority spin) of the ferromagnet. Instead, a Bloch state with arbitrary spin direction is a coherent linear combination of the spin eigenstates that in the ferromagnet (at the Fermi energy) are associated with different Fermi wave vectors $k_\uparrow^F$ and $k_\downarrow^F$. The linear coefficients of up and down spins oscillate as a function of position in the transport ($x$) direction like $\cos\left(k_{x,\uparrow}^F - k_{x,\downarrow}^F\right) x$, where $k_{x,\uparrow}^F$ and $k_{x,\downarrow}^F$ are the spin dependent wave vector components normal to the interface, which is equivalent to a precession around the exchange magnetic field with period $2\pi/\left|k_{x,\uparrow}^F - k_{x,\downarrow}^F\right|$. The total (spin) current is determined by all wave vectors at the Fermi energy, each corresponding to a different precession wave length. In high-electron-density metallic ferromagnets such as Co, Ni and Fe, a near continuum of wave vectors exists. The Fermi surface integral that determines the total currents at a distance $x$ involves a strongly oscillating integrand that cancels except for very small values of $x$ due to the destructive interference. This corresponds to an *absorption of the transverse spin current* inside the ferromagnet within the so-called transverse spin-dephasing length (also called magnetic coherence length) $\lambda_c = \pi/\left|k_{F\uparrow} - k_{F\downarrow}\right|$, which for typical transition metals is an atomistic length scale. The absorbed angular momentum is transferred to the ferromagnetic condensate and acts analogous to a mechanical torque on the magnetization that, when strong enough, may lead to a dynamic response and even a complete magnetization reversal. The spin-transfer torque thus acts on the interface of the ferromagnets and when the ferromagnetic layer is thin and its magnetization is sufficiently stiff, the interface spin-torque is transmitted to the whole ferromagnet uniformly. Part of the transverse current is reflected into the normal metal after having spent some time in the ferromagnet. Even though the penetration depth is generally small, as explained above, the exchange field is strong, and can induce a significant precession of the reflected component compared to the incident one. This is the physical origin of a proximity exchange field felt by electrons in a normal metal attached to a ferromagnet. It is usually very small in intermetallic systems.

The spin-dephasing of the transverse spin flow into the ferromagnet explains that the non-collinear perpendicular (transverse) component of the spin current can depend only on the spin accumulation in the normal metal. The non-collinear perpendicular (transverse) component of the spin current is not conserved across the normal metal-ferromagnet interface since it vanishes inside the ferromagnet unlike the Ohms law we have discussed so far. Naturally we should evaluate the transverse spin current on the normal metal side of the interface. The angular momentum of the transverse spin current is transferred as a torque on the magnetization of the ferromagnet.

In the limiting case of a normal metal interface to a half-metallic ferromagnet discussed in the Introduction, the penetration depth of transverse spins is governed by the evanescent waves corresponding to the forbidden spin direction. The analogy of the scattering of transverse spin states to the Andreev scattering at a normal metal|superconductor interface holds also for weak ferromagnets. The magnitude of the transverse spin current in the limit that the electronic structure of the majority spin of the ferromagnet is matched to that of the normal metal at an ideal interface is easily seen to be proportional to the spin accumulation component normal to the magnetization $\vec{m}_1 \times \left(\vec{V}_{S,\mathrm{N}} \times \vec{m}_1\right)$ times (twice) the number of states $N$ that ($e^2N/h$ is the Sharvin conductance per spin in the normal metal) that hit the interface

$$\vec{I}_{s1\perp} = \frac{2e^2}{h} N \vec{m}_1 \times \left(\vec{V}_{S,\mathrm{N}} \times \vec{m}_1\right). \tag{13}$$

The expressions for the spin current polarized perpendicular to the magnetization direction are less intuitive in the general case, and we give here only the final results, referring to the technical sections for the derivations. A possible spin accumulation in the ferromagnet must be collinear to its magnetization direction and can only contribute to the collinear (longitudinal) component of the spin current. The non-collinear (transverse) part of the spin current can either be in the plane with the magnetization and spin accumulation vector in the normal metal, $\vec{V}_{S,\mathrm{N}} \times \vec{m}_1$, or normal to this plane, $\vec{m}_1 \times \left(\vec{V}_{S,\mathrm{N}} \times \vec{m}_1\right)$. These two transverse linear combinations of the spin accumulations, combined with two independent conductances, the real and imaginary part of the spin mixing conductance $G_{\uparrow\downarrow}$, determine the transverse spin current:

$$\vec{I}_{s1\perp} = 2\,\mathrm{Re}\, G_{1\uparrow\downarrow} \vec{m}_1 \times \left(\vec{V}_{S,\mathrm{N}} \times \vec{m}_1\right) + 2\,\mathrm{Im}\, G_{1\uparrow\downarrow} \vec{V}_{S,\mathrm{N}} \times \vec{m}_1.$$

The expression above is the non-collinear (transverse) spin current on the normal metal side of the normal metal-ferromagnet interface. On the ferromagnetic side of the interface at a distance larger than $\lambda_c$ $\vec{I}_{sF1\perp} = 0$. The spin-transfer torque is the loss of transverse spin current, $\vec{\tau}_1 = \vec{I}_{S1\perp}$. It should be noted that the spin accumulation in the normal metal $\vec{V}_{S,\mathrm{N}}$ is, as yet, an unknown quantity that needs to be determined by circuit theory, to be discussed below. When interpreted as spin transfer to the ferromagnet, the first term in $\vec{\tau}_1$, proportional to $\vec{m}_1 \times (\vec{m}_1 \times \vec{m}_2)$ corresponds to the torque introduced first by Slonczewski. The second term, proportional to $\vec{m}_1 \times \vec{m}_2$, acts as an effective magnetic field collinear to the magnetization direction in the second ferromagnet on the first ferromagnet. We will show below that for a symmetric two-terminal device the second term can be disregarded. It is also for metallic systems often much smaller than the first term since $\mathrm{Im}\, G_{1\uparrow\downarrow} \ll \mathrm{Re}\, G_{1\uparrow\downarrow}$. These torques are source terms in the phenomenological Landau-Lifshitz-Gilbert equation for the magnetization dynamics:

$$\left(\frac{d\vec{m}_{1/2}}{dt}\right)_{bias} = \frac{\hbar}{eM_{s,1/2}} \vec{\tau}_{1/2}, \tag{14}$$

where $M_{s,1/2}$ is the total magnetization of the ferromagnetic element.

The so-called mixing conductance $G_{1\uparrow\downarrow}$ is a material parameter independent of the spin dependent conductances $G_{1\uparrow}$ and $G_{1\downarrow}$. It is a pure interface property as long as the ferromagnetic dephasing length is the smallest length scale and, in the absence of spin-flip scattering localized at the interface, can be computed microscopically from the spin-dependent transmission and reflection coefficients in the spin quantization axis of the ferromagnet, see Eq. (2). The spin-transfer torque does not depend on a possible spin accumulation in the ferromagnet and only indirectly on the voltage bias $V_1$ via the spin accumulation in the normal metal $\vec{V}_{S,\mathrm{N}}$. According to the Landauer-Büttiker formula, the spin-dependent conductances $G_{1\uparrow}$ ($G_{1\downarrow}$) are given by spin-dependent transmission probabilities. The spin mixing conductance $G_{1\uparrow\downarrow}$, is on the other hand given by the number of transport modes in the normal metal, as in the ideal half-metallic ferromagnet, that have to corrected by material-combination-dependent normal reflection processes that usually suppress the spin-transfer torque.

In the following we use the previous expressions to determine currents and spin-torques in a $\mathrm{F}_1|\mathrm{N}|\mathrm{F}_2$ spin valve. The charge accumulation $V_{C,\mathrm{N}}$ and the spin accumulation $\vec{V}_{S,\mathrm{N}}$ must be determined by the flow rates of spins and charges in the entire spin-coherent circuit. To this end we need (i) expressions for the spins and charge currents from the the normal metal into the first ferromagnet, $I_{C1}$ and $I_{S1}$, as discussed above, and (ii) similar expressions for the second ferromagnet, $I_{C2}$ and $\vec{I}_{S2}$, and (iii) conservation equations for charges and spins in the normal metal. The above expressions for the charge and spin current from the first ferromagnet into the normal metal can be rewritten slightly as

$$I_{C1}/G_1 = V_{C,\mathrm{N}} - V_1 + P_1 \vec{V}_{S,\mathrm{N}} \cdot \vec{m}_1, \tag{15a}$$

$$I_{S1}/G_1 = \vec{m}_1 \left[ \vec{V}_{S,\mathrm{N}} \cdot \vec{m}_1 + P\left(V_{C,\mathrm{N}} - V_1\right) \right] + \eta_{R1} \vec{m}_1 \times \left( \vec{V}_{S,\mathrm{N}} \times \vec{m}_1 \right) + \eta_{I1} \vec{V}_{S,\mathrm{N}} \times \vec{m}_1, \tag{15b}$$

where we have introduced the total conductance $G_1$ and polarization $P_1$

$$G_1 = G_{1\uparrow} + G_{1\downarrow},$$
$$P_1 = \frac{G_{1\uparrow} - G_{1\downarrow}}{G_{1\uparrow} + G_{1\downarrow}},$$

and the real and imaginary parts of the relative mixing conductance,

$$\eta_{R1} = \frac{2\,\mathrm{Re}\, G_{1\uparrow\downarrow}}{G_{1\uparrow} + G_{1\downarrow}},$$
$$\eta_{I1} = \frac{2\,\mathrm{Im}\, G_{1\uparrow\downarrow}}{G_{1\uparrow} + G_{1\downarrow}}.$$

of the left junction. Similarly, the charge current, $I_{C2}$, and the spin-current, $\vec{I}_{S2}$, from the right reservoir into the normal metal node is

$$I_{C2}/G_2 = V_{C,\mathrm{N}} - V_2 + P_2 \vec{V}_{S,\mathrm{N}} \cdot \vec{m}_2, \tag{16a}$$

$$\vec{I}_{s2}/G_2 = \vec{m}_2 \left[ \vec{V}_{S,\mathrm{N}} \cdot \vec{m}_2 + P_2\left(V_{C,\mathrm{N}} - V_2\right) \right] + \eta_{R2} \vec{m}_2 \times \left( \vec{V}_{S,\mathrm{N}} \times \vec{m}_2 \right) + \eta_{I2} \vec{V}_{S,\mathrm{N}} \times \vec{m}_2, \tag{16b}$$

with total conductance and polarization of the right junction, $G_2$, $P_2$, and the real and imaginary parts of the relative mixing conductance are $\eta_{R2}$ and $\eta_{I2}$.

We obtain the I-V characteristics by generalizing Kirchhoff's Laws, demanding conservation of not only charge but also spin. Disregarding spin-flip scattering, this implies that, in the stationary state,

$$\begin{aligned} I_{C1} + I_{C2} &= 0, \\ \vec{I}_{S1} + \vec{I}_{s2} &= 0. \end{aligned}$$

The spin accumulation in the normal metal can have components collinear and non-collinear to the magnetization of ferromagnet 1 and ferromagnet 2. The spin-accumulation in the orthogonal coordinate system defined by the magnetization vectors $\vec{m}_1 \neq \vec{m}_2$ and the out-of-plane vector $\vec{m}_1 \times \vec{m}_2$ reads

$$\vec{V}_{S,\mathrm{N}} = V_{S1}\vec{m}_1 + V_{S2}\vec{m}_2 + V_{S12}\vec{m}_1 \times \vec{m}_2. \tag{17}$$

The three components $V_{S1}$, $V_{S2}$ and $V_{S12}$ depend on the relative orientation of the magnetizations $\vec{m}_1 \cdot \vec{m}_2 = \cos\theta$. The first term in $\vec{\tau}_1$, proportional to $\vec{m}_1 \times (\vec{m}_1 \times \vec{m}_2)$, is similar to the Slonczewski torque. The second term, proportional to $\vec{m}_1 \times \vec{m}_2$, is the effective magnetic field collinear to the magnetization direction of the second ferromagnet when acting on the first ferromagnet and vice versa. In metallic ferromagnets $\eta_I \ll \eta_R$. For a symmetric two-terminal device (see below) this also leads to $V_{S12} \ll V_{S1}, V_{S2}$ such that it can often be disregarded.

Expressions for the spin accumulation and the current and spin-torques in the device are obtained by inserting the spin accumulation (17) into (15)

$$\begin{aligned} I_{C1}/G_1 &= V_{C,\mathrm{N}} - V_1 + P_1 \left(V_{S1} + V_{S2}\cos\theta\right), \\ \vec{m}_1.\vec{I}_{S1}/G_1 &= V_{S1} + V_{S2}\cos\theta + P_1 \left(V_{C,\mathrm{N}} - V_1\right), \\ \vec{m}_2.\vec{I}_{S1}/G_1 &= \cos\theta\left[V_{S1} + V_{S2}\cos\theta + P_1\left(V_{C,\mathrm{N}} - V_1\right)\right] + \eta_{R1}\sin^2\theta V_{S2} + \eta_{I1}\sin^2\theta V_{S1}, \\ (\vec{m}_1 \times \vec{m}_2)\cdot\vec{I}_{S1}/G_1 &= \eta_{R1}\sin^2\theta V_{S1} - \eta_{I1}\sin^2\theta V_{S2}\ . \end{aligned}$$

and for the right junction:

$$\begin{aligned} I_{C2}/G_2 &= V_{C,\mathrm{N}} - V_2 + P_2 \left(V_{S2} + V_{S1}\cos\theta\right), \\ \vec{m}_1\cdot\vec{I}_{S2}/G_2 &= \cos\theta\left[V_{S2} + V_{S2}\cos\theta + P_2\left(V_{C,\mathrm{N}} - V_2\right)\right] + \eta_{R2}\sin^2\theta V_{S1} - \eta_{I2}\sin^2\theta V_{S12}, \\ \vec{m}_2\cdot\vec{I}_{S2}/G_2 &= V_{S2} + V_{S1}\cos\theta + P_2\left(V_{C,\mathrm{N}} - V_2\right), \\ (\vec{m}_1 \times \vec{m}_2)\cdot\vec{I}_{S2}/G_2 &= \eta_{R2}\sin^2\theta V_{S12} + \eta_{I2}\sin^2\theta V_{S1}. \end{aligned}$$

We now have now a closed system of linear equations for the four unknowns $V_{C,\mathrm{N}}$, $V_{S1}$, $V_{S2}$, and $V_{S12}$. For a symmetric system, $G_1 = G_2$, $P_1 = P_2$, $\eta_{R1} = \eta_{R1}$, and $\eta_{I1} = \eta_{I2} = 0$ and a bias voltage $V_1 = V/2$ and $V_2 = -V/2$ conservation of charge and spin gives $V_{C,\mathrm{N}} = 0$, $V_{S12} = 0$, $V_{S2} = -V_{S1}$ and

$$V_{S1} = \frac{(1-\cos\theta)}{(1-\cos\theta)^2 + \eta_R\sin^2\theta}P\frac{V}{2}.$$

The angular dependence of the current through the device is thus $I_C = I_{C2} = -I_{C1}$:

$$I = \frac{G}{2}\left[1 - P^2\frac{\tan^2\theta/2}{\eta_R + \tan^2\theta/2}\right]V.$$

The magnetization torque on ferromagnet 1 reads:

$$\vec{\tau}_1 = -\vec{m}_1 \times (\vec{m}_1 \times \vec{m}_2)\,\frac{P\alpha}{(1-P^2)\sin^2\theta/2 + \eta_R\cos^2\theta/2}\frac{I}{2}$$

with modulus

$$|\vec{\tau}_1| = \frac{|\tan\theta/2|}{(1-P^2)\tan^2\theta/2 + \eta_R}\eta_R P I.$$

For small angles $\theta \ll 1$, when the magnetizations of the ferromagnets are close to the parallel configuration, the magnitude of the spin-torque is

$$|\vec{\tau}_1| \rightarrow \frac{\theta P I}{2}, \tag{18}$$

identical to the result first found by Slonczewski that for symmetric junctions turns out to be quite generally valid, e.g. for metallic, diffusive and tunnel junctions. However, in the close to antiparallel regime with $\theta - \pi \ll 1$, the torque depends explicitly on the relative mixing conductance $\eta_R$:

$$|\vec{\tau}_1| \approx \frac{\eta_R P I}{(1-P^2)\,2}\,(\theta - \pi)\,.$$

A large relative mixing conductance, $\eta_R \geq 1$, enhances the spin-torque that destabilizes the antiparallel configuration. In asymmetric spin valves, the torque depends on the mixing conductance also at small angles. This can be used advantageously to maximize the torque by judiciously engineering the device, as discussed in later sections.

In this part we intended to show that for magnetoelectronic circuits with noncollinear magnetizations resistive elements cannot be described by only two conductances for spins parallel and antiparallel to the magnetization direction. In addition, material resistances must be introduced that parameterize transverse spin currents and spin-transfer torques. This leads to a generalized circuit theory, magnetoelectronic circuit theory, which can be used to compute the angular dependent magnetoresistance as well as the spin-torques in current-induced magnetization dynamics. The relatively simple analytical expressions contain parameters that can be computed from first principles and tested most directly by experiments on the angular magnetoresistance of spin valves. In the time dependent generalization of circuit theory in which the magnetization dynamics is treated adiabatically, the mixing conductance is obtained immediately from the broadening of ferromagnetic resonance spectra of F|N bilayers, but we refer for the details of the dynamics to a different review article [33].

## III. THEORY OF CHARGE AND SPIN TRANSPORT

A theoretical physicist has often several methods at hands in order to tackle a physical problem. The best choice is not unnecessarily complex ("cracking a nut with a sledgehammer") but should still quantitatively capture the physics of the problem ahead. Sometimes different options turn out to be equivalent. Here we wish to understand magnetoelectronic effects in state-of-the-art materials, devices and circuits. For extended bulk systems one would use, of course, traditional methods to obtain the conductivity tensor [86], and these have indeed been used for magnetoelectronic hybrid systems [1]. However, other methods, like the scattering theory [87], non-equilibrium Green function theory [88], or random matrix theory [72] have distinct advantages in hybrid systems, especially at the nanometer scales.

When the samples are nearly ballistic or transport is limited by geometrical constrictions like point contacts or single tunneling barriers, direct calculation of the conductance via transmission coefficients or Green functions is the most convenient theoretical approach. The effects of disorder can *e.g.* be included by configurational averaging based on a microscopic model or random matrix theory. For inhomogeneous hybrid structures which are *dirty*, *i.e.* samples scales are larger than the mean free path, one should resort to semiclassical methods related to the Boltzmann/diffusion equation. These can most conveniently be formulated from first-principles by the Keldysh formalism which, for magnetic systems, leads to a generalization of the theory of electronics circuits as formulated by Kirchhoff as introduced in Section II.

In this Chapter we discuss elements of the electronic structure and transport in magnetoelectronics structures. Starting with the basic Stoner Hamiltonian for bulk systems, we continue with the main ingredient for a theory of hybrid systems, *viz.* the boundary conditions at interfaces between different materials. These can be implemented by first-principles for disordered systems using Green functions or random matrices. We shall see that these for the non-specialist rather inaccessible methods lead to conceptually simple boundary conditions that match the solutions of the Boltzmann or diffusion equations on both sides of a resistor.

### A. Electronic structure

Throughout this review we assume that the electronic and magnetic degrees of freedom can be described by a mean-field theory. This excludes from the outset much of the physics of strongly correlated systems associated to the colossal magnetoresistance (CMR) [89]. The spin-orbit interaction is assumed to be weak, causing spin-flip processes that can be handled by phenomenological spin-relaxation times and spin diffusion lengths. Both these assumptions are well-established for 3d-transition metal magnetoelectronics.

For ferromagnetic (including ferromagnet|paramagnet hybrid) systems the Stoner Hamiltonian in Pauli spin space reads

$$\hat{H} = \left[-\frac{1}{2m}\nabla^2 + U^C(\vec{r})\right]\hat{1} + \hat{U}^S(\vec{r}), \tag{19}$$

$$\hat{U}^S(\vec{r}) = (\vec{\boldsymbol{\sigma}} \cdot \vec{u}(\vec{r}))\Delta(\vec{r}), \tag{20}$$

where $U^C(\vec{r})$ and $\hat{U}^S(\vec{r})$ are the spin-dependent and spin-independent electronic potentials, which in density-functional theory can be formulated rigorously as functionals of the ground state spin-densities. $\hat{U}^S$ vanishes in a paramagnet. In a ferromagnet is locally diagonal in a variable direction $\vec{u}$ and proportional to the exchange-correlation potential $\Delta$. Here the $2 \times 2$ unit and Pauli spin matrices read:

$$\{\mathbf{1}, \vec{\boldsymbol{\sigma}}\} = \{\mathbf{1}, \boldsymbol{\sigma}_x, \boldsymbol{\sigma}_y, \boldsymbol{\sigma}_z\} \tag{21}$$

$$= \left\{ \begin{pmatrix} 1 & 0 \\ 0 & 1 \end{pmatrix}, \begin{pmatrix} 0 & 1 \\ 1 & 0 \end{pmatrix}, \begin{pmatrix} 0 & -i \\ i & 0 \end{pmatrix}, \begin{pmatrix} 1 & 0 \\ 0 & -1 \end{pmatrix} \right\}. \tag{22}$$

The electronic potentials, wave functions and properties have been computed with considerable success by density-functional theory in the local-density approximation [90]. In Section V we discuss first-principles calculations of transport properties. In qualitative model calculations we often assume that $U^C$ and $\hat{U}^S$ are constants in the ferromagnet and vanish abruptly at the interface to the paramagnet. It is usually a good approximation to take $\vec{u}$ constant in a piece of bulk ferromagnet and parallel to the magnetization, although it must be allowed to vary on layer index in magnetic multilayers. We then disregard the effect of a possible lateral domain structure. Since we are interested here in electron transport, both ferromagnets and paramagnets are in the following taken to be metals, with the reservation that contacts between metals may be tunnel junctions, and ferromagnetic insulators can be sinks for transverse spin currents as well as sources for effective exchange fields.

The Stoner model or the related density-functional theory is the most appropriate way to describe the transport properties of transition metals and their heterostructures sufficiently below the Curie temperatures. For model calculations, the $s-d$ model Hamiltonian is practically equivalent to the Stoner model for static properties. It is essential to realize, however, that the $s-d$ exchange parameter must be chosen large enough so that the Fermi surfaces for up and down spins are sufficiently different in order to mimic the strong spin dependent scattering at ferromagnet|normal metal interfaces that is found in experiments and first principle calculations. Furthermore, the magnetization dynamics can be very different for both models [91].

### B. Boundary conditions

We know from quantum mechanics that at interfaces between two different materials the wave functions are continuously differentiable. These boundary conditions are formally included in the scattering matrix of the interface, *i.e.* the transmission and reflection coefficients. First-principles band-structure calculations do take the microscopic boundary conditions at N|F interfaces properly into account [61], also in disordered structures [35], as discussed in Section V. In disordered structures and a semiclassical formalism we are not interested so much in wave functions, but in distribution functions. In early phenomenological treatments of collinear systems the boundary problem was circumvented by replacing the interfaces by regions of a fictitious bulk material, the resistances of which can be fitted to experiments [57]. This is not possible anymore when the magnetizations are non-collinear, however, because potential steps are essential for the description of the dephasing of the non-collinear spin current and the torque.

Semiclassical methods cannot describe processes on length scales smaller than the mean free path, thus cannot properly describe abrupt interfaces. It is possible, however, to express boundary conditions in terms of transmission and reflection probabilities. For transport, these boundary conditions translate into interface resistances arising at discontinuities in the electronic structure and disorder at the interface. This phenomenon has been also extensively studied in the quasi-classical theory of superconductivity [92], where a generalized diffusion approach can be used in the bulk of the superconductor, but proper boundary conditions at the interfaces between a superconductor and another normal or superconducting metal are essential. Finding the boundary conditions that match general distributions functions through regions in which the semiclassical approximations possibly do not hold is a non-trivial task that can be carried out by Green function or random matrix theory (RMT), as discussed below.

Electron states with spins that are not collinear to the magnetization direction are not eigenstates of a ferromagnet, but precess around the magnetization vector. In three dimensions, a non-collinear spin current is made up from many states with different Larmor frequencies that rapidly interfere destructively in a ferromagnet as a function of penetration depth. The efficient relaxation of the non-diagonal terms in the spin-density matrix is equivalent to the suppression of spin-accumulation non-collinear to the magnetization in the ferromagnet. This spin-dephasing mechanism does not exist in normal metals, in which the spin wave functions remain coherent on the length scale of the spin-diffusion length that can be of the order of microns. In ballistic systems, the spin-transfer occurs over the ferromagnetic decoherence length $\lambda_c = \pi / \left| k_F^{\uparrow} - k_F^{\downarrow} \right|$. In conventional ferromagnets the exchange energy is of the same order of magnitude as the Fermi energy and $\lambda_c$ is of the order of the lattice constant. The strongly localized regime in which the mean free path is smaller than the inverse Fermi wave-vector, $\ell < 1/k_F$, is not relevant for elemental metals. In conventional metallic ferromagnets $\ell \gg 1/k_F$, and the length scale of the spin-transfer $\lambda_c$ is necessarily smaller than the mean free path $\ell$ and therefore is not affected by disorder. This argument does not hold for gradual interfaces and domain walls. The opposite limit has been considered by Zhang *et al.* [93, 94], where $\lambda_c = \sqrt{2hD_0/J}$ (in the paper it was designated by $\lambda_J$), or with $D_0 \sim \frac{\ell^2}{\tau}$, $\lambda_c \sim \ell\sqrt{2h/J\tau}$. This implies $2h/J\tau \gg 1$ or $\lambda_c \gg \ell$ as in, *e.g.*, PdNi alloys with small Ni concentration [95]. Such conditions do not hold for ferromagnetic conductors like Fe, Co, Ni and its alloys, however. In these ferromagnets the spin accumulation must be considered parallel to the magnetization, which importantly affects the boundary conditions to normal metals in non-collinear structures. This condition is fulfilled in the approaches discussed in this Section, and has been analyzed in detail by Stiles and Zangwill [69] and Zwierzycki *et al.* [96].

### C. Scattering theory of transport

The Landauer-Büttiker scattering theory of electronic conduction [97, 98] (following [72]) provides a complete description of transport at low frequencies, temperatures, and voltages, under circumstances that electron-electron interactions beyond a mean-field approximation can be disregarded. A mesoscopic conductor is modeled by a phase-coherent disordered region connected by ideal leads, *i.e.* with negligible effects of disorder, to two electron reservoirs (see Fig. 9). Let us initially assume that the system is a normal metal without magnetic elements or spin-orbit scattering, so that the spin-degree of freedom does not play a role. Scattering in the phase-coherent region is elastic. All inelastic scattering is assumed to take place in the reservoirs, which are in thermal equilibrium (for simplicity at zero temperature) and an electrochemical potential that can be modulated by an applied voltage that is assumed to be small compared to the Fermi energy $E_F$. The ideal leads are "electron waveguides", introduced to define a basis for the scattering matrix of the disordered region. These leads should not always taken to be too literally, they are often fictitious mathematical constructs rather than real physical wires. The current limiting element is a constriction, potential barrier, and/or disordered region (shaded) that is connected by these leads to two electron reservoirs (to the left and right of the dashed lines). The scattering matrix $S$ relates the amplitudes $a^+$, $b^-$ of incoming waves to the amplitudes $a^-$, $b^+$ of outgoing waves. The wave function $\psi_n$ of an electron in a lead at $E_F$ separates into a longitudinal and a transverse part

$$\psi_n^{\pm}(\vec{r}) = \chi_n(\tilde{\boldsymbol{\rho}}, x)\, e^{\pm ik_n x}/\sqrt{k_n}. \tag{23}$$

The integer $n = 1, 2, \cdots, N$ labels the propagating modes, also referred to as scattering channels. Mode $n$ has a real wave number $k_n > 0$ and transverse wave function $\chi_n$. We assume, for simplicity of notation, that the two leads are identical. The normalization of the wave function (23) is chosen such that it carries unit current. A wave incident on the disordered region is described in this basis by a vector of coefficients

$$\vec{c}^{\,\mathrm{in}} \equiv \left(a_1^+, a_2^+, \cdots, a_N^+, b_1^-, b_2^-, \cdots b_N^-\right). \tag{24}$$

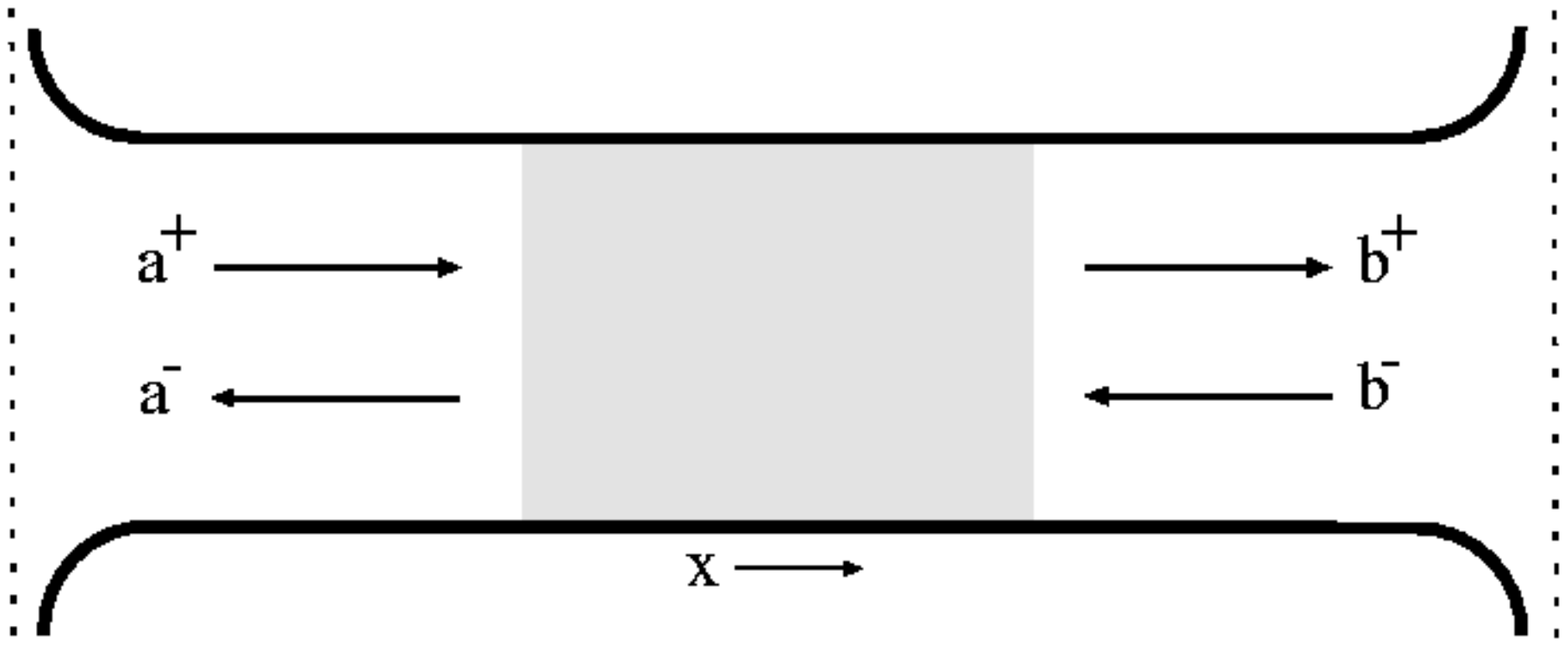

FIG. 9: the amplitudes $a^+,\ b^-$ of incoming waves to the amplitudes $a^-,\ b^+$

The first set of $N$ coefficients refers to the left lead, and the second set of $N$ coefficients to the right lead in Fig. 9. Similarly, the reflected and transmitted wave vector of coefficients reads

$$\vec{c}^{\,\mathrm{out}} \equiv \left(a_1^-, a_2^-, \cdots, a_N^-, b_1^+, b_2^+, \cdots b_N^+\right). \tag{25}$$

The scattering matrix $S$ is a $2N \times 2N$ matrix that relates these two vectors,

$$\vec{c}^{\,\mathrm{out}} = S\vec{c}^{\,\mathrm{in}} = \begin{pmatrix} r & t' \\ t & r' \end{pmatrix} \vec{c}^{\,\mathrm{in}}, \tag{26}$$

with $N \times N$ reflection matrices $r$ and $r'$ (reflection from left to left and from right to right) and transmission matrices $t$ and $t'$ (transmission from left to right and from right to left). Current conservation implies that $S$ is a unitary matrix: $S^{-1} = S^\dagger$. It is a consequence of unitarity that the four Hermitian matrices $tt^\dagger$, $t't'^\dagger$, $1 - rr^\dagger$, and $1 - r'r'^\dagger$ have the same set of eigenvalues $T_1,\ T_2, \cdots\ T_N$. Each of these $N$ transmission eigenvalues is a real number between 0 and 1. Let us now consider the spin-current through a normal metal and a ferromagnet connected by a still arbitrary junction.

A useful concept is the transfer matrix $\bar{M}$ between waves propagating to the right (left) on the right hand side of the contact $\vec{b}^+$ $(\vec{b}^-)$ and waves propagating to the right (left) on the left hand side of the contact $\vec{a}_L^+$ $(a_L^-)$

$$\begin{pmatrix} \vec{b}^+ \\ \vec{b}^- \end{pmatrix} = \bar{M} \begin{pmatrix} \vec{a}^+ \\ \vec{a}^- \end{pmatrix} . \tag{27}$$

The elements of the transfer matrix are related to the reflection and transmission coefficients by

$$\bar{M} = \begin{pmatrix} m^{++} & m^{+-} \\ m^{-+} & m^{--} \end{pmatrix} = \begin{pmatrix} t - r'(t')^{-1}r & r'(t')^{-1} \\ -(t')^{-1}r & (t')^{-1} \end{pmatrix} , \tag{28}$$

and the scattering matrix defined by Eq. (26) as

$$S = \begin{pmatrix} -(m^{--})^{-1}m^{-+} & (m^{--})^{-1} \\ m^{++} - m^{+-}(m^{--})^{-1}m^{-+} & m^{+-}(m^{--})^{-1} \end{pmatrix} , \tag{29}$$

$t_{nm}^{s\sigma}$ is the transmission matrix for incoming states from the left in mode $n$ with spin $s$ transmitted to outgoing states to the right with mode index $m$ and spin $\sigma$. The elements of the matrix of reflection coefficients $r_{nm}^{s\sigma}$ are labeled by mode index $m$ and spin $\sigma$ of incoming states from the left. $r'$ is the reflection matrix for incoming states from the right reflected to the right, and $t'$ is the transmission matrix for incoming states from the right transmitted to the left. Unitarity of the $S$-matrix implies that the transfer matrix satisfies

$$\bar{M}^\dagger \bar{\Sigma}_z \bar{M} = \bar{\Sigma}_z , \tag{30}$$

$$\bar{M} \bar{\Sigma}_z \bar{M}^\dagger = \bar{\Sigma}_z , \tag{31}$$

where $\left(\bar{\Sigma}_z\right)_{nsms'}^{\alpha\beta} = \alpha\delta_{\alpha\beta}\delta_{nsms'}$ is a Pauli matrix with respect to the direction of propagation.

In bulk systems the local electrical current $I\left(\vec{r}\right)$ as a function of an (internal) electric field distribution is well described by the linear-response Kubo formula

$$\vec{I}\left(\vec{r}\right) = \int d\vec{r}\,'\bar{\sigma}\left(\vec{r},\vec{r}\,'\right)\vec{E}\left(\vec{r}\right) \tag{32}$$

where the conductivity tensor $\bar{\sigma}$ can be calculated by standard Green function methods [86]. For the geometry in Fig. 9, and making use of the assumption that the electric field vanishes in the ideal leads, the conductance due to an applied voltage reads:

$$G = \lim_{V\to 0} I/V = \int_L d\vec{\rho} \int_R d\vec{\rho}\,'\ \bar{\sigma}\left(\vec{\rho}x, \vec{\rho}\,'x'\right) , \tag{33}$$

where the integral over the variable $\vec{\rho}$ $\left(\vec{\rho}\,'\right)$ is normal to the current direction parallel to $x\,(x')$ of the left (right) lead. Fisher and Lee [99] have shown that the Kubo formula (33) can be expressed in terms of the transmission probabilities as:

$$G = \frac{2e^2}{h}\sum_{nm}\left|t_{nm}\right|^2 = \frac{2e^2}{h}\sum_n T_n , \tag{34}$$

which is often referred to as the Landauer or Landauer-Büttiker formula. In the case of a ballistic point contact, the transmission probabilities are either zero or unity and the conductance in units of $e^2/h$ is just the number of propagating modes through the sample. In a tunnel junction, all transmission probabilities are small, whereas in diffuse wires, the conductance is dominated by a few highly-transmitting modes. Arbitrary linear statistics, like the shot noise, can conveniently be expressed in terms of the transmission matrix eigenvalues [72].

Spin can be easily reintroduced. The reservoirs may be ferromagnets, which are parameterized not only by a chemical potential, but also by a magnetization direction. By choosing for each reservoir and lead a local spin coordinate systems, the spin-up and spin-down indices $\sigma$ become good quantum numbers to label the scattering states and

$$G = \frac{e^2}{h} \sum_{n\sigma m\sigma'} |t_{n\sigma m\sigma'}|^2 \,. \tag{35}$$

Note that transitions from states $\sigma$ to $-\sigma$ represent spin-reversal of an electron passing the sample, which can be caused by spin-flip scattering due to spin-orbit scattering, spin-flop scattering at loose spins as magnetic impurities, or magnetization in the sample or reservoirs that are not collinear (parallel or antiparallel) to each other.

The scattering theory has several advantages, especially in mesoscopic physics. It is readily extended, for example, to describe transport in many-terminal samples or circuits. In combination with statistical averaging, notably random matrix theory, it can be generalized to treat dirty systems (see Section III E) as well.

### D. Keldysh Green function formalism

Non-equilibrium transport properties are most conveniently discussed in the framework of the Keldysh formalism [88]. We concentrate here on the Keldysh approach to the transport in hybrid normal|ferromagnetic (F|N) metal systems [68].

In systems involving ferromagnets electron dynamics is naturally strongly affected by the electron spin. Very generally, the elementary (retarded, advanced and Keldysh) Green functions can then be represented by $2 \times 2$ matrices in Pauli spin-space, whereas the Keldysh Green function is given by the $(4 \times 4)$ matrix

$$\check{G} = \begin{pmatrix} \hat{G}^R & \hat{G}^K \\ \hat{0} & \hat{G}^A \end{pmatrix},$$

where $\hat{G}^R$, $\hat{G}^K$ and $\hat{G}^A$ are, respectively, the retarded, Keldysh and advanced Green functions (defined below) and $\widehat{0}$ is the $(2 \times 2)$ zero matrix. The retarded Green function in spin-space is

$$\hat{G}^R(1,1') = \begin{pmatrix} G^R_{\uparrow\uparrow}(1,1') & G^R_{\uparrow\downarrow}(1,1') \\ G^R_{\downarrow\uparrow}(1,1') & G^R_{\downarrow\downarrow}(1,1') \end{pmatrix}$$

and *idem dito* for $\hat{G}^K$ and $\hat{G}^A$. Here 1 denotes the spatial and the time coordinates, $1 = \vec{r}_1 t_1$. The symbol "check" ( ˇ ) denotes $(4 \times 4)$ matrices in Keldysh space and the symbol "hat" ( ˆ ) denotes $(2 \times 2)$ matrices in spin-space. The spin-components of the Green functions indicated by the subscripts $\sigma, s$ are

$$G^R_{\sigma s}(1,1') = -i\theta(t_1 - t_{1'}) \left\langle \left[\psi_\sigma(1), \psi^\dagger_s(1')\right]_+ \right\rangle , \tag{36}$$

$$G^A_{\sigma s}(1,1') = i\theta(t_{1'} - t_1) \left\langle \left[\psi_\sigma(1), \psi^\dagger_s(1')\right]_+ \right\rangle , \tag{37}$$

$$G^K_{\sigma s}(1,1') = -i \left\langle \left[\psi_\sigma(1), \psi^\dagger_s(1')\right]_- \right\rangle , \tag{38}$$

where $\psi^\dagger_s(1)$ is the electron field operator for an electron with spin $s$ along the $z$-direction, the anticommutator is $[A,B]_+ = AB + BA$ and the commutator is $[A,B]_- = AB - BA$. The Keldysh Green functions are determined by the equation $\left(\check{H} - i\hbar\partial_t\right) \check{G}(\vec{r}t, \vec{r}\,'t') = \check{1}\delta(\vec{r}t - \vec{r}\,'t')$ with boundary conditions to be discussed below. $\check{1}$ is a $4 \times 4$ unit matrix and $\check{H}$ is block diagonal containing $\hat{H}$ twice. In the stationary state $\check{G}(1,1') = \int d(E/2\pi) \exp(iE\,(t_1 - t_{1'}))\check{G}_E(r,r')$ and the Green function on a given energy shell is determined from $\left(\check{H} - E\right) \check{G}_E(\vec{r}, \vec{r}\,') = \check{1}\delta(\vec{r} - \vec{r}\,')$. We will in the following drop the index $E$. For coordinates located in the leads, the Keldysh Green function can be expanded into quasi-one-dimensional modes as

$$\check{G}_{\sigma s}(\vec{r}, \vec{r}\,') = \sum_{nm,\alpha\beta} \tilde{G}^{\alpha\beta}_{n\sigma m s}(x,x') \chi^n_\sigma(\vec{\rho}; x) \chi^{m*}_s(\vec{\rho}\,'; x') e^{i\alpha k^n_\sigma x - i\beta k^m_s x'} , \tag{39}$$

where $x$,$\vec{\rho}$ are the longitudinal and transverse lead coordinates, $\chi^m_s(\vec{\rho}; x)$ is the transverse wave function and $k^m_s$ denotes the longitudinal wave-vector for an electron in transverse mode $m$ with spin $s$. The indices $\alpha$ and $\beta$ denote right-going $(+)$ and left-going $(-)$ modes. To denote matrices in Keldysh space, spin-space, the space spanned by the transverse modes, and the propagation direction we introduced the "tilde" ( ˜ ).

The current operator can be found from the continuity relation for the electron density. The spin-density matrix $\rho_{\sigma s}(1) = \left\langle \psi^\dagger_s(1)\psi_\sigma(1) \right\rangle$ is nothing but the $2 \times 2$ charge and spin accumulation with a direction that deviates from the quantization axis when the non-diagonal elements do not vanish. The spin-density matrix can also be written as $\rho_{\sigma s} = \left(\delta_{\sigma s}\rho_0 + \vec{\sigma}_{\sigma s} \cdot \vec{\rho}_S\right)/2$, where $\rho_0$ is the scalar charge and $\vec{\rho}_S$ the vectorial spin accumulation and we have used the elements of the Pauli matrix vector $\vec{\boldsymbol{\sigma}} = (\boldsymbol{\sigma}_x, \boldsymbol{\sigma}_y, \boldsymbol{\sigma}_z)$. $\rho_0$ and $\vec{\rho}_S$ are here densities that can be converted to the spin accumulation voltages in Chapter II via the density of states at the Fermi energy. The time-evolution of the spin-density matrix reads

$$\frac{\partial}{\partial t_1}\rho_{\sigma s} = -\frac{\partial}{\partial \vec{r}_1}\vec{J}^{\mathrm{S}}_{\sigma s} + \left(\frac{\partial \rho_{\sigma s}}{\partial t_1}\right)_{\text{prec.}} ,$$

where we have inserted the Hamiltonian (19) and found the vector spin-current

$$\vec{J}^{\mathrm{S}}_{\sigma s} = \frac{\hbar i}{2m}\left\langle \frac{\partial \psi_s^{\dagger}}{\partial \vec{r}_1}\psi_\sigma - \psi_s^{\dagger}\frac{\partial \psi_\sigma}{\partial \vec{r}_1}\right\rangle ,$$

and the spin-precession in the exchange potential $\hat{U}^S$ :

$$\left(\frac{\partial \rho_{\sigma s}}{\partial t_1}\right)_{\text{prec.}} = \frac{1}{i\hbar}\sum_{\alpha}\left[U^S_{\sigma\alpha}\rho_{\alpha s} - \rho_{\sigma\alpha}U^S_{\alpha s}\right] . \tag{40}$$

In two and three dimensions, the spin-precession (40) is an average over many states with different Larmor frequencies which averages out quickly in a ferromagnet with a large exchange splitting, leading to an efficient relaxation of the non-diagonal terms in the spin-density matrix. This means that a spin-current non-collinear to the magnetization can not penetrate a ferromagnet beyond a certain skin depth, the ferromagnetic coherence length $\lambda_c$. In normal metals (in the absence of spin-flip scattering) the spin-wave functions remain coherent and can point in any direction. In a simple Stoner model of two bands with Fermi wave vectors $k_F^{\uparrow}$ and $k_F^{\uparrow}$, the length scale over which destructive interference of the transverse component of the spin current is seen to occur on the length scale

$$\lambda_c = \frac{\pi}{\left|k_F^{\uparrow} - k_F^{\downarrow}\right|}. \tag{41}$$

In transition metal ferromagnets $\lambda_c$ is of the order of the Fermi wave vector and lattice constant, thus usually smaller than all other length scales of the problem. Stiles and Zangwill analyzed the physics of the decoherence of transverse spins injected into bulk ferromagnets via ballistic interfaces [69]. Zwierzycki *et al.* carried out numerical calculations of the transverse spin transmission through thin layers and found that the penetration depth is even more reduced when interfaces are disordered [96]. We will make use of the short ferromagnetic coherence lenght later.

The $2 \times 2$ spin-current matrix in the stationary state

$$\vec{J}^{\mathrm{S}}_{\sigma s} = \left(\frac{\partial}{\partial \vec{r}_1} - \frac{\partial}{\partial \vec{r}_{1'}}\right)\frac{\hbar i}{2m}\int \frac{dE}{2\pi}\int d(t_1 - t_{1'})e^{iE(t_1 - t_{1'})}\left\langle \psi_s^{\dagger}(1)\psi_\sigma(1')\right\rangle|_{\vec{r}_{1'}=\vec{r}_1} , \tag{42}$$

is the Keldysh component of the $4 \times 4$ current matrix in Keldysh and spin-space $\check{I}$. Along the longitudinal $x$-direction:

$$\check{I}(x) = \int d\vec{\rho} \frac{e\hbar}{m} \left(\partial_x - \partial_{x'}\right) \check{G}(\vec{r}, \vec{r}\,')|_{\vec{r}\,'=\vec{r}}.$$

In the leads the transverse wave functions $\chi_s^n(\vec{\rho}; x)$ do not depend on the $x$ coordinate, which can be exploited by the expansion

$$\check{I}_{ss'}(x) = ie \sum_{mn\alpha\beta} \left(\alpha v_s^n - \beta v_{s'}^m\right) \tilde{G}_{nsms'}^{\alpha\beta}(x, x) \int d\vec{\rho} \chi_s^n(\vec{\rho}; x) \chi_{s'}^{m*}(\vec{\rho}; x) \, , \tag{43}$$

where $v_s^n = \hbar k_s^n / m$ is the longitudinal velocity for an electron in transverse mode $n$ with spin $s$. In a normal metal, the transverse states and the longitudinal momentum are spin-independent and the Keldysh current simplifies to $\check{I}_{ss'}(x) = 2ie \sum_{n\alpha} \alpha v^n \tilde{G}_{nsns'}^{\alpha\alpha}(x, x)$ . We will subsequently use this expression to calculate the spin-current on the *normal* side of a normal metal-ferromagnet contact. In the normal metal, the 2nd term on the right hand side of the representation

$$i\tilde{G}_{nsms'}^{\alpha\beta}(x, x') = \frac{\tilde{g}_{nsms'}^{\alpha\beta}(x, x')}{\sqrt{v_s^n v_{s'}^m}} + \check{1}\delta_{ss'} \frac{\alpha \delta_{\alpha,\beta} \text{sign}(x - x')}{v_s^n} \tag{44}$$

does not contribute to the current, which therefore simplifies to

$$\check{I}_{ss'}(x) = 2e \sum_{n\alpha} \alpha \tilde{g}_{nsns'}^{\alpha\alpha}(x, x) \, . \tag{45}$$

Let us now consider the spin-current through a normal metal and a ferromagnet connected by a still arbitrary junction defined by a scattering matrix $S$, Eq. (26), and transfer matrix $\bar{M}$, Eq. (28). In order to connect the Green function to the left and to the right of the contact, we use the transfer matrix of the contact $\tilde{g}_{nsms'}^{\sigma\sigma'}(x = x_2, x') = \sum_{ls'',\sigma''} \bar{M}_{nsls''}^{\sigma\sigma''} \tilde{g}_{ls''ms'}^{\sigma''\sigma'}(x = x_1, x')$ and similarly for the $x'$-coordinate. Hence

$$\tilde{g}_2 = \bar{M} \tilde{g}_1 \bar{M}^\dagger \, , \tag{46}$$

where $\tilde{g}_{2(1)} = \tilde{g}(x = x_{2(1)}, x' = x_{2(1)})$.

The Keldysh Green functions for the Hamiltonian (19) are not unique without boundary conditions. At this point we have to introduce an essential approximation, *viz.* to assume *isotropy* of the electron distributions in the nodes, which may be done when electron propagation is chaotic or diffuse. The distribution of the incoming modes is isotropic either because of scattering within the nodes, or due to scattering off irregular boundaries of the nodes. Per definition, there are no fast spatial dependences in the leads, and the incoming modes may be described by the quasi-classical Green functions, in which the quickly varying spatial

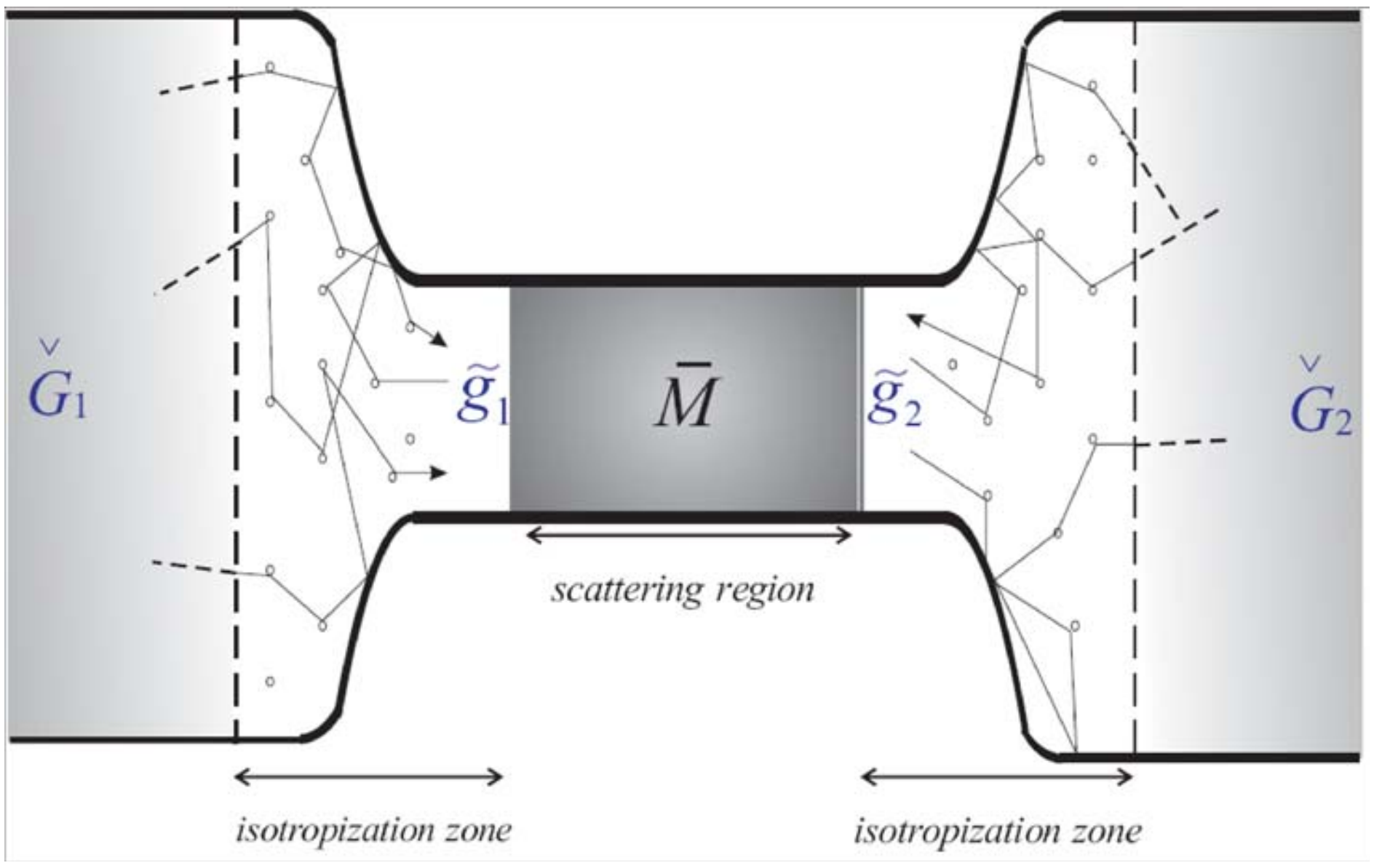

FIG. 10: Intuitive picture of the microscrocopic treatment of an unspecified contact in magnetoelectronic circuit theory [100]. Two nodes are connected by a contact/interface that constitutes the scattering region described by a transfer matrix $\bar{M}$ [Eq. (28)]. At both sides of the contact region a ballistic Green function $\tilde{g}$ [Eq. (44)] is a matrix in the space of transport channels, containing the information about the amplitudes of incoming and outgoing modes. $\tilde{g}_1$ and $\tilde{g}_2$ are connected by $\bar{M}$ according to Eq. (46). Electrons propagate between scattering regions and reservoirs through so-called "isotropization zones". These regions are supposed to have a width of the order of the mean free path, much smaller than the spin-flip diffusion length. The elastic impurity scattering redistributes the possibly anisotropic $\tilde{g}'$s without signicantly increasing the resistance, connecting them seamlessly to the isotropic quasiclassical Green functions $\check{G}_{1/2}$ that characterize the nodes.

oscillations are integrated out [88]. The outgoing modes are determined by the scattering properties of the contact. We follow the simple model of isotropization introduced by Nazarov [84] for superconductor-normal metal systems to connect the anisotropic Green functions in the scattering region, $\tilde{g}_1$ and $\tilde{g}_2$, to the isotropic Green functions $\bar{G}_1$ and $\bar{G}_2$ in the normal and ferromagnetic leads. We assume that the electrons are repeatedly scattered by point defects in the isotropization zone, as sketched in Fig. 10. Such a zone should not be interpreted too literally. It is a convenient instrument to model generic disorder that is trustworthy because the final expressions do not depend on its parameters. The nodes 1 and 2 in a circuit are connected by a scattering region such as a characterized by a transfer matrix $\bar{M}$. The Green functions $\tilde{g}_1$ and $\tilde{g}_2$ are introduced to describe the system in (possibly infinitesimally thin) ballistic strips at both sides of the scattering region that are diagonal in the space of transport channels. They contain all information about the incoming and outgoing

In the isotropization zone the quasi-classical approximation applies. On the normal metal side, the Green function is therefore determined by the well-established quasi-classical equations [88]:

$$\left(\alpha v_n \frac{\partial}{\partial x} + \check{H}_{\text{eff}}\right) G^{\alpha\beta}_{n\sigma ms}(x,x') \;=\; 0, \tag{47}$$

$$G^{\alpha\beta}_{n\sigma ms}(x,x')\left(\beta v_n \frac{\partial}{\partial x'} - \check{H}_{\text{eff}}\right) \;=\; 0, \tag{48}$$

where $\check{H}_{\text{eff}}$ is the effective quasi-classical Hamiltonian that is governed by the self-energy describing disorder scattering [84, 88]. For isotropic point scatterers the self-energy is $\check{H}_{\text{eff}} = \bar{G}_1/2\tau_{\text{imp}}$, where $\tau_{\text{imp}}$ is the impurity scattering life time. The solution of (47) and (48) is [84]

$$\check{G}(x,x') = \tilde{P}(x)\left[\tilde{g}_1 + \text{sign}(x-x')\bar{\Sigma}^z\right]\tilde{P}(-x'),$$

where $\left(\bar{\Sigma}_z\right)^{\alpha\beta}_{n\sigma ms} = \alpha\delta_{nm}\delta_{\sigma s}\delta_{\alpha\beta}$ and

$$\tilde{P}(x) = \delta_{nm}\left[\tilde{1}\cosh\left(\frac{x}{2v_n\tau_{\text{imp}}}\right) - \bar{\Sigma}^z\bar{G}_1\sinh\left(\frac{x}{2v_n\tau_{\text{imp}}}\right)\right]. \tag{49}$$

The Green function cannot continue to grow with decreasing coordinate $x$ and $x'$. From Eq. (49), we see that these conditions are compactly expressed as

$$\left(\bar{\Sigma}_z + \bar{G}_1\right)\left(\bar{\Sigma}_z - \bar{g}_1\right) \;=\; 0 \tag{50}$$

$$\left(\bar{\Sigma}_z + \tilde{g}_1\right)\left(\bar{\Sigma}_z - \bar{G}_1\right) \;=\; 0, \tag{51}$$

where $\bar{G}_1$ is the (isotropic) Green function in reservoir or node 1:

$$\left(\bar{G}_1\right)^{\alpha\beta}_{nsms'} = \delta_{nm}\delta^{\alpha\beta}(\check{G}_1)_{ss'}\,. \tag{52}$$

The importance of these effective boundary conditions is that they do not depend on the microscopic details of the isotropization, *e.g.* they are independent on the mean free path in the normal metal, and therefore likely to be universal, not depending on the specific microscopic scattering mechanism [84]. The effective boundary conditions linking the Green functions in the sample to the isotropic ones in the ferromagnetic node can be derived analogously

$$\left(\bar{\Sigma}_z - \bar{G}_2\right)\left(\bar{\Sigma}_z + \tilde{g}_2\right) \;=\; 0 \tag{53}$$

$$\left(\bar{\Sigma}_z - \tilde{g}_2\right)\left(\bar{\Sigma}_z + \bar{G}_2\right) \;=\; 0\,, \tag{54}$$

where $\bar{G}_2$ has the same form as $\bar{G}_1$, Eq. (52). In a normal metal, the retarded quasi-classical Green function is

$$\bar{G}_R = \begin{pmatrix} G_R^{++} & G_R^{+-} \\ G_R^{-+} & G_R^{--} \end{pmatrix} = \begin{pmatrix} 1 & 0 \\ 0 & 1 \end{pmatrix},$$

and the advanced one is $\bar{G}_A = -\bar{G}_R$. (Note that here 1 stands for a unit matrix in the basis of the transverse modes and spin, $1 \rightarrow \delta_{ns,ms'}\hat{1}$). The Keldysh component of the Green function is

$$\bar{G}_{K,1(2)} = \hat{h}_{1(2)} \begin{pmatrix} 1 & 0 \\ 0 & 1 \end{pmatrix}, \tag{55}$$

where the $2\times2$ distribution matrix $\hat{h}$ depends on the (non-equilibrium) distribution functions $\hat{f}(\epsilon)_{1(2)}$ in the nodes by

$$\hat{h}_{1(2)} = 2(2\hat{f}(\epsilon)_{1(2)} - 1). \tag{56}$$

$\hat{f}(\epsilon)_{1(2)}$ is a $2 \times 2$ matrix in spin-space describing the spin-accumulation in the normal metal and ferromagnet.

The isotropy conditions (50-54) connect the retarded and advanced Green function on the left and right hand side of the contact:

$$\begin{aligned} \tilde{g}_{R,1} &= \begin{pmatrix} 1 & 0 \\ \tilde{g}_{R,1}^{-+} & 1 \end{pmatrix} \\ \tilde{g}_{R,2} &= \begin{pmatrix} 1 & \tilde{g}_{R,2}^{+-} \\ 0 & 1 \end{pmatrix} \end{aligned}$$

where $\tilde{g}_{R,1}^{-+}$ and $\tilde{g}_{R,2}^{+-}$ are determined by the Green function on the right and the scattering properties of the contact. The advanced Green function is related to the retarded Green function by

$$\tilde{g}_A = -\tilde{g}_R^\dagger.$$

The Keldysh component on the left side is determined analogously, leading to

$$\tilde{g}_{K,1} = \begin{pmatrix} \hat{h}_1 1 & \hat{h}_1 r^\dagger \\ r\hat{h}_1 & \tilde{g}_{K,1}^{--} \end{pmatrix}, \tag{57}$$

$$\tilde{g}_{K,2} = \begin{pmatrix} \tilde{g}_{K,2}^{++} & r'\hat{h}_2 \\ \hat{h}_2 r'^\dagger & \hat{h}_2 1 \end{pmatrix}. \tag{58}$$

We now use the relation between the Green function on the left hand side and the right hand side of the contact (46) to obtain the retarded Green functions

$$\begin{aligned} \tilde{g}_{R,1}^{-+} &= 2r \\ \tilde{g}_{R,2}^{+-} &= 2r', \end{aligned}$$

and the Keldysh Green function

$$\tilde{g}_{K,1}^{--} = t'\hat{h}_2 (t')^\dagger + r\hat{h}_1 r^\dagger \tag{59}$$

$$\tilde{g}_{K,2}^{++} = t\hat{h}_1 t^\dagger + r'\hat{h}_2 (r')^\dagger \ . \tag{60}$$

Inserting the expression for the Keldysh component (60) into (45) and using (56), we finally arrive at the desired expression of the current through the contact on the normal metal side as:

$$\hat{I}_{F|N} = \frac{e}{h}\sum_{nm}[\hat{t}'^{nm}\hat{f}^F \left(\hat{t}'^{mn}\right)^\dagger - \delta_{nm}\hat{f}^N + \hat{r}^{nm}\hat{f}^N \left(\hat{r}^{mn}\right)^\dagger] \,, \tag{61}$$

This expression is used in Section IV A to develop a circuit theory of multi-terminal hybrid normal metal-ferromagnet systems. The Landauer-Büttiker formula is recovered when the distribution matrices $\hat{f}^F$ and $\hat{f}^N$ correspond to reservoirs, *viz.* are at equilibrium with only a bias applied. Note that this equation is very general, holding for any electrical connection between magnetic or normal nodes with arbitrary distribution functions. Eq. (61) is used in Section VII and by Kovalev et al. [70] to study phase-coherent transport through magnetic structures between normal metal nodes.

### E. Random matrix theory

Electron transport in disordered and chaotic systems can be formulated in terms of random matrix theory (RMT) [72]. In RMT, the starting point is the exact scattering matrix of the system. Instead of a configurational average over the microscopic parameters of the Hamiltonian like the defect positions, the ensemble average is obtained by averaging over all possible scattering matrices which fulfill given symmetry constraints. Waintal *et al.* [56] used that approach to calculate the spin-torque and conductance of two-terminal non-collinear hybrid normal metal-ferromagnet systems. RMT expanded to lowest order into the reciprocal numbers of modes $1/N$ turns out to be nearly equivalent to the circuit theory described in Section IV A, which is based on the Keldysh formalism sketched above, as well as the Boltzmann equation approach by Schep *et al.* [61, 62]. In contrast to the other methods, RMT can in principle be extended to include quantum effects by taking into account terms of higher order in $1/N$.

Let us follow [56] and consider a two-terminal $F_a|N|F_b$ spin valve. Four fictitious leads numbered by 0-3 are inserted into the pillar at strategic places; 0: to the right of the right ferromagnet, 1: between the normal metal node and the right ferromagnet, 2: between the left ferromagnet and the normal metal node, and 3: to the left of the left ferromagnet. The total scattering matrix of the system can be found by concatenating the scattering matrices for regions 0-3. The amplitudes of the left-going (right-going) waves $(\Psi_{iL})_{\alpha s}$ ($(\Psi_{iR})_{\alpha s}$) in node $i$ with transverse quantum number $\alpha$ and spin-component $s$ are determined by the transfer matrices $M_{iLL}$, $M_{iLR}$, $M_{iRL}$ and $M_{iRR}$, cf. Eq. (28):

$$\begin{pmatrix} \Psi_{iL} \\ \Psi_{iR} \end{pmatrix} = \begin{pmatrix} M_{iLL} & M_{iLR} \\ M_{iRL} & M_{iRR} \end{pmatrix} \begin{pmatrix} \Psi_{0L} \\ \Psi_{3R} \end{pmatrix} \tag{62}$$

The transfer matrices can be found by concatenating the scattering matrices in the different regimes with $M_{0LL} = 1$ and, cf. Eqs. (29,26):

$$M_{1LL} = \left(1 - r_a t_n r_b' \left(1 - r_n r_b'\right)^{-1} t_n' - r_a r_n'\right)^{-1} t_a' \tag{63}$$

$$M_{2LL} = \left(1 - r_\alpha r_b'\right)^{-1} t_n' \left(1 - r_a t_n r_b' \left(1 - r_n r_b'\right)^{-1} t_n' - r_a r_n'\right)^{-1} t_a' \tag{64}$$

$$M_{3LL} = t_b' \left(1 - r_\alpha r_b'\right)^{-1} t_n' \left(1 - r_a t_n r_b' \left(1 - r_n r_b'\right)^{-1} t_n' - r_a r_n'\right)^{-1} t_a'. \tag{65}$$

Analogous results are found for the other components of the transfer matrix.

The Landauer-Büttiker charge conductance reads

$$G = \frac{e^2}{h} \mathrm{Tr} M_{iLL} M_{iLL}^\dagger \ . \tag{66}$$

The torque exerted on the magnetizations equals the difference between the spin currents to the left and right of a given ferromagnetic layer. Waintal *et al.* [56] also consider the spin current response to an infinitesimal chemical potential variation $\delta\mu_3$ ($\delta\mu_0$) in the left (right) reservoir

$$\frac{\partial \vec{J}_i}{\partial \mu_3} = \frac{1}{4\pi} \mathrm{ReTr} \hat{\boldsymbol{\sigma}} \left(M_{iRL} M_{iRL}^\dagger - M_{iLL} M_{iLL}^\dagger\right) \tag{67}$$

$$\frac{\partial \vec{J}_i}{\partial \mu_0} = \frac{1}{4\pi} \mathrm{ReTr} \hat{\boldsymbol{\sigma}} \left(M_{iRR} M_{iRR}^\dagger - M_{iLR} M_{iLR}^\dagger\right) \tag{68}$$

that are not the same since spin angular momentum is transferred to the ferromagnetic order parameter. This implies that a finite magnetization torque $\vec{\tau} = \vec{J}_2 - \vec{J}_3$ may exist even when both chemical potentials are the same, reflecting the equilibrium-spin-current-induced non-local and oscillatory exchange interaction between the ferromagnets [64]. However, this equilibrium magnetization torque is sensitive to the electron phase and vanishes in the presence of disorder when the number of transport channels is large [56].

In order to carry out the ensemble average in the normal metal layer, the scattering matrix in rewritten in terms of a "polar decomposition":

$$S_n = \begin{pmatrix} r_n & t_n' \\ t_n & r_n' \end{pmatrix} = \begin{pmatrix} U & 0 \\ 0 & V' \end{pmatrix} \begin{pmatrix} \sqrt{1-T} & i\sqrt{T} \\ i\sqrt{T} & \sqrt{1-T} \end{pmatrix} \begin{pmatrix} U' & 0 \\ 0 & V \end{pmatrix} , \tag{69}$$

where $U$, $V$, $U'$ and $V'$ are $2N \times 2N$ block-diagonal unitary matrices and $T$ is a diagonal matrix with eigenvalues $0 \leq T_n \leq 1$ of the transmission matrix $tt^\dagger$ as introduced in Section III C. In the isotropic approximation the $N \times N$ blocks are assumed to be uniformly distributed in the group $U(N)$. The observables like spin-torques and the conductance (66) should be averaged not only over the eigenvalues $T_n$ as in conventional RMT, but also over the unitary matrices. For a large number of transverse channels an expansion in the small parameter $1/N$ is very useful. Without repeating calculations here, we only mention that the lowest order term is identical with the semiclassical concatenation of the scattering matrices. In this limit the quantum mechanical concatenation rules for transmission (reflection) amplitudes in spin and orbital space $t_{ns,n's'}$ are replaced by $4\times 4$ matrices of transmission (reflection) probabilities that are averaged over the transverse channels, but still reflect spin-coherence, *e.g.* the transmission probability matrix:

$$\check{T} = \frac{1}{N}\mathrm{Tr}_N \begin{pmatrix} t_{\uparrow\uparrow}t^\dagger_{\uparrow\uparrow} & t_{\uparrow\uparrow}t^\dagger_{\uparrow\downarrow} & t_{\uparrow\downarrow}t^\dagger_{\uparrow\uparrow} & t_{\uparrow\downarrow}t^\dagger_{\uparrow\downarrow} \\ t_{\uparrow\uparrow}t^\dagger_{\downarrow\uparrow} & t_{\uparrow\uparrow}t^\dagger_{\downarrow\downarrow} & t_{\uparrow\downarrow}t^\dagger_{\downarrow\uparrow} & t_{\uparrow\downarrow}t^\dagger_{\downarrow\downarrow} \\ t_{\downarrow\uparrow}t^\dagger_{\uparrow\uparrow} & t_{\downarrow\uparrow}t^\dagger_{\uparrow\downarrow} & t_{\downarrow\downarrow}t^\dagger_{\uparrow\uparrow} & t_{\downarrow\downarrow}t^\dagger_{\uparrow\downarrow} \\ t_{\downarrow\uparrow}t^\dagger_{\downarrow\uparrow} & t_{\downarrow\uparrow}t^\dagger_{\downarrow\downarrow} & t_{\downarrow\downarrow}t^\dagger_{\downarrow\uparrow} & t_{\downarrow\downarrow}t^\dagger_{\downarrow\downarrow} \end{pmatrix} . \tag{70}$$

For the normal metal node in the semiclassical regime $\check{T}_n = \frac{g_N}{N}\check{1}_4$ and $\check{R}_n = \left(1 - \frac{g_N}{N}\right)\check{1}_4$, where $g_N$ is the conductance of the normal metal layer and $\check{1}_4$ is the $4 \times 4$ unit matrix. The conductance (66) is determined by the properties of the transmission matrix $M_{3LL}M^\dagger_{3LL} = \check{T}_{b|n|a}$ and reads

$$\check{T}_{b|n|a} = \check{T}'_b \left(\check{1} - \check{R}_n\check{R}'_b\right)^{-1}\check{T}'_n \left[\check{1} - \check{R}_a\check{T}_n\check{R}'_b\left(\check{1} - \check{R}_n\check{R}'_b\right)^{-1}\check{T}'_n - \check{R}_a\check{R}'_n\right]^{-1}\check{T}'_a. \tag{71}$$

In Section IV B we demonstrate that this matrix expression can in fact be evaluated analytically and be compared with other theoretical results. Higher order terms in $N^{-1}$ correspond to quantum corrections to semiclassical transport.

### F. Boltzmann and diffusion equations

In the three previous sections we derived expressions that express currents through a resistor with a potential bias in terms of its scattering matrix. In the scattering theory (including RMT), the two nodes are at local equilibrium with a constant chemical potential bias (mostly just a voltage), whereas in the Green functions method the nodes may be in a non-equilibrium state given by a distribution function. The standard method to calculate distribution functions in a continuous bulk medium is the linearized Boltzmann equation. It can be derived microscopically from the Green function method discussed above by making a quasiclassical approximation, but not the isotropy assumption [88]. The linearized Boltzmann equation has been approximated to describe magnetic multilayers in different ways. Valet and Fert used the Boltzmann equation to include spin-flip scattering for transport in perpendicular multilayers [57], thus extending the spin-conserving two-channel resistor model. Concrete results are obtained for the diffusion equation that is equivalent to a first order expansion of the angular dependence of the distribution function in reciprocal space. Whereas collinear systems are well understood for some time now, in depth discussions of the $2 \times 2$ matrix character in Pauli spin space of the distribution function in non-collinear magnetic systems are relatively recent.

When a local non-equilibrium magnetization does not point in the direction of the spin-quantization axis, the distribution function for states at the Fermi energy $\alpha = \vec{k}\nu$ with band index $\nu$ and Bloch vector $\vec{k}$ is a matrix in Pauli spin space

$$\hat{f}_\alpha(\vec{r}) = \begin{pmatrix} f_{\uparrow\uparrow}(\vec{r}) & f_{\uparrow\downarrow}(\vec{r}) \\ f_{\uparrow\downarrow}(\vec{r}) & f_{\downarrow\downarrow}(\vec{r}) \end{pmatrix}_\alpha = f_\alpha^C(\vec{r})\,\hat{1} + \vec{\boldsymbol{\sigma}}\cdot\vec{f}_\alpha^S(\vec{r})\,. \tag{72}$$

On the right hand side the distribution has been expanded here into a basis of the unit matrix and the vector of the Pauli spin matrices (21). $f_\alpha^C$ is charge accumulation and the spin accumulation, $\vec{f}_\alpha^S$ is a vector whose direction is always parallel to the magnetization vector $\vec{m}$ in the bulk of a ferromagnet, but arbitrary in a normal metal, depending on device configuration and applied biases. $\hat{f}$ can be diagonalized by unitary rotation matrices in spin space, characterized by the polar angles $\theta$ and $\varphi$. Let us assume for the moment that these angles are piecewise constant in position space, thus disregarding magnetic domain walls [101] and magnetic field-induced spin precession in normal metals [63]. In the local spin quantization frame the distribution function is diagonal with two spin components $s = \pm 1$. In order to simplify the collision term, we introducing spin-conserving and spin-flip scattering life times $\tau_s$ and $\tau_{s,-s}^{\text{sf}}$, for the sake of argument taken to be state-independent. It is convenient to separate the distribution function into an isotropic electrochemical potential $\mu_s$ and an anisotropic term $\gamma_{\alpha s}$. For electron transport at low temperatures, the chemical potential reduces to the "spin-dependent voltage" $\mu_s = eV_s$ introduced in Section II. The anisotropic "drift" term $\gamma_{\alpha s}$ describes the deformation of the the Fermi-surface by the applied field and vanishes when averaged over the Fermi surface. The (linearized) Boltzmann equation for the stationary state reads [57]

$$\vec{v}_{\alpha s}\cdot\vec{\nabla}\left(\gamma_{\alpha s}+\mu_s\right)+\left(\frac{1}{\tau_s}+\frac{1}{\tau_{s,-s}^{\text{sf}}}\right)\gamma_{\alpha s}=\frac{\mu_{-s}-\mu_s}{\tau_{s,-s}^{\text{sf}}}. \tag{73}$$

where $\vec{v}_{\alpha s}$ is the group velocity of state $\alpha s$. Charge and spin currents read

$$j_c = \frac{e}{hA} \sum_{\alpha s} \vec{v}_{\alpha s} \gamma_{\alpha s} \ , \tag{74}$$

$$\vec{j}_s = \frac{1}{4\pi A} \sum_{\alpha s} \vec{v}_{\alpha s} s \gamma_{\alpha s} \ . \tag{75}$$

The Boltzmann equation is valid when the spatial properties do not change much on the scale of the Fermi wave length. This is not the case at atomically abrupt heterointerfaces between different material, which should be incorporated by the transmission and reflection probabilities at the interface scattering potential, that are computed by quantum mechanics. The boundary conditions between different materials or regions in which the spin quantization axes are rotated with respect to each other, are given by Eqs. (104,105) and discussed thereafter.

The Boltzmann equation is still unnecessarily complicated for most realistic systems [54, 55, 57]. Indeed, Xiao et al. [102] found good agreement between the solution of the Boltzmann equation as sketched above and a simple resistor model for a non-collinear magnetic spin valve that is based on a diffusion equation. Penn and Stiles [103] report that the solution of the diffusion equation remains close to that of the Boltzmann equation even for strong and anisotropic spin-flip scattering. The Boltzmann equation may be simplified when the disorder scattering smears out the angular dependence of the distribution function $\gamma_{\alpha s}$ in reciprocal space such that only the lowest harmonics survives. Such an "isotropy" assumption is inherent as well of Random Matrix Theory and circuit theory mentioned in the subsections above. This condition cannot be quantified in a simple manner, since bulk impurity scattering, chaotic scattering in cavities, as well as diffuse scattering at interfaces and boundaries contribute to the isotropization. In high density metal structures with very short Fermi wave lengths compared to typical sample dimensions such an approximation appears to be quite safe and justified experimentally by the successes of the series resistor model. In that limit, the Boltzmann equation reduces to the diffusion equation [54, 55, 57]

$$\nabla^2 \left[ \mu_s(\vec{r}) - \mu_{-s}(\vec{r}) \right] = \frac{\mu_s(\vec{r}) - \mu_{-s}(\vec{r})}{\ell_{sd}^2} \ . \tag{76}$$

$\ell_{sd} = \sqrt{D\tau^{\mathrm{sf}}}$ is the spin-flip diffusion length, which does not depend on spin index [104]. The spin-averaged diffusion coefficient $D$ can be written in terms of the density of states at the Fermi energy $\nu_s(E_F)$

$$\frac{1}{\left(\nu_\uparrow(E_F) + \nu_\downarrow(E_F)\right) D} = \frac{1}{\nu_\uparrow(E_F) D_\uparrow} + \frac{1}{\nu_\downarrow(E_F) D_\downarrow} \ . \tag{77}$$

in terms of the spin-dependent diffusion coefficients. In a simple two-band model $D_s = v_s^2 \tau_s / d$, where $v_s$ are the spin-dependent Fermi velocities and $d = 1, 2, 3$ is the spatial dimension. The average spin-flip relaxation time is defined as

$$\frac{1}{\tau^{\rm sf}} = \frac{1}{\tau^{\rm sf}_{\uparrow}} + \frac{1}{\tau^{\rm sf}_{\downarrow}} \ . \tag{78}$$

The currents

$$j_s(\vec{r}) = -\frac{\sigma_s}{e}\vec{\nabla}\mu_s(\vec{r}) \tag{79}$$

are governed by the spin-dependent conductivities

$$\sigma_s = \nu_s\left(E_F\right)e^2 D_s \ . \tag{80}$$

The spin-flip diffusion length in the ferromagnet is often much smaller than in normal metals and plays an important effects in the magnetotransport properties [105].

Let us now consider how the diffusion equation is modified for heterostructures and in the presence of magnetic fields [63]. For a spin valve or multilayer system invariant to translations in the lateral direction all quantities depend only on one spatial coordinate ($x$), but in the presence of a magnetic field, the diffusion equation cannot be simply written in a diagonal form. The magnetic Zeeman energy associated with the coupling between the magnetic field and the spin of the electrons is given by $g_L\mu_B\overrightarrow{\boldsymbol{\sigma}}\cdot\vec{B}/2$, where $\mu_B$ is the Bohr magneton, $g_L$ is the gyromagnetic ratio and $\vec{B}$ is the external magnetic field. Semiclassically, the spin dynamics (see *e.g.* Ref. [88]) is governed by the Bloch equations [6]

$$\frac{\partial\,\hat{f}^N(x)}{\partial t} = \frac{i}{\hbar}\left[\frac{g_L\mu_B}{2}\left(\overrightarrow{\boldsymbol{\sigma}}\cdot\vec{B}\right),\ \hat{f}^N(x)\right]_- \ . \tag{81}$$

where $[\cdots,\ \cdots]_-$ denotes a commutator. In the steady state:

$$D\frac{\partial^2\hat{f}^N(x)}{\partial x^2} = \frac{1}{\tau_{sf}}\left(\hat{f}^N(x) - \hat{\mathbf{1}}\frac{\mathrm{Tr}\left(\hat{f}^N(x)\right)}{2}\right) - \frac{i}{\hbar}\left[\frac{g\mu_B}{2}\left(\overrightarrow{\boldsymbol{\sigma}}\cdot\vec{B}\right),\ \hat{f}^N(x)\right]_- \ . \tag{82}$$

Expansion into Pauli spin matrices (21):

$$\hat{f}^N(x) = f_0^N(x)\hat{\mathbf{1}} + \vec{f}_S^N(x)\cdot\overrightarrow{\boldsymbol{\sigma}}, \tag{83}$$

where $f_0(x)$ is the local charge chemical potential and $\vec{f}_S(x)$ is a three component vector which describes the magnitude and direction of the spin-accumulation, the diffusion equation can be separated into a charge and spin-dependent part:

$$\frac{\partial^2 f_0^N(x)}{\partial x^2} = 0 \tag{84a}$$

$$\frac{\partial^2\vec{f}_S^N(x)}{\partial x^2} = \frac{1}{\left(\ell_{\rm sd}^N\right)^2}\vec{f}_S^N(x) + \frac{g\mu_B}{\hbar}\frac{\vec{B}}{D}\times\ \vec{f}_S^N(x). \tag{84b}$$

In ferromagnets, the spin accumulation is parallel to the magnetization direction $\vec{m}$ and the effect of magnetic fields can be neglected as long as they are smaller than the exchange and anisotropy fields and/or do not disturb a possible domain structure. In normal metals, the direction of the spin accumulation is arbitrary and depends in principle on the magnetization direction of all magnetic contacts.

The boundary conditions at atomically sharp interfaces for the distribution functions that obey the diffusion equation in the bulk are basically the same as that between different nodes in circuit theory that can be expressed in terms of the microscopic scattering matrices [63, 106].[4] Given the distribution functions in the bulk of the ferromagnet and normal metal, the current matrix through the contact can be calculated according to Eq. (61). The spin matrix current $\hat{I}_{F|N}$ through a contact and the non-equilibrium distribution matrix $\hat{f}^N(x)$ in the normal metal are related as

$$\hat{I}_{F|N}(x) = -S\nu_{_{DOS}} D \frac{\partial \hat{f}^N(x)}{\partial x}. \tag{85}$$

where $S$ is the surface perpendicular to the transport direction and $\nu_{_{DOS}}$ is the density of states of the normal metal. Details will be given in subsequent sections. Here we point out that the boundary conditions can be understood in a simple way when the interface resistances are small compared to the bulk resistance [108]. We then only require continuity of the particle and spin distribution functions on the normal and the ferromagnetic metal sides [12, 68]:

$$f_C^N|_{\text{N-surface}} = (f_\uparrow + f_\downarrow)/2|_{\text{F-surface}} \tag{86}$$

$$\vec{f}_S^N|_{\text{N-surface}} = \vec{m}(f_\uparrow - f_\downarrow)/2|_{\text{F-surface}} \,. \tag{87}$$

Furthermore, particle current is conserved [12, 68]:

$$[D\frac{\partial}{\partial x} f_C^N]|_{\text{N-surface}} = \frac{\partial}{\partial x}(D_\uparrow f_\uparrow + D_\downarrow f_\downarrow)|_{\text{F-surface}} \,. \tag{88}$$

We have discussed already why the non-collinear component of the spin-accumulation decays on a length scale $\lambda_c$ of the order of the lattice spacing. This leads to the third boundary condition at the F-N interface, namely that the spin-current is conserved only for the spin-component parallel to the magnetization direction [12, 68]:

$$[D\frac{\partial}{\partial x}\vec{f}_S]|_{\text{N-surface}} = \vec{m}\frac{\partial}{\partial x}(D_\uparrow f_\uparrow - D_\downarrow f_\downarrow)|_{\text{F-surface}} \,, \tag{89}$$

whereas the transverse component vanishes abruptly when going from the normal metal to the ferromagnet. Note that the ferromagnet can be here also an insulator that acts as a spin sink of transverse spin current.

[4] The relation between the boundary conditions discussed in this review and an alternative small–canting-angle expansion [107] is explaind in [106].

## IV. MAGNETOELECTRONIC CIRCUIT THEORY

State-of-the art magnetoelectronic devices are fabricated from high density metals with Fermi wave lengths on the order of the unit cell dimension. These structures are disordered on an atomic scale, especially at interfaces. Borrowing a term from the field of superconductivity these structures may be called "dirty". As a consequence, the effects of phase coherent interference of the electron waves which cause, for example, superlattice minibands in multilayer structures or quantum well states in thin layers, are small. An electron wave, which is injected as a coherent wave packet, is diffusely scattered in many direction. Even without additional inelastic scattering, the interference term of the electron waves averages out to zero (an exception is the coherent backscattering and corresponding weak localization, which can be safely disregarded in three-dimensional magnetoelectronic structures and devices of the present interest, however). The net current of electrons at the Fermi energy behaves classically and should then be described by semi- (or quasi-) classical methods, which incorporate a statistical treatment of the disorder. Note that the criterion for the validity of semiclassical approximations in heterogeneous structures is not the mean free path of the *bulk* materials. It is well established that interfaces and boundaries are the most important sources of scattering in these systems. Semiclassical methods are appropriate even in ballistic structures provided that the particle kinetics is chaotic [72]. The non-equilibrium spin accumulations are usually excited electrically, *i.e.* by applying a bias to metallic circuit with ferromagnetic elements. However, also optically excited spin-polarized plasmas [109] are accessible to similar semiclassical methods [110].

Of course, in structures of atomic dimensions like few-monolayers heterostructures, atomic break junctions, or nanoclusters semi-classical methods should be used with care. Quantum effects are relatively obvious in the non-local oscillatory exchange coupling, because without them the whole effect vanishes. In contrast, quantum effects on transport cause only small corrections on a semiclassical background electron current which remains present without quantum interference. Furthermore, in metallic structures with high electron densities many Bloch states contribute with different quantum oscillations that average out to a large extent. The two exceptions to the rule that quantum effects may be disregarded in metallic structures can be explained by a combination of wave vector selection and highly specular interfaces. By selecting essentially a single wave vector state by hot-electron injection [111] or tunneling barriers [112], the single standing wave pattern may dominate transport, even reversing the sign of the magnetoresistance. The most suitable theoretical approach for these rather exceptional cases is a numerical computation of the transport coefficients [113], that may be used in the present circuit theory if embedded as a resistive element in a larger environment (see Section VII). For completeness, we would like to add that spin injection into high-mobility semiconductor structures might become possible allowing to realize the coherent precession of the spin accumulation in a ballistic one-dimensional channel by spin-orbit interaction as envisaged by Datta [10]. The assumption of diffuse transport are then not justified.

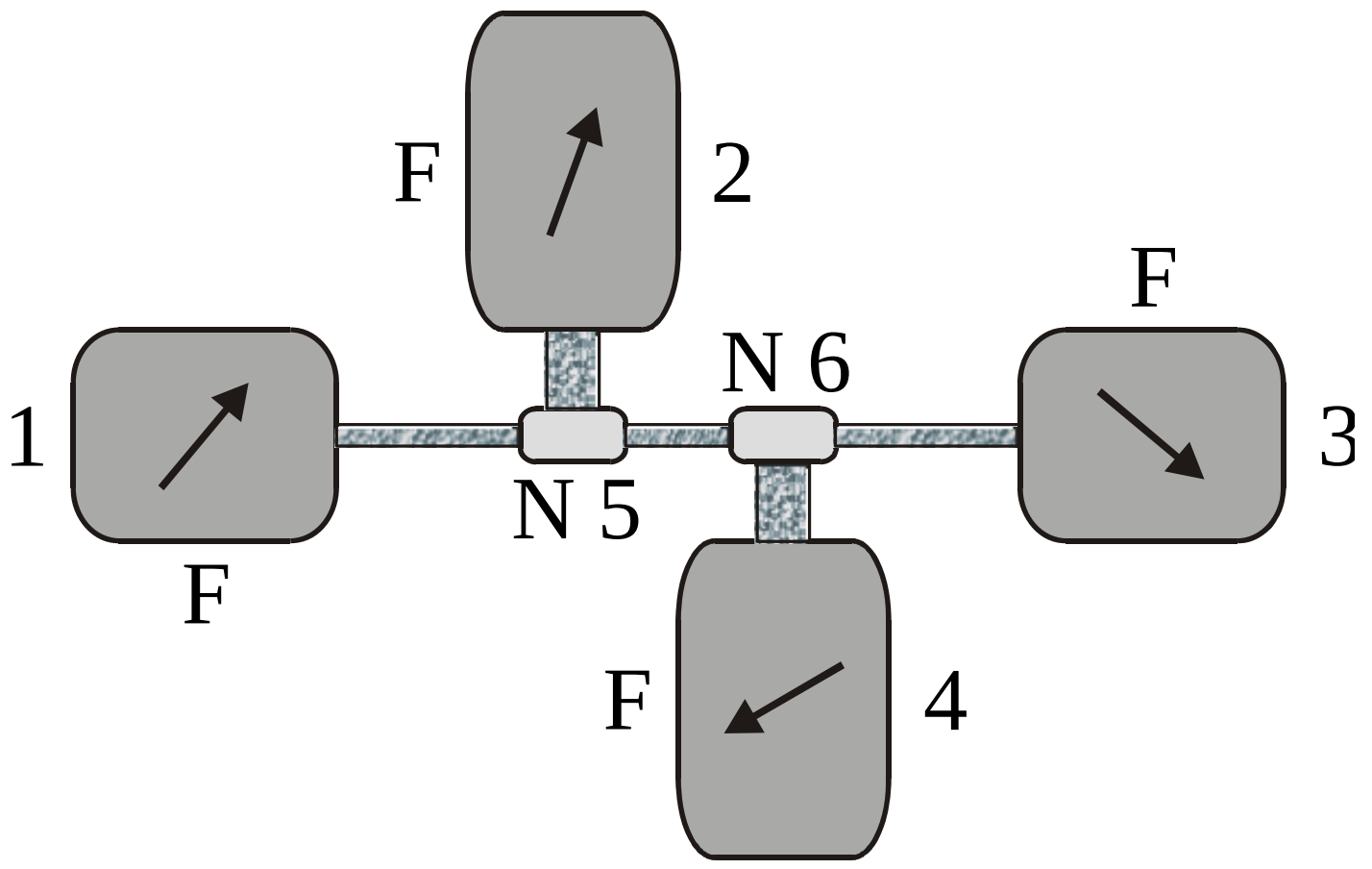

FIG. 11: A typical four terminal ferromagnet-normal metal circuit. The ferromagnetic reservoirs are connected to each other via junctions and normal metal nodes.

Here we discuss diffuse electron and spin transport in magnetic structures, referring to Section III for the basic formalisms. Magnetoelectronic circuit theory is a convenient general frame work to discuss hybrid circuits and devices. Originally, circuit theory was derived requiring complete isotropy of the distribution function in the low resistance elements, which is guaranteed when all the potential drop is over the resistive elements. A combination of the random matrix formulation of the scattering theory of transport and the approximate solution of the Boltzmann equation as suggested by Schep [61], does not suffer from this constraint, and is therefore applicable to systems with weak contact between nodes and resistors, like metallic multilayers [34]. It is also shown that circuit theory can be repaired by a renormalization of the conductance parameter. Finally, we discuss effects of spin flip scattering in the bulk.

### A. Original formulation

Transport in hybrid normal metal - ferromagnet systems can be described in the form of a circuit theory when parts of the system can be treated semiclassically. A typical (magneto)electronic circuit as schematically shown in Fig. 11 can be divided into contacts or *junctions* (resistive elements), *nodes* (low impedance interconnects) and *reservoirs* (voltage sources). The theory to be described in detail below is applicable when the junctions limit the electric current and the nodes are characterized by a local $2 \times 2$ distribution matrix in Pauli spin space that is constant in position and isotropic in momentum space. The latter condition justifies a diffusion approximation and requires that the nodes are either irregular in shape or contain a sufficient number of randomly distributed scatterers.

The current through each contact can be calculated as a function of the distribution matrices on the adjacent nodes. The spin-current conservation law then allows computation of the circuit properties as a function of the applied voltages. The recipe for calculating the current-voltage characteristics can be summarized as:

- Divide the circuit into nodes, junctions, and reservoirs.
- Specify the $2\times 2$ distribution matrix in spin-space for each node and reservoir.
- Compute the currents through contacts or junctions as a function of the distribution matrices in the adjacent nodes, which involves the spin-charge conductances specified below.
- Use the spin-current conservation law at each node, where the difference between total in and outgoing spin-currents equals the spin-relaxation rate.
- Solve the resulting system of linear equations to obtain all currents as a function of the chemical potentials of the reservoirs which are the parameters controlled by the experiments.

We denote the $2 \times 2$ distribution matrix at a given energy $\epsilon$ in the node by $\hat{f}(\epsilon)$, where hat $\hat{()}$ denotes a $2 \times 2$ matrix in spin-space. Per definition, the external reservoirs are in local equilibrium, which means that there is no spin accumulation, but a voltage bias or chemical potential difference between different reservoirs may exist. The distribution matrix is then diagonal in spin-space for normal metals as well as ferromagnets: $\hat{f} = \hat{1} f(\epsilon, \mu_\alpha)$, where $\hat{1}$ is the unit matrix, $f(\epsilon, \mu_\alpha)$ is the Fermi-Dirac distribution function and $\mu_\alpha$ is the local chemical potential in reservoir $\alpha$. The direction of the magnetization of the ferromagnetic nodes is denoted by the unit vector $\mathbf{m}_\alpha$. In ferromagnets the precession of non-equilibrium spins in the exchange field leads to an effective relaxation of the spin accumulation non-collinear to the local magnetization on the scale of the magnetic coherence length $\lambda_c$, Eq. (41). Consequently, the distribution function is limited to the form $\hat{f}^F = \hat{1} f_0^F + \hat{\boldsymbol{\sigma}} \cdot \mathbf{m} f_s^F$. Such a restriction does not hold in the normal metal node and $\hat{f}^N$ is an arbitrary Hermitian $2 \times 2$ matrix.

In the stationary state, the $2 \times 2$ non-equilibrium distribution matrices in the nodes are uniquely determined by current conservation

$$\sum_\alpha \hat{I}_{\alpha\beta} = \left( \frac{\partial \hat{f}_\beta}{\partial t} \right)_{\rm rel}, \tag{90}$$

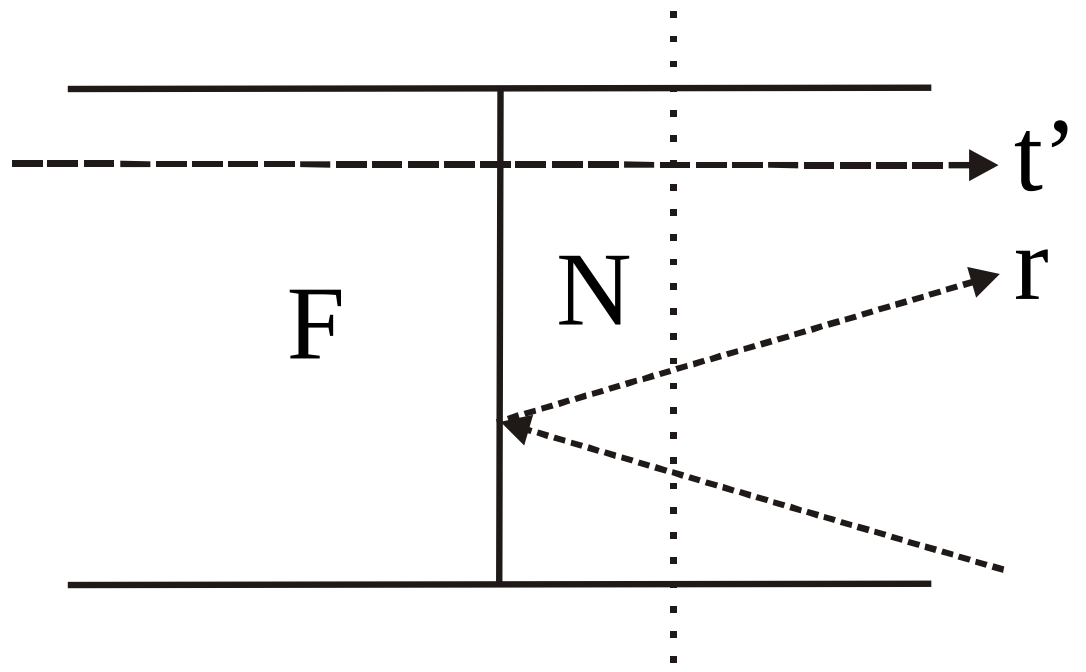

FIG. 12: A schematic picture of a ferromagnet-normal metal contact.

where $\hat{I}_{\alpha\beta}$ denotes the $2\times2$ current in spin-space which flows into the node (or reservoir) $\beta$ ( in the following taken to be normal) from all adjacent nodes (or reservoir) $\alpha$. The term on the right hand side stands for the spin-relaxation in the normal node, which vanishes when the spin-current in the node is conserved, *i.e.* when an electron spends much less time on the node than the spin-flip relaxation time $\tau_{\rm sf}$. When the spatial extension of the node in the transport direction is smaller than the spin-flip diffusion length $l_{\rm sf} = \sqrt{D\tau_{\rm sf}}$, where $D$ is the diffusion coefficient, the spin-relaxation in the node can be treated in terms of the diffusion equation for spatially independent distributions $(\partial\hat{f}^N/\partial t)_{\rm rel} = (\hat{1}{\rm Tr}(\hat{f}^N)/2 - \hat{f}^N)/\tau_{\rm sf}$. When the size of the node in the transport direction is larger than $l_{\rm sf}$ the simplest circuit theory fails and we have to solve the diffusion equation as sketched in Section III F [63].

A schematic picture of a junction between a normal metal and a ferromagnetic node is shown in Fig. 12. The current is evaluated on the normal metal side of the junction (dotted line). The current through the contact is

$$\hat{I}_{F|N} = \frac{e}{h}\sum_{nm}[\hat{t}'^{nm}\hat{f}^F\left(\hat{t}'^{mn}\right)^\dagger - (\delta_{nm}\hat{f}^N - \hat{r}^{nm}\hat{f}^N\left(\hat{r}^{mn}\right)^\dagger)]\,, \tag{91}$$

where $r^{nm}_{ss'}$ is the reflection coefficient for transverse mode $m$ with spin $s'$ at the normal metal side and reflected to transverse mode $n$ with spin $s$. $t'^{nm}_{ss'}$ is the transmission coefficient for electrons approaching from the ferromagnet in transverse mode $m$ with spin $s'$ and leaving in transverse mode $n$ with spin $s$. (Note that the Hermitian conjugate in (91) operates in both spin-space and the space spanned by the transverse modes, *e.g.* $(\hat{r}^{mn})^\dagger_{ss'} = (\hat{r}^{nm}_{s's})^*$). The relation (91) has been derived quite rigorously by the Keldysh formalism for non-equilibrium transport in Section III D but also appeals to intuitive argument in the spirit of the Landauer-Büttiker formalism (Section III C).

The relation (91) between the current and the distributions can be simplified by rotating the spin quantization axis to the local magnetization direction. Dis-regarding spin-flip processes in the contacts, the reflection matrix for an incoming electron from the normal metal transforms as

$$\hat{r}^{nm} = \sum_s \hat{u}^s r_s^{nm}\,, \tag{92}$$

where $r_\uparrow^{nm}$ ($s = \uparrow$) and $r_\downarrow^{nm}$ ($s = \downarrow$) are the spin-dependent reflection coefficients in the basis in which the spin-quantization axis is parallel to the magnetization in the ferromagnet, the spin-projection matrices are

$$\hat{u}^s = (\hat{1} + s\hat{\boldsymbol{\sigma}} \cdot \mathbf{m})/2 \tag{93}$$

and $\hat{\boldsymbol{\sigma}}$ is a vector of the Pauli spin matrices. Similarly, for the transmission matrix

$$\hat{t}'^{nm}(\hat{t}'^{mn})^\dagger = \sum_s \hat{u}^s |t_s'^{nm}|^2\,, \tag{94}$$

where $t_\uparrow^{nm}$ and $t_\downarrow^{nm}$ are the spin-dependent transmission coefficients in the basis in which the spin-quantization axis is parallel to the magnetization in the ferromagnet. Using the unitarity of the scattering matrix, we find that the general form of the relation (91) reads

$$\begin{aligned} e\hat{I}_{F|N} &= G^{\uparrow\uparrow}\hat{u}^\uparrow \left(\hat{f}^F - \hat{f}^N\right)\hat{u}^\uparrow + G^{\downarrow\downarrow}\hat{u}^\downarrow \left(\hat{f}^F - \hat{f}^N\right)\hat{u}^\downarrow \\ &\quad - G^{\uparrow\downarrow}\hat{u}^\uparrow \hat{f}^N \hat{u}^\downarrow - (G^{\uparrow\downarrow})^* \hat{u}^\downarrow \hat{f}^N \hat{u}^\uparrow \qquad (95) \\ &= \sum_{\alpha\beta} G^{\alpha\beta}\hat{u}^\alpha \left(\hat{f}^F - \hat{f}^N\right)\hat{u}^\beta, \qquad (96) \end{aligned}$$

where we have identified the spin-dependent conductances $G^{\uparrow\uparrow}$ and $G^{\downarrow\downarrow}$

$$G^{\alpha\alpha} = \frac{e^2}{h}\left[M - \sum_{nm} |r_\alpha^{nm}|^2\right] = \frac{e^2}{h}\sum_{nm} |t_\alpha^{nm}|^2\,, \tag{97}$$

and we call

$$G^{\uparrow\downarrow} = \frac{e^2}{h}\left[M - \sum_{nm} r_\uparrow^{nm}(r_\downarrow^{nm})^*\right]\,. \tag{98}$$

the "spin-mixing conductance". Physically this parameter describes the transfer of angular momentum from a spin accumulation in the normal metal to the ferromagnet (magnetization torque). The description of transmission processes through a magnetic insertion that is thinner than the magnetic coherence lengths requires a different mixing conductance of transmission [114] and is discussed in more detail in Section V.

The relation between the matrix current through a junction and the distributions in the ferromagnetic node and a normal metal node are determined by 4 parameters, the two real spin-dependent conductances ($G^{\uparrow}$, $G^{\downarrow}$) and the real and imaginary parts of the mixing conductance $G^{\uparrow\downarrow}$. These junction-specific properties can be obtained by microscopic theory or from experiments. The spin-conductances $G^{\uparrow}$ and $G^{\downarrow}$ have been used in descriptions of spin-transport for a long time [1]. The *spin-mixing conductance* is a new concept which is relevant for transport between non-collinear ferromagnets. Note that although the mixing conductance is a complex number the $2 \times 2$ current in spin-space is Hermitian and consequently the current and the spin-current in any direction given by Eq. (95) are real numbers. From the definitions of the spin-dependent conductances, Eq. (97), and the mixing conductance Eq. (98) we find $2\,\mathrm{Re}\,G^{\uparrow\downarrow} = G^{\uparrow} + G^{\downarrow} + \frac{e^2}{h}\sum_{nm}|r_{\uparrow}^{nm} - r_{\downarrow}^{nm}|^2$ and consequently the conductances should satisfy $2\,\mathrm{Im}\,G^{\uparrow\downarrow} \geq G^{\uparrow} + G^{\downarrow}$. A physical interpretation of this result is given in Ref. [68].

It is useful to rewrite the matrix current in terms of charge $I_C$ and spin current $\vec{I}_S$

$$\hat{I} = \begin{pmatrix} I_{\uparrow\uparrow} & I_{\uparrow\downarrow} \\ I_{\uparrow\downarrow} & I_{\downarrow\downarrow} \end{pmatrix} = \left(I_C\hat{1} + \vec{\boldsymbol{\sigma}} \cdot \vec{I}_S\right)/2 \tag{99}$$

in terms of the spin and charge accumulations

$$\hat{f}^N = \begin{pmatrix} f_{\uparrow\uparrow}^N & f_{\uparrow\downarrow}^N \\ f_{\uparrow\downarrow}^N & f_{\downarrow\downarrow}^N \end{pmatrix} = f_C^N\hat{1} + \vec{\boldsymbol{\sigma}}\cdot\vec{s}f_S^N \tag{100}$$

$$\hat{f}^F = f_C^F\hat{1} + \vec{\boldsymbol{\sigma}}\cdot\vec{m}f_S^N \tag{101}$$

as

$$I_C = \left(G^{\uparrow} + G^{\downarrow}\right)\left(f_C^F - f_C^N\right) + \left(G^{\uparrow} - G^{\downarrow}\right)\left(f_S^F - \vec{m}\cdot\vec{s}f_S^N\right) \tag{102}$$

$$\begin{aligned} \vec{I}_S &= \left[\left(G^{\uparrow} - G^{\downarrow}\right)\left(f_C^F - f_C^N\right) + \left(G^{\uparrow} + G^{\downarrow}\right)f_S^F + \left(2\,\mathrm{Re}\,G^{\uparrow\downarrow} - G^{\uparrow} - G^{\downarrow}\right)\vec{m}\cdot\vec{s}f_S^N\right]\vec{m} \\ &\quad -2\,\mathrm{Re}\,G^{\uparrow\downarrow}f_S^N\,\vec{s} + 2\,\mathrm{Im}\,G^{\uparrow\downarrow}f_S^N\,\vec{m}\times\vec{s}. \end{aligned} \tag{103}$$

The charge current is carried by the spin projections parallel and anti-parallel to the magnetization direction, respectively. The spin current consists of a longitudinal component parallel to the magnetization $\vec{m}$, a term parallel to $\vec{s}$ that has a transverse component in the $\vec{s}, \vec{m}$ plane that corresponds to the spin-transfer torque, and a component normal to both $\vec{s}$ and $\vec{m}$ that can be interpreted as an interface exchange field. The physics of the transverse spin current is discussed in more detail in Section IV E.

### B. Random matrix theory and Boltzmann corrections to circuit theory

It remains to discuss the relation between the random matrix theory (to leading order in the inverse number of modes) and circuit theory, as well as make connec-tions with (drift-)diffusion theory. In this subsection we find that all methods are approximate versions of the Boltzmann equation in non-collinear magnetic circuits [34, 115].

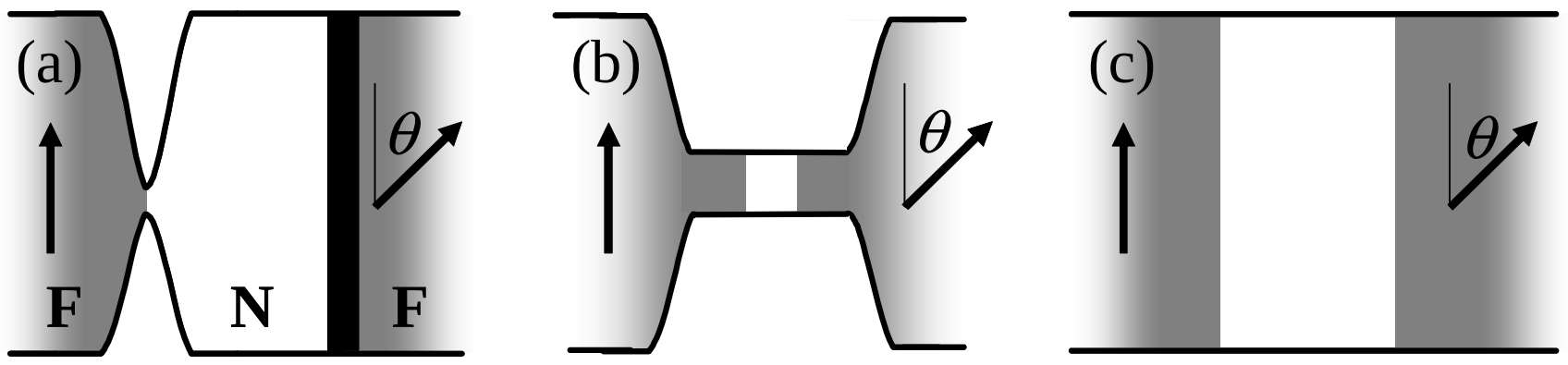

FIG. 13: Different realizations of perpendicular spin valves in which $\theta$ is the angle between magnetization directions. (a) Highly resistive junctions like point contacts and tunneling barriers limit the conductance. (b) Spin valve in a geometrical constriction amenable to the scattering theory of transport. (c) Magnetic multilayers with transparent interfaces.

#### *1. Boundary conditions*

The semiclassical approach is valid when the potential landscape varies slowly on the scale of the Fermi wave length. In heterostructures we often encounter regions in which materials change on an atomic scale, such as at intermetallic interfaces or tunnel junctions, which have to be treated quantum mechanically [63]. Let us define the "nodes" as the bulk regions, in which the semiclassical distributions are well defined. The intermediate scattering regions, or "contacts", can then be treated formally exactly by boundary conditions that link the distributions of two neighboring nodes. Consider the spin valve structures in Fig. 13, which may be part of a larger circuit. Original circuit theory described in Section IV A assumes that the nodes are in local equilibrium, an assumption that is allowed when the in and outgoing currents are not sufficiently large to disturb the distribution functions in the node such as in Fig. 13(a). In highly conductive Ohmic systems like metallic multilayers such as in Fig. 13(c), the distribution function is significantly distorted when a bias is applied. In that limit, the Landauer-Büttiker type of equations do not hold anymore. For collinear systems Schep et al. [61] showed that scattering theory is still valid when the conductances are corrected by subtracting spurious Sharvin conductances (see also [62]). Here we concentrate on non-collinear systems [34].

We denote the distribution functions in the ferromagnetic terminals by subscripts $L$ end $R$. We explicitly allow for a limited state dependence of the distributions by the superscript $\epsilon = \pm 1$ indicating whether drift is in ($\epsilon = 1$) or against ($\epsilon = -1$) the transport direction (from left to right). Taking into account the difference between the left- and right-moving distribution functions is the key generalization of the previous theories in our treatment here. In Sec. 5 it is discussed how the circuit theory can be recovered by renormalizing the conductance parameters. Here we concentrate on random matrix theory based on scattering theory.

In order to work with simple matrices instead of diadics, we follow Waintal et al. [56] and introduce the $4 \times 1$ vector representation $\left[\vec{f}_n^{\epsilon}(\vec{r})\right]^T = \left(f_{\uparrow\uparrow}(\vec{r}), f_{\uparrow\downarrow}(\vec{r}), f_{\uparrow\downarrow}(\vec{r}), f_{\downarrow\downarrow}(\vec{r})\right)_n^{\epsilon}$. The boundary conditions for the non-equilibrium distributions to the left and right of the scattering region then read:

$$\vec{f}_{R,n}^{+} = \sum_{m \epsilon L} \left(\check{T}_{L\to R}\right)_{nm} \vec{f}_{L,m}^{+} + \sum_{m \epsilon R} \left(\check{R}_{R\to R}\right)_{nm} \vec{f}_{R,m}^{-} , \tag{104}$$

$$\vec{f}_{L,n}^{-} = \sum_{m \epsilon L} \left(\check{R}_{L\to L}\right)_{nm} \vec{f}_{L,m}^{+} + \sum_{m \epsilon R} \left(\check{T}_{R\to L}\right)_{nm} \vec{f}_{R,m}^{-} . \tag{105}$$

$\check{T}$, $\check{R}$ are $4 \times 4$ transmission and reflection probability matrices and the subscripts indicate the direction of the currents ($L \to R$ denotes transmission from left to right, $R \to R$ reflection from the right, etc.). All matrix elements follow from the scattering matrix and are conveniently normalized, for example

$$\left[\left(\check{T}_{L\to R}\right)_{nm}\right]_{SS'} = \frac{1}{N_S^L} \left(\vec{t}_{nm}^{L\to R}\right)_S \left(\vec{t}_{nm}^{L\to R}\right)_{S'}^{\dagger} . \tag{106}$$

with $S \in [1,4]$. The transmission amplitudes, such as $t_{ns,ms'}^{L\to R}$ of a wave coming in from the left as mode $m$ and spin $s'$ and going out in mode $n$ and spin $s$, are here collected in vectors $\vec{t}_{nm}^{L\to R} = \left(t_{\uparrow\uparrow}^{L\to R}, t_{\uparrow\downarrow}^{L\to R}, t_{\uparrow\downarrow}^{L\to R}, t_{\downarrow\downarrow}^{L\to R}\right)_{nm}^T$. When node $L$ is a ferromagnet, which must be thicker than the magnetic coherence length $\lambda_c$ (Eq. (41)), $N_S^L = N_\uparrow^F (\delta_{S,1} + \delta_{S,2}) + N_\downarrow^F (\delta_{S,3} + \delta_{S,4})$, where $N_s^F$ is the number of transport modes for spin $s$ in the ferromagnet.

In a nutshell, this is a very general formulation of charge and spin transport, but it is not yet amenable for analytic treatment or analysis of experiments. The isotropy assumption that reduced the Boltzmann to the diffusion equation in the previous Section, enormously simplifies the results, as demonstrated in the following.

We focus here on the electrical charge current as a function of the magnetization configuration in symmetric spin valves, as in Fig. 13(b,c), in order to keep the analytical manipulations manageable. We will see later that we can derive rules from these results that are valid for general structures. $\hat{T}$ and $\hat{R}$ are functions of the magnetic configuration, which, disregarding magnetic anisotropies, can be parameterized by a single polar angle $\theta$. Asymmetric spin valves are discussed in Section VI B. In first instance, we disregard spin-flip scattering and discuss later how it can be included. Integrating over the lateral coordinates leaves a position dependence only in the transport direction ($x$). The next step is the assumption that the distribution functions for incident electrons from the left and right are isotropic in space. The distribution functions for the outgoing electrons do not have to be isotropic, as long as they are subsequently scrambled in the nodes. The isotropy assumption may be invoked when the nodes are diffuse or chaotic, such that electrons are distributed equally over all states at the (spin-dependent) Fermi surfaces (which is equivalent to replacing state dependent scattering matrix elements by its average [61]). The Fermi surface integration is then carried out easily, and the distribution functions within left and right ferromagnet nodes (at locations $x_L$ and $x_R$, respectively) are matched *via* simplified boundary conditions

$$\vec{f}^{+}(x_R) = \check{T}_{L\to R}(\theta)\,\vec{f}^{+}(x_L) + \check{R}_{R\to R}(\theta)\,\vec{f}^{-}(x_R)\,, \tag{107a}$$

$$\vec{f}^{-}(x_L) = \check{R}_{L\to L}(\theta)\,\vec{f}^{+}(x_L) + \check{T}_{R\to L}(\theta)\,\vec{f}^{-}(x_R)\,, \tag{107b}$$

where the $4\times4$ transmission and reflection probability matrices have elements like [56]:

$$\left[\check{T}_{L\to R}\right]_{SS'} = \frac{1}{N_S^L}\sum_{mn}\left(\vec{t}_{nm}^{\,R\to L}\right)_S\left(\vec{t}_{nm}^{\,R\to L}\right)_{S'}^{\dagger}. \tag{108}$$

In the coordinate systems defined by the magnetization directions, the transverse components of the spin accumulation in the ferromagnets vanish identically [12, 69]. The equations hold for any given energy. Now and in the following we assume that the scattering matrices vary only slowly on the scale of the applied voltages $eV$, a safe assumption in high density metals (except for very hot electron injection via tunneling barriers). In that case we can integrate over energies to replace the full distribution functions by the local spin current densities $\gamma_s$ and chemical potentials $\mu_s$ in the local spin quantization frame:

$$\int \vec{f}^{\pm}\left(\varepsilon; x_{L/R}\right) d\varepsilon = \left(\left(\pm\gamma_\uparrow + \mu_\uparrow\right)\left(x_{L/R}\right), 0, 0, \left(\pm\gamma_\downarrow + \mu_\downarrow\right)\left(x_{L/R}\right)\right)\ . \tag{109}$$

By this choice the explicit angle-dependence of transport is contained only in the matrices.

We can now link an arbitrary distribution on the left to compute the distributions on the right, subject to the constraint of charge current conservation. Here we focus on the simple case in which we apply a bias

$$\Delta\mu = \sum_s \left(\mu_s(x_L) - \mu_s(x_R)\right)\ , \tag{110}$$

but no spin accumulation gradient $\mu_s(x_L) - \mu_{-s}(x_L) = \mu_s(x_R) - \mu_{-s}(x_R)$ over the system. We then find that $\gamma_s(x_L) = \gamma_s(x_R)$, *i.e.* the spin current component parallel to the magnetization on left and right ferromagnets are the same. The charge current

$$I_c = \frac{e}{h} \sum_s N_s^F \gamma_s \tag{111}$$

divided by the chemical potential drop is the electrical conductance $G = eI_c/\Delta\mu$. Eqs. (107,109) then lead to

$$G = \frac{2e^2}{h} \sum_{\substack{S=1,4 \\ S'=1,4}} \left\{ N_S^F \left[ \check{1} - \check{T}_{L\to R} + \check{R}_{R\to R} \right]^{-1} \check{T}_{L\to R} \right\}_{SS'} . \tag{112}$$

When the transparency is small, all transmission probabilities are close to zero, reflection probabilities are close to unity, and the Landauer-Büttiker conductance, starting point of [56], is recovered:

$$G \to \frac{e^2}{h} \sum_{\substack{S=1,4 \\ S'=1,4}} \left\{ N_S^F \hat{T}(\theta) \right\}_{SS'} . \tag{113}$$

Indeed, in this limit the distributions to the left and right are not perturbed by the current, the nodes are genuine reservoirs, and standard scattering theory applies. Also, when $\theta = 0, \pi$, Eq. (112) is equivalent to results for the two-channel model [61].

The scattering region has still not been specified and may be interacting and/or quantum coherent. We now discuss how analytical results can be obtained in the non-interacting, diffuse limit.

### 2. *Semiclassical concatenation*

The scattering matrix of a composite system can be formulated as concatenations of the scattering matrix from separate elements, *e.g.*. the scattering matrices of bulk layers and interfaces [116]. By assuming isotropy, *i.e.* sufficient disorder or chaotic scattering, Waintal et al. [56] proved by averaging over random scattering matrices that size quantization effects like the equilibrium exchange coupling or other phase coherent phenomena are destroyed by disorder and vanish like the inverse of the number of modes. Under these conditions we are free to define nodes in the interior of the device and link them via the boundary conditions (107). This is equivalent to composing the total transport probability matrices in Eq. (112) in terms of those of individual elements by semiclassical concatenation rules [117]. The $4 \times 4$ transmission probability matrix through a F(0)|N|F($\theta$) double heterojunction as in Fig. 13 in which bulk scattering is absent, takes the form

$$\check{T}(\theta) \equiv \check{T}_{\mathrm{N}\to\mathrm{F}}(\theta) \left[ \check{1} - \check{R}_{\mathrm{N}\to\mathrm{N}}(0) \check{R}_{\mathrm{N}\to\mathrm{N}}(\theta) \right]^{-1} \check{T}_{\mathrm{F}\to\mathrm{N}}(0) , \tag{114}$$

where the interface transmission and reflection matrices as a function of magnetization angle appear. The transformations needed to obtain $\check{T}_{\mathrm{N}\to\mathrm{F}}(\theta)$ and $\check{R}_{\mathrm{N}\to\mathrm{N}}(\theta)$ require some attention. In terms of the spin-rotation

$$\hat{U} = \begin{pmatrix} \cos\theta/2 & -\sin\theta/2 \\ \sin\theta/2 & \cos\theta/2 \end{pmatrix} \tag{115}$$

and projection matrices ($s = \pm 1$)

$$\hat{u}_s(\theta) = \frac{1}{2}\begin{pmatrix} 1 + s\cos\theta & s\sin\theta \\ s\sin\theta & 1 - s\cos\theta \end{pmatrix} , \tag{116}$$

the interface scattering coefficients (omitting the mode indices for simplicity) are transformed as follows [12]

$$\hat{r}_{N\to N} = \hat{U}\hat{r}_{cN}\hat{U}^{\dagger} = \sum_s \hat{u}_s r_s^{cN} , \tag{117}$$

$$t_{ss'}^{F\to N} = U_{ss'} t_{s'}^{cF} , \tag{118}$$

$$t_{ss'}^{N\to F} = t_s^{cN} U_{ss'}^{\dagger} , \tag{119}$$

$$r_{ss'}^{F\to F} = r_s^{cF}\delta_{ss'} . \tag{120}$$

The superscript $c$ indicates that the matrices should be evaluated in the reference frame of the local magnetization, and are thus diagonal in the absence of spin-flip relaxation scattering at the interfaces. Different transformation properties for the different elements of the scattering matrix derive from our choice to use local spin-coordinate systems that may differ for each magnet. Let us, for example, inspect a transmission matrix element from the normal metal into the ferromagnet with magnetization rotated by $\theta$

$$\left[\check{T}_{N\to F}(\theta)\right]_{11} = \frac{1}{N_\uparrow^F}\sum_{mn} t_{n\uparrow m\uparrow}^{R\to L}\left(t_{n\uparrow m\uparrow}^{R\to L}\right)^{\dagger} = \frac{1}{2}(1+\cos\theta)\frac{1}{N_\uparrow^F}\sum_{mn}\left|t_{n\uparrow m\uparrow}^{cN}\right|^2 \tag{121}$$

and analogously for the other matrix elements as well as other matrices.

Transport through a more complex system can be treated by repeated concatenation of two scattering elements in terms of reflection and transmission matrices analogous to Eq. (114). In the presence of significant bulk scattering, we can represent a disordered metal $B$ with thickness $d_B$ by diagonal matrices like [56, 61]

$$\left(\check{T}_B\right)_{SS'} = \left(1 + \frac{e^2}{h}\frac{\rho_s^B d_B}{A_B}\right)^{-1}\delta_{SS'} , \tag{122}$$

where $\rho_s^B$, $A_B$ are the single-spin bulk resistivities and cross section of the bulk metal (normal or magnetic).

The *interface* parameters of the present theory are the spin-dependent Landauer-Büttiker conductances

$$g_s = \sum_{lm} \left| t^{cN}_{lm,s} \right|^2 = N_N - \sum_{lm} \left| r^{cN}_{lm,s} \right|^2 = \sum_{lm} \left| t^{cF}_{lm,s} \right|^2 = N^F_S - \sum_{lm} \left| r^{cF}_{lm,s} \right|^2 \tag{123}$$

and the real and imaginary part of the spin-mixing conductance

$$g_{s-s} = N_N - \sum_{lm} r^{cN}_{lm,s} \left( r^{cN}_{lm,-s} \right)^* ,$$

which can also be represented in terms of the total conductance $g = g_\uparrow + g_\downarrow$, polarization $p = (g_\uparrow - g_\downarrow)/g$, and relative mixing conductance $\eta = 2g_{\downarrow\uparrow}/g$.

The actual concatenation of the $4 \times 4$ matrices defined here is rather complicated even when using symbolic programming routines. This explains why in [56] analytic results were obtained only in special limiting cases. We found that final results are simple even in the most general cases, not only for Eq. (113) considered by [56], but also for Eq. (112). For the spin valves in Fig. 13, we find for the conductance as a function of angle

$$G(\theta) = \frac{\tilde{g}}{2} \left( 1 - \frac{\tilde{p}^2}{1 + \frac{|\tilde{\eta}|^2}{\operatorname{Re}\tilde{\eta}} \frac{1+\cos\theta}{1-\cos\theta}} \right) = \frac{\tilde{g}}{2} \left( 1 - \tilde{p}^2 \frac{\tan^2\theta/2}{\tan^2\theta/2 + \frac{|\tilde{\eta}|^2}{\operatorname{Re}\tilde{\eta}}} \right) , \tag{124}$$

where $\tilde{\eta} = 2\tilde{g}_{\uparrow\downarrow}/\tilde{g}$ and

$$\frac{1}{\tilde{g}_s} = \frac{1}{g_s} + \frac{e^2}{h} \frac{\rho_{F,s} d_F}{2A_F} + \frac{e^2}{h} \frac{\rho_N d_N}{2A_F} - \frac{1}{2} \left( \frac{1}{N^F_s} + \frac{1}{N_N} \right) \tag{125}$$

$$\frac{1}{\tilde{g}_{\uparrow\downarrow}} = \frac{1}{g_{\uparrow\downarrow}} + \frac{e^2}{h} \frac{\rho_N d_N}{2A_N} - \frac{1}{2N_N}. \tag{126}$$

Equation (124) is identical to the angular magnetoresistance derived by circuit theory [12] after replacement of $\tilde{g}_s$ and $\tilde{g}_{\uparrow\downarrow}$ by $g_s$ and $g_{\uparrow\downarrow}$. Physically, in Eqs. (125,126) spurious Sharvin resistances are subtracted from the interface resistances obtained by scattering theory, whereas bulk resistances are added. These corrections are large for transparent interfaces and essential to obtain agreement between experimental results of transport experiments in CPP (current perpendicular to plane) multilayers [37, 58, 118] and first-principles calculations, for conventional [35, 61, 119] as well as mixing conductances [36]. The mixing conductance parameterizes the magnetization torque due to a spin accumulation in the normal metal, governed by the reflection of electrons from the normal metal. It is therefore natural that the mixing conductance is reduced by the bulk resistance of the normal metal and we can also understand that only the normal metal Sharvin resistance has to be subtracted. The real part of the mixing conductances is often close to the number of modes in the normal metal $g_{\uparrow\downarrow} \approx N_N$, in which case $\tilde{g}_{\uparrow\downarrow} \approx N_N/2$ [114]. By letting $N_s^F \to \infty$ we are in the regime of [56]. The circuit theory is recovered when, additionally, $N_N \to \infty$. The bare mixing conductance is bounded not only from below $\mathrm{Re}g_{\uparrow\downarrow} \geqslant g/2$ [12], but also from above $|g_{\uparrow\downarrow}|^2 / \mathrm{Re}g_{\uparrow\downarrow} \leqslant 2N_N$.

### C. Generalized circuit theory

It is not obvious how these results should be generalized to more complicated circuits and devices as well as to the presence of spin-flip scattering in the normal metal. The magnetoelectronic circuit theory does not suffer from these drawbacks. Originally, it was assumed in Ref. [12] that local spin and charge currents through the contacts only depend on the generalized potential differences, and the local node chemical potentials are obtained by a spin-generalization of the Kirchhoff laws of electrical circuits. This is valid only for highly resistive contacts, such that the in and outgoing currents do not significantly disturb the quasi-equilibrium distribution of the nodes. Fortunately we are able to relax this limitation and take into account a drift term in the nodes as well. In order to demonstrate this, we construct the fictitious circuit depicted in Fig. 14.

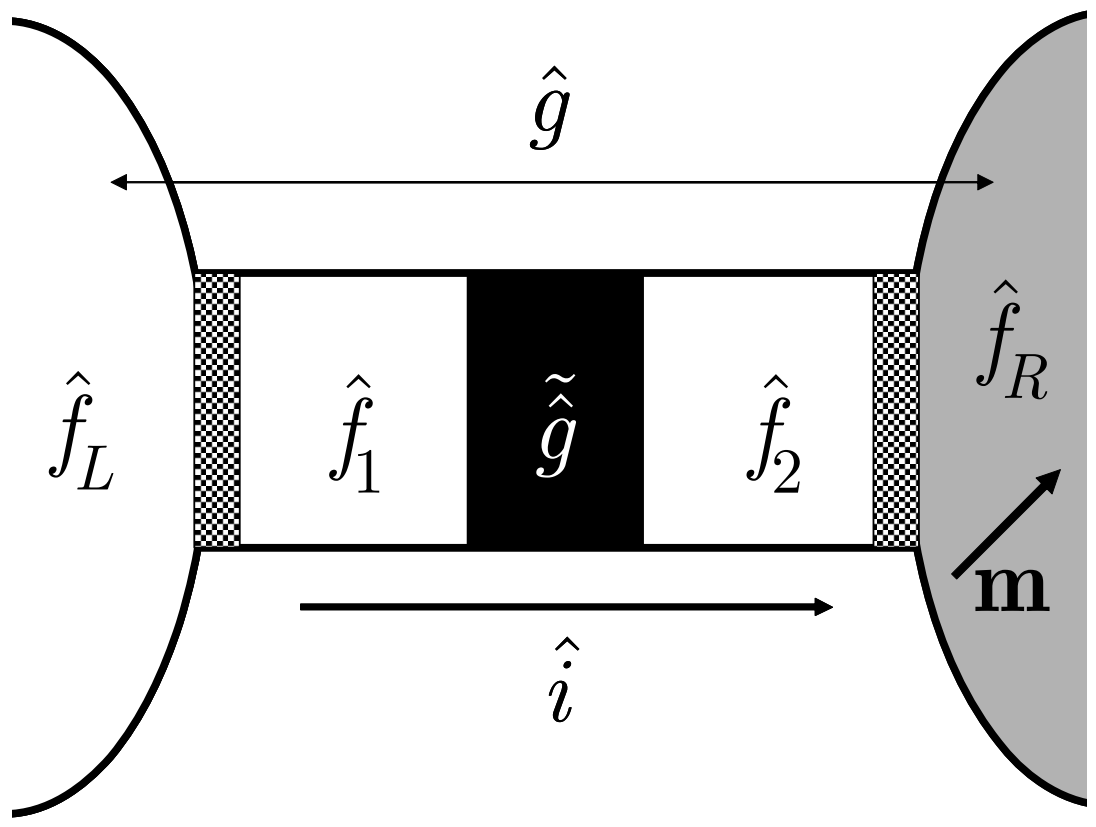

FIG. 14: Fictitious device that illustrates the generalization of circuit theory to transparent resistive elements as discussed in the text

Consider a junction that in conventional circuit theory is characterized by a matrix conductance $\hat{g}$, leading to a matrix current $\hat{\imath}$ when the normal and ferromagnetic distributions $\hat{f}_L$ and $\hat{f}_R$ are not equal. When the distributions of the nodes are isotropic, we know from circuit theory that

$$\hat{\imath} = \sum_{ss'} (\hat{g})_{ss'} \hat{u}_s \left( \hat{f}_L - \hat{f}_R \right) \hat{u}_{s'} \, , \tag{127}$$

where the projection matrices $\hat{u}_s$ are defined in Eq. (116) and $(\hat{g})_{ss} = g_s$, $(\hat{g})_{s,-s} = g_{s,-s}$. Introducing lead conductances that modify the distributions $\hat{f}_L \rightarrow \hat{f}_1$ and $\hat{f}_2 \leftarrow \hat{f}_R$, respectively, we may define a (renormalized) conductance matrix $\hat{\tilde{g}}$, which causes an identical current $\hat{\imath}$ for the reduced (matrix) potential drop:

$$\hat{\imath} = \sum_{ss'} \left( \hat{\tilde{g}} \right)_{ss'} \hat{u}_s \left( \hat{f}_1 - \hat{f}_2 \right) \hat{u}_{s'} \, . \tag{128}$$

When the lead conductances are now chosen to be twice the Sharvin conductances, and using (matrix) current conservation

$$\hat{\imath} = 2N_N \left( \hat{f}_L - \hat{f}_1 \right) \tag{129}$$

$$= \sum_s 2N_s^F \hat{u}_s \left( \hat{f}_2 - \hat{f}_R \right) \hat{u}_s \, , \tag{130}$$

straightforward matrix algebra leads to the result that the elements of $\hat{\tilde{g}}$ are identical to the renormalized interface conductances found above [Eqs. (125,126) without the bulk resistivities]. By replacing $\hat{g}$ by $\hat{\tilde{g}}$ we not only recover results for the spin valve obtained above, but we can now use the renormalized parameters also for circuits with arbitrary complexity and transparency of the contacts.

### D. Spin-flip scattering

The spin-flip scattering in ferromagnetic transition metals is quite strong because of the large d-character of many bands close to the Fermi energy that are susceptible to the spin-orbit interaction. Typical spin-flip diffusion lengths $\ell_{sd}$ range from 5 (Py)-50 (Co) nm. The spin relaxation can be weak in normal metals like Cu or Al, leading to long spin-flip diffusion lengths of the order of a micron even at room temperature, and even much longer in high-quality samples and low temperature. Spin-flip is usually not very important in magnetic multilayer structures in which the layer thicknesses are smaller than the spin diffusion lengths and can often be disregarded completely. On the other hand, the spin-flip diffusion can be of decisive importance for the transport properties in lateral structures. Even when the normal metal $\ell_{sd}^N$ is larger than the size of a normal metal island, spin-flip can dominate the transport when the dwell time is long, *e.g.* when the contacts to the ferromagnets are tunneling barriers [9]. In ferromagnets $\ell_{sd}^F$ defines the magnetoelectronically active region when in contact with a normal metal. When the resistance of the layer defined by a ferromagnet with thickness $\ell_{sd}^F$ is much larger than the interface resistance, the latter can be disregarded. In this regime the effect of the spin flip scattering can be treated with relative ease [108], since the solutions of the diffusion equation in the normal metal and the ferromagnet can be matched by requiring continuity of charge and longitudinal spin current, at the same time requiring that the transverse spin current vanishes in the ferromagnet as discussed before. This is a generalization of Valet and Fert's approach [57] to non-collinear magnetic systems, that has also been used by Stiles and Zangwill [120]. When the interface resistance cannot be neglected and the spin-flip diffusion lengths are shorter than the nodes, the spatially dependent solutions of the diffusion equation have to be matched at the interfaces in terms of the conductance parameters as before. This can often be done analytically [63, 96, 108].

The spin-flip scattering in $N$ can also be included [12] (see Section VI); it does not affect the form of Eq. (165) either, but only reduces the parameter $\tilde{\chi}$.

### E. Spin-transfer magnetization torque

We have seen that at the interface of to the ferromagnet the spin current polarized normal to the ferromagnetic order parameter is absorbed at the interface. The absorbed angular momentum is transferred to the ferromagnetic order parameter as a torque. This is the microscopic basis of Slonczewski's magnetization torque [26]. Given a chemical potential imbalance, the normal metal only, we can easily compute the torque via the spin current (103) as a function of the spin accumulation $\vec{s}f_S^N$ on the normal metal node. After integration over energies energies the total spin torque $\vec{\tau} = \frac{\hbar}{2e}\left(\vec{I}_S - \vec{I}_S \cdot \vec{m}\right)$ reads

$$\vec{\tau} = \frac{\hbar}{2e}\left(-2\,\mathrm{Re}\,G^{\uparrow\downarrow}\,\vec{m}\times\vec{\mu}_s\times\vec{m} - 2\,\mathrm{Im}\,G^{\uparrow\downarrow}\,\vec{\mu}_s\times\vec{m}\right), \tag{131}$$

where $\vec{\mu}_s = \int d\epsilon \vec{s} f_s(\epsilon)$ is the total spin accumulation in the adjacent normal metal.

In Chapter I we pointed out that the Slonczewski torque can be interpreted qualitatively in terms of Andreev scattering for a half-metallic ferromagnet. Here we see that this is a simplification, since it is only the real part of the mixing conductance that can be interpreted as a torque. The imaginary part of the mixing conductance acts as an effective field [69, 121]: although the spin current penetrates the ferromagnet typically only a couple of Angstroms, during the brief interaction the spin can precess a finite angle around the exchange field. The transferred angular momentum transferred in this way has an effect on the ferromagnet equal to an effective magnetic field parallel to the spin accumulation. There is mounting evidence that this effective field can be very significant in rather than good intermetallic interfaces [122–124].

### F. Overview

At this point it seems appropriate to lean back and review the instruments available to discuss the magnetoelectronic properties of non-collinear circuits and devices. We have two formally exact and therefore equivalent formalisms to express electrons and spin transport, i.e. the Landauer-Büttiker scattering theory and the Keldysh Green function method. If properly handled, both therefore give the correct results, in principle, but in practice there are distinct advantages and disadvantages. Random matrix theory is based on the scattering formalism and proceeds from the assumption of diffuse or chaotic scattering. Under this hypothesis a systematic expansion in the reciprocal number of modes $N^{-1}$ of the normal metal spacer allows the systematic inclusion of quantum interference effects, thus including, at least in principle, effects like Anderson localization. The linearized Boltzmann equation follows from the Keldysh Green function method in the quasiclassical approximation. It can take into account arbitrary anisotropy effects, the importance of which is not fully established yet. The diffusion equation is obtained by the additional assumption of isotropy. Circuit theory is a hybrid theory, in which an isotropy assumption is made for the distribution functions for strategically chosen nodes, that are then connected by quantum mechanical boundary conditions given by scattering theory. The original version of circuit does not hold when the nodes are connected by very transparent interfaces, but this can be repaired easily. The two-terminal transport according to the updated version of circuit theory is identical to random matrix theory for a two-terminal spin valve to lowest order in $N^{-1}$. Circuit theory subsequently turned out to be very convenient to generalize results to include spin-flip scattering, many-terminal devices, time-dependent effects, etc., and will therefore be in the focus of the discussions in the following, as well an in [73].

## V. CALCULATING THE SCATTERING MATRIX FROM FIRST PRINCIPLES

In previous chapters transport properties were formulated in terms of scattering matrices or in terms of Green functions. The transport properties of a given circuit or device could be summarized in terms of a few and measurable parameters, *viz.* the spin-dependent conductances of interfaces and bulk materials. We derived expressions that relate the conductances directly to the (density-functional) Hamiltonian of the system, which, where necessary, were estimated in the simple parabolic-band Stoner model. While the Stoner model yields valuable insights, many transport properties depend sensitively on the composition of specific samples and on the details of how they were made. To understand such aspects and ultimately to be able to guide experimentalists in their choice of materials, we need to go beyond the free-electron picture and simple models of interfaces and try to take into account the complex nature of transition metal electronic structures and to describe interfaces realistically on an atomic scale.

Various methods have been developed for calculating the transmission of electrons through an interface (or a more extended scattering region) from first principles [35, 47, 61, 125–136], or using as input electronic structures which were calculated from first principles [137–143]. Most are based upon a formulation for the conductance in terms of non-equilibrium Green's functions [144] which reduces in the appropriate limit to the well known Fisher-Lee linear-response form [99] for the conductance of a finite disordered wire embedded between crystalline leads. An alternative technique, suitable for Hamiltonians that can be represented in tight-binding form, has been formulated by Ando [145] and is based upon direct matching of the scattering-region wave function to the Bloch modes of the leads[5]. A third approach based upon "embedding" [148, 149] has been combined with full-potential linearized augmented plane wave method to yield what is probably the most accurate scheme to date [132, 133] but is numerically very demanding.

Our main purpose in this chapter is to outline a scheme suitable for studying mesoscopic transport in inhomogeneous, mainly layered, transition metal magnetic materials which is
(i) physically transparent
(ii) first-principles, requiring no free parameters,
(iii) capable of handling complex electronic structures characteristic of transition metal elements and
(iv) very efficient in order to be able to handle lateral supercells to study layered systems with different lattice parameters and to model disorder very flexibly.
A tight-binding (TB) muffin-tin-orbital (MTO) implementation of the Landauer-Büttiker formulation of transport theory within the local-spin-density approximation (LSDA) of density-functional-theory (DFT) satisfies these requirements.

As discussed in Section III, the formulation by Landauer and Büttiker of electronic transport in terms of scattering matrices where the transmission matrix element $t_{nm}$ is the probability amplitude that a state $|n\rangle$ in the left-hand lead incident on the scattering region from the left (see Fig. 15) is scattered into a state $|m\rangle$ in the right-hand lead is intuitively very appealing because wave transport through interfaces is so naturally described in terms of transmission and reflection. Usually, explicit calculation of the scattering states is avoided by making use

[5] The relationship between the wave function matching (WFM) [145] and Green function [99, 144] approaches is not immediately obvious. It was suggested recently that WFM was incomplete [146] but the complete equivalence of the two approaches could be proven [147].

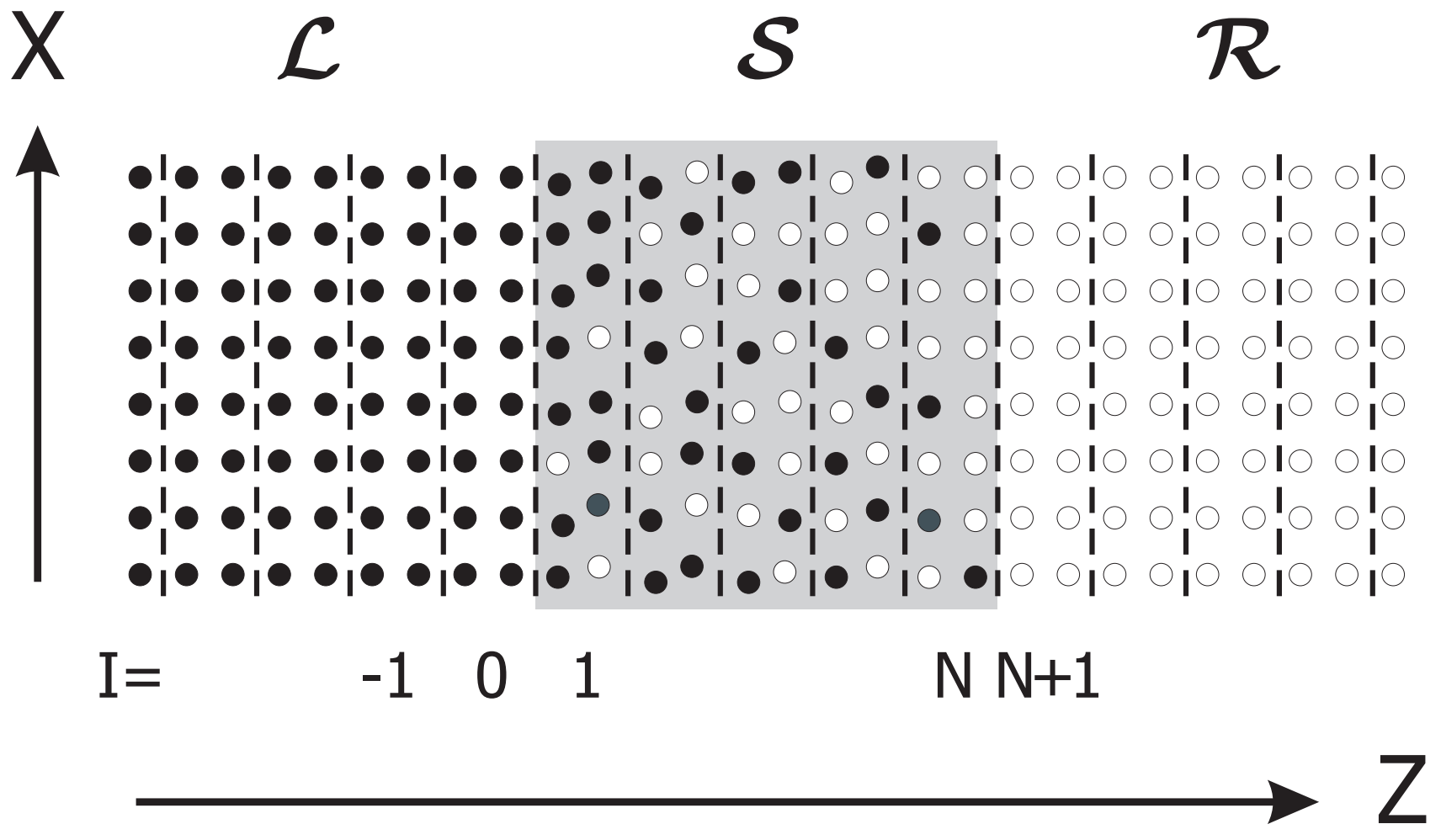

FIG. 15: Sketch of the configuration used in the Landauer-Büttiker transport formulation to calculate the two-terminal conductance. A (shaded) scattering region ($\mathcal{S}$) is sandwiched by left- ($\mathcal{L}$) and right-hand ($\mathcal{R}$) leads which have translational symmetry and are partitioned into principal layers perpendicular to the transport direction. The scattering region contains $N$ principal layers but the structure and chemical composition are in principle arbitrary.

of the invariance properties of the trace in Eq. (1)

$$G = \frac{e^2}{h} \sum_{nm} |t_{nm}|^2 \tag{132}$$

to calculate the conductance directly from Green functions expressed in some convenient localized orbital representation [144]. However, we want to make contact with the large body of theoretical literature on mesoscopic physics and address a wider range of problems in the field of spin-dependent transport as covered by the magnetoelectronic circuit theory. As evident from the previous sections, this requires calculation of the full microscopic transmission and reflection matrices $\mathbf{t}$ and $\mathbf{r}$ making use of the explicit knowledge of the scattering states to analyse our numerical results. Our first requirement of physical transparency is satisfied by choosing a computational scheme which yields the full scattering matrix and not just the conductance.

In developing a scheme for studying transport in transition metal multilayers, a fundamental difference between semiconductors and transition metals must be recognized. Transition metal atoms have two types of electrons with different orbital character. The $s$ electrons are spatially quite extended and, in solids, form broad bands with small effective masses; they conduct well. The $d$ electrons are much more localized in space, form narrow bands with large effective masses and are responsible for the magnetism of transition metal elements. The “magnetic” electrons, however, become itinerant by hybridization with $s$-electrons and do contribute to electrical transport. The appropriate framework for describing metallic magnetism, even for the late $3d$ transition metal elements, is band theory [90]. An extremely successful framework exists for treating itinerant electron systems from first-principles and this is the Local Density Approximation (LDA) of Density Functional Theory (DFT). For band magnetism, the appropriate extension to spin-polarized systems, the local spin-density approximation (LSDA) satisfies our second requirement of introducing no free parameters.[6]

Oscillatory exchange coupling in layered magnetic structures was discussed by Bruno in terms of generalized reflection and transmission matrices [150] which were calculated by Stiles [77, 78] for realistic electronic structures using a scheme [125, 126] based on linearized augmented plane waves (LAPWs). At an interface between a non-magnetic and a magnetic metal, the different electronic structures of the majority and minority spin electrons in the magnetic material give rise to strongly spin-dependent reflection [151, 152]. Schep *et al.* used transmission and reflection matrices calculated from first-principles with an embedding surface Green function method [153] to calculate spin-dependent interface resistances for specular Cu|Co interfaces embedded in diffusive bulk material [61]. The resulting good agreement with experiment indicated that interface disorder is less important than the spin-dependent reflection and transmission from a perfect interface. Calculations of domain wall resistances as a function of the domain wall thickness illustrated the usefulness of calculating the full scattering matrix [47, 154]. However, the LAPW basis set used by Stiles and Schep was computationally too expensive to allow repeated lateral supercells to be used to model interfaces between materials with very different, incommensurate lattice parameters or to model disorder. This is true of all plane-wave based basis sets which typically require of order 100 plane waves per atom in order to describe transition metal atom electronic structures reasonably well.

Muffin-tin orbitals (MTO) form a flexible, minimal basis set leading to highly efficient computational schemes for solving the Kohn-Sham equations of DFT [155–157]. For the close packed structures adopted by the magnetic materials Fe, Co, Ni and their alloys, a basis set of 9 functions ($s$, $p$, and $d$ orbitals) per atom in combination with the atomic sphere approximation (ASA) for the potential leads to errors in describing the electronic structure which are comparable to the absolute errors incurred by using the local density approximation. This should be compared to typically 100 basis functions per atom required by the more accurate LAPW method. MTOs thus satisfy our third and fourth requirements of being able to treat complex electronic structures efficiently.

[6] Because the magnetism of transition metals depends very sensitively on atomic structure [90], it is important to know this structure quite accurately. The current drive to make devices whose lateral dimensions approach the nanoscale means that it is becoming increasingly important to know the atomic structures of these small systems microscopically while at the same time it is more difficult to do this characterization experimentally. It has become a practical alternative to determine minimum-energy structures theoretically by minimizing as a function of the atomic positions the total energy obtained by solving the Schrödinger equation self-consistently within the local density approximation (LDA) of Density Functional Theory (DFT), thereby avoiding the use of any free parameters.

The tight-binding linearized muffin tin orbital (TB-LMTO) surface Green function (SGF) method was developed to study the electronic structure of interfaces and other layered systems. When combined with the coherent-potential approximation (CPA), it allows the electronic structure, charge and spin densities of layered materials with substitutional disorder to be calculated self-consistently very efficiently [158]. To calculate transmission and reflection matrices from first principles, we combined the wave-function matching (WFM) formalism described by Ando [145] for an empirical tight-binding Hamiltonian, with an ab-initio TB-MTO basis [157]. The method which results was applied to a number of problems of current interest in spin-transport: to the calculation of spin-dependent interface resistances where interface disorder was modelled by means of large lateral supercells [35]; to the first principles calculation of the spin mixing conductance parameter entering theories with non-collinear magnetizations, relevant for the spin-transfer torque [36] and the related problem of Gilbert damping enhancement in the presence of interfaces [96]; to a generalized scattering formulation of the suppression of Andreev scattering at a ferromagnetic|superconducting interface [19]; to the problem of how spin-dependent interface resistances influence spin injection from a metallic ferromagnet into a III–V semiconductor [83]. These examples amply demonstrate that the fourth requirement is well satisfied.[7]

### A. Formalism

In the layered magnetic structures which are encountered in the study of spin transport, Bloch translational symmetry is broken so that the Kohn-Sham equations have to be solved for the infinite system represented by Fig. 15. This problem is reduced to finite size (Fig. 16) by replacing the semi-infinite leads by an appropriate energy dependent boundary condition as follows. First a local orbital basis $|i\rangle$ is introduced in which the Hamiltonian $\mathcal{H}$ has block tridiagonal form. To study transport, it is convenient to group the atoms in layers labelled $I$ (see Fig. 15) and to collect all of the expansion coefficients $c_{Ii}$ for orbital $i$ in layer $I$ in a vector $\mathbf{C}_I \equiv c_{Ii}$. For electrons with energy $\varepsilon$ the resulting equation of motion (EoM) relating the vectors of coefficients $\mathbf{C}_I$ for layers $I-1$, $I$, and $I+1$ is

$$\mathcal{H}_{I,I-1}\mathbf{C}_{I-1} + (\mathcal{H}-\varepsilon)_{I,I}\mathbf{C}_I + \mathcal{H}_{I,I+1}\mathbf{C}_{I+1} = 0 \tag{133}$$

forming an infinite set of homogeneous linear equations for $I = -\infty \ldots + \infty$ . For the muffin-tin orbitals we will use, $i$ is a combined index $\mathbf{R}lm$, where $l$ and $m$ are the azimuthal and magnetic quantum numbers, respectively, of the MTO defined for an atomic-spheres-approximation (ASA) potential on the site $\mathbf{R}$. (In the case of magnetic materials, the effective potentials depend on the spin $\sigma$ of the orbital which must consequently be labelled $\mathbf{R}lm\sigma$. To prevent the notation becoming too cumbersome, the spin index will not be shown unless strictly necessary.) The EoM does not restrict us to only considering nearest neighbour interactions since atoms can be simply grouped into layers defined to be so thick that the interactions between layers $I$ and $I \pm 2$ are negligible (see Fig. 15). Such layers are called *principal layers*. Their thickness depends on the range of the interactions which in turn partly depends on the spatial extent of the orbital basis. It can be minimized by using the highly localized tight-binding MTO representation [155–157].

[7] A version of the method implemented for a real-space grid has also been developed and applied to the calculation of the conductance of atomic wires [136].

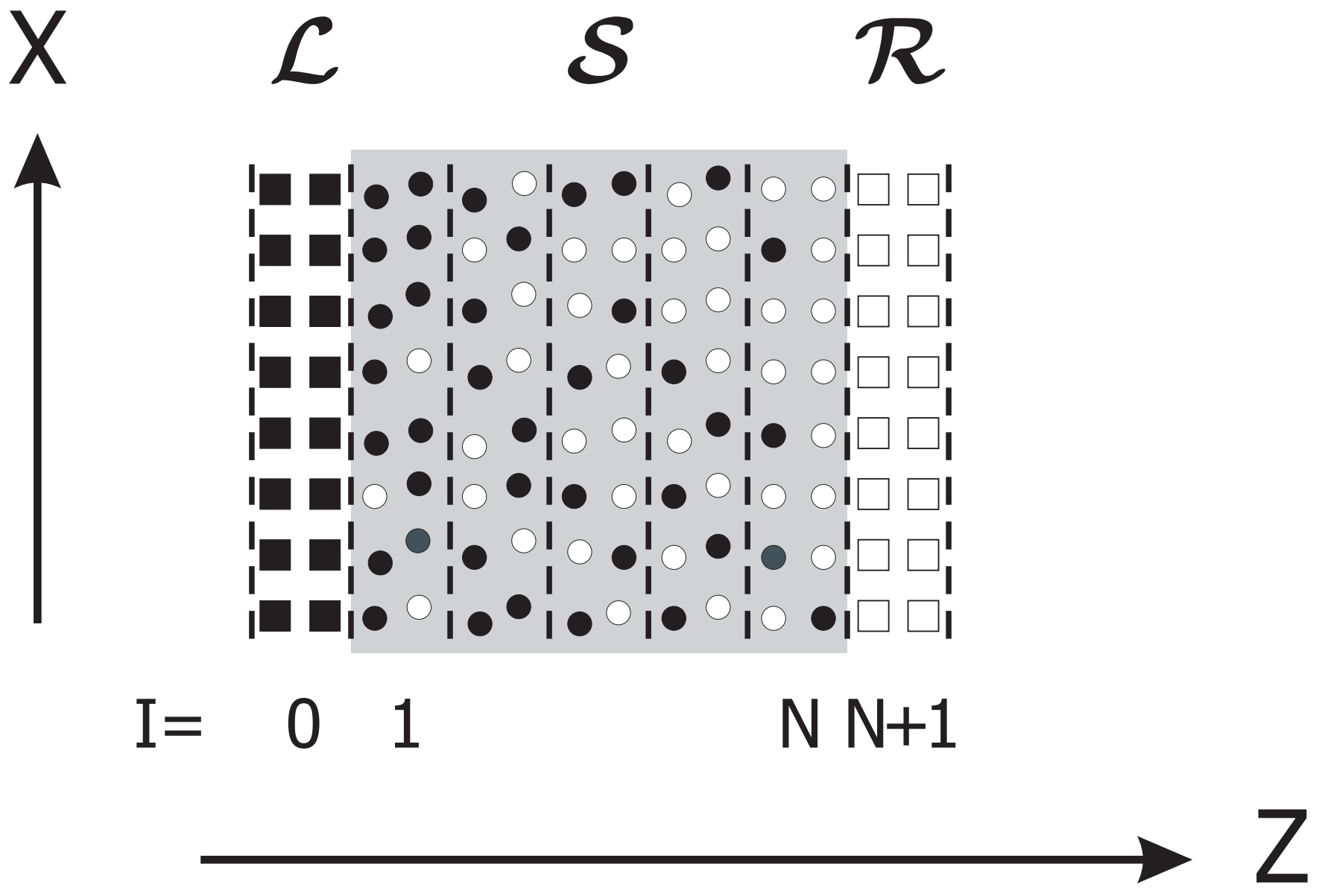

FIG. 16: The semi-infinite leads and corresponding Hamiltonian problem with an infinite number of layers from Fig. 15 is replaced by finite leads and an effective Hamiltonian. The effective Hamiltonian is simply constructed by adding the energy-dependent self-energies $\Sigma_{\mathcal{L}}$ and $\Sigma_{\mathcal{R}}$ to the $\mathcal{H}_{0,0}$ and $\mathcal{H}_{N+1,N+1}$ Hamiltonian blocks, respectively, where the indices are principal layer indices. The lead atoms modified by the self energy are depicted as squares. The left and right self-energies are constructed in terms of generalized Bloch matrices and blocks of lead hopping matrices for the left-hand and right-hand leads, respectively.

The infinite set of equations (133) is solved by splitting the problem into (i) a part involving only the semi-infinite leads which have Bloch translational symmetry outside the shaded scattering region in Fig. 15 and (ii) making the problem for the scattering region finite by replacing the infinite leads with an appropriate boundary condition in the layers $I = 0$ and $I = N + 1$. The infinite set of homogeneous linear equations (133) is reduced to a finite set of inhomogeneous linear equations for $0 \leq I \leq N + 1$ whose solution basically requires inverting the modified Hamiltonian matrix. The inhomogeneous part is determined by the boundary condition we impose. This depends on which element of the scattering matrix we require - transmission or reflection from the left- or the right-hand side. The modification to the Hamiltonian consists of adding an energy-dependent self-energy to the $I = 0$ and $I = N + 1$ diagonal blocks of the Hamiltonian. This self-energy is constructed from two ingredients. The first is the off-diagonal (in layer indices) block of the Hamiltonian which describes hopping in the leads. The second is a generalized "Bloch" matrix which describes the change left- or right-going electrons undergo as a result of a translation taking us from one principal layer to another. This is the essence of the wave-function matching (WFM) method for calculating the transmission and reflection matrices due to Ando [145].

To combine the WFM method with muffin-tin orbitals, it turns out to be convenient to use the so-called "tail-cancellation" condition[8]

$$\sum_{R',l'm'} \left[ P^{\alpha}_{Rlm}(\varepsilon) \delta_{RR'} \delta_{ll'} \delta_{mm'} - S^{\alpha}_{RlmR'l'm'} \right] c^{\alpha}_{R'l'm'} = 0, \tag{134}$$

in terms of *potential functions* $P^{\alpha}_{Rlm}(\varepsilon)$ which characterize the AS potentials and the potential-independent screened *structure constant* matrix $S^{\alpha}_{RlmR'l'm'}$ whose range in real space depends on a set of *screening* parameters $\{\alpha_l\}$. The set of parameters which minimize the range of hopping is denoted $\alpha = \beta$. The equation analogous to (133) which we use to solve the scattering problem is then

$$-S^{\beta}_{I,I-1} \mathbf{C}_{I-1} + \left( P^{\beta}_{I,I}(\varepsilon) - S^{\beta}_{I,I} \right) \mathbf{C}_I - S^{\beta}_{I,I+1} \mathbf{C}_{I+1} = 0. \tag{135}$$

---

[8] This equation is nothing other than the well known KKR equation of electronic structure theory (discussed in many solid state physics textbooks) in the so-called atomic spheres approximation in which the kinetic energy in the interstitial region is taken to be zero and the volume of the interstitial region is made to vanish by replacing the muffin tin spheres with space filling atomic spheres. This choice leads to structure constants which are energy and scale independent, unlike the KKR structure constants. The potential function $P_l(\varepsilon)$ is simply related to the logarithmic derivative $D_l(\varepsilon)$ as $P_l(\varepsilon) = 2(2l+1)\,(D_l(\varepsilon)+l+1)\,/\,(D_l(\varepsilon)-1)$.

$\mathbf{C}_I \equiv c_{Ii} \equiv c_{IRlm}$ is a $(l_{\max}+1)^2 N \equiv M$ dimensional vector describing the amplitudes of the $I$-th layer consisting of $N$ atom sites with $(l_{\max}+1)^2$ orbitals per site. $P_{I,I}$ and $S_{I,J}$ are $M \times M$ matrices.

In the actual calculation of the transmission matrix, we adopt the following procedure. First of all two separate self-consistent field calculations are performed for the left ($\mathcal{L}$) and right ($\mathcal{R}$) leads making use of their perfect lattice periodicity to calculate the electron densities (for magnetic materials, the spin-densities) as well as corresponding Fermi energies. Next, a self consistent field calculation is carried out for the scattering region $\mathcal{S}$ between the leads subject to the requirement that the Fermi energies in the right- and left-hand leads are equal. We now have a charge (spin) density in all space as well as the corresponding Kohn-Sham effective potential and can proceed to the solution of the transport problem.

The calculation of the scattering matrix is split into two distinct parts. In the first stage, the eigenmodes of the leads, of which there are $2M$, are calculated using the EoM (135) and making use of the lattice periodicity. By calculating their $\mathbf{k}$ vectors (which are in general complex) and velocities $v_{\mathbf{k}}$, the eigenstates can be classified as being either left-going (-) or right-going (+). They form a basis in which to expand any left- and right-going waves and have the convenient property that their transformation under a lattice translation in the leads is easily calculated using Bloch's theorem (with $\mathbf{k}$ complex).

In the second stage, the scattering region $\mathcal{S}$ which mixes left- and right-going lead eigenmodes is introduced in the layers $1 \leq I \leq N$ . The scattering region can be a single interface, a complex multilayer or a tunnel junction, and the scattering can be introduced by disorder or simply by discontinuities in the electronic structure at interfaces. The $n \rightarrow m$ element of the reflection matrix, $r_{mn}$, is defined in terms of the ratio of the amplitudes of left-going and right-going solutions in the left lead (in layer 0 for example) projected onto the $n^{\text{th}}$ right-going and $m^{\text{th}}$ left-going propagating states ($\mathbf{k}$ vector real) renormalized with the velocities so as to have unit flux. The scattering problem is solved by direct numerical inversion of a matrix with the leads included as a boundary condition so as to make finite the matrix which has to be inverted. By using real energies throughout we avoid problems distinguishing propagating and evanescent states which are encountered when a small but finite imaginary part of the energy is used.

Even though the theoretical scheme outlined above contains no adjustable parameters, its practical implementation does involve numerous approximations, some physical, others numerical, which need to be evaluated. Any workable scheme must be based upon an independent particle approximation. Our confidence in the corresponding single particle electronic structures is derived from the agreement (or lack thereof !) with reliable experimental gauges such as Fermi surfaces determined using methods such as de Haas-van Alphen measurements or the occupied and unoccupied electronic states close to the Fermi energy determined by, for example, photoelectron spectroscopy.

### B. Calculations

In this section we illustrate the formalism sketched in the previous section by calculating the transmission matrix for an ideal ordered Cu|Co(111) interface (V B 1), then describe how interface disorder can be modelled using large lateral supercells and analyse the results (V B 2).

#### 1. *Ordered Interfaces*

Cu and Co have slightly different atomic volumes. The equilibrium lattice constant of Cu is 3.614 $\AA$ and of Co 3.549 $\AA$, assuming an *fcc* structure. Even in the absence of interface disorder, the lattice spacing will not be homogeneous and will depend on the lattice constant of the substrate on which the sample was grown, on the global and local concentrations of Cu and Co, and on other details of how the structure was prepared. In principle we could calculate all of this by energy minimization. However, we judge that the additional effort needed is not justified by current experimental knowledge. Instead, we content ourselves with estimating the uncertainty which results from plausible variations in the (interface) structure by considering two limiting cases and one intermediate case. In each case an *fcc* structure is assumed, with lattice constants corresponding to (i) the atomic volume of Cu, (ii) the atomic volume of Co, (iii) an intermediate case with arithmetic mean of Cu and Co atomic volumes.

Our starting point is a self-consistent TB-LMTO SGF calculation [158] for the interface embedded between semi-infinite Cu and Co leads whose potentials and spin-densities were determined self-consistently in separate "bulk" calculations. The charge and spin-densities are allowed to vary in $n_{\rm Cu}$ layers of Cu and $n_{\rm Co}$ layers of Co bounding the interface. The results of these calculations for Cu|Co(111) interfaces and the three different lattice constants detailed above are given in Table I for $n_{\rm Cu}$=4, $n_{\rm Co}$=4. In the Cu layers, only tiny moments are induced. Only four layers away from the interface on the Co side, the magnetic moments are seen to be very close to the bulk values. At the interface, where the $d$-bandwidth is reduced as a result of the lower coordination number, the moments are suppressed rather than enhanced. This occurs because the majority-spin $d$ bands are full, preventing the conversion of minority to majority $d$ electrons. While the number of majority-spin electrons remains essentially constant, the number of minority-spin $d$ electrons is enhanced at the expense of the free-electron like $sp$ electrons whose bandwidth is less strongly affected by the reduction in coordination number. A 2% change in lattice constant changes the bulk magnetic moment of *fcc* Co by less than 3% and the effect of changing the basis (*spd* to *spdf*) has a comparable effect. From Table I we see that the interface moments behave in a similar fashion. The magnetic moment of the interface Co atoms increases by $\sim$ 3%, from 1.58 $\mu_B$/atom for $a = 3.549\,\mathring{A}$ to $1.63\,\mu_B$/atom for $a = 3.614\,\mathring{A}$ for an *spd* basis compared to a 2.3% increase for bulk atoms. Thus the *sp* to $d_{\rm min}$ conversion is enhanced at the interface by the reduced $d$-bandwidth.

| $a(\mathring{A})$ | 3.549 | | 3.581 | 3.614 |
|---|---|---|---|---|
| Basis | *spdf* | *spd* | *spd* | *spd* |
| $m_{Cu}$(bulk) | 0.000 | 0.000 | 0.000 | 0.000 |
| $m_{Cu}$(int-4) | 0.001 | 0.001 | 0.001 | 0.001 |
| $m_{Cu}$(int-3) | -0.001 | 0.000 | 0.000 | 0.000 |
| $m_{Cu}$(int-2) | -0.005 | -0.005 | -0.005 | -0.005 |
| $m_{Cu}$(int-1) | 0.002 | 0.004 | 0.003 | 0.001 |
| $m_{Co}$(int+1) | 1.53 | 1.58 | 1.61 | 1.63 |
| $m_{Co}$(int+2) | 1.62 | 1.66 | 1.67 | 1.69 |
| $m_{Co}$(int+3) | 1.60 | 1.64 | 1.66 | 1.68 |
| $m_{Co}$(int+4) | 1.61 | 1.65 | 1.67 | 1.68 |
| $m_{Co}$(bulk) | 1.609 | 1.646 | 1.67 | 1.684 |
| $G^{\rm maj}(111)$ | 0.41 | 0.43 | 0.43 | 0.43 |
| $G^{\rm min}(111)$ | 0.38 | 0.38 | 0.37 | 0.36 |

TABLE I: Variation of the layer-resolved magnetic moments (in Bohr magnetons) for Cu|Co(111) interfaces with basis set and lattice constant. These results were obtained with von Barth-Hedin's exchange-correlation potential. In the last two rows, the interface conductances are given in units of $10^{15} : \Omega^{-1}\mathrm{m}^{-2}$.

Once the interface potential has been obtained, the transmission matrix can be calculated and the BZ summation carried out. The convergence of this summation, shown in Fig. 17 for a lattice constant of $a = 3.614\mathring{A}$ and an *spd* basis, closely parallels the behaviour found on calculating the Sharvin conductance of the leads and does not represent a limitation in practice. Converged transmission probabilities

$$G^\sigma(\hat{n}) = \frac{e^2}{h} \sum_{m,n,\vec{k}_\parallel} T^\sigma_{mn}(\vec{k}_\parallel) = \frac{e^2}{h} \sum_{m,n,k_\parallel} |t^\sigma_{mn}(\vec{k}_\parallel)|^2 \tag{136}$$

are given in the last two rows of Table I. The apparently modest spin-dependence of "bare" interface conductances ($\sim 20\%$) can lead to spin-dependent interface resistances differing by a factor of $\sim 3-5$. To obtain estimates of the interface resistance for highly transparent interfaces, the "bare" transmissions cannot be used. $R^{LB} = 1/G^{LB}$ results in a finite "interface" resistance, even for a fictitious interface between identical materials. Schep *et al.* [61] derived an expression for the resistance of transparent interfaces in terms of the interface transmission, which takes into account the finiteness of the conductance of the perfect leads:

$$R_\sigma^{Schep}(A|B) = \frac{h}{e^2}\left[\frac{1}{\sum T_{mn}^\sigma} - \frac{1}{2}\left(\frac{1}{N_A^\sigma} + \frac{1}{N_B^\sigma}\right)\right] \tag{137}$$

where $N_A^\sigma$ and $N_B^\sigma$ are the Sharvin conductances of the materials A and B forming the interface, in units of $e^2/h$ . A more detailed discussion of this "Boltzmann correction" was given in Section IV B.

| N\|M | Au\|Fe | | Cu\|Co | |
|---|---|---|---|---|
| Layer | clean | dirty | clean | dirty |
| $m_N$(bulk) | 0.000 | 0.000 | 0.000 | 0.000 |
| $m_N$(int-3) | 0.000 | 0.000 | 0.001 | 0.000 |
| $m_N$(int-2) | 0.001 | -0.003 | -0.000 | -0.003 |
| $m_N$(int-1) | -0.002 | 0.010 | -0.004 | -0.003 |
| $m_N$(int) | 0.064 | 0.026 | 0.006 | 0.010 |
| $m_F$(int) | - | 2.742 | - | 1.410 |
| $m_N$(int) | - | 0.128 | - | 0.036 |
| $m_F$(int) | 2.687 | 2.691 | 1.545 | 1.540 |
| $m_F$(int-1) | 2.336 | 2.396 | 1.635 | 1.596 |
| $m_F$(int-2) | 2.325 | 2.363 | 1.621 | 1.627 |
| $m_F$(int-3) | 2.238 | 2.282 | 1.627 | 1.624 |
| $m_F$(bulk) | 2.210 | 2.210 | 1.622 | 1.622 |

TABLE II: Layer-resolved magnetic moments in Bohr magnetons for single N|M interfaces (N=Au,Cu; F=Fe,Co). The magnetic moments for the Cu|Co(111) interface differ slightly from those given in Table I because a different exchange-correlation potential (Perdew-Zunger) was used.

The majority-spin case can be readily understood in terms of the geometry of the Fermi surfaces of Cu and Co so we begin by discussing this simple case before examining the more complex minority-spin channel.

*a. Clean Cu|Co (111) Interface: Majority Spins* In the absence of disorder, crystal momentum parallel to the interface is conserved. If, for a given value of $\vec{k}_\parallel$, there is a propagating state in Cu incident on the interface but none in Co, then an electron in such a state is completely reflected at the interface. Conversely, $\vec{k}_\parallel$'s for which there is a propagating state in Co but none in Cu also cannot contribute to the conductance. To determine the existence of such states, it is sufficient to inspect projections of the Fermi surfaces of *fcc* Cu and majority-spin Co onto a plane perpendicular to the transport direction $\hat{n}$, shown in Fig. 18 for $\hat{n} = (111)$.

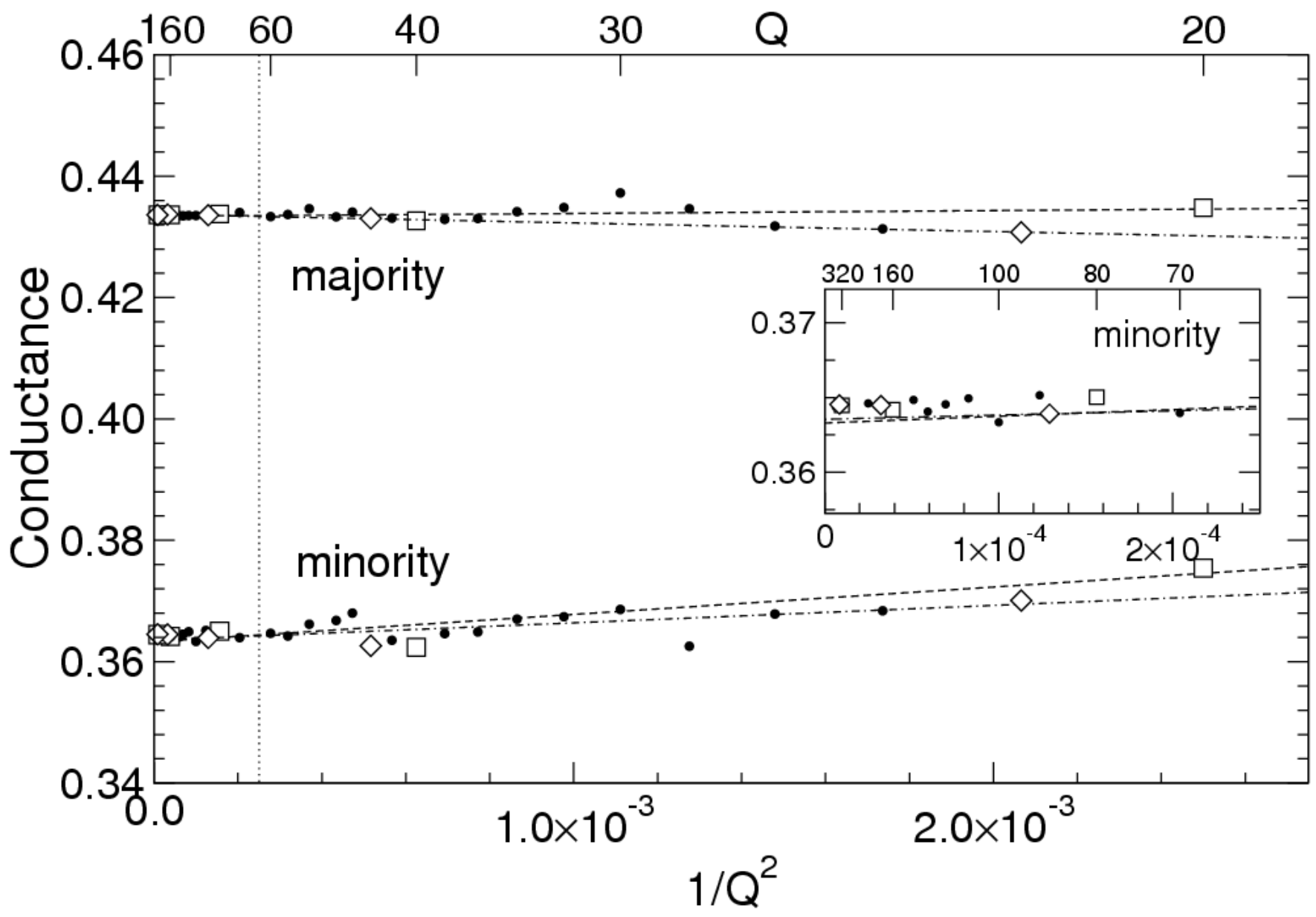

FIG. 17: Interface conductance $G^{\sigma}(111)$ (in units of $10^{15}$ : $\Omega^{-1}\mathrm{m}^{-2}$) for an *fcc* Cu|Co(111) interface for majority and minority spins plotted as a function of the normalized area element used in the Brillouin zone summation, $\Delta^2\mathbf{k}_{\parallel}/A_{BZ} = 1/Q^2$. $Q$, the number of intervals along the reciprocal lattice vector is indicated at the top of the figure. The squares represent the series ($Q = 20, 40, 80, 160, 320$) least-squares fitted by the dashed line; the diamonds, the series ($Q = 22, 44, 88, 176, 352$) least-squares fitted by the dash-dotted line. The part of the curve for the Co minority spin case to the left of the vertical dotted line is shown on an expanded scale in the inset. An *fcc* lattice constant of $a = 3.614\mathring{A}$ and von Barth-Hedin exchange correlation potential were used.

The first feature to note in the figure (left-hand and middle panels) is that per $\vec{k}_{\parallel}$ there is only a single channel with positive group velocity so that the transmission matrix in (136) is a complex number whose modulus squared is a transmission probability with values between 0 and 1. It is plotted in the right-hand panel and can be interpreted simply. Regions which are depicted blue correspond to $\vec{k}_{\parallel}$'s for which there are propagating states in Cu but none in Co. These states have transmission probability 0 and are totally reflected. For values of $\vec{k}_{\parallel}$ for which there are propagating states in both Cu and Co, the transmission probability is very close to one, depicted red. These states are essentially free electron-like states which have the same symmetry in both materials and see the interface effectively as a very low potential step. Close to the centre of the figure there is an annular region where there are propagating states in Co but none in Cu so they do not contribute to the conductance. Performing the sum in (136), we arrive at an interface conductance of $0.43 \times 10^{15}$ : $\Omega^{-1}\mathrm{m}^{-2}$ to be compared to the Sharvin conductances for Cu and Co of 0.58 and 0.47 respectively, in the same units for $a = 3.549\,\AA$ and an *spd* basis. The interface conductance of 0.43 is seen to be essentially the Sharvin conductance of the majority states of Co reduced because the states closest to the $\Lambda$-axis (corresponding to the symmetry axis of the figures, the $\Gamma L$ line in reciprocal space) do not contribute.

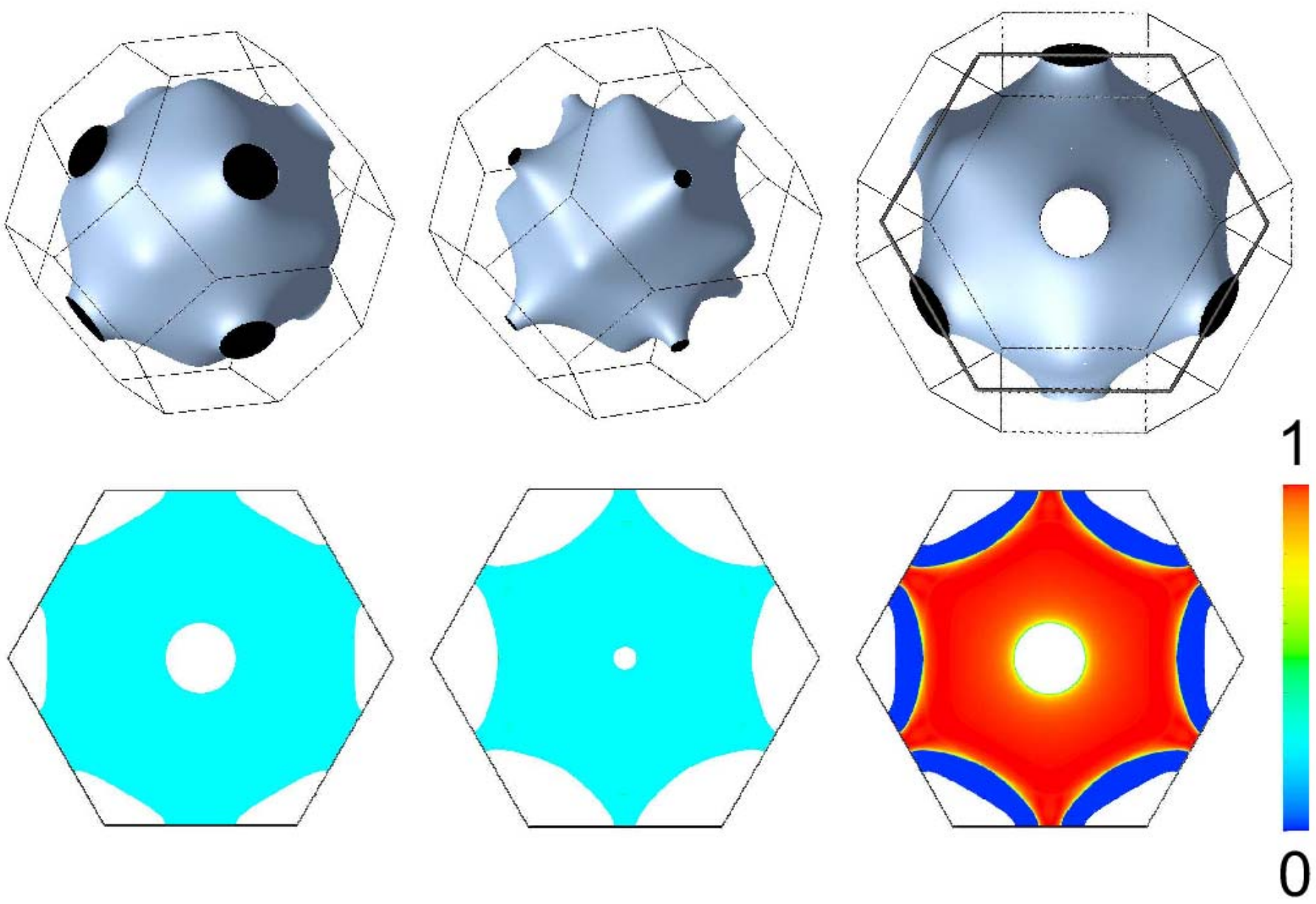

FIG. 18: Top row: Majority-spin fcc Fermi surfaces of Cu (left), Co (middle), and of Cu viewed along the (111) direction (right). Bottom row: projection of the majority-spin fcc Fermi surfaces onto a plane perpendicular to the (111) direction for Cu (left) and Co (middle). Transmission probability for majority-spin states as a function of transverse crystal momentum, $T(\mathbf{k}_{||})$ for an *fcc* Cu|Co(111) interface (right).

*b. Clean Cu|Co (111) Interface: Minority Spins* The minority-spin case is considerably more complex because the Co minority-spin $d$ bands are only partly filled, resulting in multiple sheets of Fermi surface. These sheets are shown in Fig. 19 together with their projections onto a plane perpendicular to the (111) transport direction. Compared to Fig. 18, one difference we immediately notice is

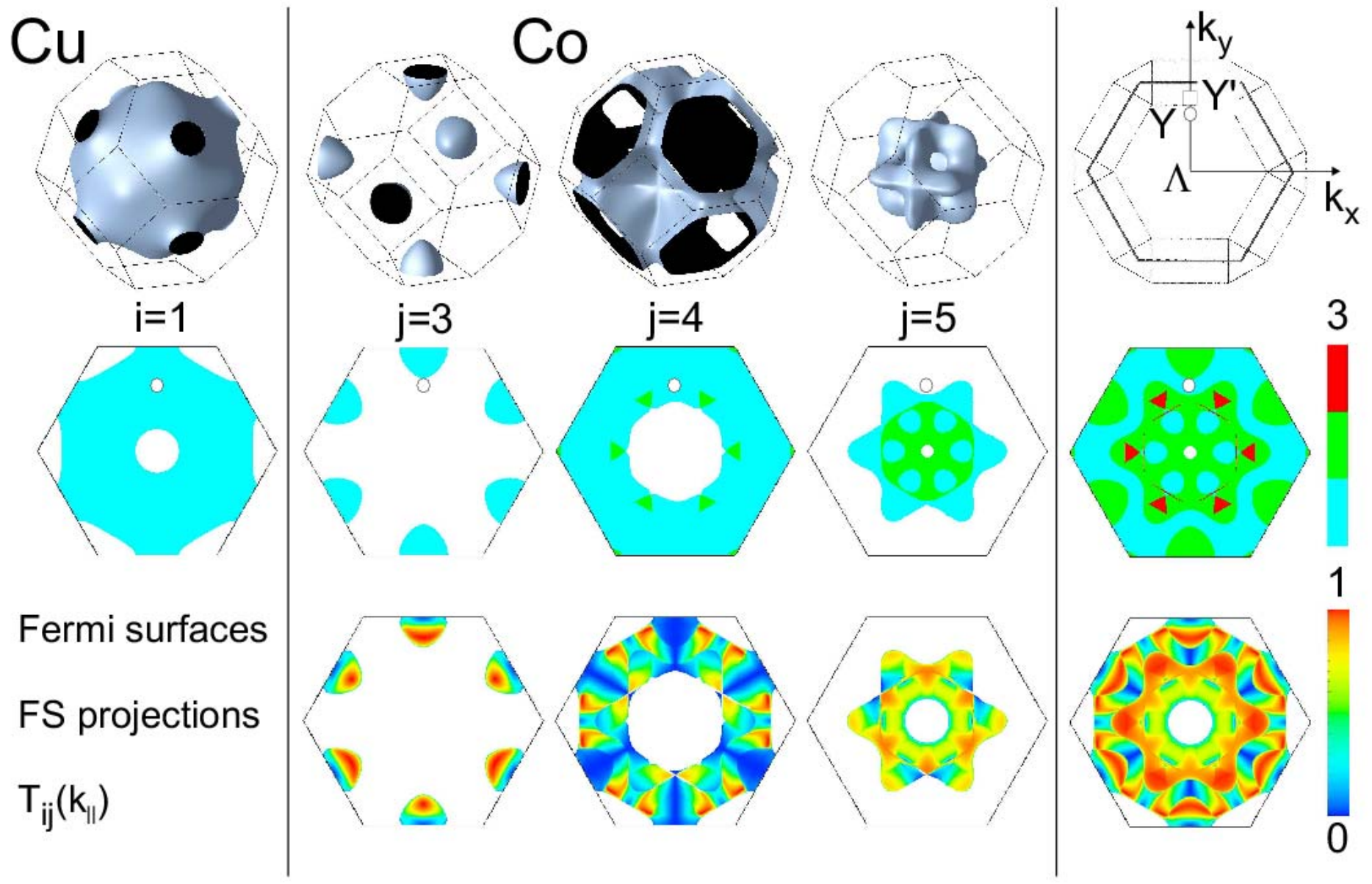

FIG. 19: Top row: Fermi surface (FS) of fcc Cu; of the third, fourth and fifth FS sheets of minority-spin fcc Co; projection of the Brillouin zone (BZ) on a plane perpendicular to the (111) direction and of the two dimensional BZ. Middle row: corresponding projections of individual FS sheets and of Co total. Bottom row: probability $T_{ij}(\mathbf{k}_{||})$ for a minority-spin state on the single FS sheet of Cu to be transmitted through a Cu|Co(111) interface into FS sheet $j$ of *fcc* Co as a function of the transverse crystal momentum $\mathbf{k}_{||}$. The point Y is such that there are only propagating states in Cu and in the fourth FS sheet of Co. For the point Y' slightly further away from $\Lambda$ and indicated by a small open square there is, in addition, a propagating state in the third FS sheet of Co.

that even single Fermi surface (FS) sheets are not single valued: for a given $\vec{k}_{||}$ there can be more than one mode with positive group velocity. The areas depicted green in the projections of the FS sheets from the fourth and fifth bands are examples where this occurs.

An electron incident on the interface from the Cu side, with transverse crystal momentum $\vec{k}_{||}$, is transmitted into a linear combination of all propagating states with the same $\vec{k}_{||}$ in Co; the transmission matrix $t^{\sigma}_{mn}(\vec{k}_{||})$ is in general not square but rectangular. The transmission probabilities $T_{mn}(\vec{k}_{||})$ are shown in the bottom row of Fig. 19. Because there is only a single incident state for all $\vec{k}_{||}$, the maximum transmission probability is one. Comparison of the total minority-spin transmission probability $T_{\mathcal{LR}}(\vec{k}_{\|})$ (Fig. 19, bottom right-hand panel) with the corresponding majority-spin quantity (right-hand panel of Fig. 18) strikingly illustrates the spin-dependence of the interface scattering, much more so than the integrated quantities might have led us to expect; the interface conductances, 0.38 and $0.43 \times 10^{15} : \Omega^{-1}\mathrm{m}^{-2}$ from Table I, differ by only $\sim 15\%$.

Three factors contribute to the large $\vec{k}_{\|}$-dependence of the transmission probability: first and foremost, the complexity of the Fermi surface of both materials but especially of the minority spin of Co; secondly and inextricably linked with the first because of the relationship $\hbar v_{\vec{k}} = \nabla_k \varepsilon(\vec{k})$, the mismatch of the Fermi velocities of the states on either side of the interface. Thirdly, the orbital character of the states $m$ and $n$ which varies strongly over the Fermi surface and gives rise to large matrix element effects.

The great complexity of transition metal Fermi surfaces, clear from the figure and well-documented in standard textbooks, is not amenable to simple analytical treatment and has more often than not been neglected in theoretical transport studies. Nevertheless, as illustrated particularly well by the ballistic limit [59, 76], spin-dependent band structure effects have been shown to lead to magnetoresistance ratios comparable to what are observed experimentally in the current-perpendicular-to-plane (CPP) measuring configuration and cannot be simply ignored in any quantitative discussion. Most attempts to take into account contributions of the $d$ states to electronic transport do so by mapping the five $d$ bands onto a single tight-binding or free-electron band with a large effective mass.

Fermi surface topology alone cannot explain all aspects of the transmission coefficients seen in Fig. 19. For example, there are values of $\vec{k}_{\|}$, such as that labelled $Y$ in the figure, for which propagating solutions exist on both sides of the interface yet the transmission probability is zero. This can be understood as follows. We begin by choosing a $(1 \times 1)$ interface unit cell so that the atoms which occupy the ABC sites characteristic of the stacking in the *fcc* structure all lie along the $y$-axis. At $\vec{k}_{\|} = Y$, the propagating states in Cu have $\{s, p_y, p_z, d_{yz}, d_{3z^2-r^2}, d_{x^2-y^2}\}$ character and are even with respect to reflection in the plane formed by the $y$-axis and the transport direction perpendicular to the (111) plane which we choose to be the $z$-axis. For this $\vec{k}_{\|}$ the only propagating state in Co is in the fourth band. It has $\{p_x, d_{xy}, d_{xz}\}$ character which is odd with respect to reflection in the $yz$ plane. Consequently, the corresponding hopping matrix elements in the Hamiltonian (and in the Green function) vanish and the transmission is zero.

Along the $k_y$ axis the symmetry of the states in Cu and those in the fourth band of Co remain the same and the transmission is seen to vanish for all values of $k_y$. However, at points further away from $\Lambda$, we encounter states in the third band of Co which have even character whose matrix elements do not vanish by symmetry and we see substantial transmission probabilities. Similarly, for points closer to $\Lambda$, there are states in the fifth band of Co with even character whose matrix elements also do not vanish and again the transmission probability is substantial. Because it is obtained by superposition of transmission probabilities from Cu into the third, fourth and fifth sheets of the Co FS, the end result, though it may appear very complicated, can be straightforwardly analysed in this manner k-point by k-point.

Though the underlying lattice symmetry is only threefold, the Fermi surface projections shown in Fig. 19 have six-fold rotational symmetry about the line $\Lambda$ because bulk $fcc$ structure has inversion symmetry (and time-reversal symmetry). The interface breaks the inversion symmetry so $T_{mn}(\vec{k}_\parallel)$ has only threefold rotation symmetry for the individual FS sheets. However, in-plane inversion symmetry is recovered for the total transmission probability $T_{\mathcal{LR}}(-\vec{k}_\parallel) = T_{\mathcal{LR}}(\vec{k}_\parallel)$ which has full sixfold symmetry.

### *2. Interface Disorder*

Instructive though the study of perfect interfaces may be in gaining an understanding of the role electronic structure mismatch may play in determining giant magnetoresistive effects, all measurements are made on devices which contain disorder, in the diffusive regime. Because there is little information available from experiment about the nature of this disorder, it is very important to be able to model it in a flexible manner, introducing a minimum of free parameters. To model interfaces between materials with different lattice constants and disorder, we use *lateral supercells.*

*a. Lateral Supercells* The TB-MTO scheme is computationally very efficient and allows us to use large lateral supercells to model in a simple fashion interface disorder, interfaces between materials whose underlying lattices are incommensurate, or quantum point contacts. This treatment becomes formally exact in the limit of infinitely large supercells. In practice, satisfactory convergence is achieved for supercells of quite moderate size.

The use of lateral supercells makes it possible to analyse diffuse scattering particularly simply. We consider an $H_1 \times H_2$ lateral supercell defined by the real-space lattice vectors

$$\vec{A}_1 = H_1 \vec{a}_1 \quad \text{and} \quad \vec{A}_2 = H_2 \vec{a}_2 \tag{138}$$

where $\vec{a}_1$ and $\vec{a}_2$ are the lattice vectors describing the in-plane periodicity of a primitive unit cell (Fig. 20). The cells contained within the supercell are generated by the set of translations

$$\left\{ \vec{T}_\parallel = h_1 \vec{a}_1 + h_2 \vec{a}_2 \, ; 0 \leq h_1 < H_1, : 0 \leq h_2 < H_2 \right\} . \tag{139}$$

In reciprocal space the supercell Brillouin zone is defined by the reduced vectors

$$\vec{B}_1 = \vec{b}_1 / H_1 \quad \text{and} \quad \vec{B}_2 = \vec{b}_2 / H_2 \tag{140}$$

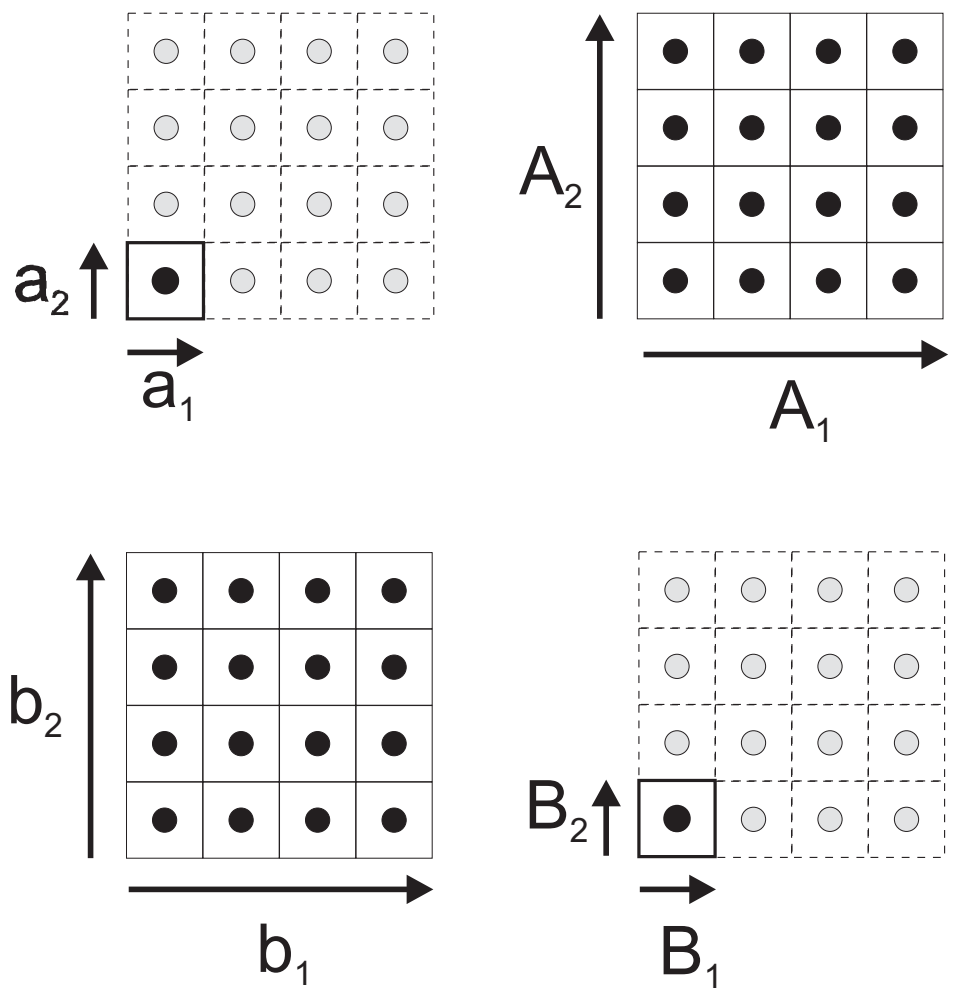

FIG. 20: Illustration of lateral supercells and corresponding 2D interface Brillouin zones. Top panel: lattice vectors for a primitive unit cell containing a single atom (lhs) and a $4 \times 4$ supercell (rhs). Bottom panel: a single k-point in the BZ (rhs) corresponding to the $4 \times 4$ real-space supercell is equivalent to $4 \times 4$ k-points in the BZ (lhs) corresponding to the real-space primitive unit cell.

where $\vec{b}_1$ and $\vec{b}_2$ are the reciprocal lattice vectors corresponding to the real space primitive unit cell. As a result the Brillouin zone (BZ) is folded down, as shown schematically in Fig. 20 (bottom rhs), and the single $\vec{k}_{\|}^{\mathbb{S}}$ point ($\mathbb{S}$ is used to label supercell quantities) in the supercell BZ corresponds to the set of $H_1 \times H_2$ $\vec{k}$ points in the original unfolded BZ

$$\left\{\vec{k}_{\|} = h_1\vec{B}_1 + h_2\vec{B}_2 \; ; 0 \leq h_1 < H_1, : 0 \leq h_2 < H_2\right\}. \tag{141}$$

Solutions associated with different $\vec{k}_{\|}$ in the primitive unit cell representation become different "bands" at the single $\vec{k}_{\|}$ in the supercell representation.

The lead states are calculated using the translational symmetry of the primitive unit cell so that the computational effort scales linearly with the size of the supercell *i.e.* as $(H_1 \times H_2)$ rather than as $(H_1 \times H_2)^3$ which is the scaling typical for matrix operations. Another advantage is that it enables us to analyse the scattering. By keeping track of the relation between supercell "bands" and equivalent eigenmodes at different $\vec{k}_{\|}^{\mathbb{S}}$ (Fig. 20) we can straightforwardly obtain $t_{mn}(\vec{k}_{\|1}, \vec{k}_{\|2})$ and other scattering coefficients. In other words the "interband" specular scattering in the supercell picture translates, in the presence of disorder in the scattering region, into the "diffuse" scattering between the $\vec{k}_{\|}$ vectors belonging to the set (141). Since this approach is formally only valid if sufficiently large supercells are used, we begin by studying how the interface conductance depends on the lateral supercell size.

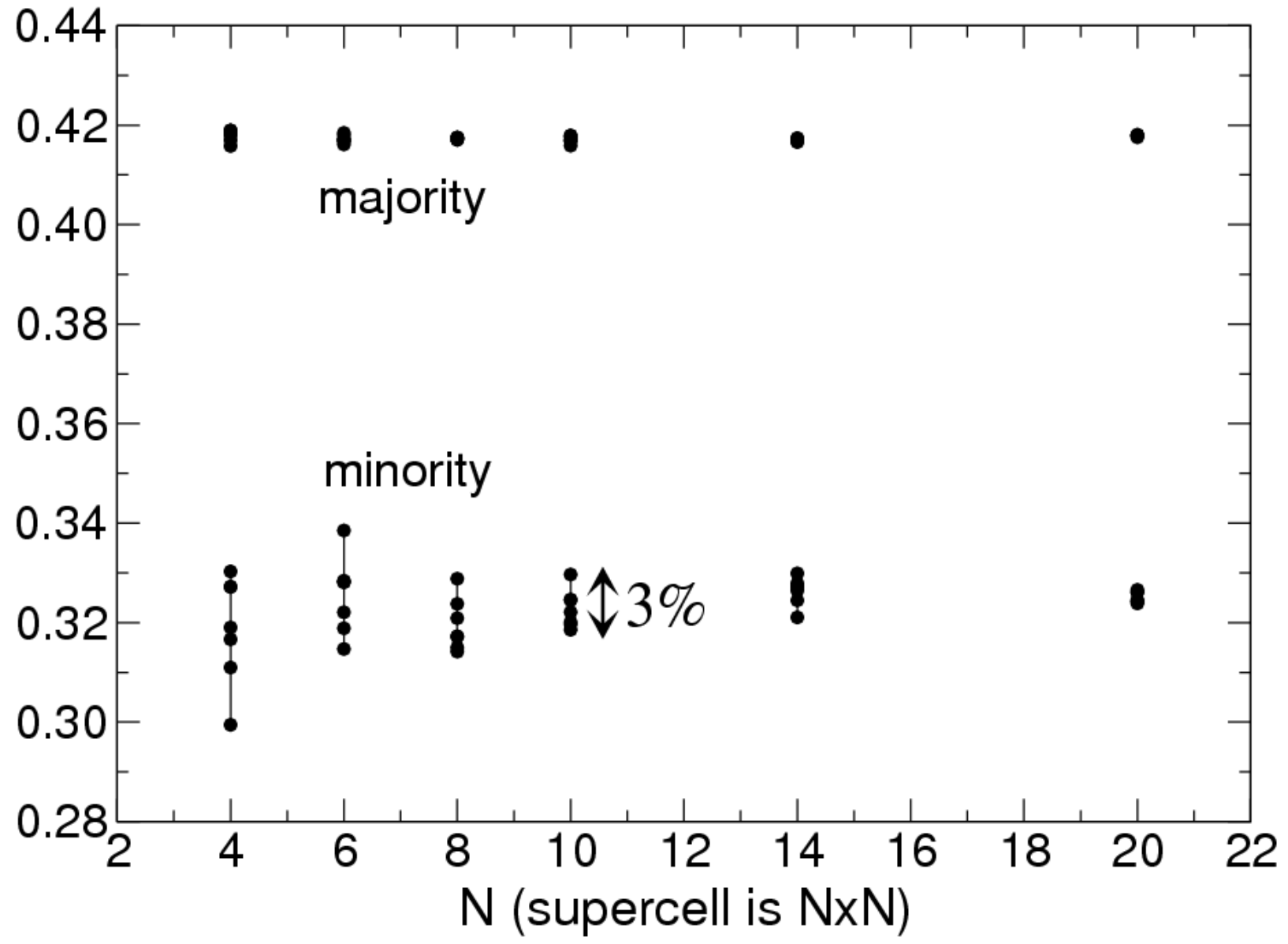

FIG. 21: Interface conductance of a Cu|Co(111) interface for different configurations of disorder as a function of N where the lateral unit cells contain $N \times N$ sites. These results are for an *fcc* lattice constant of $a = 3.549\mathring{A}$, an *spd* basis and Perdew-Zunger exchange-correlation potential.

To perform fully self-consistent calculations for a number of large lateral supercells and for different configurations of disorder would be prohibitively expensive. Fortunately, the coherent potential approximation (CPA) is a very efficient way of calculating charge and spin densities for a substitutional disordered $A_xB_{1-x}$ alloy with an expense comparable to that required for an ordered system with a minimal unit cell [159]. The output from such a calculation are atomic sphere potentials for the two sites, $\upsilon_A$ and $\upsilon_B$. The layer CPA approximation generalizes this to allow the concentration to vary from one layer to the next [158].

Once $\upsilon_A$ and $\upsilon_B$ have been calculated for some concentration $x$, an $N \times N$ lateral supercell is constructed in which the potentials are distributed at random, maintaining the concentration for which they were calculated self-consistently. For a given value of $N$, a number of such random distributions is generated. The conductances calculated for $4 \leq N \leq 20$ are shown in Fig. 21 for a Cu|Co(111) interface in which the Cu and the Co layers forming the interface are totally mixed

to give two layers of 50%-50% interface alloy. The sample-to-sample variation is largest for the minority spin case, ranging from $\pm 5\%$ for a modest $4 \times 4$ unit cell and decreasing to about $\pm 0.5\%$ for a $20 \times 20$ unit cell. For $N \sim 10$, the spread in minority spin conductances is $\sim 3\%$ which is comparable to the typical uncertainty we associated with the LDA error, the uncertainty in lattice constants or the error incurred by using the ASA.

Comparing the conductances without and with disorder, we see that disorder has virtually no effect on the majority spin channel (0.42 versus $0.42 \times 10^{15} \Omega^{-1}\text{m}^{-2}$) which is a consequence of the great similarity of the Cu and Co majority spin potentials and electronic structures. However, in the minority-spin channel the effect (0.38 versus $0.32 \times 10^{15}\, \Omega^{-1}\text{m}^{-2}$) is much larger. Equation (137) can be used to obtain estimates of the interface resistances $R_{\sigma}^{Schep}$, which can be compared to values extracted from experiment [58, 160]. A comparison of theoretical values derived in this way with experiment is reproduced from Ref. [62] in Table III.

The results in Table III were obtained for disordered interfaces comprising two monolayers (2ML) of 50%-50% alloy, a model derived from X-ray [162] NMR [163, 164], and magnetic EXAFS [165] studies. Though plausible, it does contain large uncertainties. In [166] the effect of changing the interface alloy concentration and the number of layers containing disorder was studied for the Cu|Co(111) interface. As expected, the minority spin electrons are much more sensitive to the details of the interface disorder than the majority spin electrons. Nevertheless, the interface resistances calculated for the minority spin channel using Equation (137) lie within the range of values extracted from experiment for sputtered and MBE grown multilayers and tabulated in Table I of Ref. [160].

The differences between $AR^{Schep}$ and $AR^{LB}$ are very significant for highly transparent interfaces. The agreement of the computed interface resistances with experiments, which was already found to be good for specular Cu|Co interfaces [61, 119], is improved for minority spins by including interface disorder [35]. However, for the majority-spin case, the calculated interface resistance is *larger* than the value extracted from experiment, whether or not interface disorder is included. We will return to this again briefly after discussing how the interface scattering can be analysed.

### *3. Analysis of Interface Disorder Scattering: Cu|Co*

When disorder is modelled using lateral supercells, the transmission matrix elements can be categorized as being either *ballistic* or *diffuse*, depending upon whether or not transverse momentum is conserved. The scattering induced by two layers of 50%-50% alloy is illustrated in Figs. 22 and 23 for the majority and minority spins, respectively, of a Cu|Co(111) interface. The calculations were performed for a single $\mathbf{k}_{\|}$ point, $\Gamma$, and a $20 \times 20$ lateral supercell, equivalent

| Interface | Roughness | $AR_{maj}^{Schep}$ | $AR_{min}^{Schep}$ | $AR_{maj}^{LB}$ | $AR_{min}^{LB}$ |
|---|---|---|---|---|---|
| Au/Ag(111) | Clean | 0.094 | 0.094 | 2.41 | 2.41 |
| Au/Ag(111) | 2 layers 50-50 alloy | 0.118 | 0.118 | 2.43 | 2.43 |
| Au/Ag(111) | Exp. [160] | $0.100 \pm 0.008$ | $0.100 \pm 0.008$ | | |
| Cu/Co(100) | Clean | 0.33 | 1.79 | 2.27 | 3.11 |
| Cu/$Co_{hcp}$(111) | Clean | 0.60 | 2.24 | 2.67 | 3.65 |
| Cu/Co(111) | Clean | 0.39 | 1.46 | 2.39 | 2.80 |
| Cu/Co(111) | 2 layers 50-50 alloy | 0.41 | 1.82±0.03 | 2.40 | 3.14 |
| Cu/Co(111) | Exp. [160] | 0.26±0.06 | 1.84±0.14 | | |
| Cr/Fe(100) | Clean | 2.82 | 0.50 | 3.51 | 1.45 |
| Cr/Fe(100) | 2 layers 50-50 alloy | 0.99 | 0.50 | 1.68 | 1.45 |
| Cr/Fe(110) | Clean | 2.74 | 1.05 | 4.22 | 3.17 |
| Cr/Fe(110) | Clean [119] | 2.11 | 0.81 | | |
| Cr/Fe(110) | 2 layers 50-50 alloy | 2.05 | 1.10 | 3.53 | 3.22 |
| Cr/Fe(110) | Exp. [161] | 2.7±0.4 | 0.5±0.2 | | |

TABLE III: Interface resistances, in units of ($\mathrm{f\Omega m^2}$), for a number of commonly studied interfaces. $AR_{\sigma}^{Schep}$ are theoretical values calculated using Eq. (137); $AR_{\sigma}^{LB}$ are theoretical values obtained using "bare" transmission probabilities. Some of the values differ from those which can be calculated using interface and Sharvin conductances quoted in the text because other lattice constants, exchange-correlation potentials, etc. were used.

to using a $1 \times 1$ interface cell and k-space sampling with $20 \times 20$ points in the corresponding BZ. Figs. 22(a) and 22(b) show the majority-spin Fermi surface projections of *fcc* Cu and Co, respectively, and the coarse $20 \times 20$ grid is seen to yield a good representation of the detailed Fermi surface projections shown in Fig. 18. The probability, $T(\mathbf{k}_{\|}, \mathbf{k}'_{\|})$, that a state in Cu with transverse momentum $\mathbf{k}_{\|}$ is scattered on transmission through the disordered interface into the state in Co with transverse momentum $\mathbf{k}'_{\|}$, is shown in Fig. 22(c) for $\vec{k}_{\|} = Y$ on the $k_y$ axis (see the inset in Fig. 19) and is dominated by the $\mathbf{k}_{\|}$-conserving forward scattering, the *specular transmission*: $T(\vec{k}_{\|} = Y, \vec{k}_{\|} = Y) = 0.93$. Indeed, the diffuse scattering is so weak that it cannot be seen on a scale of $T$ from 0 to 1. To render it visible, a magnification by a factor 500 is needed, shown in Fig. 22(d). The total diffuse scattering, $T_{diff}(Y) = \sum_{k'_{\|} \neq k_{\|}} T(\vec{k}_{\|} = Y, \vec{k}_{\|} \neq Y) = 0.04$ can be seen from the

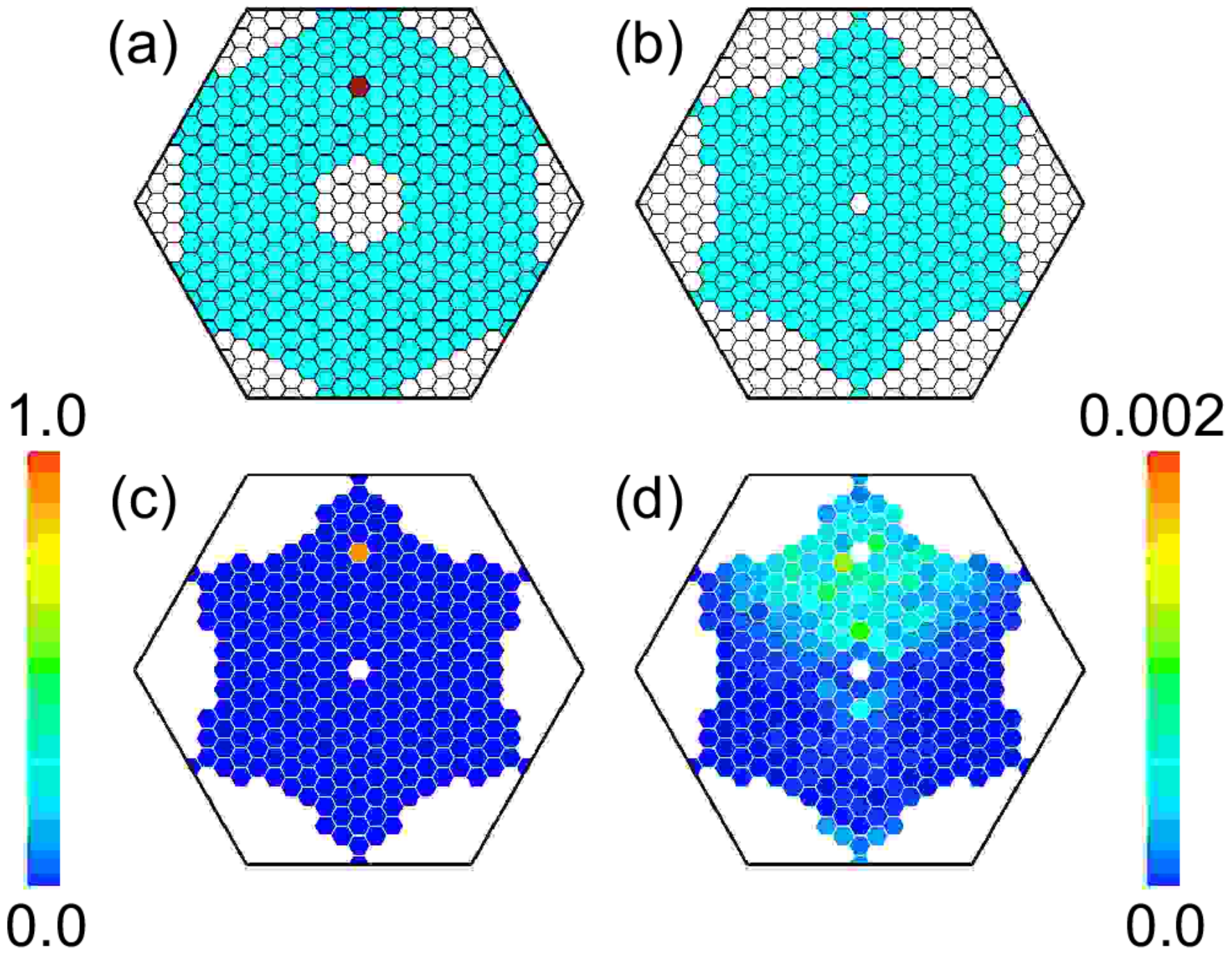

FIG. 22: Fermi surface projections of majority-spin *fcc* Cu (a) and Co (b) derived from a single k-point using a $20 \times 20$ lateral supercell. The honeycomb lattice represents the corresponding $20 \times 20$ mesh of $\mathbf{k}_{\|}$ points in reciprocal space. The red point in the Cu Fermi surface projection corresponds to the point $Y$ in Fig. 19. $T(Y, \mathbf{k}'_{\|})$ is shown in (c) and magnified by a factor 500 in (d) where the ballistic component $T(Y, \mathbf{k}'_{\|} = Y)$ is indicated by a white point because its value goes off the scale.

figure to be made up of contributions of $T \sim 0.0004$ per $\vec{k}_{\|}$-point from roughly a quarter of the BZ (100 $\vec{k}_{\|}$ points) centred on $\vec{k}_{\|} = Y$. The total transmission, $T_{total} = T_{spec} + T_{diff} = 0.93 + 0.04 = 0.97$, compared to a transmission of 0.99 in the absence of disorder. In the majority case, there is thus a strong specular transmission peak surrounded by a weak diffuse background.

The minority-spin Fermi surface projections of *fcc* Cu and Co are shown in Figs. 23(a) and 23(b), respectively. Compared to the corresponding panels in Fig. 19, the $20 \times 20$ point representation is seen to be sufficient to resolve the individual Fermi surface sheets of Co. To study the effect of scattering, we consider two different situations. In the first, we again consider $\vec{k}_{\|} = Y$, for which the transmission in the absence of disorder was zero as a result of the symmetry of the states along the $k_y$ axis. $T(\vec{k}_{\|} = Y, \vec{k}_{\|})$ is shown in Fig. 23(c). By contrast with the majority-spin case just examined, there is now scattering to all other k-points in

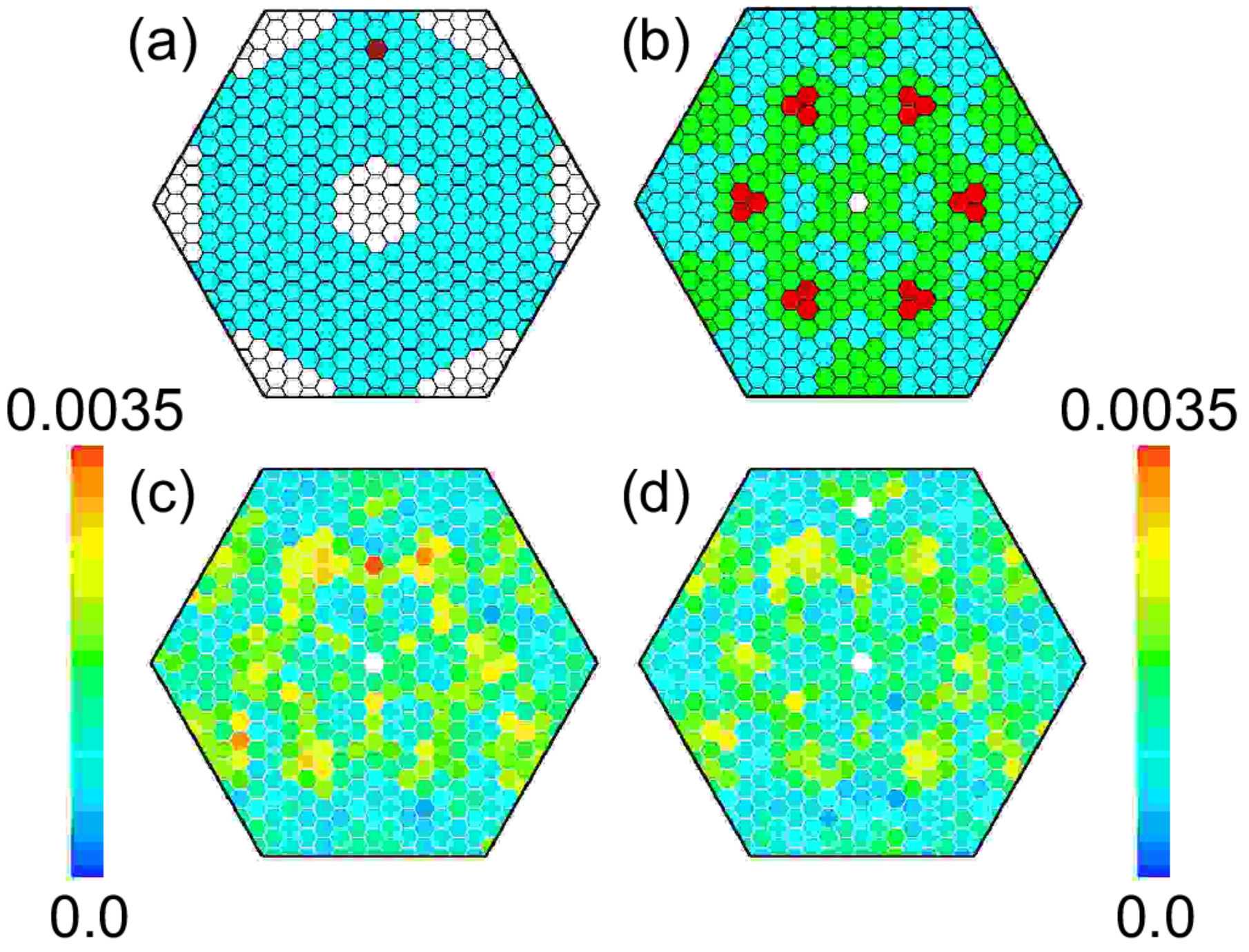

FIG. 23: Fermi surface projections of minority-spin *fcc* Cu (a) and Co (b) derived from a single k-point using a $20 \times 20$ lateral supercell. The red circle in the Cu Fermi surface projection (a) corresponds to the point $Y'$ in Fig. 19. (c) $T(Y, \mathbf{k}'_{\|})$ and (d) $T(Y', \mathbf{k}'_{\|})$; the ballistic component $T(Y', \mathbf{k}'_{\|} = Y')$ is indicated by a white point because its value goes off the scale.

the 2D BZ, $\sum_{k'_{\|} \neq k_{\|}} T(\vec{k}_{\|} = Y, \vec{k}_{\|} \neq Y) = 0.58$. The specular transmission $T(Y, Y)$ has increased from 0.00 in the clean case, to only 0.01 in the presence of disorder. The effect of disorder is to increase the total transmission, $T_{total}(Y) = \sum_{k'_{\|}} T(\vec{k}_{\|} = Y, \vec{k}_{\|})$ from 0.00 to $T_{spec}(Y) + T_{diff}(Y) = 0.01 + 0.58 = 0.59$. Disorder thus *increases* the transmission for states which were strongly reflected in its absence.

The second case we consider is that of a k-point slightly further away from the origin $\Lambda$ along the $k_y$ axis which had a high transmission, $T(Y') = 0.98$, in the absence of disorder. For this k-point, $T(\vec{k}_{\|} = Y', \vec{k}_{\|})$, shown in Fig. 23(d), looks very similar to Fig. 23(c). There is strong diffuse scattering with $\sum_{k'_{\|} \neq k_{\|}} T(\vec{k}_{\|} = Y', \vec{k}_{\|} \neq Y') = 0.54$ while $T(Y', Y')$ has been drastically decreased from 0.98 in the clean case, to 0.06 as a result of disorder. The total transmission, $T_{total}(Y') = T_{spec}(Y') + T_{diff}(Y') = 0.06 + 0.54 = 0.60$, is almost identical to what was found for the $Y$ point. The effect of disorder has been to *decrease* the transmission of states which, in its absence, were weakly reflected. The strong k-dependence of the transmission found for the minority-spin channel in the specular case is largely destroyed by a small amount of disorder.

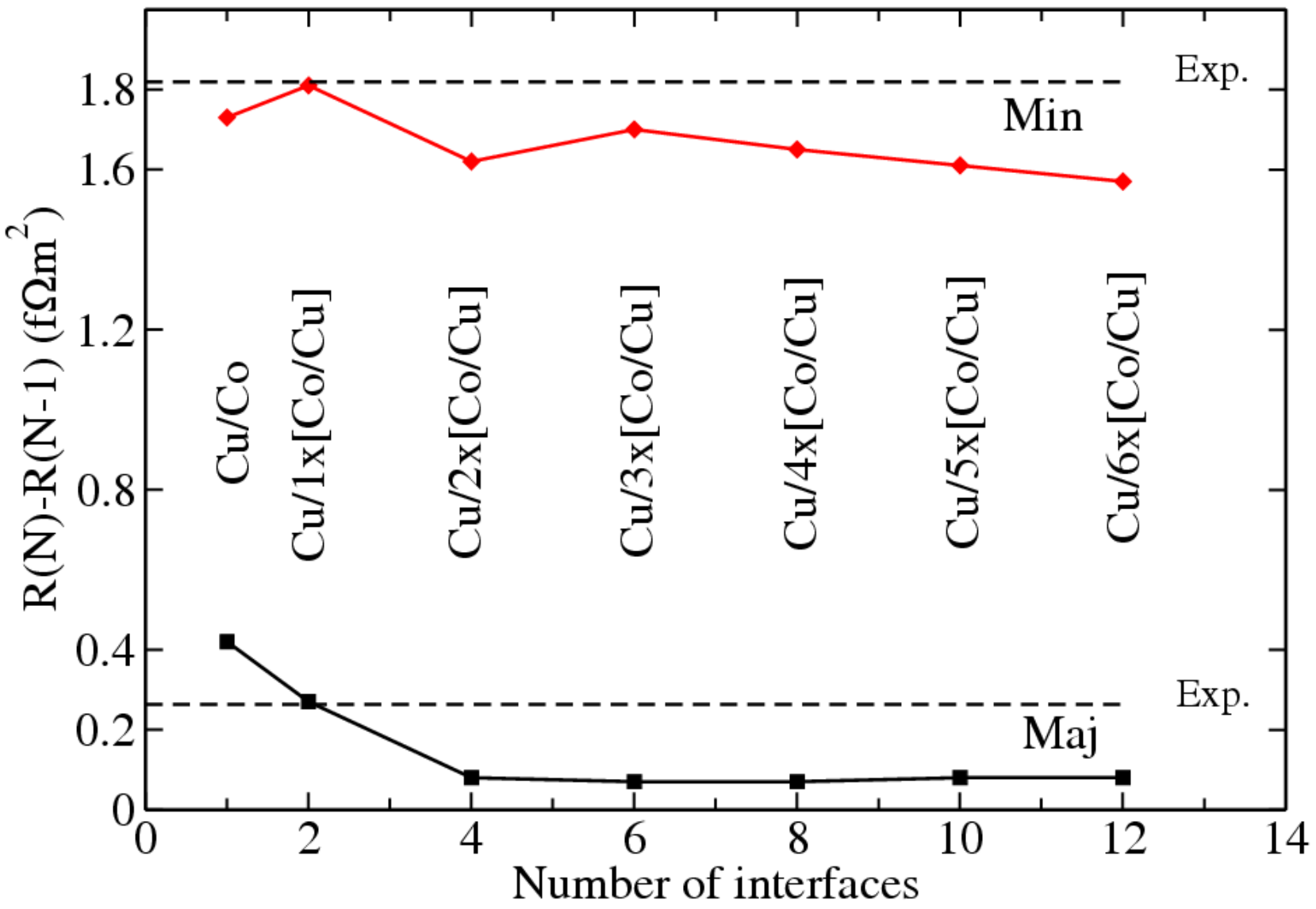

FIG. 24: Differential interface resistance as the number of interfaces increase for a disordered Cu|Co(111) multilayer embedded between Cu leads. An $8 \times 8$ lateral supercell was used and the interface was modelled as two layers of 50%-50% alloy.

To derive (137), it was assumed that there is loss of phase coherence between adjacent interfaces. The scattering of majority-spin electrons at the Cu|Co(111) interface is so weak that this assumption is not obviously valid. It can be examined by seeing whether or not the total resistance of a Cu|Co multilayer containing N interfaces scales linearly with N. If this is so, then the incremental interface resistance $R(N) - R(N-1)$ should be independent of $N$. Note that the lead correction in (137) is independent of N and thus drops out of the incremental resistance.

The results of an extensive series of calculations for a disordered $Cu_{10}|Co_{10}(111)$ multilayer attached to Cu leads are shown in Fig. 24 which includes experimental values [160] for comparison. An $8 \times 8$ lateral supercell was used so that, according to Fig. 21, the error on the interface transmission due to configuration averaging is negligible for the majority-spin case and of order 5% for the minority spins. In the largest calculation represented in Fig. 24, the scattering region contained about 7200 atoms. For the strongly scattered minority spins, the interface resistance is essentially constant for all N (within the error bar of the calculation set by the configuration averaging, choice of exchange-correlation potential etc.). For the majority spin case there is a sharp drop before the incremental resistance levels off to a constant value for $N \geq 4$ which is a factor 3 to 4 lower than the experimental value. This implies a breakdown of the series resistor model for the majority spins or else there is some other source of majority-spin scattering which has not been included in the calculation. Including bulk scattering does not change this result significantly. Other materials specific studies of the transport properties of Cu|Co multilayers carried out with other methods [75] also find that substitutional disorder alone cannot account for the reported resistivities in the majority-spin channel.

*a. Cr|Fe* An extreme example of how interface disorder can *enhance* interface transmission is found for the bcc(100) orientation of Cr|Fe. Whereas the majority-spin band structures were well matched in the case of Cu|Co, the situation is reversed for Cr|Fe and it is the minority-spin electronic structures which match well. Calculating the interface resistance using (137), we find values of 2.82 f$\Omega$m$^2$ and 0.50 f$\Omega$m$^2$ for majority and minority spin, respectively, in the absence of disorder and values of 0.99 f$\Omega$m$^2$ and 0.50 f$\Omega$m$^2$, respectively when interface disorder is modelled as two layers of 50-50 alloy as was done for Cu|Co. Thus disorder reduces the majority-spin interface resistance by almost a factor of three while having only a small effect on the well-matched minority spin channel just as in the Cu|Co majority spin case.

The qualitative difference between Cr|Fe and Cu|Co can be understood in terms of their Fermi surface projections and transmission probabilities $T_{\mathcal{LR}}(\vec{k}_\parallel)$. In the Cu|Co(111) majority-spin case (Fig. 18), there was a large area of the 2D BZ where states on both sides matched very well and interface disorder led to mainly forward scattering with virtually no reduction of the total transmission. In the minority-spin case (Fig. 19), the situation was more complicated because the average transmission was much lower in the absence of disorder ($\sim$ 60%) and the disorder-induced reduction of the interface transmission ($\sim$ 20%) resulted from two competing effects: transmission enhancement by symmetry-breaking for channels which were closed for reasons of symmetry and transmission reduction by diffuse scattering for channels which were very transparent in the absence of disorder. On balance, defect scattering reduced the transmission probability and thus *increased* the interface resistance of the Cu|Co minority spin channel.

In spite of there being multiple sheets of Cr and majority-spin Fe Fermi surfaces which overlap in large regions of the 2D Brillouin zone (see Figs. 25(a-c)), the average majority-spin transmission probabilities in the absence of disorder, shown in Fig. 25(d), are very small throughout the BZ while those for the minority-spin case, shown in Fig. 25(e), are quite substantial. As in the Cu|Co case, there are two mechanisms by which interface disorder *increases* the interface transmission.

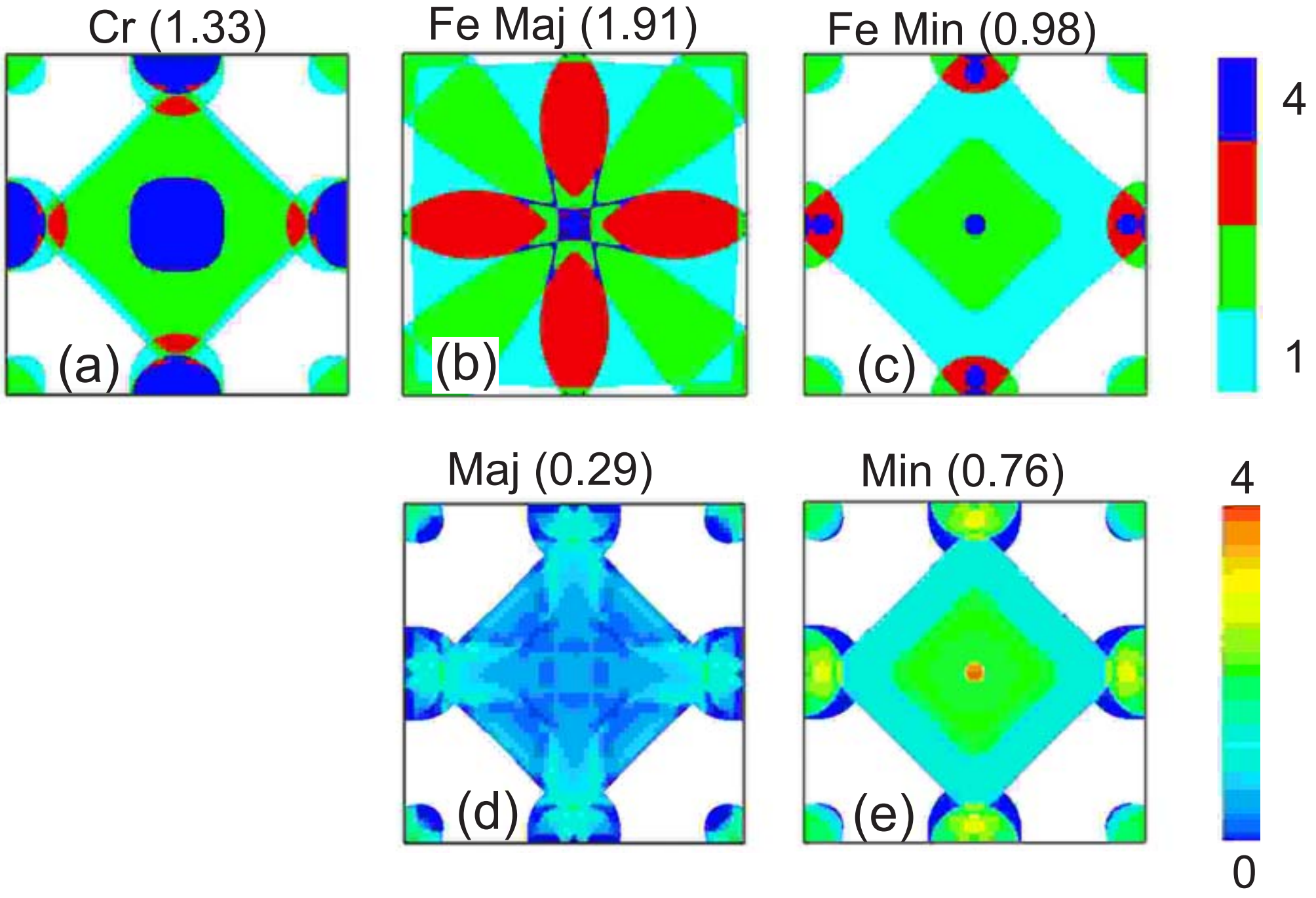

FIG. 25: Fermi surface projections of *bcc* (a) non-magnetic Cr, (b) Fe, majority-spin and (c) Fe, minority-spin. (d) and (e) show the majority- respectively, minority-spin interface transmissions for a clean lattice-matched *bcc* Cr|Fe interface. The result of integrating the number of channels over the whole Brillouin zone is given in brackets at the top of each panel.

Majority-spin electrons with small $\vec{k}_{\|}$ are almost completely reflected at the clean Cr|Fe interface because the electronic states on both sides of the interface do not match well. Defect scattering is found to increase the transmission of these electrons strongly. Furthermore, for large areas of the two-dimensional Brillouin zone, there are no propagating states on the Cr side. Propagating modes in Fe with these values of $\vec{k}_{\|}$, which were inaccessible to Cr electrons in the ordered case, can be reached by diffusive scattering. This opens up a large number of new channels and this increase in transmission for the Cr|Fe majority-spin channel translates into a reduction by a factor 3 of the interface resistance. Recent CPP experiments on Fe/Cr(110) [161] show that the spin-averaged resistance agrees very well with the theoretical prediction, but not the polarization dependence.

## C. Mixing conductances and spin torque

The real and imaginary parts of the complex mixing conductance introduced in Section IV for an N|F interface describe how the interface reflects the different components of a vector spin current incident from the non-magnetic side. If, instead of a single interface, we consider an N|F|N sandwich structure containing a magnetic layer of finite thickness $d$, then it also becomes necessary to consider how a spin current may be transmitted from the non-magnetic layer on the left into that on the right. To do this, we consider how a spin current flows through the entire system in response to an externally applied spin accumulation $\vec{\mu}$ induced in the left lead only and the ferromagnet is magnetized along the $z$ axis. The spin current incident on the interface from the left is then proportional to the number of incoming channels in the lead $\vec{I}^{\rm L}_{\rm in} = g^{\rm Sh}_N \vec{\mu}$. The reflected spin current is given by

$$\vec{I}^{\rm L}_{\rm out} = \frac{1}{2\pi} \begin{pmatrix} g^{\rm Sh}_N - {\rm Re} g^r_{\uparrow\downarrow} & -{\rm Im} g^r_{\uparrow\downarrow} & 0 \\ {\rm Im} g^r_{\uparrow\downarrow} & g^{\rm Sh}_N - {\rm Re} g^r_{\uparrow\downarrow} & 0 \\ 0 & 0 & g^{\rm Sh}_N - (g^\uparrow + g^\downarrow)/2 \end{pmatrix} \vec{\mu} \tag{142}$$

where

$$g^\sigma = \sum_{nm} |t^\sigma_{nm}|^2 \tag{143}$$

are the conventional Landauer-Büttiker conductances and

$$g^r_{\uparrow\downarrow} = S^{-1} \sum_{mn} \left( \delta_{mn} - r^\uparrow_{mn} r^{\downarrow\star}_{mn} \right), \tag{144}$$

is the *reflection* mixing conductance introduced in Section IV. The transverse component of the reflected spin current is seen to be determined by the real and imaginary parts of $g^{\rm Sh}_N - g^r_{\uparrow\downarrow} = \sum_{mn} r^\uparrow_{mn} r^{\downarrow\star}_{mn}$. The transmitted spin current is given by [167]

$$\vec{I}^{\rm R}_{\rm out} = \frac{1}{2\pi} \begin{pmatrix} {\rm Re} g^t_{\uparrow\downarrow} & {\rm Im} g^t_{\uparrow\downarrow} & 0 \\ -{\rm Im} g^t_{\uparrow\downarrow} & {\rm Re} g^t_{\uparrow\downarrow} & 0 \\ 0 & 0 & (g^\uparrow + g^\downarrow)/2 \end{pmatrix} \vec{\mu} \tag{145}$$

where the *transmission* mixing conductance $g^t_{\uparrow\downarrow}$, which describes the transverse component of the transmitted spin current which is subject to precession and absorption within the magnetic layer, is given by

$$g^t_{\uparrow\downarrow} = S^{-1} \sum_{mn} t'^{\uparrow}_{mn} t'^{\downarrow\star}_{mn}. \tag{146}$$

Here, $S$ is the N|F contact area, conductances are given per unit contact area, and $m$ and $n$ denote scattering states at the Fermi energy of the normal-metal

leads. Thus we see that the reflection and transmission mixing conductances, which determine how much of an incoming spin angular momentum flux is lost at an interface with a magnetic layer of finite thickness, are defined in terms of the spin-dependent reflection and transmission matrices we calculated in the last section. We now proceed to determine typical values of $g^t_{\uparrow\downarrow}$ and $g^r_{\uparrow\downarrow}$ and to study their dependence on magnetic layer thickness and on interface disorder.

In order to do so, we consider two representative N|F interfaces: the Cu|Co(111) interface considered in the last chapter and Au|Fe(001). Because both of these systems are nearly ideally lattice matched, we assume common lattice constants for both metals of a given structure: $a_{\mathrm{Cu|Co}} = 3.549$ Å and $a_{\mathrm{Au|Fe}} = \sqrt{2} \times 2.866$ =4.053 Å. The same two step procedure of performing self-consistent LSDA calculations for the potentials and subsequent calculation of the scattering matrix using the potentials as input was followed. The atomic-sphere (AS) potentials of 4 monolayers on either side of the magnetic layer (or interface) were iterated to self-consistency while the potentials of more distant layers were held fixed at their bulk values. Disordered interfaces were modelled with one atomic layers of a 50%-50% alloy (for N|F|N systems) or two (for single N|F interfaces). While this model is probably reasonable for $\mathrm{Cu_{fcc}|Co_{fcc}}$ because of the nearly perfect lattice match and structural compatibility, the situation is more complicated for $\mathrm{Au_{fcc}|Fe_{bcc}}$ because of the large difference in AS sizes for Au and Fe with Wigner-Seitz radii of 2.99 and 2.67 Bohr atomic units, respectively. We have assumed here that the disorder is only substitutional and that the diffused atoms occupy the AS of the same size as that of the host element. In the Au|Fe|Au case, where the alloy is only one atomic monolayer (ML) thick, we assume that the Fe atoms diffuse into Au. While the validity of this model can be questioned, the insensitivity of the final results to the details of the disorder (e.g. one versus two monolayers of alloy) indicate that this is not a critical issue. The layer-resolved magnetic moments for single interfaces are given in Table IV.

The two-dimensional Brillouin zone (2D BZ) summation required to calculate the mixing conductances using Eqs. (144) and (146) was performed with $k_{||}$-mesh densities equivalent to $10^4$ points in a 2D BZ of a $1 \times 1$ interface unit cell. The uncertainties resulting from this BZ summation and from impurity ensemble averaging are of the order of a few times $10^{12}\ \Omega^{-1}\mathrm{m}^{-2}$, which is smaller than the size of the symbols used in the figures.

### *1. Calculated Mixing Conductances*

Figures 26 to 29 show how $G^r_{\uparrow\downarrow} = (e^2/h)g^r_{\uparrow\downarrow}$ and $G^t_{\uparrow\downarrow} = (e^2/h)g^t_{\uparrow\downarrow}$ depend on the thickness $d$ of the magnetic layer (measured in atomic layers) for specular ($\vec{k}_{||}$-preserving) Au|Fe|Au(001) and Cu|Co|Cu(111) systems. Both quantities exhibit quasi-oscillatory behaviour with, however, noticeably different periods and ampli-

| N\|M | Au\|Fe | | Cu\|Co | |
|---|---|---|---|---|
| Layer | clean | dirty | clean | dirty |
| $m_N$(bulk) | 0.000 | 0.000 | 0.000 | 0.000 |
| $m_N$(int-3) | 0.000 | 0.000 | 0.001 | 0.000 |
| $m_N$(int-2) | 0.001 | -0.003 | -0.000 | -0.003 |
| $m_N$(int-1) | -0.002 | 0.010 | -0.004 | -0.003 |
| $m_N$(int) | 0.064 | 0.026 | 0.006 | 0.010 |
| $m_F$(int) | - | 2.742 | - | 1.410 |
| $m_N$(int) | - | 0.128 | - | 0.036 |
| $m_F$(int) | 2.687 | 2.691 | 1.545 | 1.540 |
| $m_F$(int-1) | 2.336 | 2.396 | 1.635 | 1.596 |
| $m_F$(int-2) | 2.325 | 2.363 | 1.621 | 1.627 |
| $m_F$(int-3) | 2.238 | 2.282 | 1.627 | 1.624 |
| $m_F$(bulk) | 2.210 | 2.210 | 1.622 | 1.622 |

TABLE IV: Layer-resolved magnetic moments in Bohr magnetons for single N|M interfaces (N=Au,Cu; F=Fe,Co). The magnetic moments for the Cu|Co(111) interface differ slightly from those given in Table I because a different exchange-correlation potential (Perdew-Zunger) was used.

tudes. The values of both $G^r_{\uparrow\downarrow}$ and $G^t_{\uparrow\downarrow}$ are determined by two factors: the matching at the interface of the normal metal and ferromagnetic metal states (described by the scattering coefficients of the single interface) and the phases accumulated by electrons on their passage through the magnetic layer (quantum-size effect). The first factor determines the amplitudes of the oscillations and (for $G^r_{\uparrow\downarrow}$) the asymptotic values, while the second is responsible for the observed quasi-periodicity. In order to better understand this, it is instructive to interpret the transmission and reflection coefficients of the finite-size magnetic layer in terms of multiple scattering at the interfaces. We first note that both Cu and Au have only one left- and one right-going state at the Fermi level for each value of $\vec{k}_{||}$ and spin so that the summations in Eqs. (144) and (146) reduce to integrations over the 2D BZ involving the complex-valued functions $r^\sigma(\vec{k}_{||})$ and $t^\sigma(\vec{k}_{||})$. Retaining only lowest-order thickness-dependent terms, dropping explicit reference to $\vec{k}_{||}$, and dropping the primes on $t^{'}$, we then have

$$t^\sigma \approx t^\sigma_{F\to N}\Lambda^\sigma t^\sigma_{N\to F} \tag{147}$$

$$r^\sigma \approx r^\sigma_{N\to N} + t^\sigma_{F\to N}\Lambda^\sigma r^\sigma_{F\to F}\Lambda^\sigma t^\sigma_{N\to F} \tag{148}$$

where $t^\sigma_{N\to F} = (t^\sigma_1, \ldots, t^\sigma_n)^T$ is a vector of transmission coefficients between a single propagating state in the normal metal and a set of states in the ferromagnet, $\Lambda^\sigma$ is a diagonal matrix of phase factors $e^{ik^{\sigma}_{j\perp}d}$ ($j$ is an index of the states in the ferromagnet), $r^{\sigma}_{N\to N}$ is a scalar reflection coefficient for states incoming from the normal metal and $r^{\sigma}_{F\to F}$ is a square matrix describing reflection on the ferromagnetic side. The set of states in the ferromagnet consists of both propagating and evanescent states. The contribution of the latter decreases exponentially with the thickness of the layer.

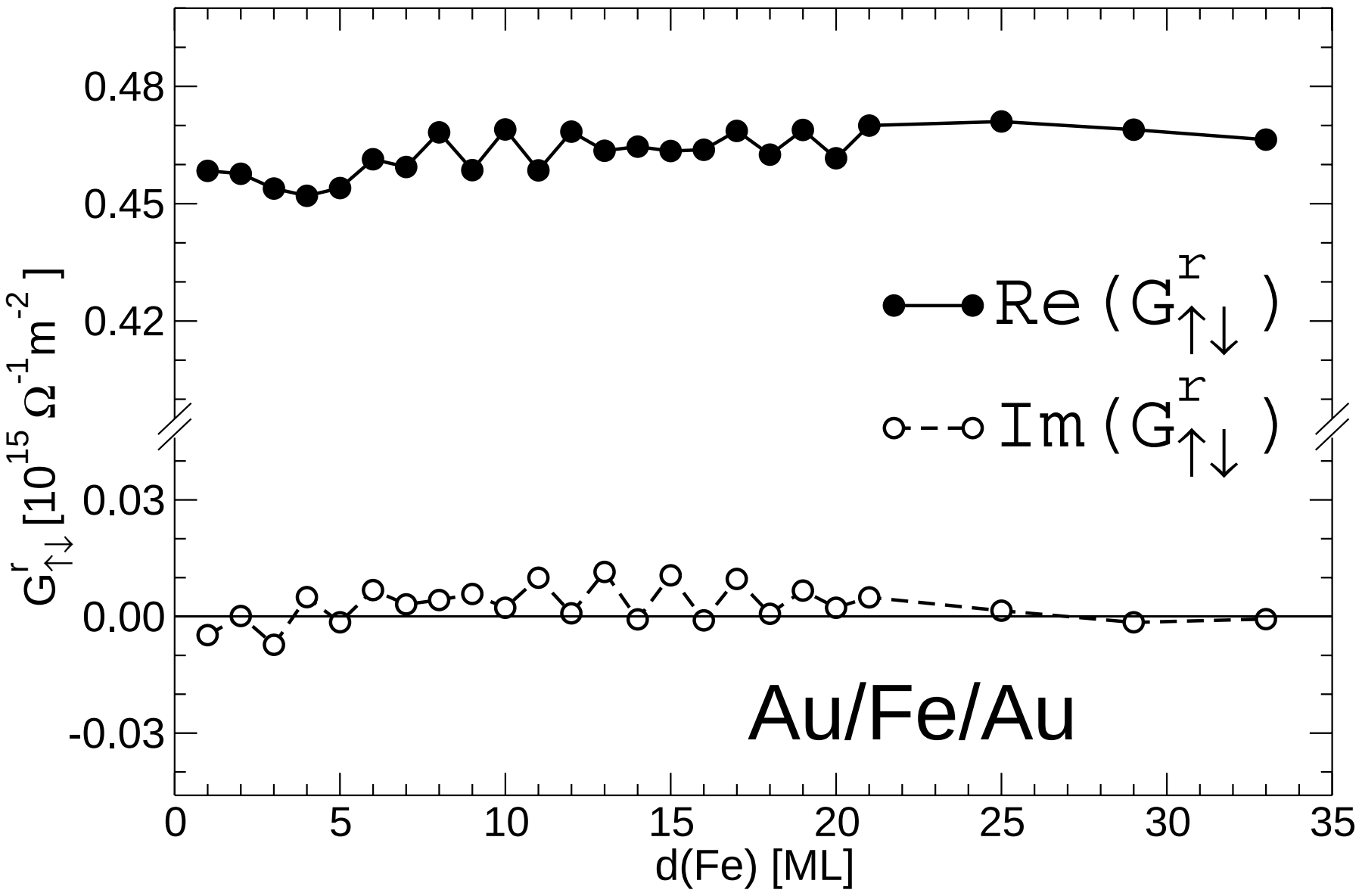

FIG. 26: Reflection spin-mixing conductance (per unit area) of a Au|Fe|Au(001) trilayer with perfect interfaces as a function of the thickness $d$ of the Fe layer. In this and subsequent plots, mixing conductances expressed in terms of number of conduction channels per unit area are converted to $\Omega^{-1}\mathrm{m}^{-2}$ using the conductance quantum $e^2/h$, i.e. $G_{\uparrow\downarrow} = (e^2/h)g_{\uparrow\downarrow}$.

Concentrating first on the thickness dependence of $g^t_{\uparrow\downarrow}$, we notice that, in view of Eq. (147), the summation in Eq. (146) is carried out over terms containing phase factors $e^{i(k^{\uparrow}_{i\perp}-k^{\downarrow}_{j\perp})d}$. Because of the large differences between majority and minority Fermi surfaces of the ferromagnet, this typically leads to rapidly oscillating terms which mostly cancel out on summing over $\vec{k}_{||}$. It can be argued [69] in the spirit of the theory of interlayer exchange coupling [65] that the only long-range contributions originate from the vicinity of points for which $\nabla_{k_{||}}(k^{\uparrow}_{i\perp} - k^{\downarrow}_{j\perp}) = 0$, corresponding to the stationary phase of the summand in Eq. (146). These contributions will then exhibit damped oscillations around zero value as seen in Figs. 27 and 29.

Turning to $g^r_{\uparrow\downarrow}$, we find on substituting Eq. (148) into Eq. (144) that there are two thickness-independent contributions. The first comes from summing the $\delta_{nm}$ term in Eq. (144) and is nothing other than the number of states in the normal metal (i.e. the Sharvin conductance). The second comes from the $r^{\uparrow}_{N\to N}r^{\downarrow *}_{N\to N}$ term and provides an interface-specific correction to the first. Superimposed on these two is the contribution from the thickness-dependent terms which, to lowest order, contain phase factors $e^{i(k^{\sigma}_{i\perp}+k^{\sigma}_{j\perp})d}$ and $e^{-i(k^{\sigma}_{i\perp}+k^{\sigma}_{j\perp})d}$. Just as in the case of $g^t_{\uparrow\downarrow}$, one can argue that the integral over these terms will have oscillatory character. However, the oscillations will have different periods and occur around the constant value set by the first two contributions. It is clear that the value approached asymptotically by $g^r_{\uparrow\downarrow}$ is simply the reflection mixing conductance evaluated for a single interface.

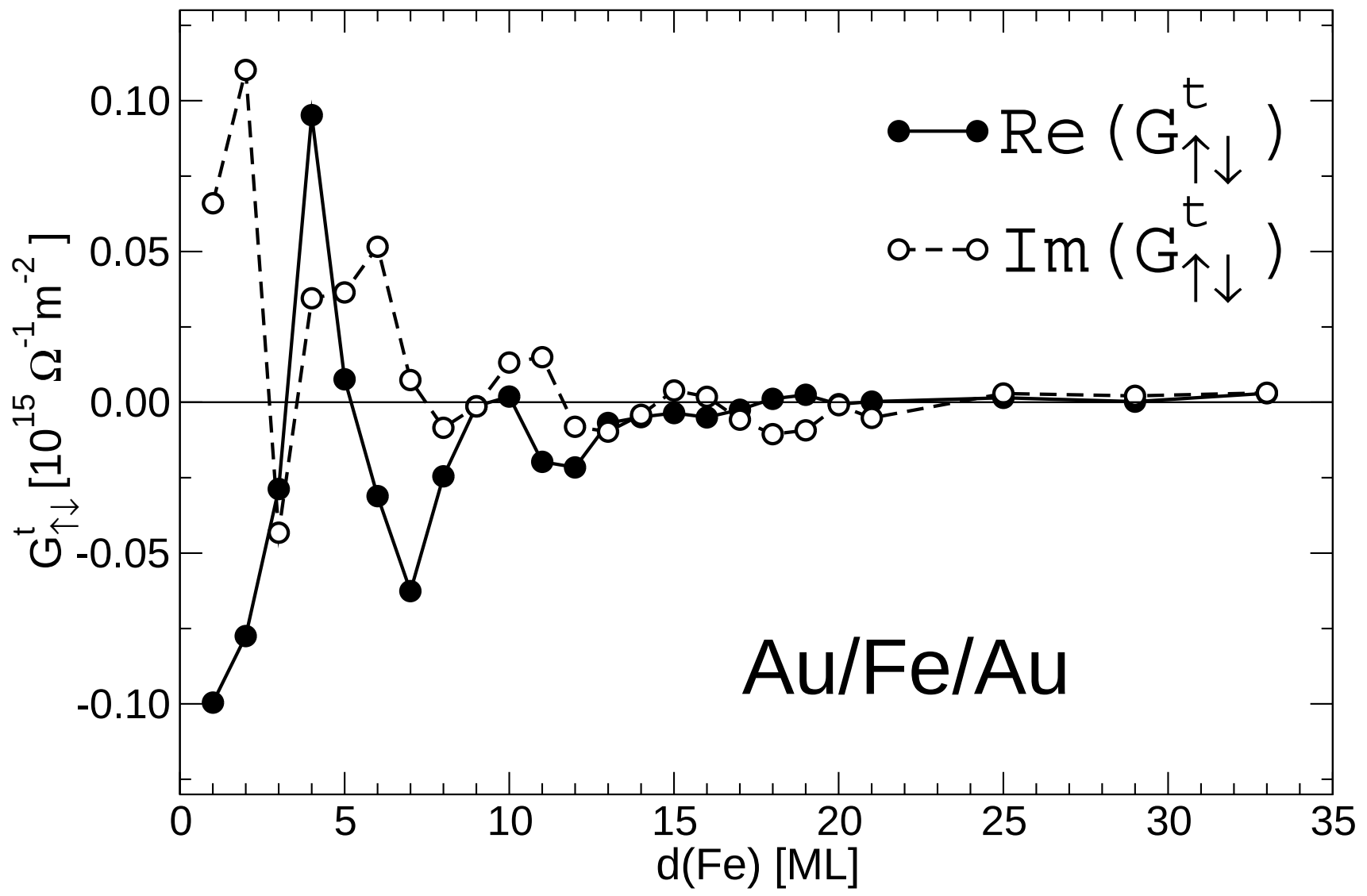

FIG. 27: Transmission spin-mixing conductance of a Au|Fe|Au (001) trilayer with perfect interfaces as a function of the thickness $d$ of the Fe layer.

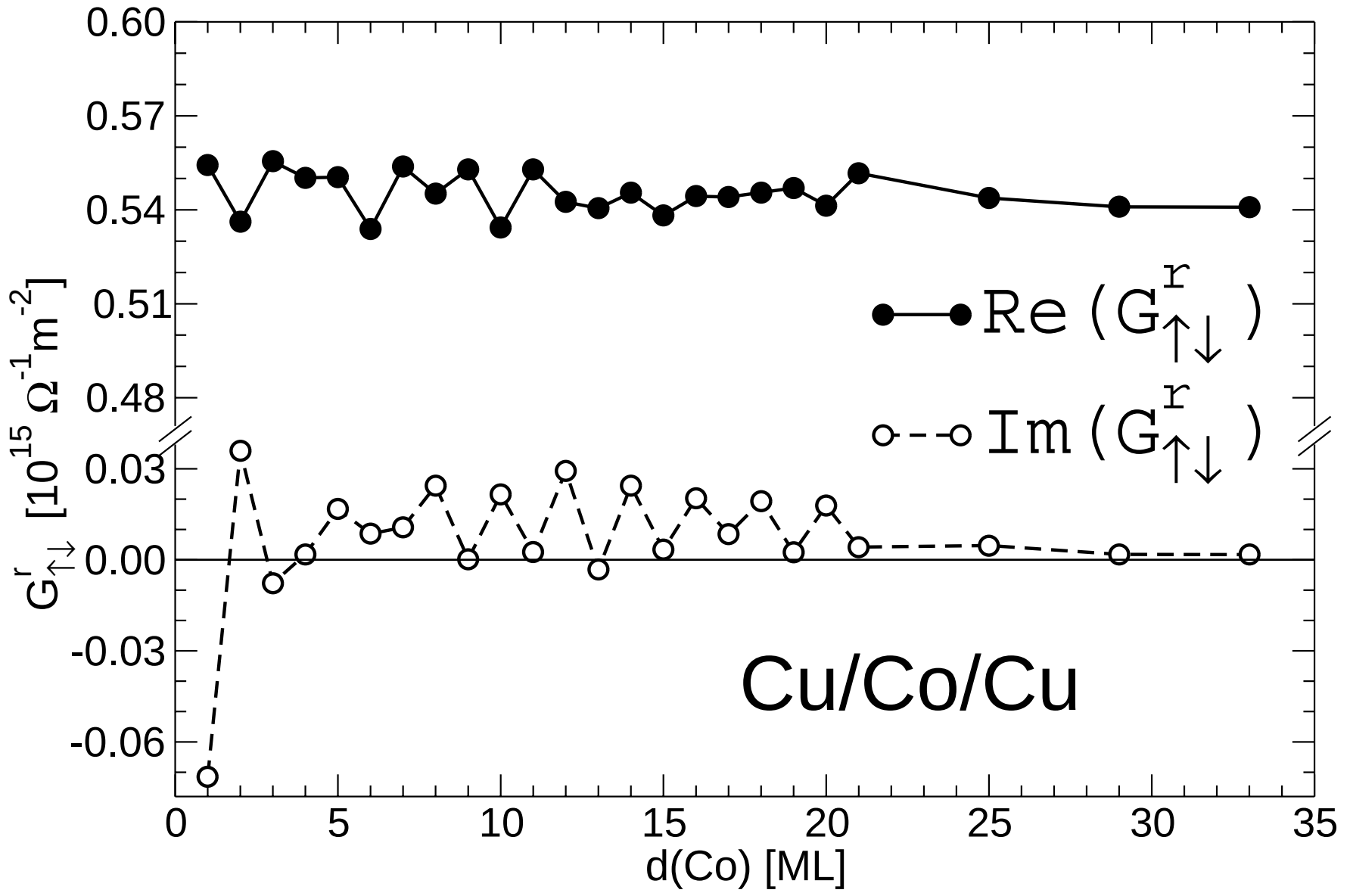

FIG. 28: Reflection spin-mixing conductance of a Cu|Co|Cu (111) trilayer with perfect interfaces as a function of the thickness $d$ of the Co layer.

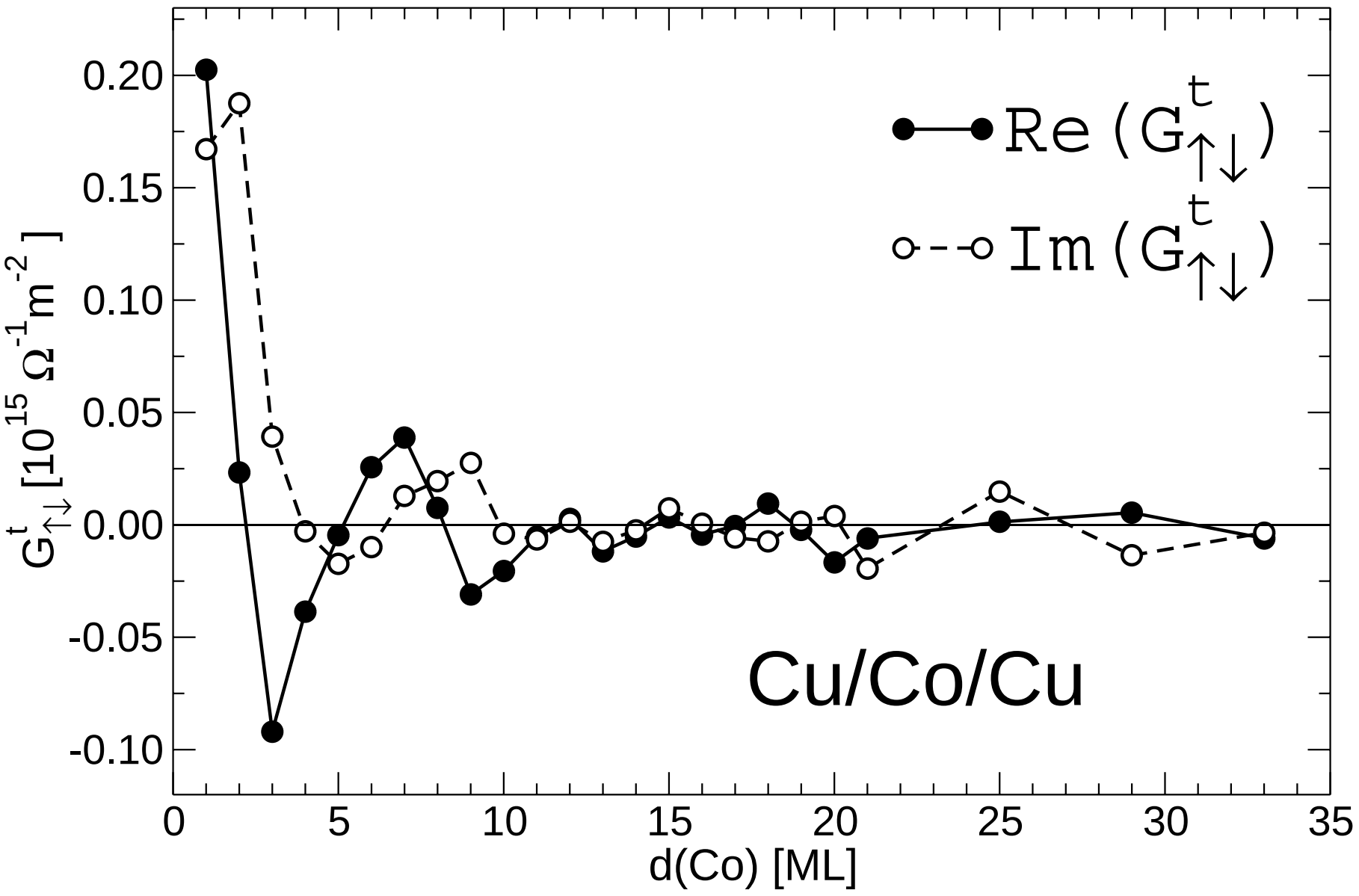

FIG. 29: Transmission spin-mixing conductance of a Cu|Co|Cu (111) trilayer with perfect interfaces as a function of the thickness $d$ of the Co layer.

The (quasi)periodicity of $g^r_{\uparrow\downarrow}$ and $g^t_{\uparrow\downarrow}$ as a function of the magnetic-layer thickness $d$ clearly depends (through the $\Lambda^{\sigma}$) on the electronic structure of the internal part of the magnetic layer, which for metallic systems is practically identical to that of the bulk material. The amplitudes, on the other hand, are related to the interfacial scattering coefficients introduced in Eqs. (147) and (148). Analyzing the scattering properties of the single interface enables us in the following to understand why the amplitudes of oscillation of $g^t_{\uparrow\downarrow}$ are substantially larger than those of $g^r_{\uparrow\downarrow}$ for the two systems considered. We begin by noting that the transmission probability for states in the majority-spin channel assumes values close to one over large areas of the Brillouin zone for both Cu|Co and Au|Fe, as illustrated in

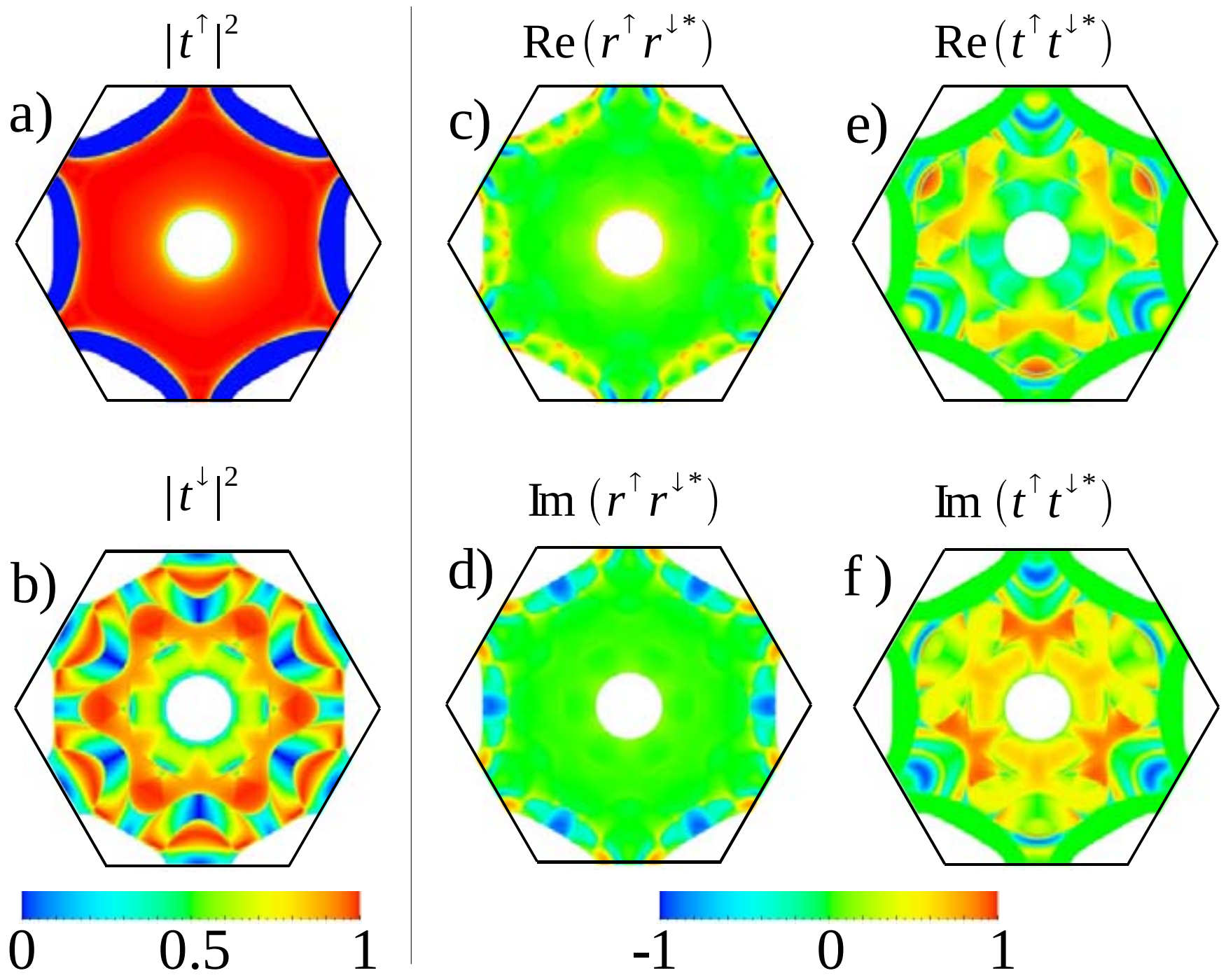

FIG. 30: Plotted within the first Brillouin zone for the Cu|Co(111) interface: transmission probability for (a) majority- and (b) minority-spins. (c) Real and (d) imaginary parts of $r^{\uparrow}_{N\to N} r^{\downarrow\star}_{N\to N}$. (e) Real and (f) imaginary parts of $t^{\uparrow}_{\rm int} t^{\downarrow\star}_{\rm int}$ where $t\sigma_{\rm int} = t\sigma_{F\to N} \cdot t\sigma_{N\to F}$ as discussed in the text. Note the different scales for panels (a), (b) and for (c) - (f).

Fig. 30(a) for the Cu|Co(111) interface. For Cu|Co, this results from the close similarity of the corresponding Cu and Co electronic structures. The situation is more complicated for Au|Fe because the majority-spin Fermi surface of Fe consists of several sheets, unlike that of Au. However, one of these sheets is made up of states which match well with the states in Au. In the minority-spin channel, on the other hand, the transmission probability varies between 0 and 1; see Fig. 30(b). The average magnitude of the "spin-mixing" products of Eqs. (144) and (146) are therefore determined mostly by the majority-spin scattering coefficients with the modulation, as a function of $\vec{k}_{||}$, provided by the corresponding minority-spin coefficients.

The small reflectivity for the majority-spin states has a direct consequence for the values of the mixing conductances. In the case of $g^{r}_{\uparrow\downarrow}$, the second term under the sum in Eq. (144) will typically have a negligible magnitude. This follows directly from $r^{\uparrow}_{N\to N} \approx 0$ [and consequently also $(r^{\uparrow}_{F\to F})_{i,j} \approx 0$ ] and Eq. (148) and is illustrated in Fig. 30 (c) and (d) for the $r^{\uparrow}_{N\to N} r^{\downarrow *}_{N\to N}$ term. As we can see, the only non-zero contributions in this case come from the outer regions of the Brillouin zone, where states from the normal metal are perfectly reflected because of the absence of propagating majority-spin states in the ferromagnet. Independently varying phases (as a function of $\vec{k}_{||}$) for "up" and "down" reflection coefficients lead, in the course of integration over $\vec{k}_{||}$, to additional cancellation of already small contributions. The final outcome is that the values of $g^{r}_{\uparrow\downarrow}$ are determined mostly by the first term in the Eq. (144), i.e. the Sharvin conductance of the lead.

Because the interface transmission in the majority-spin channel is uniformly large almost everywhere in the Brillouin zone, the transmission through the magnetic layer also remains large for arbitrary thicknesses and its magnitude (but not its phase) is only weakly modulated by the multiple scattering within the layer. The magnitude of the $t^{\uparrow} t^{\downarrow *}$ product is then modulated mostly by the variation of the transmission in the minority-spin channel, as a function of $\vec{k}_{||}$. To demonstrate the effect of the interface scattering on $g^{t}_{\uparrow\downarrow}$, values of the product $t^{\uparrow}_{\rm int} t^{\downarrow *}_{\rm int}$ are shown in Figs. 30(e) and (f) for a Cu|Co (111) interface. Here, $t^{\uparrow}_{\rm int}$ is defined as the scalar product of the interface transmission vectors: $t^{\sigma}_{\rm int} = t^{\sigma}_{F\to N} \cdot t^{\sigma}_{N\to F}$. As one can see, the values assumed by the real and imaginary part of this product vary strongly throughout the Brillouin zone. Unlike the case of $g^{r}_{\uparrow\downarrow}$, however, the values span the entire range from -1 to +1. An imbalance of positive and negative contributions is therefore more likely to produce a sizeable integrated value. The complex values of $t^{\uparrow} t^{\downarrow *}$ are further modified by thickness- and $\vec{k}_{||}$-dependent phase factors discussed above, which leads to the oscillatory damping seen in Figs. 27 and 29.

Figures 31 and 32 show the same quantities ($G^{r}_{\uparrow\downarrow}$ and $G^{t}_{\uparrow\downarrow}$) calculated in the presence of disorder modelled by 1 monolayer of 50 % alloy added on each side of the magnetic layer. For both systems we have used $10 \times 10$ lateral supercells. The thickness $d$ in this case is that of the clean ferromagnetic layer. For both material systems, the effect of disorder is to strongly reduce the amplitudes of the (quasi)oscillations. The reflection mixing conductance becomes practically constant at the level of its asymptotic (i.e. interfacial) value listed in Table V for clean and disordered interfaces.[9] The disorder here was modelled by 2 ML of 50 % alloy. In spite of this difference, the values are practically identical to the asymptotic ones seen in Figs. 26, 28, 31, and 32. In particular, Im$G^r_{\uparrow\downarrow}$ assumes prohibit the transfer of spin angular momentum which can be mediated by evanescent states, for example in the case of a magnetic insulator. The contribution from such $\vec{k}_{\parallel}$ points *should* be included in the 2D-BZ integration.

[9] The values given in Table V differ somewhat from ones reported previously in [36]. There are two reasons for this. Firstly, these calculations were performed using energy-independent muffin-tin orbitals linearized about the centres of gravity of the occupied conduction states. The current implementation [83, 168] uses energy-dependent, (non-linearized) MTO's, calculated exactly at the Fermi energy which improves the accuracy of the method. Secondly, on performing the 2D-BZ integration in Eq. (144), it was assumed by [36] that the contribution to the sum of $\vec{k}_{||}$ points for which there are no propagating states in the ferromagnet should be neglected. However, the lack of propagating states in the ferromagnet does not necessarily

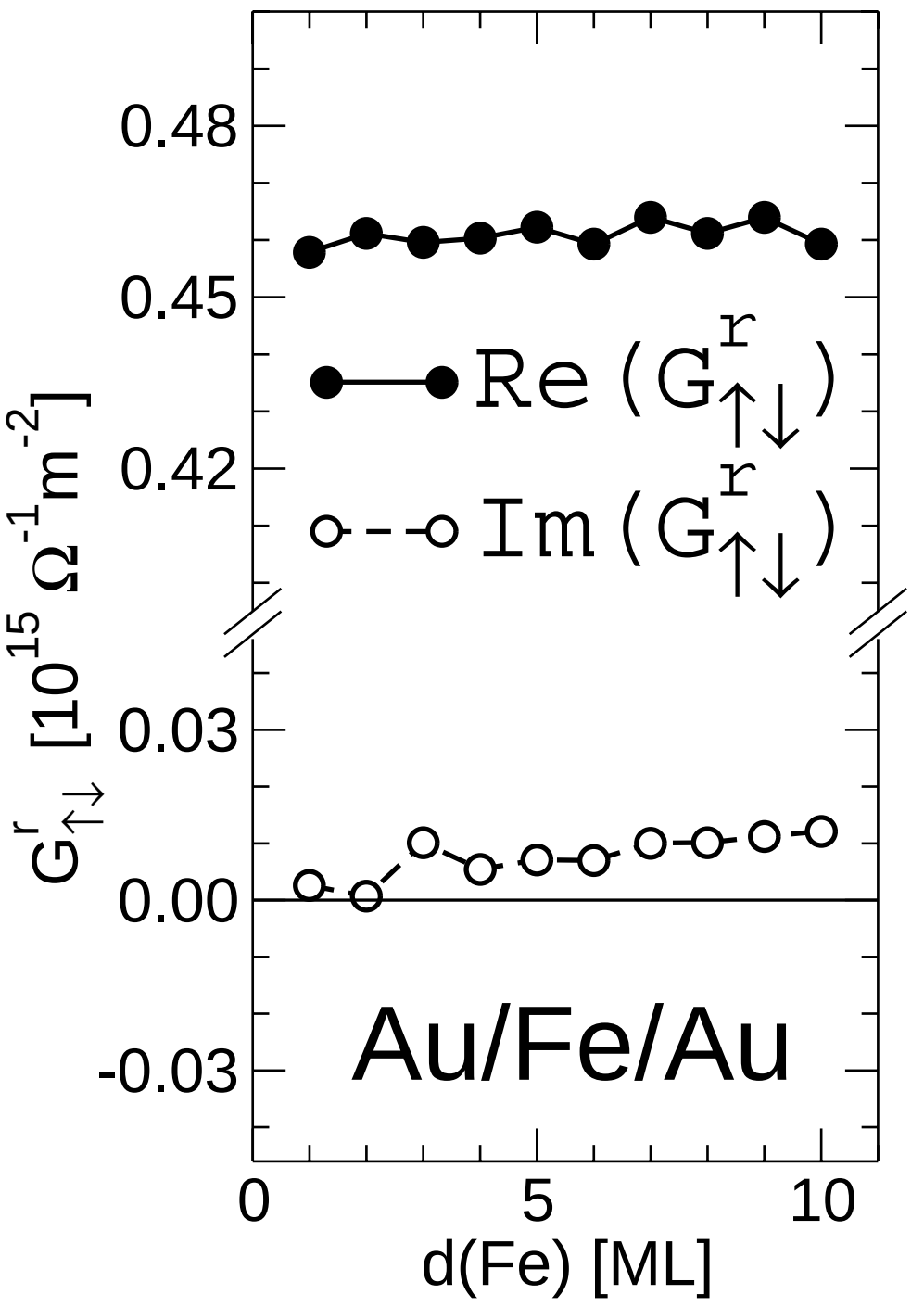

FIG. 31: Spin-mixing conductances of a Au|Fe|Au (001) trilayer with disordered interfaces as a function of the thickness $d$ of the Fe layer.

| System | Interface | $G^{\uparrow}$ | $G^{\downarrow}$ | Re$G^r_{\uparrow\downarrow}$ | Im$G^r_{\uparrow\downarrow}$ | $G^{\rm Sh}_N$ | $G^{\rm Sh}_{F\uparrow}$ | $G^{\rm Sh}_{F\downarrow}$ |
|---|---|---|---|---|---|---|---|---|
| Au\|Fe | clean | 0.40 | 0.08 | 0.466 | 0.005 | 0.46 | 0.83 | 0.46 |
| (001) | alloy | 0.39 | 0.18 | 0.462 | 0.003 | | | |
| Cu\|Co | clean | 0.42 | 0.38 | 0.546 | 0.015 | 0.58 | 0.46 | 1.08 |
| (111) | alloy | 0.42 | 0.33 | 0.564 | -0.042 | | | |
| Cr\|Fe | clean | 0.14 | 0.36 | 0.623 | 0.050 | 0.63 | 0.90 | 0.46 |
| (001) | alloy | 0.26 | 0.34 | 0.610 | 0.052 | | | |

TABLE V: Interface conductances in units of $10^{15}$ $\Omega^{-1}$m$^{-2}$.

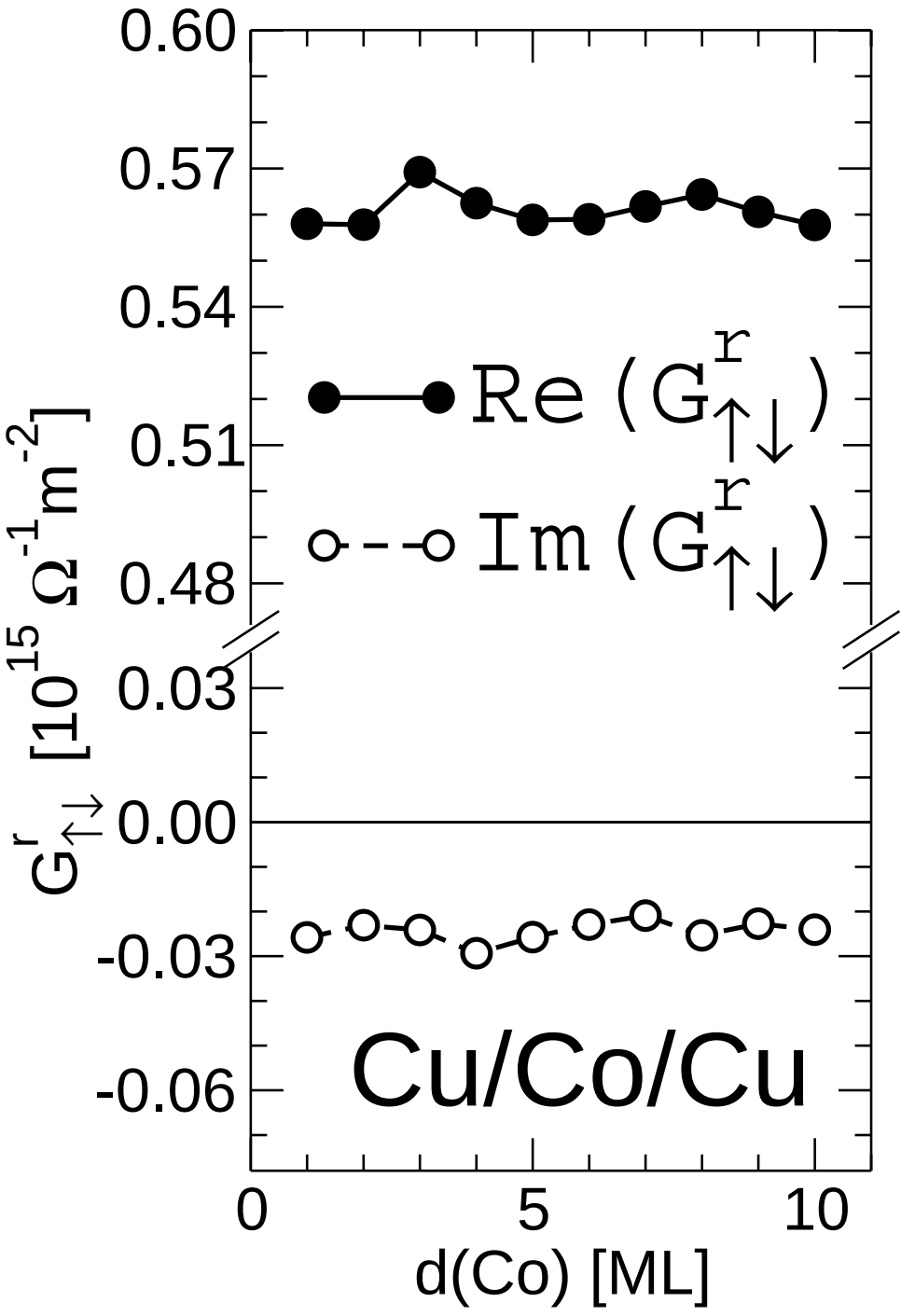

FIG. 32: Spin-mixing conductances of a Cu|Co|Cu (111) trilayer with disordered interfaces as a function of the thickness $d$ of the Co layer.

values two orders of magnitude smaller than Re$G^r_{\uparrow\downarrow}$, with the latter being close to the Sharvin conductance of the normal metal. This approximate equality results once again from a combination of amplitude (small $|r^\uparrow|$) and uncorrelated spin-up and spin-down phase effects.

For $G^t_{\uparrow\downarrow}$, the oscillations are not entirely damped out but their amplitude is substantially reduced. In fact, the values of $G^t_{\uparrow\downarrow}$ become negligible compared to Re$G^r_{\uparrow\downarrow}$ for all but the thinnest magnetic layers. In addition, we expect that diffusive scattering in the bulk of the magnetic layer, which for simplicity has not been included here, will have a similar effect. This means that for all practical purposes, spin transport is determined by just $G^\uparrow$, $G^\downarrow$ and $G_{\uparrow\downarrow} \equiv G^r_{\uparrow\downarrow}$ calculated for a single interface.

### *2. Cr|Fe*

Expression (144) suggests that having a large number of propagating channels on the non-ferromagnetic side of the interface should lead to a large mixing conductance. For Cr, particularly in the centre of the Brillouin zone where there are four propagating states, this is seen to be the case; see 25(a). For the Cr|Fe interface the band structure matching and the effect of interface disorder are quite different compared to Cu|Co. Whereas the majority spin states of Cu|Co match very well, it is the minority spin-states in Cr and Fe which match best; see Figs. 25(d-e). For a perfect Cr|Fe(001) interface, the mixing conductance (in units of $e^2/h$) is $g^{\rm Sh}_{\rm Cr} - {\rm Tr}(\mathbf{r}^{\uparrow}\mathbf{r}^{\downarrow\dagger}) \sim 1.3$, compared to the "normal" conductance $g^{\uparrow} + g^{\downarrow} = 0.29 + 0.76$.

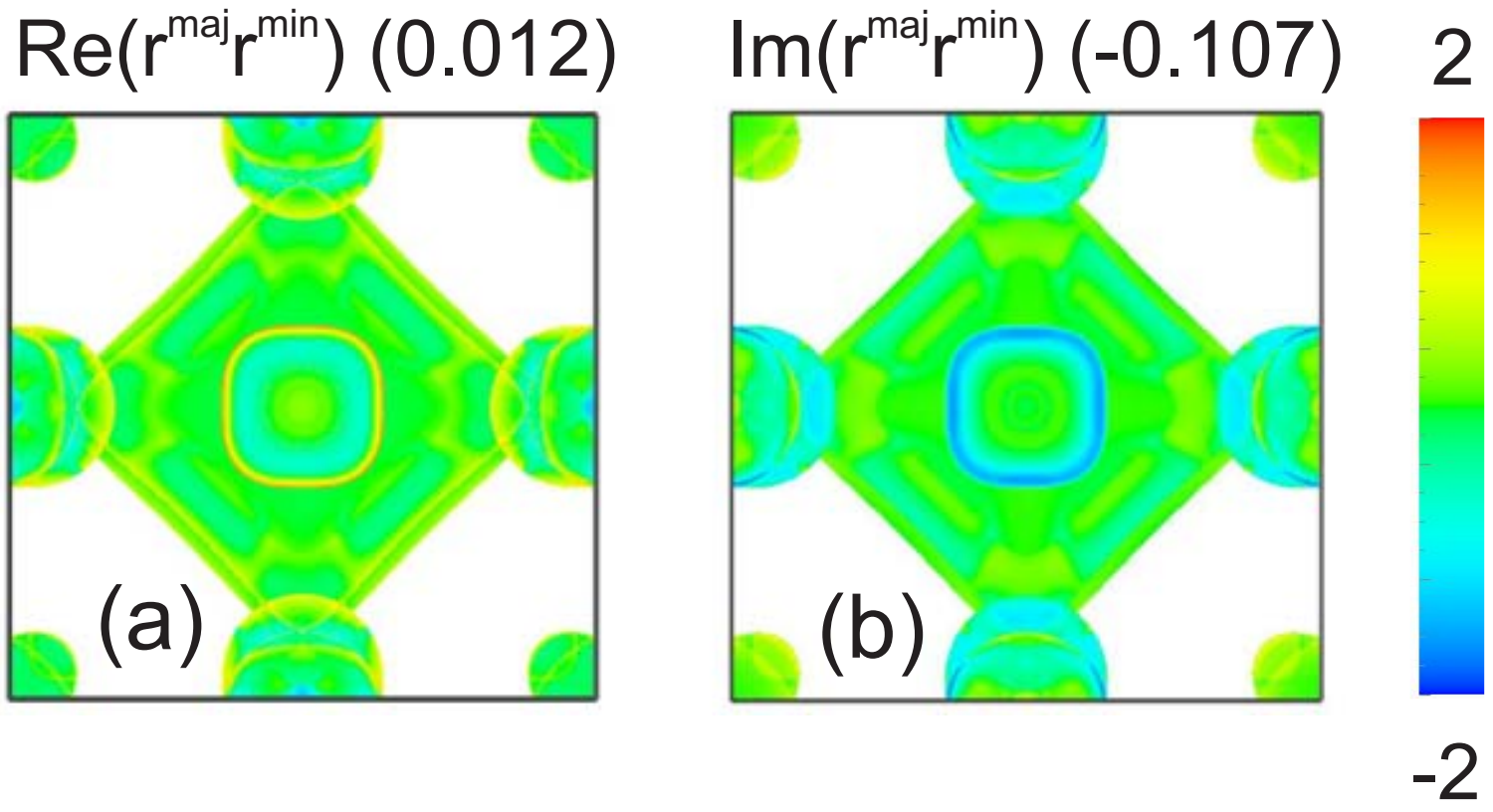

FIG. 33: For a clean Cr|Fe interface (a) and (b) show the real and imaginary parts, respectively, of the mixing conductance as a function of transverse crystal momentum. Units of conductance are $e^2/h$. The result of integrating over the whole Brillouin Zone is given in brackets at the top of each panel.

It is of interest to have a closer look at the $\vec{k}_{\|}$ resolved mixing conductance $g_{\uparrow\downarrow}$. Just as in the case of Cu|Co, the term ${\rm Tr}(\mathbf{r}^{\uparrow}\mathbf{r}^{\downarrow\dagger})$, shown in Fig. 33 is very small throughout most of the Brillouin zone (BZ) so that $g_{\uparrow\downarrow}$ is in essence given by the Cr Sharvin conductance which takes the value 4 in the region surrounding the centre of the BZ so that the real part of $g_{\uparrow\downarrow}$ is very large. Starting from the origin, the real part of the correction term ${\rm Tr}(\mathbf{r}^{\uparrow}\mathbf{r}^{\downarrow\dagger})$ in Fig. 33(a) is first almost zero, then turns moderately negative in an annular region centred on the origin, then becomes strongly positive in a very narrow band before returning to zero. In an annulus around the centre where the real part changes sign, the imaginary part is large and negative. In this region, where Re $tr(r^{\uparrow}r^{\downarrow\dagger})$ is negative, $g_{\uparrow\downarrow}$ is even larger than the number of channels in Cr at the same $\vec{k}_{\|}$-points, shown in Fig. 25(a); at the same $\vec{k}_{\|}$-points the transmission of majority spin electrons is very low; Fig. 25(d). Thus at some $\vec{k}_{\|}$ the mixing conductance can be much larger than the normal conductance. Although $g_{\uparrow\downarrow}$ is thus large for some values of $\vec{k}_{\|}$, in most parts of the BZ, the minority-spin reflection is very low, Fig. 25(e) so that after averaging over the BZ, $g_{\uparrow\downarrow}$ is not very large compared with the normal conductance and the imaginary part is quite small, typically an order of magnitude lower than the real part for the interfaces for which values are given in Table V.

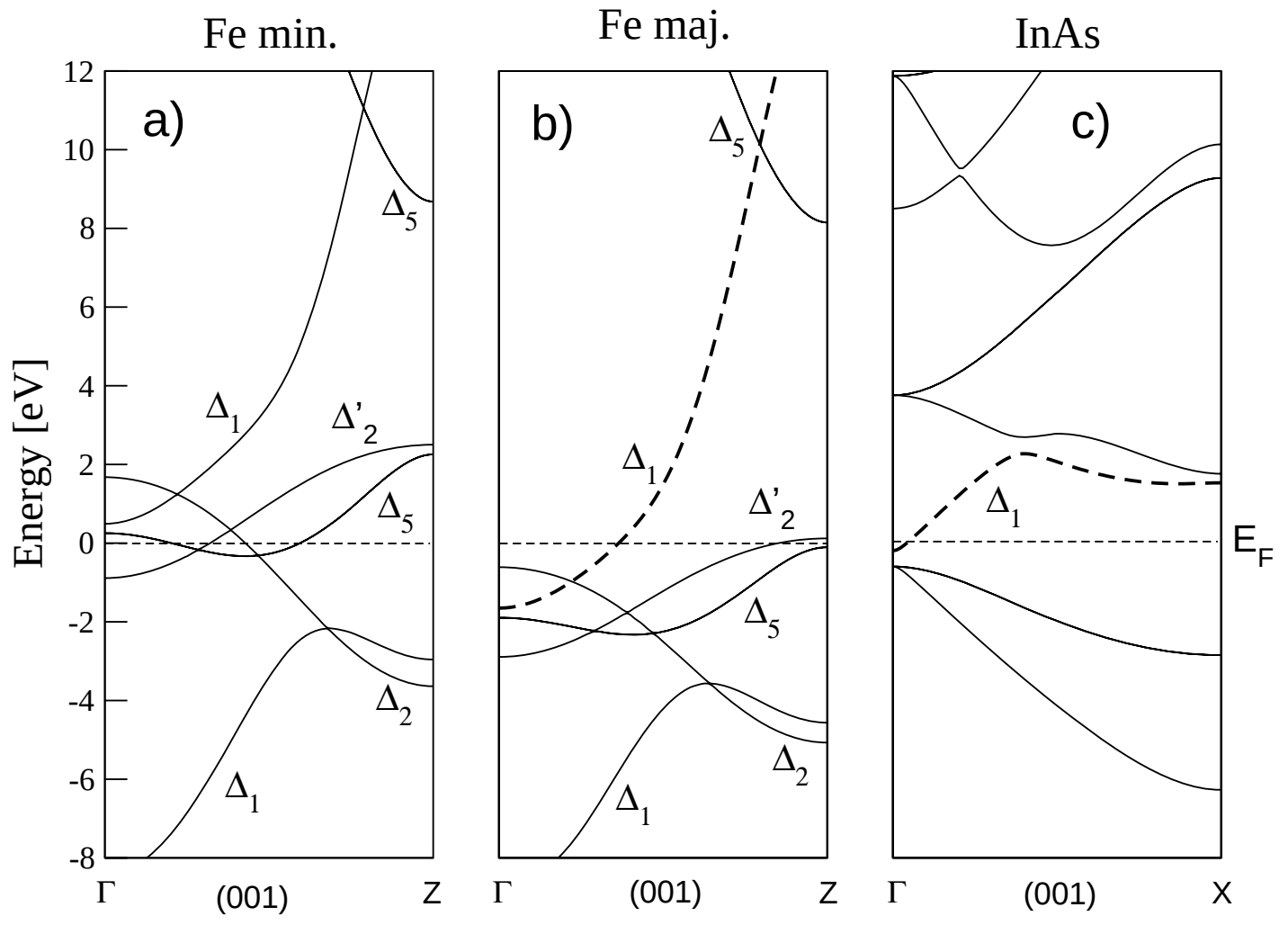

FIG. 34: Energy band structures of tetragonal Fe minority spin states (a), majority spin states (b), and InAs states (c) at $\mathbf{k}_{\parallel} = 0$ for $\mathbf{k} = (00k_z)$ perpendicular to the interface.

One way of circumventing the cancellation which occurs on averaging over the BZ is to choose a non-magnetic material with a very small Fermi surface so that cancellation can less easily occur. The ideal situation would be where the size and position of the Fermi surface could be modified in a simple fashion. An example where the size, but not the position, of the Fermi surface of the non-magnetic material can be modified is when there is an Ohmic contact between a ferromagnet (such as Fe) and a doped semiconductor (such as InAs). The band structures of these materials are shown in Fig. 34 where Fe has been tetragonally distorted so its in-plane lattice constant matches that of bulk InAs [83]. In this situation the magnitude of the real and imaginary parts of the mixing conductance are comparable in size though obviously their absolute values are small, being limited by the restricted conductance of a doped semiconductor. For InAs/Fe, the imaginary part of the mixing conductance is quite sensitive to interface disorder [110]. The imaginary part of $G_{\uparrow\downarrow}$ is related to the spin precession which results from the non-collinear alignment of the spins of the injected electrons and the magnetization (or an external magnetic field). A non-vanishing imaginary part of the mixing conductance, $\mathrm{Im}\, G_{\uparrow\downarrow}$, should result in antisymmetry with respect to time reversal [63, 121].

### *3. Spin current induced torque*

The rapid decay of $g^t_{\uparrow\downarrow}$ (and $g^r_{\uparrow\downarrow} - g^{\mathrm{Sh}}_N$) discussed in previous paragraphs as a function of increasing magnetic layer thickness implies that the absorption of the transverse component of the spin current occurs within a few monolayers of the N|F interface. The limit $g^t_{\uparrow\downarrow} \to 0$ and $g^r_{\uparrow\downarrow} \to g^{\mathrm{Sh}}_{\mathrm{N}}$ corresponds to the situation where all of the incoming transverse polarized spin current is absorbed in the magnetic layer. The torque is then proportional to the Sharvin conductance of the normal metal. As demonstrated in Figs. 26-29, 31, and 32, this is the situation for all but the thinnest (few monolayers) and cleanest magnetic layers. This is the basic justification for the expression for the torque given in Eq. 4.42 which contains only the reflection spin-mixing conductance parameter, $G_{\uparrow\downarrow}$.

Because of the $\vec{k}_{\|}$ and spin-dependence of the phase of the reflection matrices, the mixing conductance need not be small even when $|r^\sigma| \sim 1$. Indeed, the spin-mixing can remain substantial, even when the conventional conductance is made vanishingly small, as in the case of a tunnel junction. This means that if it is possible to create a spin-accumulation in the absence of an electrical current on one side of a tunnel junction, this would still lead to the exertion of a torque on the magnetization. This leads us to consider the three-terminal device ("spin-flip transistor") [12, 68] sketched in Fig. 35 and discussed in more detail in Section VI E. By passing a current from one ferromagnetic element FM1 into another, FM2, through a non-magnetic node NM, a spin-accumulation is induced in NM that is in electrical contact with a third ferromagnetic element FM. Once there is a spin-accumulation in NM, the torque on FM is determined only by the mixing conductance even if FM is a magnetic insulator or the top magnetic element of a magnetic tunnel junction. In the latter case it is possible to independently determine the orientation of FM by measuring the TMR [169]. The spin torque is that of the metallic junction, but without the energy dissipation caused by the particle current. In practical memory devices it may be advantageous to be able to achieve this separation of particle and spin injection.

## D. Relationship to calculations of spin-dependent transport in metallic systems

So far we have been concerned with the quantitative and qualitative characteristics of the transmission and reflection of electron states at *single* interfaces between real materials, one of which is an itinerant ferromagnet. The advantage of focussing on the full scattering matrix, rather than simply calculating the conductance, is that it provides us with greater insight and is a very convenient point of contact with other theories, such as the circuit theory described in the previous section; with it we can make contact relatively easily with more complex transport problems.[10] We wish to argue that such an approach may be more fruitful than a frontal, brute force approach to calculating transport properties entirely from first principles; we are not aiming at a comprehensive review of realistic approaches to calculating GMR; for this, we refer to a number of existing reviews [1, 37, 40].

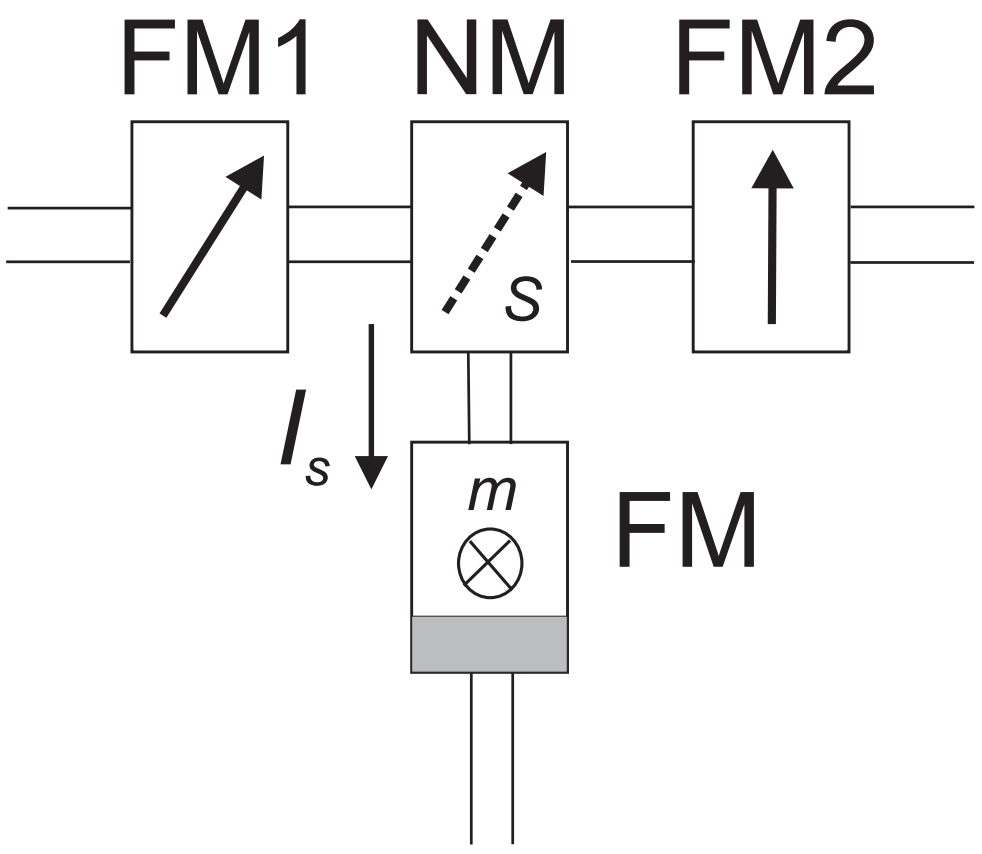

FIG. 35: Sketch of a three-terminal device where a normal metal (NM) element is connected to three ferromagnetic elements FM1, FM2 and FM. An applied bias causes a current to flow between FM1 and FM2 creating a spin accumulation in NM. If FM is a tunnel junction or a magnetic insulator then the particle flow into FM will be vanishingly small.

#### 1. *History*

In spite of the fact that a large body of theoretical research on mesoscopic transport had been formulated in terms of scattering matrices (see Section III C) and that oscillatory exchange coupling was successfully formulated in terms of reflection matrices [65, 150], a considerable time passed before it was realized that spin-dependent electronic structure mismatch plays an essential role in GMR. To understand why, it is perhaps useful to recall the situation at the time. The origin of the residual resistivity in crystalline materials was qualitatively well understood as resulting from the scattering from deviations from full translational symmetry such as intrinsic or extrinsic point defects, dislocations, stacking faults etc. However, even for simple metals whose electronic structures are well described by nearly free electron (NFE) models and only a small number of plane waves, it was still quite cumbersome to calculate the residual resistivity for point defects quantitatively [170]. For transition metals, with partly filled $d$ shells, the complexity of the electronic structure made the description of transport properties much more difficult and attention focussed on understanding (i) the dilute limit where the resistivity is proportional to the (known) concentration of impurities or intrinsic defects and (ii) substitutional alloys in which there is a well defined underlying crystal structure but the lattice sites are occupied randomly by two or more different types of atoms.

[10] A good example of this is the study of the materials dependence of the suppression of Andreev scattering at a ferromagnetic|superconducting interface. This is a problem which had been studied phenomenologically [72] without taking into account details of the electronic structure of materials which might be used in an actual experiment. Because it had been formulated in terms of the scattering matrix for the F|S interface with the superconducting material in its normal state, it was straightforward to introduce and study the dependence on the constituent materials [19].

In practice, first-principles materials-specific studies of the dilute limit were based on iterative solution of the quasiclassical Boltzmann equation [171, 172] combined with scattering $T$-matrices which describe how the host electronic states are scattered by an impurity atom. Early calculations were performed for host materials such as Cu which have simple, nearly free electron like electronic structures and for transition metal impurities. The phase shifts describing the scattering by the impurities were either derived from experiment [173] or calculated from first-principles [174]. The bottleneck was the determination of the scattering matrix for point defects in host materials with complex electronic structures, in a form suitable for transport calculations. Once the problem of calculating the electronic structure of point impurities self-consistently was resolved [175, 176], the way was open to combine the two parts of the problem. This was done first for non-magnetic systems [177] and then for magnetic systems [178]. For elemental solids with one type of atom per unit cell, this problem can now be solved routinely [179] and yields results in impressively good agreement with experiments performed in the appropriate dilute limit.

The study of the resistivity of substitutional, concentrated alloys was based on the fully quantum mechanical Kubo-Greenwood formalism in the linear-response regime [180, 181]. Though the formalism itself is quite general, evaluation of the conductivity tensor requires averaging over configurations of disorder. To do this, many practical implementations made use of the coherent potential approximation (CPA) treatment of disorder [159] which is capable of describing (all concentrations of) substitutional random alloys. Qualitative studies of the electronic structure of disordered alloys frequently made use of tight-binding implementations of the CPA but first-principles studies were based either on a multiple-scattering, Korringa-Kohn-Rostoker (KKR) implementation [182] or on a closely related LMTO formulation [158]. Updating the charge density in a self-consistent electronic structure calculation within density functional theory using the CPA requires a configura-tion averaged (single particle) Green's function $\langle G \rangle$. To study transport, however, a configuration-averaged *product* of Green's functions $\langle GG^* \rangle$ is required, a more formidable undertaking. Worked out by Butler [183] and applied to binary alloys in a single site approximation, the multiple-scattering theory implementation [184] of transport theory in the CPA gave results in very good agreement with experiment and later on with supercell calculations [185]. The application to more complex systems, however, appeared daunting.

This, then, was the state-of-the-art of materials-specific transport calculations at the time GMR was discovered. The challenge it presented for the Boltzmann equation approach, was to extend the procedures for calculating the electronic structure self-consistently for point impurities in a simple host to one for a much more complex host characterized by a translational unit cell containing a period of a superlattice, an order of magnitude more atoms than could be handled at the time by state of the art methodology. Straightforward extension of matrix operations which scale as $\sim$(rank of matrix)$^3$ in the number of numerical operations and as $\sim$(rank of matrix)$^2$ for memory requirements to a system an order of magnitude larger was not possible; it was necessary to develop new algorithms and/or wait for the arrival of more powerful computers with more memory. The challenge facing the KKR-CPA was similar.

Not surprisingly, therefore, the first theoretical studies of the GMR effect were based upon simple, empirical models for the electronic structure: either free electron or tight-binding descriptions of a single electron band. For reasonable choices of the parameters in such models (band centres and widths, exchange splitting etc.) the electronic structure doesn't exhibit any striking spin-dependence. In the absence of a strong spin dependence in the underlying electronic structure, the GMR effect was attributed essentially entirely to spin-dependent disorder scattering [186–190].

#### *2. Ballistic GMR*

In retrospect, the most striking "disorder" effect is the spin-dependent scattering which results from the breaking of translational symmetry induced by a single interface, a point made clear by transmission probabilities such as those shown in Figs.(18, 19 and 25), first calculated by Stiles [77, 78] to study oscillatory exchange coupling or calculated by van Hoof [153] and applied to spin-dependent transport [47, 61, 154]. The single interface scattering matrix could be concatenated in a manner familiar from optics [see equations (147) and (148) in Sect.V C 1 ] to construct scattering matrices for an F|N|F sandwich or any multilayer com-

posed of individual F|N interfaces.[11] Alternatively, a finite multilayer could be inserted between the leads of Fig. 15 in which case the effects of multiple scattering and evanescent states are included automatically and exactly; this has been done for $s$-band tight-binding model Hamiltonians [190–194], for more realistic *spd* tight-binding Hamiltonians [138, 195, 196] and for first-principles (DFT-LDA) tight-binding Hamiltonians [35]. As the number of multilayer periods is increased, it becomes interesting to consider the limiting case of transmission through a perfectly periodic multilayer. In this case full translation periodicity is recovered and the electronic structure $\varepsilon_{i\sigma}(\vec{k})$ is obtained by performing a conventional electronic structure calculation for the corresponding superlattice. The resulting eigenstates are, by construction, propagating states with transmission probability unity. The conductance in spin channel $\sigma$ in the direction $\hat{n}$ through an area $A$ of material is

$$G^{\sigma}(\hat{n}) = \frac{e^2}{h}\frac{A}{4\pi^2}\frac{1}{2}\sum_i \int_{FS(i\sigma)} \frac{dS}{|\vec{v}_{i\sigma}(\vec{k})|}|\hat{n}\cdot\vec{v}_{i\sigma}(\vec{k})|. \tag{149}$$

where

$$\vec{v}_{i\sigma}(\vec{k}) = \frac{1}{\hbar}\nabla_{\vec{k}}\,\varepsilon_{i\sigma}(\vec{k}) \tag{150}$$

is the group velocity of the corresponding Bloch state and the integration is carried out over all sheets $i$ of the Fermi surface. To determine the conductance of this system, all that is needed is to count the number of propagating modes in the transport direction. This is most easily done by determining the area of the projection of the Fermi surface in the transport direction [59, 76]. When these calculations were carried out for *fcc* Co/Cu and *bcc* Fe/Cr multilayers, Schep *et al.* found that conductances in the current perpendicular to the multilayer plane (CPP) configuration depended strongly on whether the magnetizations of adjacent magnetic layers were aligned parallel (P) or antiparallel (AP). In the absence of any disorder, he found values for the magnetoresistance, defined as MR $= (G_{P\uparrow} + G_{P\downarrow} - 2G_{AP})/2G_{AP}$, as large as 120% for Co/Cu and 230% for Fe/Cr multilayers. These values are comparable to measured values. Interface disorder modifies this picture quantitatively [35] but not qualitatively [190]. It can increase or decrease the magnetoresistance depending on how transparent the clean interfaces are. It was suggested that a large CPP effect in the ballistic regime was an important condition for a GMR, either CPP or CIP.

---

[11] In practice, this would be quite inefficient because of the need to take account of multiple reflections. Indeed, to be entirely correct, such a procedure would need to include explicitly the transmission (and reflection) probability amplitudes for scattering between evanescent states ("tunnelling") as well as the phase changes incurred on propagating through the "bulk" F and N layers.

We conclude from this that the giant magnetoresistance in the CPP configuration is in zeroth order determined by the spin-dependent electronic structure mismatch which is intrinsic to a particular interface and the complex electronic structures of the materials from which the interface is formed. This result is only changed in detail when finite, rather than infinite multilayers are considered [195]. The spin-dependent matching described by the scattering matrix is obviously to be found in related quantities such as atomic or muffin tin potentials, densities of states, phase-shifts and derived quantities such as bulk conductivities [197] Disorder plays a secondary role giving rise to an "extrinsic" resistivity whose spin-dependence is "intrinsic". Any model which contains realistic electronic structures will result in reasonable values for the MR but the values of the resistivities will depend on how disorder is modelled [198, 199]. The advantage of considering the ballistic limit is that it is at one and the same time an exact result in which all of the complexity of the Fermi surfaces of transition metal elements is included. It is "exact" in the sense that the ballistic limit can in principle be realized experimentally, for example, in a point contact measuring configuration. The failure up till now of efforts to observe GMR in this limit [200, 201] does not detract from this.

In the current-in-plane (CIP) configuration, the situation is more complicated. The calculated MR values in the CIP geometry in the ballistic regime [76] were much smaller than the highest experimental values of 115% for Co/Cu multilayers [202] and 220% for Fe/Cr multilayers [203] indicating that some additional scattering mechanism is important. Using a single-band tight-binding model, Asano et al. [190] found that, unlike in the CPP case, the CIP MR did depend critically on interface defect scattering. In the ballistic limit, Schep found a strong dependence of the MR on the orientation of Fe/Cr multilayers: large for "rough" (100) and small for "smooth" (110) orientations, indicating that the CIP MR was very sensitive to the interface morphology down to the monolayer level. This is in agreement with the widespread assumption that interface disorder is responsible for CIP-GMR (though the interpretation of the origin of the spin-dependence is different). Experimentally the CIP MR in Fe/Cr multilayers is found to depend strongly on the structural properties of the interface [204].

Study of the ballistic regime was an important simplification which allowed parameter-free calculations to be applied to the study of giant magnetoresistance. Doing so "merely" required performing electronic structure calculations for large supercells (containing as many as 32 atoms) and determining the area of the Fermi surface projected onto a plane in the transport direction. Otherwise, as stated above, it is an exact treatment of a well-defined limiting case.

In the case of the Boltzmann equation, the difficulty lay in carrying out the necessary impurity calculations for systems with similar periodicity, a project only achieved recently [75]. As a result, the earliest calculations made recourse to simplifying assumptions varying in sophistication from using state and spin independent relaxation times [205, 206]; through spin dependent but state independent relaxation times [60]; a variety of approximations to construct spin and state dependent relaxation times [207–209]; until full iterative solution of the Boltzmann equation including vertex corrections was achieved and used to study the influence of impurity type and position on the magnetoresistance in Cu|Co multilayers [75] without the assumptions made in a qualitatively similar, previous study by the same authors [210]. The fully iterated Boltzmann equation also represents a well-defined physical limiting case, but one which would be even more difficult to realize experimentally than the ballistic limit (since there is no equivalent of the "point contact" configuration which might allow it to be studied). Not only would it require a periodic superlattice; it would require that the superlattice be so lightly doped that the mean free path would be much shorter than the superlattice period and that coherent scattering between impurities can be neglected.

The CPA is in this respect more realistic in principle, especially implementations where the alloy concentration can be allowed to vary from layer to layer [158]. In the context of magnetic overlayers and multilayers it has been a very useful tool for studying properties such as magnetic moments which are related to the configuration-averaged Green's functions $\langle G \rangle$. However, calculating the conductivity requires a configuration-averaged product of Green's functions, $\langle GG^* \rangle$, and this remains an intractable problem for inhomogeneous systems such as magnetic multilayers. Instead, $\langle GG^* \rangle$ is usually approximated by $\langle G \rangle \langle G^* \rangle$ [44, 211], i.e. neglecting so-called vertex corrections (corresponding to the "scattering-in" terms in the semiclassical Boltzmann equation). It is difficult to determine *a priori* how serious an approximation this is.[12] Even then, additional approximations must be made in order to evaluate $\langle G \rangle$ such as: a "single-site" approximation;[13] complex rather than real energies;[14] arbitrariness in the boundary conditions, all of which can obscure the physical results.

To attempt a detailed discussion of work in this direction [211, 213–216] would take us too far afield and we refer instead to a number of recent review articles [40, 44]. We simply note that the motion of electrons under the influence of an electric field parallel to the plane of a multilayer will be very strongly influenced by the specular and diffuse scattering at individual interfaces described by the scattering matrix discussed in the previous section. A detailed description of CIP transport requires taking the different geometry into account explicitly [217].

[12] A recent comparison [212] of interface conductances obtained using a tight binding Hamiltonian and modelling interface disorder in a supercell approach or using the CPA (neglecting vertex corrections) concludes that "without vertex corrections, ..., CPA cannot be used to describe phenomena such as CPP GMR".

[13] A more accurate "single-cell" approximation including vertex corrections has been implemented but only for a single band tight binding model [191].

[14] The calculations presented in the previous section were performed for real energies.

When this is done, then a result emerges which is common to most of the approaches based on ab-initio description of the electronic structure which have been explored so far, for both Co/Cu and Fe/Cr multilayers: the calculated values of GMR are generally found to be (much) higher than the measured values. The resistivity of the high resistivity channel (minority spin channel for Co/Cu; majority for Fe/Cr) is described reasonably but that of the low resistivity channel (majority for Co/Cu; minority for Fe/Cr) is substantially lower than the values extracted from experiment. It is possible to include a spin-independent scattering which changes the resistivity of the low resistivity channel proportionately much more than that of the high resistivity channel but this is essentially an exercise in fitting since the microscopic origin of the scattering is not known [196, 199].

### *4. Spin-injection and spin tunneling*

Spin-dependent matching of electronic structures not only plays a role at interfaces between metallic ferromagnets and non-magnetic metals. It also occurs at the interface between itinerant ferromagnets and semiconductors or insulators, where the electronic structure of the semiconductor/insulator in a very small region of reciprocal space dominates the injection/tunneling [as in Fig. 34] when there is lattice matching and in the absence of disorder so transverse crystal momentum is conserved. Most work has focussed on systems containing Fe(001)-related interfaces because in this orientation the lattice constant of Fe is reasonably well matched to those of a number of inorganic semiconductors and to MgO. Vacuum barriers have also been considered in a number of studies [137, 139, 218]. Almost all of the transport studies in which realistic multi-band electronic structures were used were based on transmission probabilities calculated either with real space Green's function techniques [99, 144] or the wave-function matching scheme [145]. The electronic structures were either calculated directly from first-principles [44, 83, 141, 218–225] or obtained by fitting to first-principles electronic structures [137, 139, 142, 143, 226]. For disorder-free interfaces, very large polarizabilities of the injection or tunnel currents are predicted, much larger than those predicted by the densities-of-states related formula [227] which has been used quite successfully to interpret experiments where the barrier was an amorphous oxide. Recent experiments on single-crystal MgO-based tunnel junctions [228–230] find much larger TMR values but they are still substantially lower than the predicted values. This can be interpreted as indicating that the quality of the tunnel junctions used in the recent experiments are beginning to approach the ideal considered theoretically. The only calculations to consider disordered interfaces [83] find that the polarization depends sensitively on the detailed interface structure. As in the case of GMR in metallic systems, much more work is needed to characterize the interface disorder experimentally.

## VI. APPLICATIONS

With the basic theory as sketched in the previous chapter we are in a position to describe arbitrary magnetoelectronic circuits and devices. In Section VI A the modelling of different resistive elements of practical interest are discussed. In Sections VI B and VI C we review simple magnetoelectronic devices and their characteristics.

### A. Resistive elements

#### *1. Diffuse wire connections*

When a ferromagnetic and a normal metal reservoir are connected by a normal metallic wire which is much longer than the elastic impurity-scattering mean-free path, Eq. (95) can be derived by the diffusion equation. On the normal metal side of such a diffuse wire the distribution matrix $\hat{f}$ equals that in the node $\hat{f}^N$. On the ferromagnet side of the junction, the distribution matrix approaches the equilibrium distribution in the reservoir, $f^F\hat{1}$. As mentioned above, spins noncollinear to the local exchange field relax very fast since electrons with different spins are not coherent. The associated additional resistance (and torque) and other interface related excess resistances are assumed here to be small compared to the diffuse resistance of the long wire (see also [108]). Sufficiently far from the interface the distribution in the ferromagnet consists therefore of only two components. Only the spin-current parallel to the magnetization of the ferromagnet is conserved. We denote the cross-section of the contact $A$, the length of the ferromagnetic part of the contact $L^F$, the length of the normal part of the contact $L^N$, the (spin dependent) resistivity in the ferromagnet $\rho^{Fs}$, and the resistivity in the normal metal $\rho^N$. The (spin-dependent) conductance of the ferromagnetic part of the contact $G^{DFs} = A/(\rho^{Fs}L^F)$ and the conductance of the normal part of the contact $G^{DN} = A/(\rho^N L^N)$. Solving the diffusion equation $\nabla^2\hat{f} = 0$ on the normal and ferromagnetic side and the boundary conditions mentioned above, we find the current through a diffuse junction:

$$\hat{\imath}^D = G^{D\uparrow}\hat{u}^\uparrow\left(\hat{f}^F - \hat{f}^N\right)\hat{u}^\uparrow + G^{D\downarrow}\hat{u}^\downarrow\left(\hat{f}^F - \hat{f}^N\right)\hat{u}^\downarrow \tag{151}$$

$$- G^{DN}\left(\hat{u}^\uparrow\hat{f}^N\hat{u}^\downarrow + \hat{u}^\downarrow\hat{f}^N\hat{u}^\uparrow\right), \tag{152}$$

where the total spin dependent conductance is $1/G^{Ds} = 1/G^{DFs} + 1/G^{DN}$. This result can be understood as a specific case of the generic Eq. (95) with $G^\uparrow = G^{D\uparrow}$, $G^\downarrow = G^{D\downarrow}$, and $G^{\uparrow\downarrow} = G^{DN}$. The mixing conductance in the diffuse limit therefore depends on the conductance of the normal part of the junction only, which is a consequence of the relaxation of spins non-collinear to the magnetization direction in the ferromagnet.

### *2. Ballistic junction*

In the semiclassical model proposed in Refs. [71, 231] the channels are either completely reflected or transmitted, with $N^\uparrow$ and $N^\downarrow$ being the number of transmitted channels for different spin directions. We find that the spin conductance $G^{B\uparrow} = (e^2/h)N^\uparrow$, $G^{B\downarrow} = (e^2/h)N^\downarrow$ and the mixed conductance is determined by the lowest number of reflected channels, $G^{B\uparrow\downarrow} = \max(G^{B\uparrow}, G^{B\downarrow})$ and is real.

### *3. Tunnel junction*

In tunnel junctions we can expand Eqs. (97) and (98) in terms of the small transmission. Let us illustrate this for a single channel

$$\frac{h}{e^2}G_T^s = |t_s|^2 = 1 - |r_s|^2 ; \ \frac{h}{e^2}G_T^{\uparrow\downarrow} = 1 - r_\uparrow r_\downarrow^*. \tag{153}$$

$G_T^s$ is exponentially small for tunnel junctions and the modulus of the reflection coefficient is close to one:

$$r_s = e^{i\phi_s} - \delta r_s. \tag{154}$$

To lowest order in $\delta r_s$:

$$G_T^s = 2\,\mathrm{Re}[e^{i\phi_s}\delta r_s^*], \tag{155}$$

$$\frac{h}{e^2}G_T^{\uparrow\downarrow} = 1 - e^{i\phi_\uparrow - \phi_\downarrow} + \delta r_\uparrow e^{-i\phi_\downarrow} + e^{i\phi_\uparrow}\delta r_\downarrow^* \tag{156}$$

$$\frac{h}{e^2}\,\mathrm{Re}\,G_T^{\uparrow\downarrow} = 1 - \cos(\phi_\uparrow - \phi_\downarrow) + \mathrm{Re}[e^{i\phi_\uparrow}\delta r_\downarrow^*] + \mathrm{Re}[e^{-i\phi_\downarrow}\delta r_\uparrow], \tag{157}$$

$$\frac{h}{e^2}\,\mathrm{Im}\,G_T^{\uparrow\downarrow} = -\sin(\phi_\uparrow - \phi_\downarrow) + \mathrm{Im}[e^{i\phi_\uparrow}\delta r_\downarrow^*] + \mathrm{Im}[e^{-i\phi_\downarrow}\delta r_\uparrow]. \tag{158}$$

For tunnel junctions, the conductances are exponentially small. When $\phi_\uparrow = \phi_\downarrow$ :

$$\frac{h}{e^2}\operatorname{Re}G_T^{\uparrow\downarrow} = [G_T^{\uparrow} + G_T^{\downarrow}]/2 \tag{159}$$

$$\frac{h}{e^2}\operatorname{Im}G_T^{\uparrow\downarrow} = \operatorname{Im}[e^{i\phi_\downarrow}\delta r_\downarrow^*] + \operatorname{Im}[e^{-i\phi_\uparrow}\delta r_\uparrow] \tag{160}$$

so that both $\operatorname{Im}G_{\uparrow\downarrow}$ and $\operatorname{Re}G_{\uparrow\downarrow}$ are of the same order of magnitude as the diagonal conductances, thus small as well.

For N|F|I junctions in which F is thinner than $\ell_{sd}^F$, significantly different phase shifts for the spin-up and spin-down electrons cause large mixing conductances in spite of negligible diagonal ones [36]. This reflects the fact that the mixing conductance is an interface property involving only a penetration depth of the order of the magnetic coherence length. When the barrier itself is exchange split as in magnetic insulators like the Europium chalcogenides [232], the imaginary part of $G_{\uparrow\downarrow}$ can be much larger than the spin-dependent conductances themselves [121]! Evidence is mounting that the imaginary part of the mixing conductance, also called "effective field" or "spin-dependent interface phase shift", can be quite large also for conventional ferromagnets [122–124].

## B. Perpendicular spin valves

We will now illustrate circuit theory of spin-transport by computing the transport properties of the archetypal two-terminal device, the perpendicular spin valve.[15] It consists basically of three layers, *viz.* a normal metal sandwiched by two ferromagnetic reservoir. Whereas the spin valve has in the beginning been mainly employed in the current in plane (CIP) configuration, it recently became a focus of the field of magnetoelectronics because of the current-induced magnetization reversal, which has been very convincingly demonstrated in these devices, see [8] and references in there. We focus here on the resistance of the device as a function of the magnetization angle difference and the spin-current induced torque felt by the magnetizations when an electric current is driven through the device.

### 1. Angular magnetoresistance

The normal metal node in these devices is smaller than the spin-diffusion length so that the spatial distribution function is homogeneous within the node. The elementary manipulations are shown in more detail in the Appendix. Let us first consider a normal metal node attached to two ferromagnetic reservoirs with identical junctions, *e.g.* $G_1^{\uparrow} = G_2^{\uparrow} = G^{\uparrow}$, $G_1^{\downarrow} = G_2^{\downarrow} = G^{\downarrow}$ and $G_1^{\uparrow\downarrow} = G_2^{\uparrow\downarrow} = G^{\uparrow\downarrow}$. The relative angle $\theta$ between the magnetization in the two ferromagnetic reservoirs is arbitrary and the details of the junctions leading to the conductances given above do not have to be specified. With the aid of (90) and (95) we find the current

$$G(\theta) = \frac{G}{2}\left(1 - \frac{p^2}{1+g_{\mathrm{sf}}}\frac{\tan^2\theta/2}{\tan^2\theta/2+\zeta}\right) = G(-\theta)\,, \tag{161}$$

where[16]

$$\zeta = \frac{(\eta_R + g_{\mathrm{sf}})^2 + \eta_I^2}{(1+g_{\mathrm{sf}})\,(\eta_R + g_{\mathrm{sf}})} \geqslant 1 \tag{162}$$

[15] An elementary discussion of the physics and the calculations of the spin valve is given in Section II and Appendix B.

Here $G = G^{\uparrow} + G^{\downarrow}$ is the total conductance of one contact, $p = (G^{\uparrow} - G^{\downarrow})/G$ the polarization, $\eta = 2G^{\uparrow\downarrow}/(G^{\uparrow}+G^{\downarrow})$ the relative mixing conductance and $g_{\mathrm{sf}} = 2g_{\mathrm{sf}}/G$ the relative 'spin-flip conductance' $g_{\mathrm{sf}} = \mathcal{D}e^2/(2\tau_{\mathrm{sf}})$ [233] ($\mathcal{D}$ is the energy density of states). Note that the inequality $\alpha \geqslant 1$ follows from (98). In the absence of spin-flip scattering and small interface transparencies compared to the Sharvin conductances, Eq. (161) reduces to Eq. (124). In the limit of strong spin-flip scattering $\zeta \to 1 + (\eta_R - 1)/g_{\mathrm{sf}}$

$$g(\theta) = \frac{g}{2}\left(1 - \frac{p^2}{g_{\mathrm{sf}}}\left(1 - \frac{1}{g_{\mathrm{sf}}}\right)\frac{\tan^2\theta/2}{\tan^2\theta/2+1}\left(1 - \frac{(\eta_R - 1)/g_{\mathrm{sf}}}{\tan^2\theta/2+1}\right)\right) \tag{163}$$

$$= \frac{g}{2}\left(1 - \frac{p^2\,(1-\cos\theta)}{2g_{\mathrm{sf}}}\left[1 - \frac{1 + (\eta_R - 1)\,(1+\cos\theta)}{2g_{\mathrm{sf}}}\right]\right) \tag{164}$$

we see that not only the magnetoresistance decreases, but that to leading order the aMR assumes a simple cosine dependence.

We plot in Fig. 36 the current (161) as a function of $\theta$. When the magnetizations are parallel ($\theta = 0$), there is no spin-accumulation on the normal metal node and the current is given by Ohm's law $I_{\mathrm{P}} = I(\theta = 0) = GV/2$. The anti-parallel magnetization configuration ($\theta = \pi$) generates the largest spin-accumulation, reducing the particle current to $I_{\mathrm{AP}} = I(\theta = \pi)) = G(1 - p^2/(1 + g_{\mathrm{sf}}))V/2$. In this case the magneto-resistance ratio $(I_{\mathrm{P}} - I_{\mathrm{AP}})/I_{\mathrm{P}}$ is $p^2/(1 + g_{\mathrm{sf}})$, irrespective of the mixing conductance. For long spin-flip relaxation times ($g_{\mathrm{sf}} = 0$) we generalized here previous results for two tunnel junctions [234]. For intermediate $\theta$ the current increases with increasing $\alpha$ (Fig. 36), since a large relative mixing conductance implies that spins orthogonal to the magnetization in the reservoirs escape the normal metal node easily. This suppresses the spin accumulation and allows a higher charge current. When $\alpha \gg 1$ and $\theta$ is not close to $\pi$ the current approaches Ohm's law $I = GV/2$ irrespective of the magnetization angle. When the magnetizations are nearly anti-parallel and the mixing conductance becomes irrelevant. For large $\alpha$ the current has a narrow dip at $\theta \lesssim \pi$ which is governed by the polarization $p$.

[16] This result, derived by W. Wetzels (unpublished), slightly corrects the original one in [68].

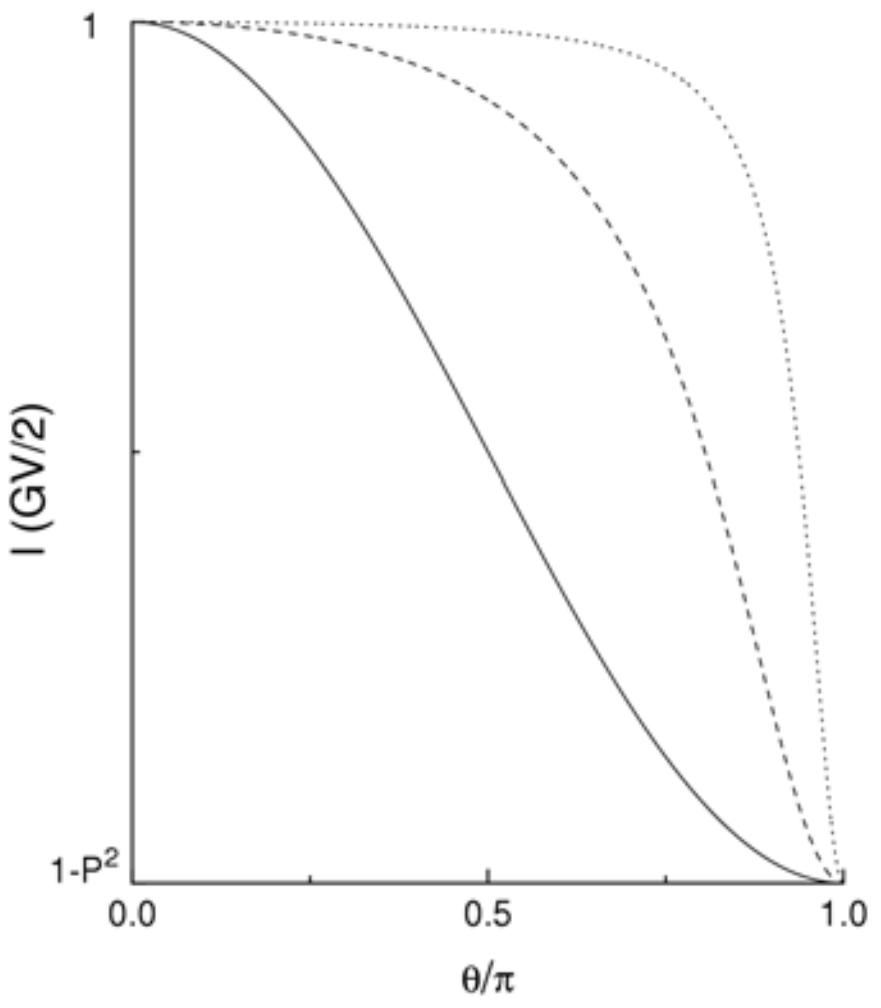

FIG. 36: The current as a function of the relative angle between the magnetizations (angular magnetoconductance) in a F|N|F spin valve for $\zeta = 1$ (full curve), 10 (dashed), 100 (dotted), where $\varsigma$ has been defined in Eq. (**??**).

The current non-collinear spin-valve has a universal property which does not depend of the nature of the contacts as long as they do not flip spins. By scaling the conductance change to the difference between the current in the parallel and anti-parallel configuration $(G/2)p^2/(1+g_{\mathrm{sf}})V$ the current change for *any* two terminal device (in the semiclassical regime) should be higher than its universal value at the minimum value of $\eta$, $|\eta| = 1$. Thus, our theory predicts that measured points on the current *vs.* magnetization angle are all above the universal curve for $|\eta| = 1$ (see also Fig. 36)).

Intermetallic interfaces in a diffuse environment (see Fig. 1c) have been studied thoroughly by the Michigan State University collaboration [39] and others [37, 38, 118, 235, 236] in perpendicular (CPP) spin valves. These experiments provided a large body of evidence for the two-channel (*i.e.* spin-up and spin-down) series resistor model and a wealth of accurate transport parameters such as the spin-dependent interface resistances for various material combinations [35, 37, 61, 119]. In exchange-biased spin valves, it is possible to measure the electric resistance as a function of the angle between magnetizations, which has been analyzed exper-imentally and theoretically [237]. Pratt *c.s.* observed that experimental magnetoresistance curves [238] could accurately be fitted by the form [12]

$$\frac{R(\theta)-R(0)}{R(\pi)-R(0)}=\frac{1-\cos\theta}{\chi(1+\cos\theta)+2} \tag{165}$$

According to the new insights described above, the free parameter $\chi$ is a function of renormalized microscopic parameters

$$\chi=\frac{1}{1-p^2}\frac{|\tilde{\eta}|^2}{\mathrm{Re}\tilde{\eta}}-1 \tag{166}$$

in terms of the relative mixing conductance $\tilde{\eta}=2\tilde{g}_{\uparrow\downarrow}/\tilde{g}$, the polarization $p=(\tilde{g}_{\uparrow}-\tilde{g}_{\downarrow})/\tilde{g}$, and the average conductance $\tilde{g}=\tilde{g}_{\uparrow}+\tilde{g}_{\downarrow}$. Slonczewski [239] rederived Eq. (165) by a phenomenological version of magnetolectronic circuit theory, approximating the mixing conductance by the Sharvin conductance multiplied by a factor $2/\sqrt{3}$.Others also found the form(165) giving different interpretations to the parameter $\chi$ for special situations [94, 240, 241].

Experimental values for the parameters for Cu/Permalloy (Py) spin valves are $\tilde{\chi}=1.2$ and $p=0.6$ [238]. Disregarding a very small imaginary component of the mixing conductance [36], using the known values for the bulk resistivities, the theoretical Sharvin conductance for Cu ($0.55\cdot 10^{15}\,\Omega^{-1}\mathrm{m}^{-2}$/spin [61]), and the spin-flip length of Py as the effective thickness of the ferromagnet $\left(\ell_{sd}^{F}=5\text{ nm [39]}\right)$, we arrive at the bare Cu/Py interface mixing conductance $G_{\uparrow\downarrow}=0.39\,(3)\cdot 10^{15}\,\Omega^{-1}\mathrm{m}^{-2}$. This value may be compared with the calculated mixing conductance for a disordered Co/Cu interface ($0.55\cdot 10^{15}\,\Omega^{-1}\mathrm{m}$ [36]). The agreement is reasonable, but leaves some room for material and device dependence that deserves to be investigated in the future. The mixing conductance can also be determined from the excess broadening of ferromagnetic resonance spectra. A larger mixing conductance in Pt/Py can be explained by the larger density of conduction electrons in Pt compared to Cu [114]. Reasonable agreement between experiment and theory has been also found for Fe/Au multilayers [242].

Eq. (161) has been derived assuming that the islands are made from metals that behave like a normal Fermi liquid [243]. The angular magnetoresistance in spin valves with a Luttinger liquid in the center turns out to have the same functional form [244, 245]. However, in a Luttinger liquid the current depends nonlinearly on the applied source-drain bias voltage [244]. Coulomb charging reduces the spin-accumulation in the linear response regime [234] and modifies the angular magnetoresistance as well [246].

Interesting new effects are predicted when the magnetically active region of the junction is asymmetric (Fig. 37) [247]. The non-monotonicity can be explained by the spin-accumulation which exists in the presence of asymmetry even for a parallel configuration $\theta=0$. When the polarization of the first and mixing conductance of the second contact are sufficiently large, the increase of the spin accumulation by the diagonal conductance is more than offset by the additional channel opened by the mixing conductance, leading to the minimum observed in Fig. 37. Such a non-monotonicity has been observed by Urazhdin et al. [248] in Py|Cu|Py exchange-biased CPP spin valves in which one of the magnetic contacts was made thinner than the spin-flip diffusion length of 5 nm. The experimental results plotted in Fig. 38 compare favorable with the circuit theory model [249].

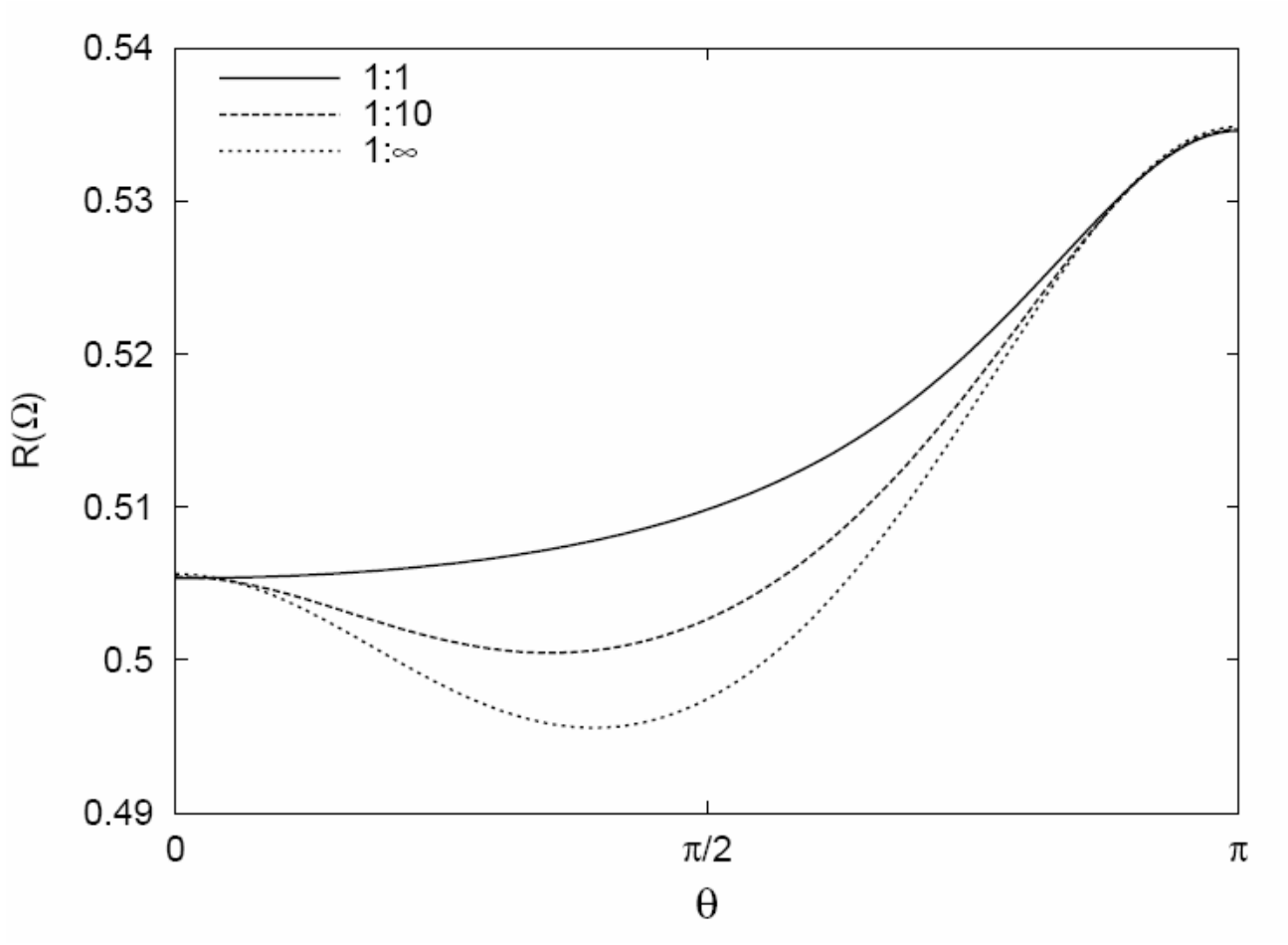

FIG. 37: Angular magnetoresistance of a spin valve with asymmetrically distributed resistances of the magnetically active region of the two ferromagnetic contacts.

### *2. Spin-transfer torque*

We have seen that the magnetization torque [26] is caused by the absorption of the spin current polarized normal to the magnetic order parameter [12, 56]. Flow of such a spin current requires a bias of the distribution functions between the normal metal and the ferromagnet. This can be generated most easily in a perpendicular spin valve, with a polarizing ferromagnet as source and the ferromagnet to be switched as an analyzer. The magnetization torque as an integral property of the device can be computed easily by circuit theory.

The magnetization torque or spin transfer on a ferromagnet equals the spin current through the interface with vector component normal to the magnetization direction and its evaluation is closely related to the charge conductance. An analytical expression for the spin valve torque as a function of the applied potential

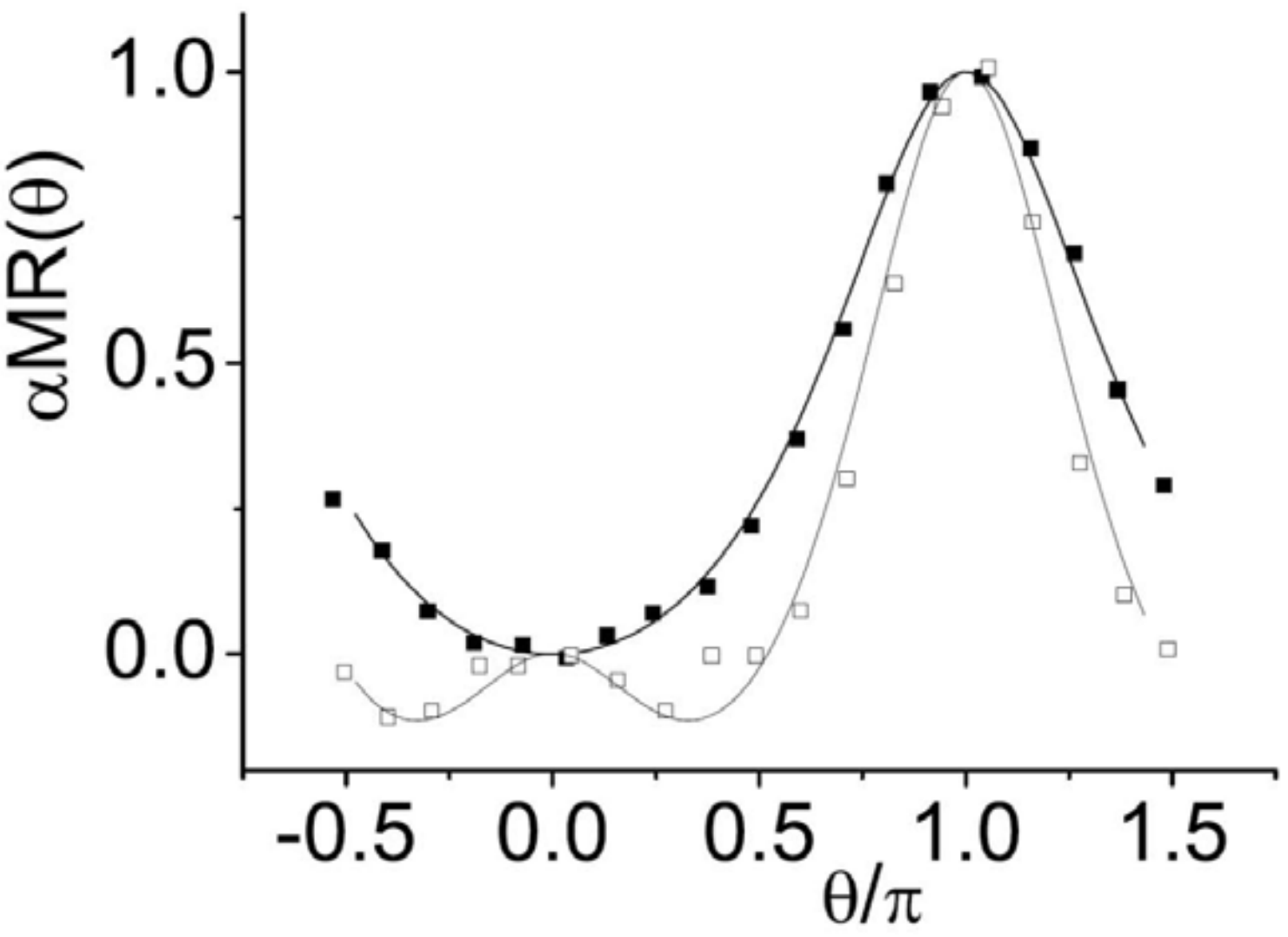

FIG. 38: Angular magnetoresistance (aMR) of Py|Cu|Py spin valves with a symmetric and asymmetric resistance distribution in the magnetically active region. The data points are experimentals results from Urazhdin *et al.* [248] and the curves have been computed by circuit theory [249].

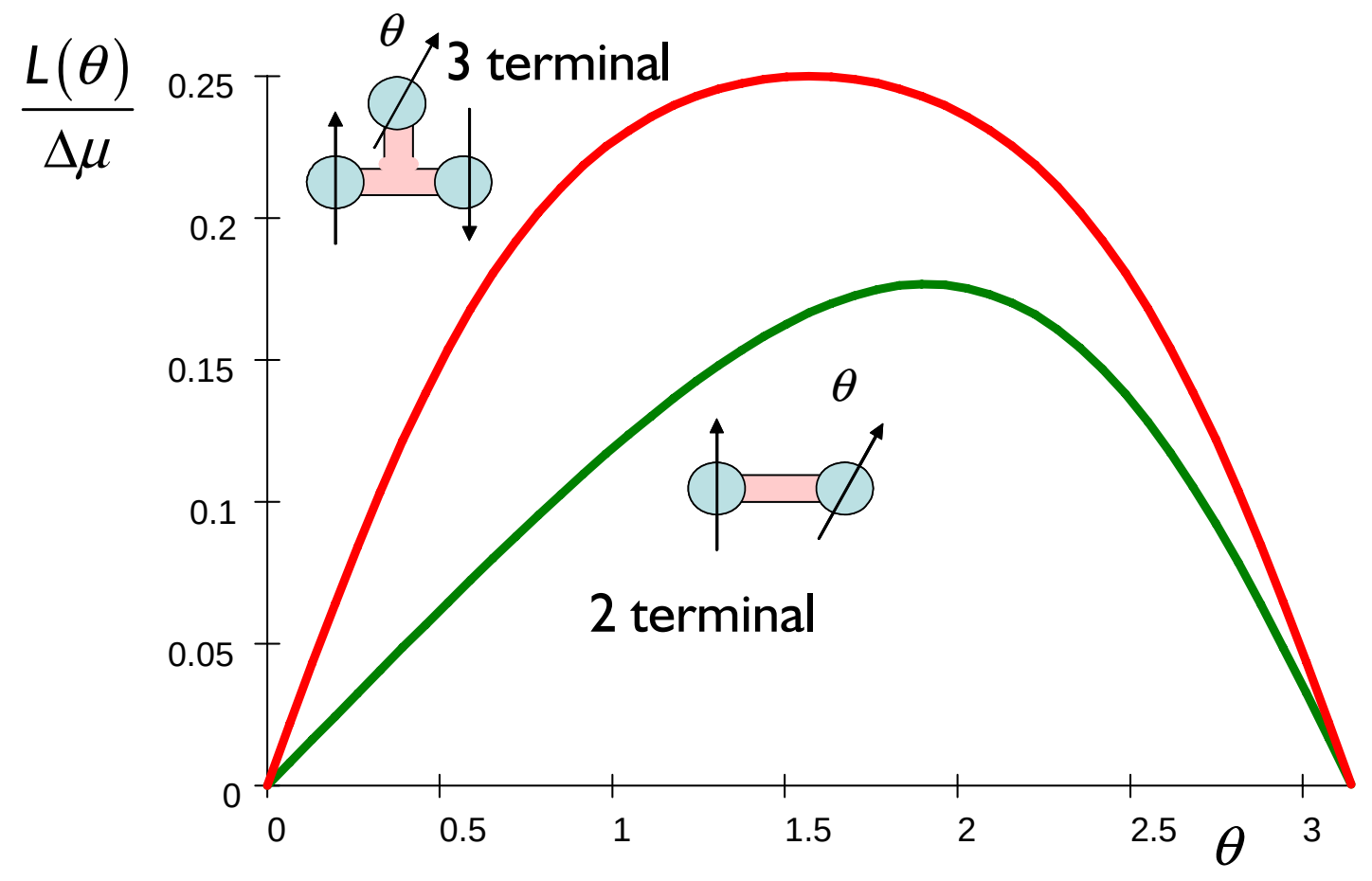

FIG. 39: The spin-accumulation induced magnetization torque for a two-terminal spin valve and a three-terminal spin-flip transistor. $\Delta\mu$ is the source-drain bias and all contact parameters are taken to be the same, with $\mathrm{Re}\,\eta = 2$ and $\mathrm{Im}\,\eta = 0$

bias:

$$\left|\frac{\vec{L}(\theta)}{\Delta\mu}\right| = \frac{gp}{8\pi}\frac{|\eta|^2}{\operatorname{Re}\eta}\frac{\sin\theta}{1+\frac{|\eta|^2}{\operatorname{Re}\eta}-\left(1-\frac{|\eta|^2}{\operatorname{Re}\eta}\right)\cos\theta} \tag{167}$$

and particle current bias:

$$\left|\frac{\vec{L}(\theta)}{hI_c/e}\right| = \frac{p}{4\pi}\frac{|\eta|^2}{\eta_r}\frac{\sin\theta}{1-p^2+\frac{|\eta|^2}{\operatorname{Re}\eta}-\left(1-p^2-\frac{|\eta|^2}{\operatorname{Re}\eta}\right)\cos\theta} \tag{168}$$

The torque is a vector normal to the plane defined by the magnetization vectors. The sign should be chosen such that at a positive particle current (from left to right) the torque rotates the magnetization vectors to the left. Note that the imaginary part of the mixing conductance is here taken into account explicitly, but the torque remains coplanar to the magnetization of the contacts, *i.e.* the effect of an out-of-plane "effective" field vanishes. Previous results [26, 56] are recovered in the limit that $\eta \to 2$ and $p \to 1$. The torque is maximized by the polarization of the injecting contact and the mixing conductance of the interface to the ferromagnet that should be rotated.

Stiles and Zangwill [120] directly solved the Boltzmann equation for spin valves to obtain angular magnetoresistance and spin torque, approximating the mixing conductance by the number of modes (note that in a direct solution of the Boltzmann equation this parameters should not be renormalized). The numerical results agree well with the functional form (165) [102]. Slonczewski [239] obtained a similar result with a simplified circuit theory and also pointed out the relation between the angular magnetoresistance and the spin torque. Shpiro *et al.* [94] found the form (165) to be valid in the limit of vanishing exchange splitting, thus in a regime different from the transition metal ferromagnets considered here [69].

When the spin valve is asymmetric, the minima of in the non-monotonous angular magnetoresistance of Fig. 37 correspond to zero torque as plotted in Fig. 40, which has important consequences for the switching dynamics in spin valves structures under an applied bias [250]. Note that similar results can be obtained as well when bulk scattering is dominant [108] and when the exchange splitting is small [94]. In [241] an analytic expression is derived by circuit theory for the torques in asymmetric junctions.

The effect of the magnetization torque on the dynamics of the ferromagnetic order parameter will be reviewed elsewhere [33].

### C. Johnson's spin transistor

We now proceed to discuss many three terminal system with a single node. We assume in the following that the typical transport time of a spin into the node is shorter than the spin-flip relaxation time, so that the right hand side of (90) effectively vanishes.

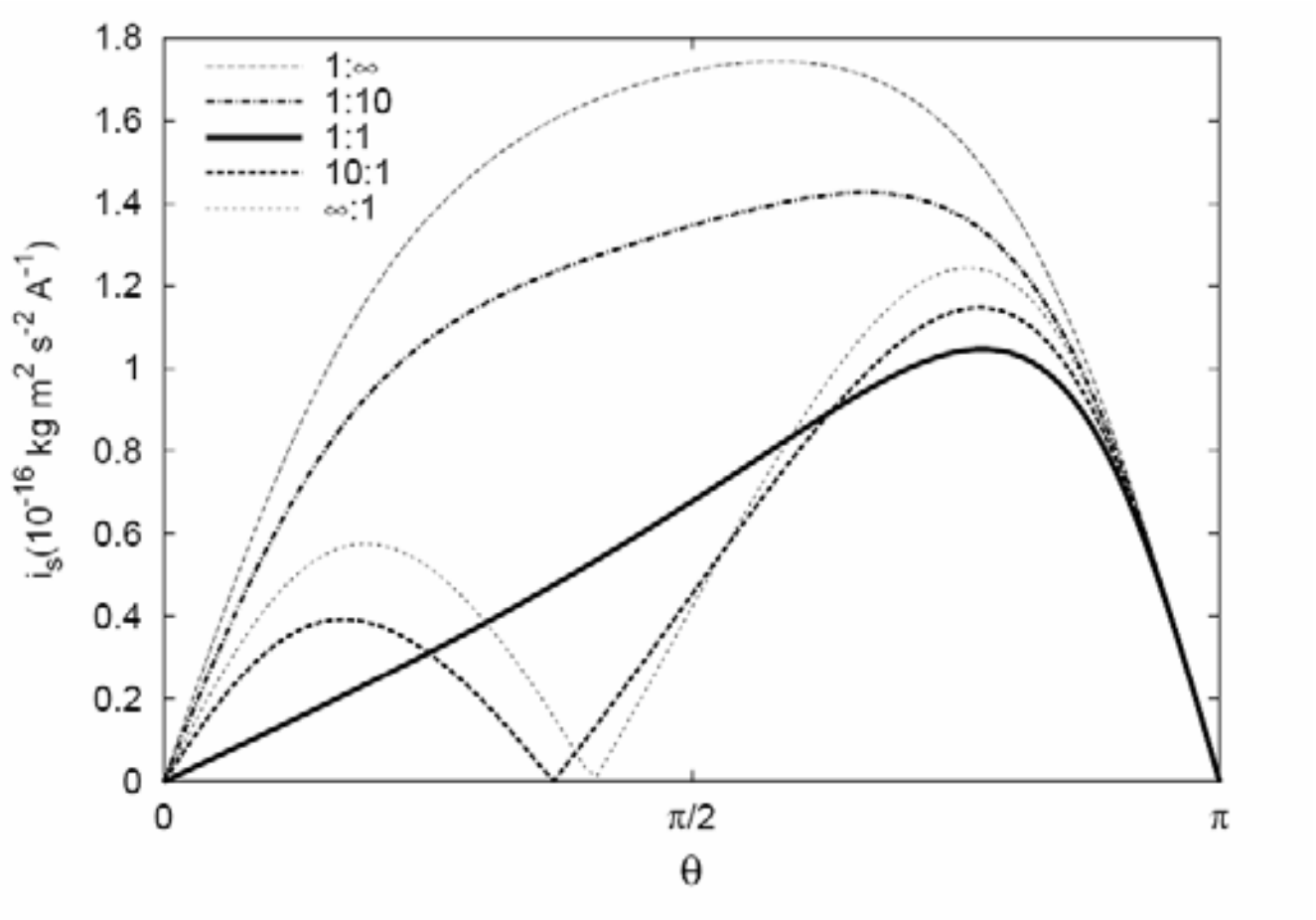

FIG. 40: Magnetization torque of a spin valve with asymmterically distributed resistances of the magnetically active region of the two ferromagnetic contacts.

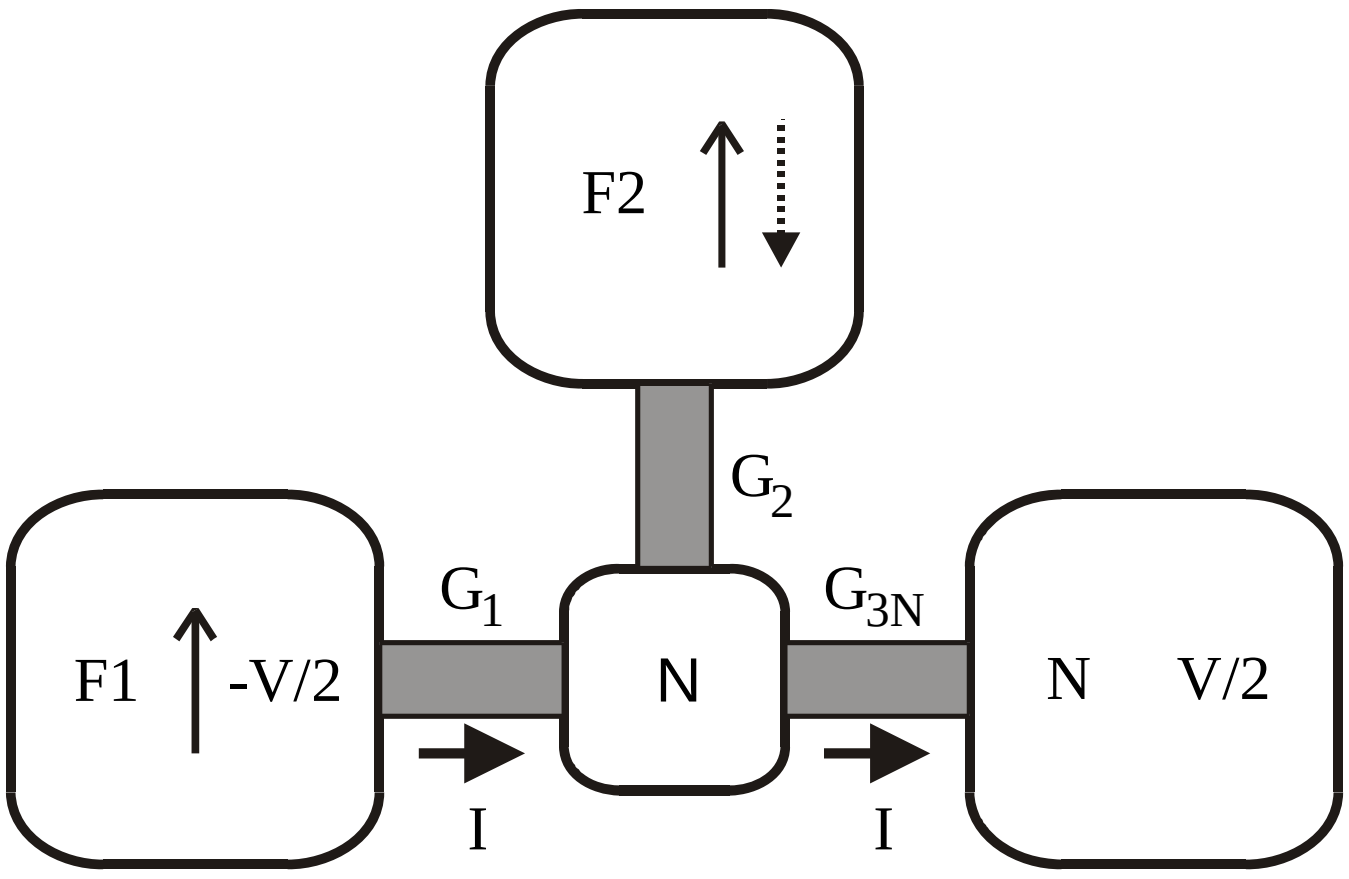

FIG. 41: The Johnson [11] transistor consisting of a normal metal film with two ferromagnetic, F1 and F2, and one normal metal contact N. F1 and N are soure and drain, and the voltage at F2 is a function of the relative magnetic orientation of F1 and F2.

Let us now consider a set-up similar to the (pedagogical model of) Johnson's spin-transistor [11] as shown in Fig. 41 which was also discussed by Geux *et al.* [251] in the context of a multi-terminal Landauer-Büttiker formalism for collinear magnetization configurations. A small normal metal node $N$ is attached to two ferromagnetic reservoirs and one normal metal reservoir by three junc-tions. A voltage bias applied to the ferromagnetic reservoir F1 and the normal metal reservoir N causes a current between the same reservoirs passing through the normal metal node. The spin-accumulation on the normal metal node injected by F1 affects the chemical potential of ferromagnet F2, which is adjusted such that the charge current into F2 vanishes. We characterize the junction between the first (second) ferromagnet and the normal metal node by the total conductance $G_1 = G_1^{\uparrow} + G_1^{\downarrow}$ ($G_2 = G_2^{\uparrow} + G_2^{\downarrow}$), the polarization $p_1 = (G_1^{\uparrow} - G_1^{\downarrow})/G_1$ ($p_2 = (G_2^{\uparrow} - G_2^{\downarrow})/G_2$) and the relative mixing conductance $\eta_2 = 2G_2^{\uparrow\downarrow}/(G_2^{\uparrow} + G_2^{\downarrow})$ ($\eta_1 = 2G_1^{\uparrow\downarrow}/(G_1^{\uparrow} + G_1^{\downarrow})$). The junction between the normal metal reservoir and the normal metal node is characterized by a single conductance parameter, $G_{3N}$. $\theta$ is the relative angle between the magnetization of ferromagnet F1 and ferromagnet F2. We assume that the typical rate of spin-injection into the node is faster than the spin-flip relaxation rate, so that the right hand side of (90) can be set to zero.

The current through the normal metal node is insensitive with respect to a flip in the magnetization direction of ferromagnet F1 or F2 when the interface conductance $G_2$ is small: $I(\theta) \simeq I(\theta + \pi)$. The chemical potential of ferromagnet F2 is modified, however, since it is sensitive to the magnitude and direction of the spin-accumulation on the normal metal node: $\mu_2(\theta) \neq \mu_2(\theta+\pi)$. A spin-'resistance' $R_s(\theta)$ can be defined as the ratio between the difference in the chemical potential of ferromagnetic F2 when ferromagnet F1 or ferromagnet F2 is flipped: $R_S = (V_2(\theta + \pi) - V_2(\theta))/I(\theta)$. In the collinear configuration ($\theta = 0$) the spin-resistance $R_s(\theta = 0)$ is thus the ratio between the difference in the chemical potential of ferromagnetic F2 when its magnetization is parallel ($\mu_2^{\mathrm{P}}$) and anti-parallel ($\mu_2^{\mathrm{AP}}$) to the magnetization of ferromagnet F1 and a current ($I$) passes from ferromagnet F1 to the normal metal reservoir N. With the aid of the general conductances, valid for arbitrary junctions, we solve for the non-equilibrium distribution function on the normal metal node (90) under the condition that no particle current enters ferromagnet F2. Using the solution for the non-equilibrium distribution function we find the current (95) through the system and subsequently the non-equilibrium chemical potential of ferromagnet F2.

Let us first discuss the results in the collinear configuration, $\theta = 0$ and $\pi$. The spin-resistance

$$R_S(\theta = 0) = \frac{2p_1 p_2}{G_1(1 - p_1^2) + G_{3N} + G_2(1 - p_2^2)}\,, \tag{169}$$

does not depend on the relative mixing conductances $\eta_1$ and $\eta_2$ that are only relevant for the transport properties in systems with non-collinear magnetization configurations. The spin-resistance is proportional to the product of the polarizations of the junctions to ferromagnet F1 and ferromagnet F2. In order to measure a large effect of the spin-accumulation, *e.g.* a large spin-resistance, highly resistive junctions are advantageous. On the other hand, the resistance has to be small enough so that the electron dwell time is shorter than the spin-flip relaxation. Eq.

(169) covers a large class of experiments, since we have not specified any details about the junctions between the reservoirs and the normal metal node but is valid only for a normal metal node that is smaller than the spin-diffusion length and can therefore not be applied directly to Johnson's experiment [252].

The general formulation in terms of transmission probabilities of 251 is exact, but, in order to include the effects of spin-relaxation the transmission probabilities were treated as pair-wise resistors between the reservoirs. This corresponds to an equivalent circuit in which resistors connect the three reservoirs in a "ring" topology. The present model, on the other hand, can be described by a "star" configuration circuit, in which all resistors point from the reservoirs to a single node. The present model is more accurate when the contacts dominate the transport properties, whereas Geux's model is preferable when the resistance of the normal metal island is important. Effectively, Johnson's thin film device appears to be closer to the star configuration.

Let us now proceed to discuss the results when the magnetization directions are non-collinear. The analytical expression for the spin-resistance is much simpler when the two contacts F1-N and F2-N are identical, $G_1 = G_2 \equiv G$, $p_1 = p_2 \equiv p$ and $\eta_1 = \eta_2 \equiv \eta$. Furthermore we disregard the imaginary part of the mixing conductance which is very small or zero in the model calculations of tunnel, ballistic and diffusive contacts presented in this paper as well as in first-principle band-structure calculations [36]. The resulting spin-resistance

$$R_S(\theta) = \frac{2(G_{3N} + 2G\eta)p^2\cos\theta}{\left[G_{3N} + G(1+\eta-p^2)\right]^2 - G^2(1-\eta-p^2)^2\cos^2\theta} \tag{170}$$

is an even function of the relative angle between the magnetization directions $\theta$ and we recover (169) when $\theta = 0$. The spin-resistance vanishes when the magnetizations are perpendicular $\theta = \pi/2$ as expected from symmetry considerations. The angular dependence is approximately proportional to $\cos\theta$ when the relative mixing conductance is not too large, $\eta \approx 1$. For larger mixing conductances, $\eta \gg 1$, the spin-accumulation on the normal metal island is suppressed in the perpendicular configuration $\theta = \pi/2$ due to the increased transport rates for spins between the normal metal node and ferromagnet F2. Consequently the spin-resistance is small and only weakly dependent on the relative angle around $\theta = \pi/2$.

### D. Laterally structured multi-terminal devices

The Groningen Group did pioneering work in the transport studies of lateral metallic magnetic structures based on the Cu/Py with Ohmic contacts [23] and Al/Co tunneling contacts [24]. Compared to multilayers in which the interfaces often dominate the transport properties, bulk scattering is often more important in these lateral structures. This requires solving the diffusion equations with boundaries conditions governed by the mixing conductances [63]. More recently, a single normal metal Al island smaller than the spin diffusion length and four Co tunneling contacts have been experimentally realized [9]. Since now the spin accumulation in the normal metal is constant, one may disregard the resistances of the normal metal node and the ferromagnetic bulk, the most basic form of circuit theory applies. Zaffalon and van Wees [253] also analyzed the current-voltage characteristics in this system of a metallic dot with four contacts taking into account the spin precession due to an applied magnetic field and derived analytical results for the spin accumulation are obtained. By detailed analysis of the experimental data they most parameters of their sample, such as the magnetization direction of the contacts can thereby be obtained. Unfortunately, the relative low conductances of the tunnel junctions causes long residence times of the electrons on the dot, making them susceptible to bulk spin flip scattering that in the experiments dominate the effects of the contact mixing conductances that could therefore not yet be measured in these lateral samples. Recently, increased spin-related signals in multi-terminal lateral structures with Ohmic Py/Cu contacts have been measured [254, 255]. Improved lateral structures allow controlled experiments on the energy dependence of spin-selective tunneling through magnetic junctions [256].

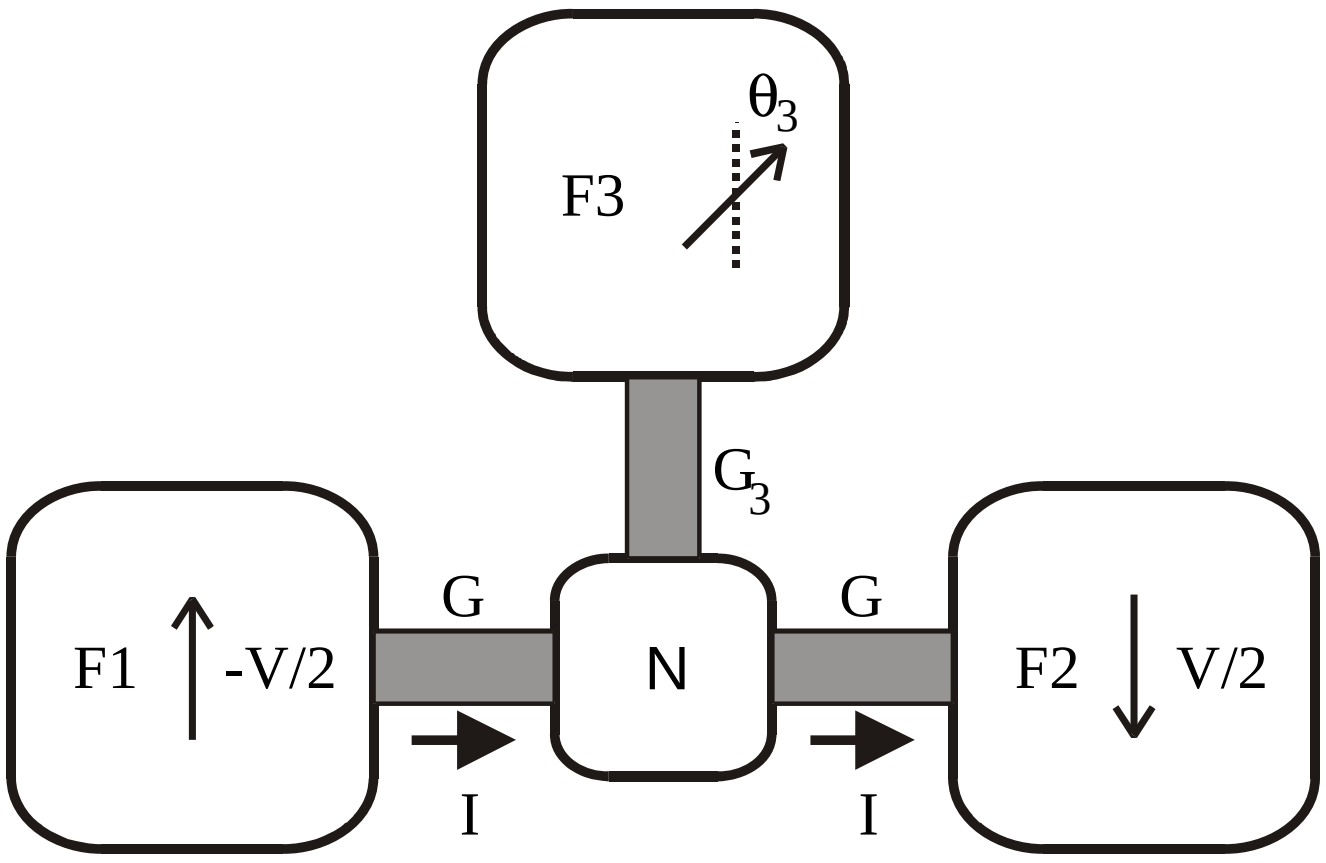

FIG. 42: A three terminal device comprising of three ferromagnetic reservoirs. F1 and F2 are source and drain contacts with fixed antiparallel magnetizations, whereas F3 is magnetically soft with a variable magnetization direction.

### E. Spin-flip transistor

Here we compute the characteristics of the novel 3-terminal 'spin-flip transistor' device shown in Fig. 42. A normal metal node (N) is connected to 3 ferromagnetic reservoirs (F1, F2 and F3) by arbitrary junctions parameterized by our spin-conductances. A source-drain bias voltage $V$ applied between reservoir 1 and 2 causes an electric current $I$ between the same reservoirs. The charge flow into reservoir 3 is adjusted to zero by the chemical potential $\mu_3$. Still, the magnetization direction $\vec{m}_3$ influences the current between reservoir 1 and 2. We assume that spin relaxation in the normal node can be disregarded so that the right hand side of (90) is set to zero. Furthermore, we assume that the voltage bias $V$ is sufficiently small so that the energy dependence of the transmission (reflection) coefficients can be disregarded. To further simplify the results the junctions 1 and 2 are taken to be identical, $G_1^{\uparrow} = G_2^{\uparrow} \equiv G^{\uparrow}$, $G_1^{\downarrow} = G_2^{\downarrow} \equiv G^{\downarrow}$ and $G_1^{\uparrow\downarrow} = G_2^{\uparrow\downarrow} \equiv G^{\uparrow\downarrow}$. Contact 3 is characterized by the conductances $G_3^{\uparrow}$, $G_3^{\downarrow}$ and $G_3^{\uparrow\downarrow}$. We find the distribution in the normal node by solving the four linear Eqs. (90). The current through the contact between reservoir 1 (2) and the node is obtained by inserting the resulting distribution for the normal node into Eq. (95).

When the magnetizations in reservoir 1 and 2 are parallel there is no spin-accumulation since contacts 1 and 2 are symmetric and consequently ferromagnet 3 does not affect the transport properties. The current is then simply a result of two total conductances $G = G^{\uparrow} + G^{\downarrow}$ in series, $I = GV/2$. The influence of ferromagnet 3 depends on the spin accumulation in the normal metal node and largest when the magnetizations of the source and drain reservoirs are antiparallel, $\vec{m}_1 \cdot \vec{m}_2 = -1$. We denote the relative angle between the magnetization in reservoir 3 and reservoir 1 (reservoir 2) $\theta_3$ $(\pi - \theta_3)$. The current is an even function of $\theta_3$ and symmetric with respect to $\theta_3 \to \pi - \theta_3$ as a result of the symmetry of the device, e.g. the current when the magnetizations in reservoir 1 and 3 are parallel equals the current when the magnetizations in reservoir 1 and 3 are antiparallel. Due to the finite mixing conductance at non-collinear magnetization the third contact acts as a drain for the spin-accumulation in the node, thus allowing a larger charge current between reservoir 1 and 2. The relative increase of the current due to the reduced spin-accumulation $\Delta_3(\theta_3) = [I(\theta_3) - I(\theta_3 = 0)]/I(\theta_3 = 0)$, is plotted in Fig. 43 as a function of $\theta_3$. The maximum of $\Delta_3$ is achieved at $\theta_3 = \pi/2$ $(\theta_3 = 3\pi/2)$ and equals $(\mathrm{Im}G_3^{\uparrow\downarrow} = 0)$

$$\Delta_3 = p^2 \frac{2GG_3}{2G + G_3\eta_3} \cdot \frac{\eta_3 - 1 + p_3^2}{2G(1-p^2) + G_3(1-p_3^2)} \tag{171}$$

introducing the total conductance of the contact $G_i = G_i^{\uparrow} + G_i^{\downarrow}$, the polarization of the contact $p_i = (G_i^{\uparrow} - G_i^{\downarrow})/(G_i^{\uparrow} + G_i^{\downarrow})$ and the relative mixing conductance $\eta = 2G_i^{\uparrow\downarrow}/(G_i^{\uparrow} + G_i^{\downarrow})$. The influence of the direction of the magnetization of the reservoir 3 increases with increasing polarization $p$ and increasing relative mixing conductance $\eta_3$ and reaches its maximum when the total conductances are of the same order $G_3 \sim G$. Note that the physics of this three terminal device is very different from that of Johnson's spin transistor [257]; the latter operates with collinear magnetizations of two ferromagnetic contacts whereas the third may be normal.

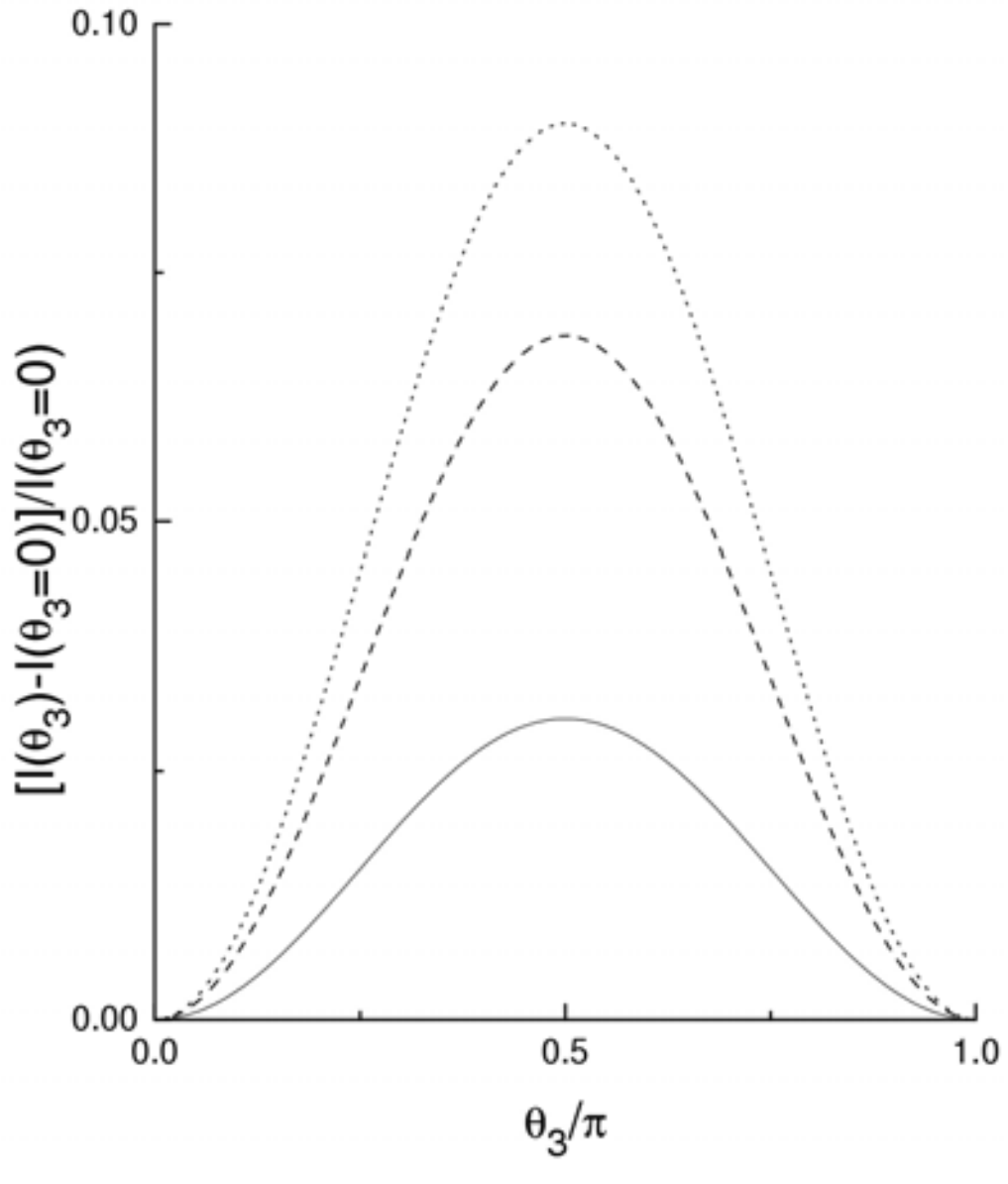

FIG. 43: Current *vs.* relative magnetization direction in ferromagnet 2.

By circuit theory it is straightforward to compute the torque on the base contact of the spin-flip transistor with antiparallel source-drain magnetizations [36]. Let us assume that the three contacts are identical, and the base contact magnetization lies in the plane of the source and drain magnetizations. Assuming that we may disregard spin-flip scattering in the base contact, the in-plane torque $L_b$ turns out to be always larger than the spin valve torque $L$ in the two-terminal spin valve, Eq. (167), with a symmetric and flatter dependence on the angle of the base magnetization direction $\theta$ (see Fig. 39)

$$L_b\left(\theta\right) = -\frac{gp\,\mathrm{Re}\,\eta\sin\theta}{\left(1-\mathrm{Re}\,\eta\right)\cos^2\theta + \mathrm{Re}\,\eta + \frac{6}{2+\left|\eta\right|^2/\left(\mathrm{Re}\,\eta\right)^2}}\frac{\Delta\mu}{4\pi}. \tag{172}$$

In the presence of a significant imaginary part of the mixing conductance, we also find an out-of-plane (effective field) torque $L_{\perp}\left(\theta\right)$ with the same angular dependence and

$$\frac{L_{\perp}}{L_b} = -2\frac{\mathrm{Im}\,\eta\,\mathrm{Re}\,\eta}{\left|\eta\right|^2 + 2\,\mathrm{Re}\,\eta}. \tag{173}$$

The spin-flip transistor might become an element of an alternative MRAM architecture that efficiently employs the spin-transfer effect.

### F. Spin-torque transistor

We have discussed above that in the "spin-flip transistor" the source-drain current is modulated by the base magnetization direction via the spin accumulation in the conducting channel. In the following, we investigate the device parameters of the spin-flip transistor operated as an amplifier by controlling the base magnetization by a second spin valve in an integrated device that we call "spin-torque transistor" (Fig. 44). The lower part of this device consists of source and drain contacts made from high-coercivity metallic magnets with antiparallel magnetizations that are biased by an electrochemical potential $\mu_S$. The source-drain electric current $I_{SD}$ induces a spin accumulation in the normal metal node *N1.* We attach an electrically floating base (or gate) electrode *B,* which is magnetically very soft and has good electric contact to *N1.* When the magnetization angle $\theta$ is not 0 or $\pi$ a spin current flows into the base that decreases the spin accumulation and increases $I_{SD}$ with $\theta$ up to $\pi/2$. On the other hand, the spin accumulation in *N1* exerts a torque on $B$ which strives to lower $\theta$. $\theta$, and thus $I_{SD}$ could be modulated, *e.g.*, by the Ørsted magnetic field generated electrically by the "write line" of an MRAM element, but this does not appear viable. We therefore propose the transistor in Fig. 1, which integrates a second spin valve with magnetizations rotated by $\pi/2$ from the lower one. An applied bias $\mu_B$ creates a another torque which pulls the magnetization into the direction collinear to the upper contacts. The base electrode then settles into a configuration at which both torques cancel each other. A variation in $\mu_B$ then modulates $\theta$ and consequently $I_{SD}$. In the following we discuss the figures of merit of the transistor action, *viz.* the transconductance and the current gain of this device.

For most transition metal based structures exchange splittings are large, Fermi wavelengths short, and interfaces disordered. Electron propagation is therefore diffuse and ferromagnetic (transverse spin) coherence lengths are smaller than the mean-free path [69]. In these limits magnetoelectronic circuit theory is a convenient formalism [12, 34]. Spin-flip relaxation can be disregarded in the normal metal node of small enough structures, since Al and Cu have spin-flip diffusion lengths of the order of a micron [23, 24]. Spin-flip in the source and drain electrodes can simply be included by taking their magnetically active thickness as the smaller of the spin-flip diffusion length and physical thickness. The base electrode is assumed to be magnetically soft and the thickness is taken to be smaller than the spin-flip diffusion length. These assumption are not necessary, since magnetic anisotropies and spin-flip in the base can readily be taken into account, but these complications only reduce the device performance and will be treated elsewhere. The source-drain

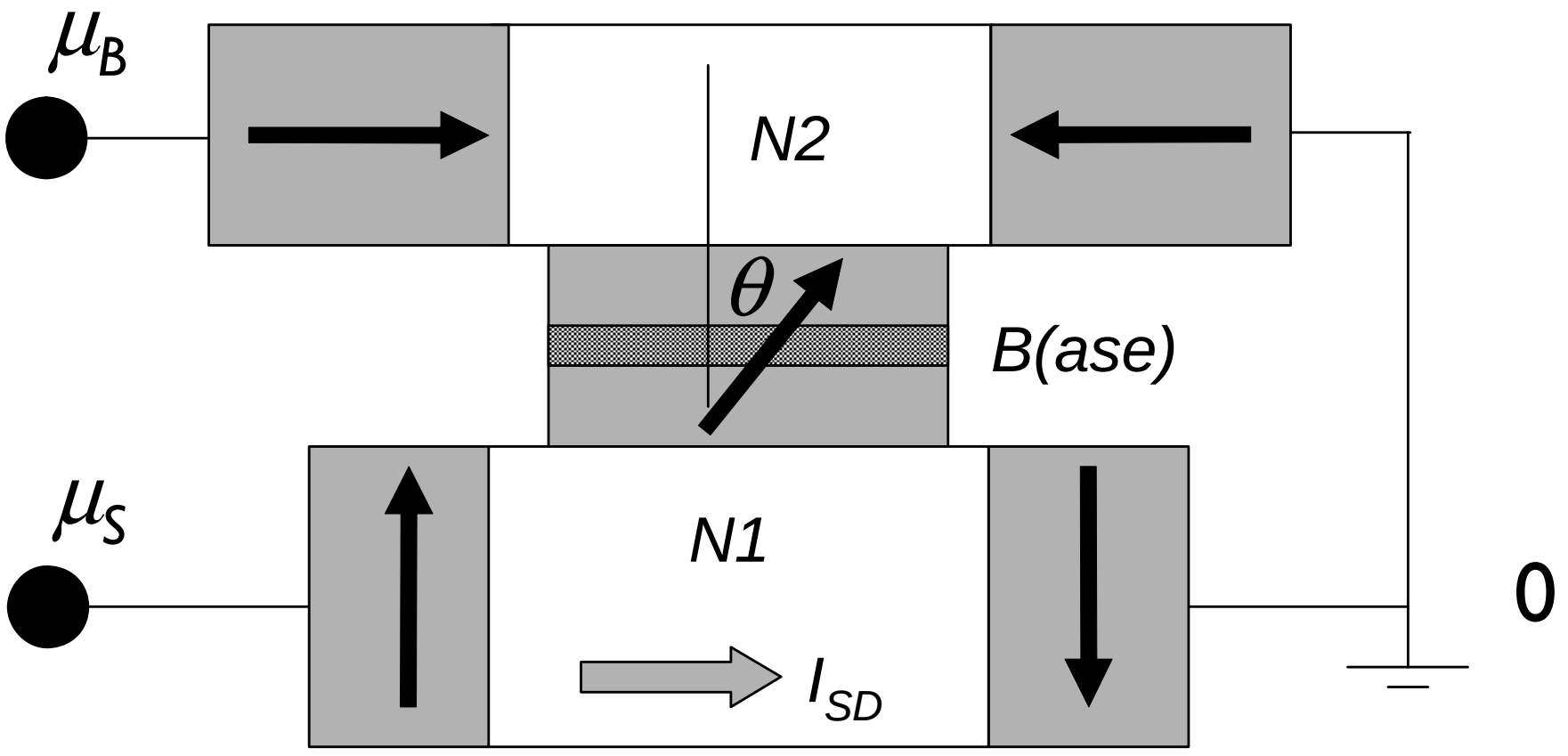

FIG. 44: Schematic sketch of the spin-transfer transistor consisting of two spin-flip transistors with a common base contact $B$ and source drain contact magnetizations which are rotated by 90°. The magnetization direction of the base $B$ is controlled by the chemical potentials $\mu_B$ and $\mu_S$.

current dependence on the base magnetization angle $\theta$ then reads [12]:

$$I_{SD}(\theta) = \frac{e}{h}\frac{g_S\mu_S}{2}\frac{2\left(g_B^{\uparrow\downarrow} + g_S\left(1-p_S^2\right)\right)g_S^{\uparrow\downarrow} + g_B^{\uparrow\downarrow}\left(-2g_S^{\uparrow\downarrow} + g_S\left(1-p_S^2\right)\right)\cos^2\theta}{2\left(g_S + g_B^{\uparrow\downarrow}\right)g_S^{\uparrow\downarrow} + g_B^{\uparrow\downarrow}\left(g_S - 2g_S^{\uparrow\downarrow}\right)\cos^2\theta}, \tag{174}$$

where $g_S = g_S^{\uparrow} + g_S^{\downarrow}$ and $p_S = \left(g_S^{\uparrow} - g_S^{\downarrow}\right)/g_S$ are the normal conductance and polarization of the source, and $g_S^{\uparrow\downarrow}$ and $g_B^{\uparrow\downarrow}$ are the "mixing conductances" of the source and base contacts, respectively. Drain and source contact conductances are taken to be identical. All conductance parameters are in units of the conductance quantum $e^2/h$, contain bulk and interface contributions [34], can be computed from first-principles and are taken to be real [36]. The torque on the base magnetization created by the spin-accumulation is proportional to the transverse spin-current [34] into $B$:

$$L_B(\theta) = \frac{1}{2\pi}\frac{p_S g_S g_S^{\uparrow\downarrow} g_B^{\uparrow\downarrow}\sin\theta\mu_S}{2\left(g_S + g_B^{\uparrow\downarrow}\right)g_S^{\uparrow\downarrow} + g_B^{\uparrow\downarrow}\left(g_S - 2g_S^{\uparrow\downarrow}\right)\cos^2\theta}, \tag{175}$$

A steady state with finite $\theta$ exists when $L_B(\theta)$ equals an external torque, either from an applied magnetic field, or a spin accumulation from the upper side in Fig. 1. The differential source-drain conductance $\tilde{G}_{SD}$ subject to the condition of a constant external torque reads:

$$\tilde{G}_{SD} \equiv \left(\frac{\partial I_{SD}(\theta)}{\partial \mu_S}\right)_{L_B} = \frac{I_{SD}}{\mu_S} + \left(\frac{\partial I_{SD}}{\partial \theta}\right)_{\mu_S} \left(\frac{\partial \theta}{\partial \mu_S}\right)_{L_B} \tag{176}$$

$$= \frac{I_{SD}}{\mu_S} - \left(\frac{\partial I_{SD}}{\partial \theta}\right)_{\mu_S} \frac{L_B(\theta)}{\mu_S \left(\frac{\partial L_B(\theta)}{\partial \theta}\right)_{\mu_S}}, \tag{177}$$

where the first term on the right hand sides is the derivative with respect to $\mu_S$ for constant $\theta$ and the second term arises from the source-drain bias dependence of $\theta$. The general equations are unwieldy and not transparent. The most important parameter turns out to be the spin-polarization $p_S$ of the source and drain contacts. We therefore choose a model system with $p_S$ variable, but other parameters fixed for convenience, *viz.* the same parameters for both spin-flip transistors and $g_B^{\uparrow\downarrow} = g_S^{\uparrow\downarrow} = g_S$, which holds approximately for metallic interfaces with identical cross sections [36]. We find that

$$\tilde{G}_{SD} = \frac{e^2}{h}\frac{g_S}{2}\left(1 - p_S^2 \frac{2 + \cos^2\theta + \frac{4\sin^2\theta}{2-\cos^2\theta}}{4 - \cos^2\theta}\right) \tag{178}$$

may become negative, since an increased source-drain bias tends to rotate the angle to smaller values, thus reducing the source-drain current. At the sign change of $\tilde{G}_{SD}$, the output impedance of the spin valve becomes infinite, which can be useful for device applications.

We now demonstrate that it is attractive to modulate $I_{SD}$ by the spin-transfer effect [26, 27, 56]. In contrast to previous work that was focussed on magnetization reversal by large currents, we envisage controlled rotations by small voltages. The base is supposed to be highly resistive, consisting of a magnetic insulator, or, alternatively, of two magnetically coupled ultrathin magnetic films separated by a thin insulator. The device might be realized in the lateral thin-film geometry by van Wees *c.s.* [9, 23, 24], using a soft magnet with a circular disk shape for the base, sandwiched in a cross configuration of normal metal films with ferromagnetic contacts. The device characteristics can be computed for the complete parameter space by circuit theory, but the important features are retained by proceeding as above and also assuming the same parameters for the upper and lower sections. The stationary state of the biased spin-transfer transistor is described by the angle $\theta_0$ at which the two torques on the base magnet cancel each other. For the present model this is the solution of the transcendental equation

$$\frac{\mu_B}{\mu_S} = \frac{7 + \cos 2\theta_0}{7 - \cos 2\theta_0} \tan\theta_0. \tag{179}$$

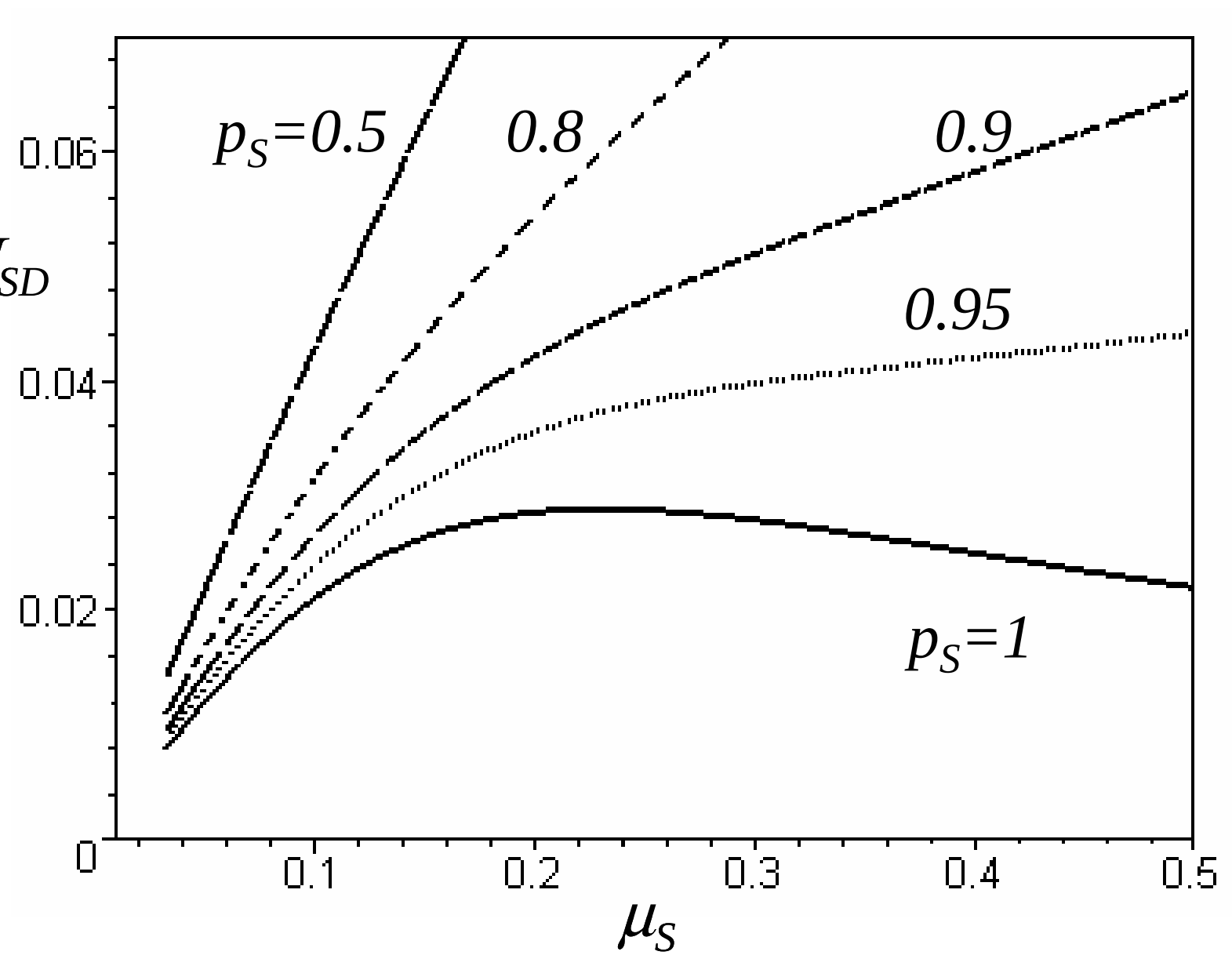

FIG. 45: Source-drain current Eq. (174) of the spin-transfer transistor, divided by the contact conductance $e^2 g_S/h$ , *i.e.* in (voltage) units of $\mu_S/e$, as a function of $\mu_S$ and the polarization $p_S$ of the source and drain contacts. A constant $\mu_B = 0.2$ (in the same units as $\mu_S$) is applied.

The calculated source-drain differential conductance (now without tilde) has to be computed now under condition of constant $\mu_B$ rather then a constant torque

$$G_{SD} \equiv \left(\frac{\partial I_{SD}(\theta)}{\partial \mu_S}\right)_{\mu_B} = \frac{I_{SD}}{\mu_S} + \left(\frac{\partial I_{SD}}{\partial \theta}\right)_{\mu_S} \left(\frac{\partial \theta}{\partial \mu_S}\right)_{\mu_B} \tag{180}$$

which is plotted as a function of $\mu_S$ and polarization $p_S$ in Fig. 2. Note that with increasing $p_S$ strong non-linearities develop, which for large polarizations lead to a zero and negative differential resistance at $\mu_B \approx \mu_S$. The physical reason is the competition between the Ohmic current, which for constant resistance increases with the bias, and the increasing torque, which at constant $\mu_B$ decreases the current, as noted above.

The differential transconductance measures the increase of the source-drain current (at constant $\mu_S$) induced by an increased chemical potential of the base electrode $T(\theta) \equiv (\partial I_{SD}(\theta)/\partial \mu_B)_{\mu_S}$. We focus discussion here on the differential current gain, *i.e.* the ratio between differential transconductance and channel conductance $\Gamma = T/G_{SD}$, as a representative figure of merit. In the regime $\mu_B \ll$

$\mu_S$ and thus small $\theta_0 \to 3\mu_B/(4\mu_S)$, the current gain becomes

$$\lim_{\mu_B \to 0} \Gamma = \frac{\frac{1}{2}\theta_0}{\frac{1-p_S^2}{1+p_S^2} - \frac{1}{3}\theta_0^2}. \tag{181}$$

For small polarizations the $\theta_0^2$ in the denominator may be disregarded and $\Gamma \sim \theta_0$, thus is proportional to the control potential $\mu_B$. When the polarization is close to unity, however, we see that $\Gamma$ becomes singular at small angles and changes sign. This behavior reflects the negative differential resistance found above for $\mu_B$ and $\theta$. For complete polarization ($p_S = 1$) $\Gamma = -3/(2\theta_0)$. For polarizations (slightly) smaller than unity we may tune the transistor close to the optimal operation point of infinite output impedance

$$\theta_{0,c} = \sqrt{3\frac{1-p_S^2}{1+p_S^2}} \tag{182}$$

at which $\Gamma \sim (\theta_0 - \theta_{0,c})^{-1}$.

The working principle of this spin-transfer transistor is entirely semiclassical, thus robust against, for example, elevated temperatures. The derivations assumed absence of phase coherence and electron correlation, but the physics most likely survives their presence. The base contact is preferably a magnetic insulator or contains a thin insulating barrier (F|I|F), but the contact to the normal metal should be good (for a large mixing conductance). Tunnel junctions may be used for the source-drain contacts, but this will slow down the response time. It should be kept in mind as well that the dwell time of electrons in the device must be larger than the spin-flip relaxation time. The basic physics, such as the non-linearity of the source-drain conductance in Fig. 45, should be observable for conventional ferromagnetic materials. Large current gains exist for incomplete polarization close to unity of the source and drain ferromagnets, but at the cost of non-zero "off" currents. A useful device should therefore be fabricated with (nearly) half-metallic ferromagnets for sources and drains. As base magnet, a thin film of any soft ferromagnet is appropriate as long as it is thicker than the ferromagnetic (transverse spin) coherence length, but not too thick in order to keep the response to torques fast. We recommend a couple of monolayers of permalloy on both sides of a very thin alumina barrier. A possible schematic sample geometry is sketched in Fig. VI F.

In conclusion, we propose a robust magnetoelectronic three terminal device which controls charge currents via the spin-transfer effect. It can be fabricated from metallic thin films in a lateral geometry, but its usefulness will derive from the availability of highly polarized (half-metallic) ferromagnets.

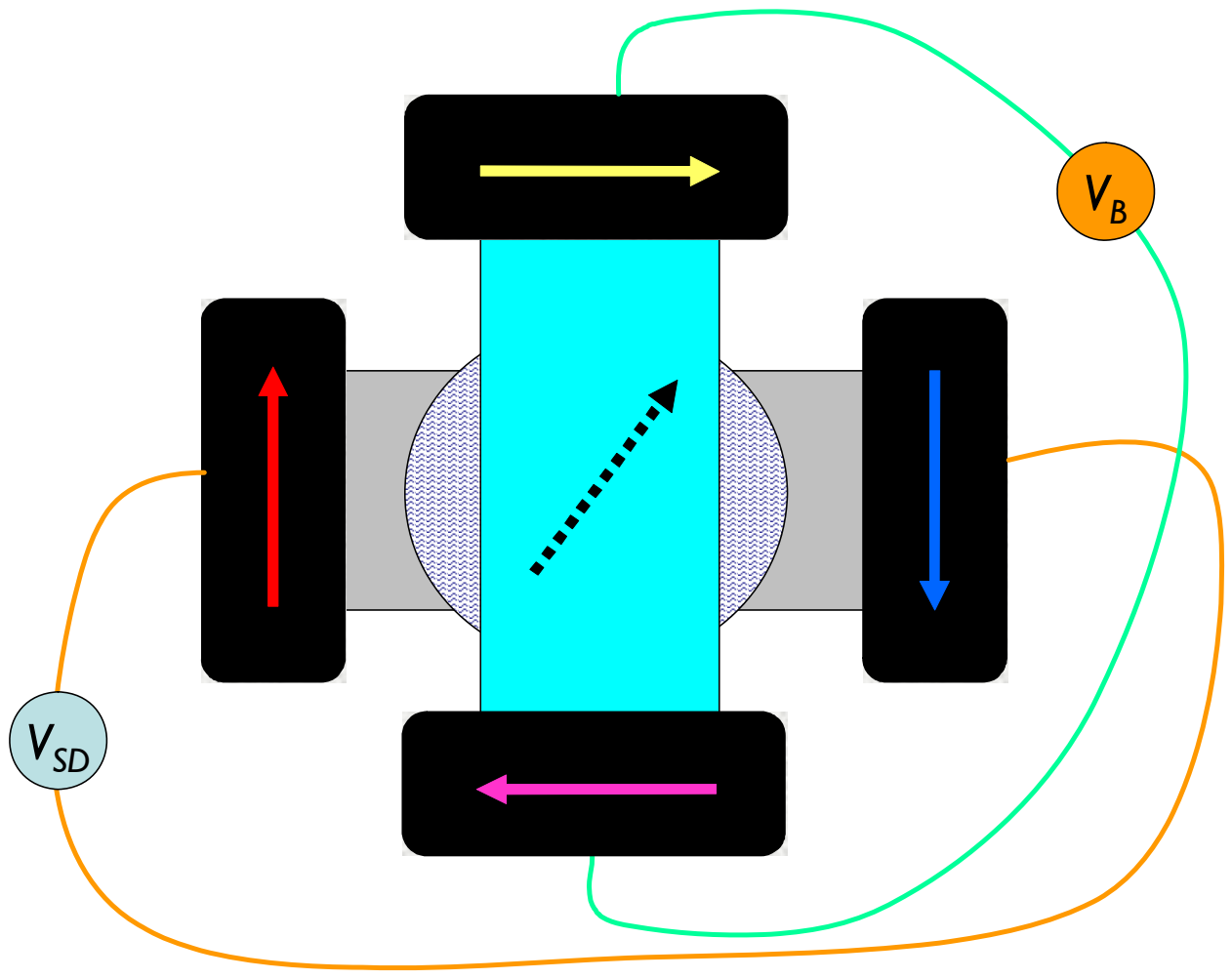

## VII. BEYOND SEMICLASSICAL TRANSPORT

Previous sections focussed on the regime, in which the circuit properties are uniquely specified by the circuit elements. We then take for granted that two single resistors in series cause a resistance that is just the sum of both. This is called semiclassical since quantum non-localities are supposed to be averaged out by disorder or chaotic scattering. This approach is valid for conductors with a sufficient amount of scattering, either in the bulk or at the boundary of the system that effectively scramble the electron distribution functions. This is the case for most state-of-the-art hybrid ferromagnet-normal metal systems because of their very high electron density and metallic bonding that is tolerant to lattice imperfections and interface disorder. Deviations from the semiclassical approximation such as resonant tunneling, quantum well states and weak localization have been studied in great detail in semiconductor heterostructures like quantum wells and the two-dimensional electron gas in which the scattering mean free path can reach macroscopic length scales. In very thin metallic multilayers, quantum interference effects beyond the semiclassical approximation are well known as the non-local oscillatory exchange coupling that can be understood in terms of a magnetic-configuration dependent population of quantum wells states [64]. In dedicated metallic structures it has been demonstrated experimentally that also in transport ballistic [111] and quantum-interference effects [112] can be resolved. Theoretically, transport in magnetic multilayers has often been treated quantum mechanical as well, be it only for the reason that it is sometimes easier to completely disregard effects of disorder [59, 258]. These calculations are useful to helps understand transport in inhomogeneous systems, but are not always experimentally relevant an except for strongly sample-dependent transport properties do not reveal much new physics. The regime between the ballistic and diffuse can be investigated by quantum statistical methods [237] or by brute force numerical simulations [35, 113, 259]. Here we would like to emphasize that the additional degrees of freedom in hybrid ferromagnet-normal metal systems may give rise to novel spin effects that have not yet been studied very much. In this Section we present just one example of a new phenomenon that might become into experimental reach with improved sample quality, *viz.* a magnetoelectronic spin echo [167]. Kovalev et al. [70] discuss coherence effects in ultrathin ferromagnetic layers that might be somewhat easier to observe than the spin echo explained in the following.

### A. Magnetoelectronic circuit theory with phase coherent elements

Magnetoelectronic circuit theory as discussed in previous sections is a useful tool even when a part of the circuit cannot be treated semiclassically. It simply means we are not allowed to insert nodes into those regions that support ballistic or phase coherent transport, meaning that the resistive elements connecting semiclassical nodes become more complex. The conductance tensors are still governed by the scattering matrices and the rest of the formalism is unchanged.

Consider therefore an arbitrary compound ferromagnet-normal metal section cut out of a magnetoelectronic circuit, as depicted in Fig. 46. The scattering region is connected to a normal metal on the left ($i = 1$) and to a normal metal on the right ($i = 2$). We assume that in the left and right normal metals in Fig. 46 charge spin accumulations are excited by external circuits, as before. These nodes are characterized by a non-equilibrium distribution function $\hat{f}^{(i)}(\epsilon)$ for electrons at energy $\epsilon$ that is a $2 \times 2$ matrix in Pauli spin space. The chemical potential matrix $\hat{\mu}^{(i)} = \int d\epsilon \hat{f}^{(i)}(\epsilon)$ contains a scalar charge chemical potential component $\mu_0^{(i)} = \mathrm{Tr}\left[\hat{\mu}^{(i)}\right]/2$ and the vector spin accumulation $\vec{\mu} = \mathrm{Tr}\left[\hat{\mu}^{(i)}\boldsymbol{\vec{\sigma}}\right]/2$, $\boldsymbol{\vec{\sigma}}$ being the vector of Pauli matrices. Note that with these definitions the equilibrium values are $\mu_0^{(1)} = \mu_0^{(2)}$ and $\vec{\mu}^{(1,2)} = 0$.

We are interested in the $2 \times 2$ matrix currents $\hat{\imath}^{(i)}$ at the left- and the right-hand side of the scattering region that respond to the non-equilibrium spin accumulations. When spin currents and ferromagnetic magnetization directions are not collinear, $\hat{\imath}^{(1)} \neq \hat{\imath}^{(2)}$, *i.e.*, spin current is not conserved but absorbed by the ferromagnetic order parameters. These currents can be calculated by the rules of circuit theory in terms of the scattering matrix of the connection, albeit that the general expression for the current between two normal modes have not been given in the original papers. We find the following expression for the matrix current to the left (1)

$$\hat{\imath}^{(1)} = \frac{1}{2}\sum_{nm}\left[\hat{t}_{nm}\hat{\mu}^{(2)}\hat{t}_{nm}^{*} - \hat{\mu}^{(1)} + \hat{r}_{nm}\hat{\mu}^{(1)}\hat{r}_{nm}^{*}\right]. \tag{183}$$

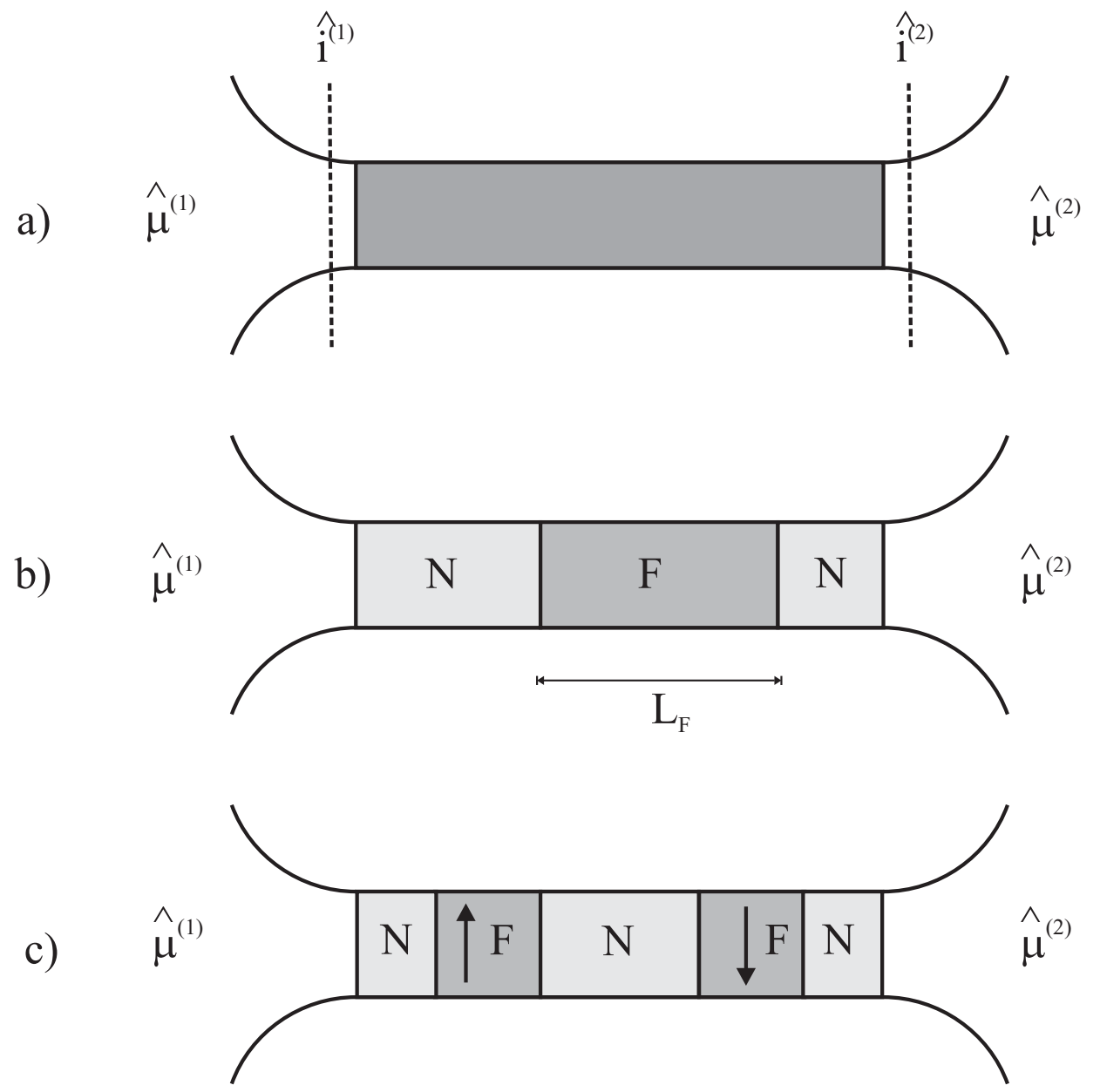

FIG. 46: a) The scattering region couples the left (1) and right (2) normal-metal nodes with nonequilibrium chemical potentials $\hat{\mu}^{(1)}$ and $\hat{\mu}^{(2)}$. b) The scattering region consists of one uniform ferromagnet and normal metals. c) The scattering region consists of two identical antiferromagnetically-aligned ferromagnets separated by a normal metal. A partial electron spin echo can be seen in transport in this configuration.

Similarly, for the right side (2) the matrix current is

$$\hat{\imath}^{(2)} = \frac{1}{2} \sum_{nm} \left[ \hat{t}'_{nm} \hat{\mu}^{(1)} \hat{t}'^{*}_{nm} - \hat{\mu}^{(2)} + \hat{r}'_{nm} \hat{\mu}^{(2)} \hat{r}'^{*}_{nm} \right] , \tag{184}$$

where $\hat{r}_{n,m}$, $\hat{t}_{n,m}$, $\hat{r}'_{n,m}$, and $\hat{t}'_{n,m}$ are reflection and transmission matrices of the scattering region both in the space spanned by the transverse channels (momentum labeled by $n$ and $m$) and the spin space (denoted by the hats). Possible quantum interference effects within the structure are automatically included by a proper evaluation of the scattering matrices. When the magnetizations are collinear, we obtain from the matrix currents, Eq. (183) and Eq. (184), expressions that are similar to Eq. (97) and Eq. (98) found previously. In the following we present a system in which quantum interference effects lead to measurable consequences in the current-voltage characteristics.

## B. Magnetoelectronic spin echo

In previous sections we discussed that a spin current polarized perpendicular to the magnetization direction decays on the scale of the ferromagnetic spin-coherence length when penetrating a metallic ferromagnet with constant order parameter. The magnetoelectronic spin echo in ferromagnetic-normal-ferromagnetic spin valves comprises the prediction that this apparently lost spin current may partially reappear when adding a second identical but antiparallel ferromagnet [167]. This is a re-entrant phenomenon that resembles the time-dependent spin-echo effect in the collective magnetization of nuclei under pulsed rf excitation [260]. However, unlike this conventional spin echo effect, the magnetoelectronic spin echo effect is governed by quantum interference and is quenched by disorder. A similar quantum interference effect has been predicted independently by Blanter and Hekking [261]. Let us first discuss the origin of the spin echo in more detail before we discuss our proposed experimental setup to measure the spin echo.

The spin echo [260] is the following phenomenon: Subjecting a collections of magnetized non-interacting spins in an inhomogeneous environment to two transverse rf pulses in succession with a time delay $\tau$, the initially destroyed magnetic order reappears at time $\tau$ after the second pulse seemingly out of nowhere. Hahn [260] explained the spin echo with the following simple picture: Imagine runners at a race track that start at the same time but with different running speeds. After waiting a sufficiently long time, the runners are apparently randomly distributed over the race track. However, when we order the runners to suddenly turn around at time $t = \tau$ (rf pulse) and run with the same speed in the opposite direction, they all meet again at $t = 2\tau$.

The magnetoelectronic analogue to this spin echo can be observed in ferromagnetic-normal-ferromagnetic (F|N|F) spin valves with antiparallel magnetization alignment. These are the devices recently used to study current induced magnetization reversal by the spin transfer (see [8] and reference therein) and the present effect bares direct connection to this phenomenon. As mentioned earlier, most experiments appear to confirm Slonczewski's magnetization torque as the physical mechanism of this effect. It is equivalent to a transfer of spin angular momentum by the absorption by the magnetization order parameter of the spin-current components transversely polarized to the magnetization direction. This absorption is an effect of quantum interference on atomic length scales that lead to a vanishing transverse spin current and, as discussed below, the spin current can be recovered in analogy with the nuclear spin echo.

In order to understand the magnetoelectronic spin echo in more detail consider a spin current which is injected into a ferromagnetic metal with a polarization perpendicular to its magnetization. As discussed on several occasions in this review, the transverse spin state is not an eigenstate of the ferromagnet, but a coherent linear combination of the majority and minority spin eigenstates with different

Fermi wave vectors $k_{F\uparrow}$ and $k_{F\downarrow}$. The spin therefore precesses on a length scale depending on the perpendicular component of the wave-vector difference at the Fermi energy that depends on the angle of incidence. In elemental ferromagnets like Co, Ni and Fe, a large number of modes with different spin precession lengths in the ferromagnet contribute to the total current. Their destructive interference leads to a *loss of transverse spin current* inside the ferromagnet on a length scale of $\lambda_c = \pi/|k_{F\uparrow} - k_{F\downarrow}|$ [12, 69] (the so-called transverse spin-coherence length), which for typical transition metals is only a few Angstroms. The lost angular momentum is transferred to the ferromagnetic condensate, which thus experiences a torque that can lead to a magnetization reversal, as predicted by Slonczewski and Berger [26, 27, 85]. We investigate (anti)symmetric F|N|F spin valves, in which the magnetization of the second ferromagnet points into the opposite direction of the first one such that the spin precession of a propagating spin in the first ferromagnet is time reversed in the second ferromagnet, the lost transverse spin current is recovered, and the total spin torque on the magnetizations, *i.e.*, the absorbed spin angular momentum, is significantly reduced. A transverse spin current therefore can propagate through the (anti)symmetric F|N|F spin valve even though it apparently vanishes in the center.

For a specific scattering region with fixed ferromagnetic magnetization directions parallel or antiparallel to the spin-quantization axis ($z$ direction) and disregarding spin-flip scattering, the transmission matrices are spin diagonal, $(\hat{t}_{nm})_{s,s'} = \delta_{s,s'} t^s_{nm}$, where $t^s_{nm}$ is the transmission coefficient for an electron with spin $s$ ($s =\uparrow$, parallel, and $\downarrow$, antiparallel). The same holds for the transmission coefficients $t'^s_{nm}$ as well as the reflection coefficients $r^s_{nm}$ and $r'^s_{nm}$. Spin and charge currents through the scattering region depend on both the charge and the spin chemical potentials. In order to simplify the discussion, we assume in the following that the outer normal metal nodes are spin but not charge biased, $\mu_0^{(1)} = \mu_0^{(2)}$. This is a perfectly realistic situation, as outlined in Refs. [32, 262], and does not imply any loss of generality for our purposes. We first compute the spin current $\mathbf{I}^{(2)} = \mathrm{Tr}\left[\hat{\imath}^{(2)}\boldsymbol{\sigma}\right]$ on the right of the scattering region and find its components to be:

$$
\begin{aligned}
I_x^{(2)} &= \left[g_R^{t'}\mu_x^{(1)} + g_I^{t'}\mu_y^{(1)}\right] - \left[g_R^{r'}\mu_x^{(2)} + g_I^{r'}\mu_y^{(2)}\right] , \\
I_y^{(2)} &= \left[g_R^{t'}\mu_y^{(1)} - g_I^{t'}\mu_x^{(1)}\right] - \left[g_R^{r'}\mu_y^{(2)} - g_I^{r'}\mu_x^{(2)}\right] , \\
I_z^{(2)} &= g\left[\mu_z^{(1)} - \mu_z^{(2)}\right] , \qquad (185)
\end{aligned}
$$

where the conductances associated with transmission and reflection are $g^{t'} = \sum_{nm} t'^{\uparrow}_{nm} t'^{\downarrow *}_{nm}$, $g^{r'} = \sum_{nm}\left[\delta_{nm} - r'^{\uparrow}_{nm} r'^{\downarrow *}_{nm}\right]$, $g = (1/2)\sum_{nm}\left[\left|t'^{\uparrow}_{nm}\right|^2 + \left|t'^{\downarrow}_{nm}\right|^2\right]$, and the subscripts $R$ and $I$ refer to real and imaginary parts, respectively. Similarly, the current on the left of the scattering region $\mathbf{I}^{(1)}$ is expressed as Eq. (185) with the conductances $g^t$ and $g^r$ denoting the substitutions $t' \to t$ and $r' \to r$ and with the substitutions $1 \leftrightarrow 2$.

The longitudinal component of the spin current of Eq. (185) (*i.e.*, the component of the spin current collinear to the magnetization in the scattering region, which is the $z$ direction) is conserved, as expected, $I_z^{(1)} + I_z^{(2)} = 0$. If the scattering region would be a normal metal, the transmission and reflection coefficients are spin independent and consequently the $x$ and $y$ components of the spin current are conserved as well. When the magnetization directions in the scattering region are noncollinear with the spin accumulations, the transverse spin current is not necessarily conserved, so that in general $I_x^{(1)} + I_x^{(2)} \neq 0$, $I_y^{(1)} + I_y^{(2)} \neq 0$. As explained above, for a single ferromagnetic layer thicker than the ferromagnetic spin-coherence length $\lambda_{\mathrm{sc}}$, we know that $g^t$ and $g^{t'}$ vanish rapidly. The transverse spin current is then exclusively determined by the (reflection) spin-mixing conductance $g^r$ [12]. For high-density metallic systems, the phases of $r^{\uparrow}$ and $r^{\downarrow}$ are large and uncorrelated, and therefore $g^r \approx g_{\mathrm{S}} \approx g^{r'}$, where $g_{\mathrm{S}}$ is the Sharvin conductance given by the number of propagating channels in the normal-metal leads. The conventional conductance $g$ for the longitudinal spin component is determined by the transmission probability for spin-up and spin-down electrons. The spin currents are then simply $I_z^{(2)} = g(\mu_z^{(1)} - \mu_z^{(2)}) = -I_z^{(1)}$, and $I_{x,y}^{(1,2)} = -g_{\mathrm{S}}\mu_{x,y}^{(1,2)}$. The latter expression represents the loss of spin current that is directly proportional to the spin torque [36].

Let us now turn to the main subject of this chapter, the F|N|F spin valve in Fig. 46c. The lengths of the left and right leads are irrelevant for the transport properties. We define a scattering matrix for the left half of the scattering region (consisting of the left normal-metal lead with a length of half of the central normal metal spacer, the left ferromagnet, and again one half of the normal metal spacer) and the an analogous scattering matrix for the right half with opposite magnetization direction. The total scattering matrix can be found by concatenation, most conveniently for a spin-quantization axis collinear to the magnetizations. The (spin-dependent) transmission and reflection matrices for the entire scattering region then read

$$t^s = t_2^s[1 - r_1^{'s} r_2^s]^{-1} t_1^s, \tag{186}$$

$$r^s = r_1^s + t_1^{'s} r_2^s [1 - r_1^{'s} r_2^s]^{-1} t_1^s, \tag{187}$$

where $s$ denotes the spin of the electron and subscripts 1 and 2 left and right regions of the scatterer, respectively. Similar expressions hold for $t'$ and $r'$.

As mentioned above a spin valve is usually opaque for the transverse spin component. In order to observe the spin echo, the device must be structurally *symmetric* and *clean*, with antiparallel magnetization configurations. In general, the transmission (186) and reflection (187) matrices are nondiagonal in the space spanned by the transverse waveguide modes. For clean systems with specular scattering, reflection and transmission matrices in Eqs. (186) and (187) are diagonal in the transverse wave vector. Here it is implied that the outermost normal metals have simple Fermi surfaces (like Cu), but the principal arguments hold also for multiband normal metals. Furthermore, for mirror-symmetric ferromagnetic layers and an antiparallel and symmetric F|N|F spin valve for a suitable gauge choice symmetry dictates that $t^{\uparrow}_{1,ii} = t'^{\uparrow}_{1,ii} = t^{\downarrow}_{2,ii} = t'^{\downarrow}_{2,ii}$, $t^{\downarrow}_{1,ii} = t'^{\downarrow}_{1,ii} = t^{\uparrow}_{2,ii} = t'^{\uparrow}_{2,ii}$, $r^{\uparrow}_{1,ii} = r'^{\uparrow}_{1,ii} = r^{\downarrow}_{2,ii} = r'^{\downarrow}_{2,ii}$, and $r^{\downarrow}_{1,ii} = r'^{\downarrow}_{1,ii} = r^{\uparrow}_{2,ii} = r'^{\uparrow}_{2,ii}$, so that

$$t^{\uparrow} = t^{\downarrow} = t^{\downarrow}_1 t^{\uparrow}_1 [1 - r^{\uparrow}_1 r^{\downarrow}_1]^{-1} \,. \tag{188}$$

We find that $g^t_I = 0$ and $g^{t'}_I = 0$, but in contrast to the single ferromagnetic layer, $g^t_R$ and $g^{t'}_R$ are nonzero, which implies that a transverse spin current is permitted now. This is the *spin echo* in electron transport: The transverse spin coherence apparently lost on traversing the first ferromagnet, reappears after passing through the second ferromagnet, since $g^t = g^{t'} = g$. On the other hand there is no connection between $r^{\uparrow}$ and $r^{\downarrow}$. Consequently, the reflection mixing conductance is approximately equal to the Sharvin conductance, $g^r \approx g_{\mathrm{S}}$, and is not affected by the second ferromagnet. The current in the right lead is thus

$$I^{(2)}_{\alpha} = g\mu^{(1)}_{\alpha} - g^r \mu^{(2)}_{\alpha} \,, \qquad (\alpha = x,\ y) \,, \tag{189}$$

$$I^{(2)}_z = g\left[\mu^{(1)}_z - \mu^{(2)}_z\right] \tag{190}$$

and the current in the left lead becomes

$$I^{(1)}_{\alpha} = g\mu^{(2)}_{\alpha} - g^r \mu^{(1)}_{\alpha} \,, \qquad (\alpha = x,\ y) \,, \tag{191}$$

$$I^{(1)}_z = g\left[\mu^{(2)}_z - \mu^{(1)}_z\right] \,. \tag{192}$$

The first terms on the right-hand side of Eqs. (189) and (191) provide a quantitative expression for our spin echo in clean systems. Note that, although a transverse spin current is allowed through the sample, it is *not conserved*, $I^1_x + I^2_x \neq 0$, and $I^1_y + I^2_y \neq 0$, as $g^r \approx g_{\mathrm{S}} > g$, and the total magnetization torque on the spin valve is not completely quenched. Whereas the transmission mixing conductance is now strongly enhanced, the reflection mixing conductance is essentially unmodified, as noted above, and continues to exert a torque on the spin valve. The transmission of the transverse spin current suppresses an important contribution to the torque present in single-layer ferromagnets [56, 69], however.

### C. Conditions for observation

In order to observe the spin echo, the transverse wave vector must be conserved, because otherwise the electron precession through the second ferromagnet is not exactly time-reversed compared to that of the first ferromagnet. Disorder such as impurity scattering, interface alloying, and layer thickness fluctuations can be detrimental to the spin echo. The coherent propagator or transmission coefficient is exponentially damped on the length scale of the mean free path. This does not have to be a great concern, because the mean free path in bulk materials can be much larger than the film thicknesses in multilayers. More critical is the interface quality, since monolayer fluctuations of the layer thickness can already significantly dephase the spin echo. One might therefore question the observability of the spin echo in state-of-the-art transition-metal structures in which transport is usually well described by the classical series-resistor model [37]. There are several reasons to be optimistic, however. For one, the nonlocal exchange coupling in magnetic multilayers is a robust quantum-interference effect routinely observed in multilayers, oscillating not only as a function of the normal- but also of the magnetic-layer thickness [263, 264]. Furthermore, transport experiments with focused electrons, either by tunneling barriers [112] or by hot carriers [111] are indicative of specular scattering at high-quality interfaces. Indeed, conventional transport in high-density metallic structures consists of a large semiclassical background with relatively small quantum corrections [40]. In contrast, the spin-echo signal vanishes in the semiclassical approximation. The suppression of the spin echo by disorder thus provides direct information about the degree of quantum interference in electron transport.

First-principles band-structure calculations can provide quantitative estimates for the magnitude of the spin echo [36]. Such calculations can provide important information on the effect of interband scattering, impurity scattering at interfaces and bulk disorder on the suppression of the spin echo. We expect that a thickness difference between the two ferromagnetic layers would roughly suppress the spin echo as a single ferromagnetic layer of width equal the difference would suppress a transverse spin current. We thus estimate that the spin echo signal decays to 10-20 % of its original value by one atomic layer mismatch, and that the decay is algebraic[69]. A small misalignment $\theta$ of the magnetizations is not detrimental to the spin echo since the difference in phase shifts traversing the ferromagnets is small, $\Delta\phi = k_F L_F[\cos(\theta) - 1] \approx k_F L_F \theta^2/2 > 2\pi$, which gives $\theta^2 < 2(\lambda_F/L_F)$. This is easily satisfied e.g. for $L_F = 20\lambda_F$ resulting in $\theta < 18$ degrees.

We suggest a double-layer thin-film arrangement shown in Fig. 47 a for experimental observation of the spin echo. The $F_3|N|F_4$ spin valve should be thinner than the mean free path for impurity scattering. A Co/Cu/Co structure with a copper-layer thickness corresponding to the first antiparallel-coupling energy minimum should be a good choice because the spin flip in these materials is weak. Besides, this system, in particular its fabrication, is thoroughly investigated. The transverse spin accumulation driving the spin current through layers $F_3$ and $F_4$ can be excited by a spin battery operated by ferromagnetic resonance [262] or by a current-biased antiparallel spin valve with ferromagnets $F_1$ and $F_2$ whose magnetizations are rotated by 90 degrees with respect to $F_3$ and $F_4$, analogously to the spin-torque transistor [15]. The biased ferromagnets $F_1$ and $F_2$ create a spin accumulation in the normal metal between them. This spin accumulation can traverse the $F_3|F_4$ spin valve only when conditions for the spin echo are fulfilled. In that case, an induced spin accumulation in the upper normal metal, and hence the spin echo, can be detected as a voltage depending on the magnetization angle $\theta$ of ferromagnet $F_5$ as sketched in Fig. 2b. A similar set-up can be used to measure coherent transmission of a transverse spin current through a ultrathin magnetic layer [70].

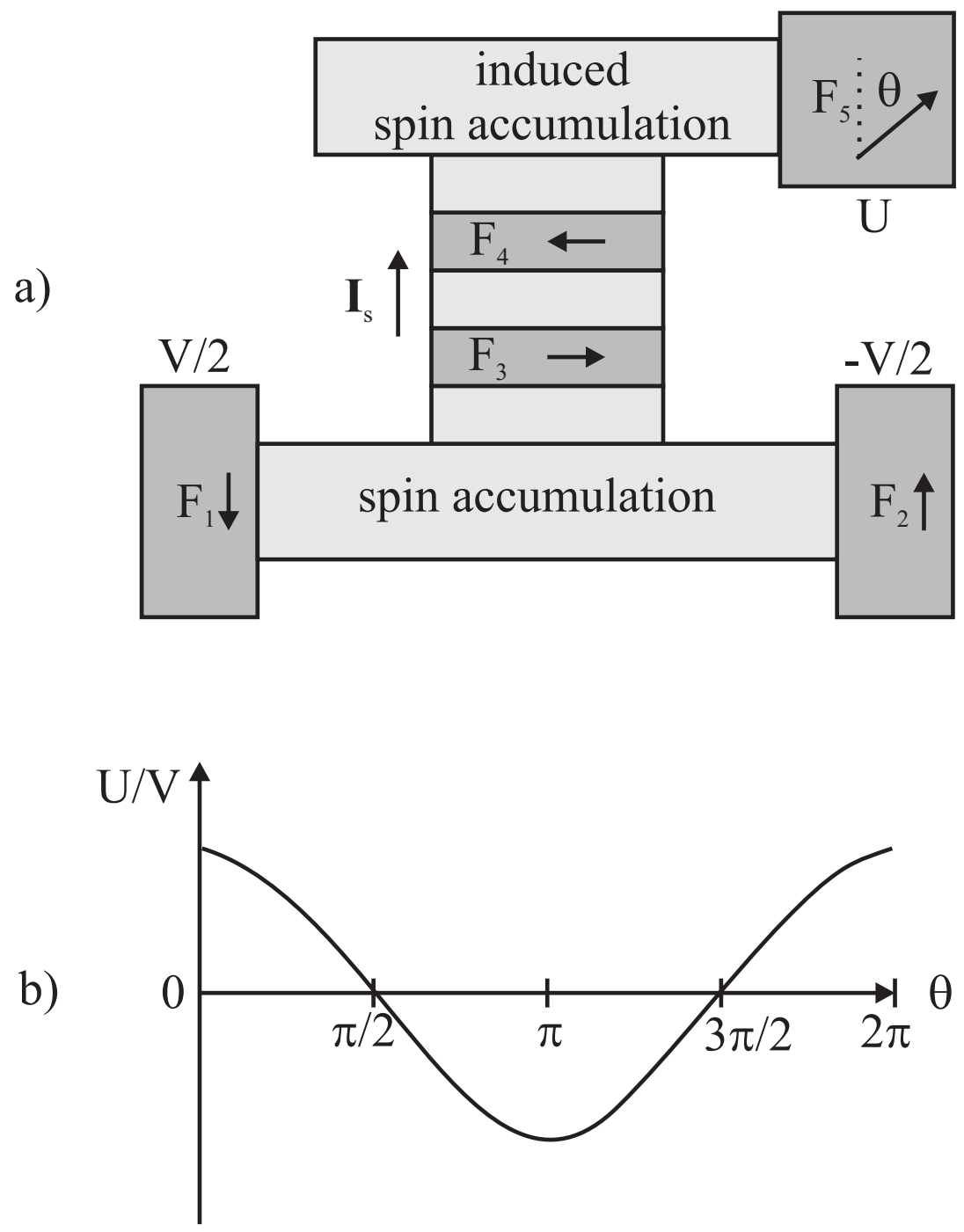

FIG. 47: Proposed experiment to observe the spin echo. Grey areas denote ferromagnetic layers. The spin echo is measured via the angular dependence of the potential $U(\theta)$ (or simply by its drop when $\theta$ is changed from 0 to $\pi$).

In conclusion, a magnetoelectronic spin echo in spin valves is a sensitive measure of quantum coherence in metallic magnetic multilayers. Its observation would create a break of the paradigm defended in this review, namely that quantum interference effects on transport in ferromagnetic systems may be disregarded.

## VIII. CONCLUSION AND OUTLOOK

We argued in this review that the main physics of the magnetoelectronic properties of metallic devices is understood by semiclassical concepts with parameters that are accessible to quantum mechanical calculations. Combined with state of the art first principles calculations DC and selected AC magnetoelectronic transport properties can be predicted with (semi)quantitative accuracy. Much of the evidence is derived from the success of the two-channel resistor model for the giant magnetoresistance in perpendicular spin valves with collinear magnetization directions. Experimental confirmation of our predictions in non-collinear systems is as yet limited to the angular resistance of CPP spin valves, the critical current for spin-wave excitations and magnetization reversal, and excess damping of the magnetization dynamics in normal-ferromagnetic hybrid structures, but we are optimistic that further support will be forthcoming.

The formalism of circuit theory is a finite element analysis of the differential equation that govern the semiclassical distribution functions including the spin accumulation. It is easy to use, also for non-experts and experimentalists, and we hope that it will be adopted by other researchers. The formalism should be useful for engineering applications like the test of new device designs that, *e.g.*, put effects like the magnetization torque to good use.

Of course, we do not pretend to promote a theory for everything. Novel directions that require extension of the present theory or completely different approaches, are indeed numerous. A straightforward extension would be a treatment of spin-orbit interactions that could help to get to grips with spin-flips at interfaces that have been measured quite recently [265]. Spin-orbit interaction also causes the anomalous Hall effects [266] that is also beyond the applicability of the present approach, in spite of being semiclassical in nature. The coupling between mechanical and ferromagnetic degrees of freedom is another classic topic of magnetism that may turn out to produce surprises in very small structures [267, 268].

Tunnel junctions that separate two ferromagnets require a different treatment. Transport through non-collinear F|I|F should in principle be computed for each angle. However, the projections of the two magnetizations appear to dominate the angular magnetoresistance (cosine dependence) and torque (sine dependence) with constant coefficients [269]. Unclear are as yet also the reasons for observed strong non-linear effects at higher applied bias [270, 271]. The recent develepment of MgO single crystal tunnel junctions [229, 230, 272] should help to better understand these systems.

The present mean-field theory needs to be improved when Coulomb interaction induced correlations start to play a role. The simplest case would be the Coulomb blockade in small normal metal grains that are contacted by ferromagnets with variable magnetization directions. Previous work concentrated on collinear structures [233, 273, 274], and first results have been published for very small islands (quantum dots) in the electric quantum limit [275] and for metallic islands [124]. Interesting systems that recently attract much attention are carbon nanotubes [22, 276–278] or conducting molecules/polymers with ferromagnetic contacts [279, 280]. Interesting correlation effects may be expected [244].

In Chapter VII we already gave on example of phase coherent effects that are clearly distinguishable from the semiclassical ones that have been discussed in previous sections. The large international effort in spintronics of nano-scale semiconducting structures based on the high-mobility two-dimensional electron gas should also reveal many as yet not anticipated effects beyond the semiclassical paradigm. A convergence between the up to now separate fields of metal-based magnetoelectronics and semiconductor-based spintronics is already in full action.

If surprises are the most delightful aspects of physics, practitioners of magnetoelectronics and spintronics may look forward to a pleasant future.

#### Acknowledgments

We acknowledge important contributions by our collaborators Yaroslav Tserkovnyak, Yuli Nazarov, Daniel Huertas-Hernando, Jan Manschot, Alex Kovalev, Wouter Wetzels, Maciej Zwierzycki, Ke Xia, Bart van Wees, and Mohand Talanana to the work reviewed here and for permission to publish their unpublished results. We are grateful to Bert Halperin, Andy Kent, Mark Stiles, Axel Hoffmann, and Peter Levy for numerous discussions. This work is part of the research program for the "Stichting voor Fundamenteel Onderzoek der Materie" (FOM) and the use of supercomputer facilities was sponsored by the "Stichting Nationale Computer Faciliteiten" (NCF), both financially supported by the "Nederlandse Organisatie voor Wetenschappelijk Onderzoek" (NWO). It was also supported by the EU Commission FP6 NMP-3 project 505587-1 "SFINX".and the Research Council of Norway, NANOMAT Grants No. 158518/431 and 158547/431.

## APPENDIX A: SPIN-ROTATION TRANSFORMATION

We consider the transmission and reflection matrices between a normal metal and a uniform ferromagnet in which the magnetization direction $\mathbf{m} = (\sin\theta\cos\varphi, \sin\theta\sin\varphi, \cos\theta)$ is a spatially independent unit vector in terms of the polar angles $\theta$ and $\varphi$. The Schrödinger equation reads

$$\hat{H}(\vec{r})\psi(\vec{r}) = E\psi(\vec{r}),$$

where $\psi(\vec{r})$ is a two-component spinor. In the Hamiltonian

$$\hat{H}(\vec{r}) = \hat{U}\left[-\frac{1}{2m}\nabla^2\hat{1} + V_s(\vec{r})\sigma_z + \hat{V}_c(\vec{r})\right]\hat{U}^+ \tag{A1}$$

$V_s(\vec{r})$ denotes the spin-dependent potential and $\hat{V}_c(\vec{r})$ is the scattering potential of the contact. The unitary matrix $\hat{U}$

$$\hat{U} = \begin{pmatrix} \cos\frac{\theta}{2} & -e^{-i\phi/2}\sin\frac{\theta}{2} \\ e^{i\phi/2}\sin\frac{\theta}{2} & \cos\frac{\theta}{2} \end{pmatrix}$$

diagonalizes the spin-dependent potential which vanishes in the normal metal, $V_s(\vec{r}) = 0$ for $x < x_l$ and attains its bulk value in the ferromagnet for $x > x_r$. The F|N contact is represented by the scattering potential

$$\hat{V}_c(\vec{r}) = \begin{pmatrix} V_\uparrow(\vec{r}) & V_{\text{sf}}(\vec{r}) \\ V_{\text{sf}}^\dagger(\vec{r}) & V_\downarrow(\vec{r}) \end{pmatrix} \tag{A2}$$

connecting the bulk values in the intermediate region between the normal metal and the ferromagnet for $x < x_l$ and $x > x_r$, respectively. The off-diagonal terms in (A2) represent the exchange potentials due to a non-collinear magnetization in the contact, spin-orbit interaction or spin-flip scatterers. The Hamiltonian (A1) can be diagonalized in spin-space by

$$\psi(\vec{r}) = \hat{U}\phi(\vec{r})\,, \tag{A3}$$

where the spinor $\phi(\vec{r})$ is governed by the Schrödinger equation

$$\left[-\frac{1}{2m}\nabla^2\hat{1} + V_s(\vec{r})\sigma_z + \hat{V}_c(\vec{r}) - E\right]\phi(\vec{r}) = 0\,.$$

Let us first consider an incoming wave from the normal metal in the transverse mode $n$ and with spin $s$ collinear to the magnetization in the ferromagnet. The wave function in the normal metal is

$$\phi_s^n(\vec{r}) = \sum_{ms'} \frac{\chi_N^m(\vec{\rho})}{\sqrt{k_m}} [\delta_{s's}\delta^{mn}\xi_s e^{ik^n x} + r_{cN,s's}^{mn}\xi_{s'} e^{-ik^m x}], \tag{A4}$$

where $\xi_\uparrow^\dagger = (1,0)$ and $\xi_\downarrow^\dagger = (0,1)$ are the spin-up and spin-down basis sates, $\chi_N^m(\vec{\rho})$ is a transverse wave function, $k_m$ is the longitudinal wave vector for mode $m$ and $r_{cN,s's}^{mn}$ is the reflection matrix from state $ns$ to state $ms'$. We would like to transform the result for the reflection matrix into a basis with arbitrary spin quantization axis. To this end we introduce the spinor wave function, of which the incoming component is spin-up and down in the general coordinate system:

$$\psi_s^n(\vec{r}) = \sum_{ms'} \frac{\chi_N^m(\vec{\rho})}{\sqrt{k_m}} \left[\delta_{s's}\delta^{mn}\xi_s e^{ik_n x} + r_{s's}^{mn}\xi_{s'} e^{-ik_m x}\right]. \tag{A5}$$

In order to satisfy the incoming-wave boundary conditions we cannot simply use the transformation (A3) but have to expand the wave function spinor in terms of the basis states $\phi(\vec{r})$ as

$$\psi_s^n(\vec{r}) = \hat{U}\sum_\sigma \phi_\sigma^n(\vec{r}) a_{\sigma s}, \tag{A6}$$

where $a_{\sigma s}$ are expansion coefficients to be determined by equating (A5) and (A6). For the incoming spin-up/down electron in the general coordinate system we require

$$a_{\sigma s} = \xi_\sigma^\dagger \hat{U}^\dagger \xi_s = \left(U^\dagger\right)_{\sigma s};\ \hat{a} = \hat{U}^\dagger \tag{A7}$$

and

$$r_{s's}^{mn} = \sum_{\sigma'\sigma} U_{s'\sigma'} r_{\mathrm{C},\sigma'\sigma}^{mn} U_{\sigma s}^\dagger. \tag{A8}$$

When spin-flip in the contacts may be disregarded, the reflection matrix

$$\hat{r}^{nm} = \sum_\sigma U_{s'\sigma} r_{cN,\sigma\sigma}^{mn} U_{\sigma s} = \hat{u}_\uparrow r_{cN,\uparrow}^{mn} + \hat{u}_\downarrow r_{cN,\downarrow}^{mn}, \tag{A9}$$

can be conveniently expressed in terms of the spin-projection matrices $\hat{u}_\uparrow$ and $\hat{u}_\downarrow$ (93)

$$\hat{u}_\pm(\theta) = \frac{1}{2}\left(\hat{1} \pm \boldsymbol{\sigma}\cdot\boldsymbol{m}\right).$$

We can understand the $N \to N$ reflection as first transforming the incoming (spin-up) electron from the general frame to the one parallel to the magnetization, let them be reflected normally and subsequently transforming back to the general frame.

The other coefficients of the scattering matrix are transformed differently because we chose the spin quantization axes in the ferromagnet invariantly collinear with the magnetization. In that basis the incoming wave from the ferromagnet reads

$$\phi_s^n(\vec{r}) = \sum_{ms'} \frac{\chi_{Fs}^m(\rho)}{\sqrt{k_{ms}}} \left[\delta_{s's}\delta^{mn}\xi_s e^{ik_s^n x} + r_{cF,s's}^{mn}\xi_{s'} e^{-ik_{s'}^m x}\right], \tag{A10}$$

where $\chi_m^{Fs}(\vec{\rho})$ is the (spin-dependent) transverse wave function and $k_{ms}$ is the spin-dependent Fermi wave-vector. The outgoing wave into the normal metal is

$$\phi_s^n(\vec{r}) = \sum_{ms'} \frac{\chi_N^m(\vec{\rho})}{\sqrt{k_m}} t_{\mathrm{cF},s's}^{m,n} \xi_{s'} \exp(ik_{s'}^m x). \tag{A11}$$

In the general coordinate system the transmitted wave is a linear combination of spin-up and spin-down states in the normal metal, which is given by the transformation Eq. (A3) directly and therfore

$$\hat{t}_{F\to N}^{nm} = \hat{U}\hat{t}_{cF}^{nm};\ t_{F\to N,s's}^{nm} = \sum_\sigma \hat{U}_{s'\sigma} t_{cF,\sigma s}^{nm} \tag{A12}$$

In the absence of spin-flip scattering in the contact $t_{F\to N,s's}^{nm} \to \hat{U}_{s's} t_{cF,s}^{nm}$.

The transmitted states from the normal metal into the ferromagnet are again spin-diagonal but the incoming states in the normal metal have to be transformed. So we get

$$t_{\sigma s}^{N\to F} = t_{cN,\sigma}\left(\hat{U}^{\dagger}\right)_{\sigma s};\ \hat{t}_{\sigma s}^{N\to F} = \hat{t}_{cN,\sigma}\hat{U}^{\dagger} \tag{A13}$$

The reflection process of an electron in the ferromagnet at a single interface to a normal metal, on the other hand, is independent of the spin quantization axes in the normal metal $\hat{r}_{F\to F} = \hat{r}_{cF}$. In the absence of spin-flips at the interface $(\hat{r}_{F\to F})_{ss'} \to \delta_{ss'} r_{cF,s}$.

The current in the normal metal Eq. (91) is, for a given energy shell and associating the first index with $\Psi$ and the second index with $\Psi^{\dagger}$,

$$\frac{h}{e}\hat{I} = -M\hat{f}^N + \sum_{nm}\left[\hat{r}^{mn}\hat{f}^N\left(\hat{r}^{nm}\right)^{\dagger} + \hat{t}'^{mn}\hat{f}^F\left(\hat{t}'^{nm}\right)^{\dagger}\right]. \tag{A14}$$

The contribution from the transmission probability to the spin-current is therefore ($\hat{f}^F$ is diagonal)

$$\begin{aligned}
eI_{\alpha\delta}^F &= \frac{e^2}{h}\sum_{nm\beta} t'^{mn}_{\alpha\beta} f_\beta^F \left(t'^{nm}_{\beta\delta}\right)^{\dagger} \\
&= \frac{e^2}{h}\sum_{nm\beta} U_{\alpha\beta} f_\beta^F \left|t'^{mn}_{\beta}\right|^2 U^{+}_{\beta\delta} = \frac{e^2}{h}\sum_{nm\beta}\hat{u}_\beta\left|t'^{mn}_{\beta}\right|^2 f_\beta^F \\
e\hat{I}^F &= G^{\uparrow\uparrow}\hat{u}_\uparrow f_\uparrow^F + G^{\downarrow\downarrow}\hat{u}_\downarrow f_\downarrow^F\,,
\end{aligned} \tag{A15}$$

with spin-dependent conductances

$$G^{ss} = \frac{e^2}{h}\left(M - \sum_{nm}|r_s^{nm}|^2\right) = \frac{e^2}{h}\sum_{nm}|t'^{mn}_s|^2\,.$$

Similarly, the contribution from the normal metal is

$$\begin{aligned}
-M\hat{f}^N + \sum_{nm}\hat{r}^{mn}\hat{f}^N\left(\hat{r}^{nm}\right)^{\dagger} &= -\sum_{nm}\left[\hat{f}^N\delta_{nm} - \sum_{ss'}\hat{u}_s\hat{f}^N\hat{u}^{+}_{s'}r_s^{mn}\left(r_{s'}^{mn}\right)^*\right] && \text{(A16)}\\
&= -\sum_{ss'}\hat{u}_s\hat{f}^N\hat{u}_{s'}\sum_{nm}\left(\delta_{nm} - r_s^{mn}\left(r_{s'}^{mn}\right)^*\right) && \text{(A17)}\\
&= -\sum_{ss'}\hat{u}_s\hat{f}^N\hat{u}_{s'}G_{ss'} && \text{(A18)}
\end{aligned}$$

where according to the unitarity of the scattering matrix $M - \sum_{nm}|r_s^{nm}|^2 = \sum_{nm}|t'^{nm}_s|^2$ and the mixing conductance is introduced as

$$G^{\uparrow\downarrow} = \frac{e^2}{h}\left[M - \sum_{nm}\left(r_\uparrow^{mn}\right)^* r_\downarrow^{nm}\right].$$

By adding both currents we recover Eq. (95):

$$\frac{h}{e}\hat{I} = \sum_{\alpha\beta} G^{\alpha\beta}\hat{u}_\alpha \left(\hat{f}^F - \hat{f}^N\right)\hat{u}_\beta \tag{A19}$$

## APPENDIX B: HOW TO USE CIRCUIT THEORY

Here we demonstrate how to compute transport properties of simple devices by magnetoelectronic circuit theory. We consider the limit of long $\tau_{sf}$, which in favorable cases is experimentally relevant even at room temperature [23]. Although the matrix manipulations are tedious in general, analytical formulas can be derived in terms of algebraic expressions of trigonometric functions of the magnetization angles. Here we concentrate on collinear and 90° degree magnetization configurations for which results are obtained more easily. The following results are derived with conventional transition metals in mind. Half-metallic (fully polarized) ferromagnetic metals (HMF) are interesting systems that maximize magnetoelectronic effects and simplify manipulation and results, however. We therefore also quote results in the limit of HMF's using the following model for the interface with a normal metal:

$$\hat{G} = \begin{pmatrix} G^{\uparrow} & G^{\uparrow\downarrow} \\ \left(G^{\uparrow\downarrow}\right)^* & G^{\downarrow} \end{pmatrix} \overset{HMF}{\rightarrow} \begin{pmatrix} G & G \\ G & 0 \end{pmatrix}. \tag{B1}$$

The particle current at the normal side of an F|N contact and directed into the normal metal is according to Eq. (96) [68]

$$\hat{I}^{(\vec{m})}(\varepsilon) = \sum_{\alpha\beta} G^{\alpha\beta}\hat{u}_\alpha(\vec{m})\left(\hat{f}^F(\varepsilon) - \hat{f}^N(\varepsilon)\right)\hat{u}_\beta(\vec{m}),$$

where the spin-$\frac{1}{2}$ projection matrices are according to Eq. (93)

$$\hat{u}_\uparrow(\vec{m}) = \frac{1}{2}\left(\hat{1} + \vec{m}\cdot\hat{\boldsymbol{\sigma}}\right);\ \hat{u}_\downarrow(\vec{m}) = \frac{1}{2}\left(\hat{1} - \vec{m}\cdot\hat{\boldsymbol{\sigma}}\right).$$

It is implied in the following that the energy integration of the occupation functions and spectra currents has been carried out. We take the ferromagnets as reservoirs at equilibrium $\int \hat{f}^F(\varepsilon)\, d\varepsilon = \mu^F\hat{1}$. Some insight can be gained by re-writing the current and the distribution function in the form of a scalar particle and a vectorial spin contribution. The distribution function on the normal node can be written as

$$\int \hat{f}^N(\varepsilon)\, d\varepsilon = \begin{pmatrix} f^{\uparrow\uparrow} & f^{\uparrow\downarrow} \\ \left(f^{\uparrow\downarrow}\right)^* & f^{\downarrow\downarrow} \end{pmatrix} = \mu^N\hat{1} + \vec{S}\cdot\hat{\sigma}, \tag{B2}$$

where $\mu^N = \left(f^{\uparrow\uparrow} + f^{\downarrow\downarrow}\right)/2$ is the local chemical potential and

$$\vec{S}=\frac{1}{2}\begin{pmatrix} f^{\uparrow\downarrow} + \left(f^{\uparrow\downarrow}\right)^* \\ f^{\uparrow\downarrow} - \left(f^{\uparrow\downarrow}\right)^* \\ f^{\uparrow\uparrow} - f^{\downarrow\downarrow} \end{pmatrix} = \begin{pmatrix} \operatorname{Re} f^{\uparrow\downarrow} \\ i \operatorname{Im} f^{\uparrow\downarrow} \\ \frac{1}{2}\left(f^{\uparrow\uparrow} - f^{\downarrow\downarrow}\right) \end{pmatrix} \equiv \begin{pmatrix} M_R \\ M_I \\ \Sigma \end{pmatrix} \tag{B3}$$

the spin accumulation vector. The junction parameters can be rewritten as $G = G^{\uparrow} + G^{\downarrow}$, $p = \left(G^{\uparrow} - G^{\downarrow}\right)/G$, $\eta_R = 2\operatorname{Re} G^{\uparrow\downarrow}/G$, and $\eta_I = 2\operatorname{Im} G^{\uparrow\downarrow}/G$. The matrix-current through an F|N interface $\hat{I} = (I_C \hat{1} + \vec{I}_S \cdot \hat{\boldsymbol{\sigma}})/2$ can then be expanded into vector components in terms the scalar particle current:

$$I_C = G\left(\mu^F - \mu^N\right) - pG\vec{m}\cdot\vec{S}$$

and the vector spin current:

$$\vec{I}_S = G[p(\mu^F - \mu^N) - (1-\eta_R)\,\vec{S}\cdot\vec{m}]\vec{m} - \eta_R\vec{S} + \eta_I(\vec{S}\times\vec{m}) \tag{B4}$$

with component perpendicular to the magnetization direction

$$\vec{I}_\perp = \vec{I}_S - \left(\vec{I}_S\cdot\vec{m}\right)\vec{m} = -G\eta_R\left[\vec{S} - \left(\vec{S}\cdot\vec{m}\right)\vec{m}\right] + G\eta_I(\vec{S}\times\vec{m}) \tag{B5}$$

that is proportional to the spin-torque exerted by the polarized current on the ferromagnet $\vec{L} = -\hbar\vec{I}_\perp/\left(2e\right)$ [56].

When the magnetization is parallel to the quantization axis, $\vec{m}_{\pm z} = (0,0,\pm 1)$ :

$$\hat{u}_z^{\uparrow} = \left(\hat{1} + \hat{\sigma}_z\right)/2 = \begin{pmatrix} 1 & 0 \\ 0 & 0 \end{pmatrix};\ \hat{u}_z^{\downarrow} = \left(\hat{1} - \hat{\sigma}_z\right)/2 = \begin{pmatrix} 0 & 0 \\ 0 & 1 \end{pmatrix} \tag{B6}$$

and $\hat{u}_{-z}^{\uparrow} = \hat{u}_z^{\downarrow}$. It is easy to see that

$$\hat{I}^{(z)} = \begin{pmatrix} G^{\uparrow}\left(\mu^F - f_{\uparrow\uparrow}\right) & -G^{\uparrow\downarrow} f_{\uparrow\downarrow} \\ -\left(G^{\uparrow\downarrow} f_{\uparrow\downarrow}\right)^* & G^{\downarrow}\left(\mu^F - f_{\downarrow\downarrow}\right) \end{pmatrix};\ \hat{I}^{(-z)} = \begin{pmatrix} G^{\uparrow}\left(\mu^F - f_{\downarrow\downarrow}\right) & -G^{\uparrow\downarrow}\left(f_{\uparrow\downarrow}\right)^* \\ -\left(G^{\uparrow\downarrow}\right)^* f_{\uparrow\downarrow} & G^{\downarrow}\left(\mu^F - f_{\uparrow\uparrow}\right) \end{pmatrix} \tag{B7}$$

Non-diagonal elements become important when the magnetization is not collinear to the quantization axis. For $\vec{m}_x = (1,0,0)$ :

$$\hat{u}_x^{\uparrow} = \left(\hat{1} + \hat{\sigma}_x\right)/2 = \frac{1}{2}\begin{pmatrix} 1 & 1 \\ 1 & 1 \end{pmatrix};\ \hat{u}_x^{\downarrow} = \left(\hat{1} - \hat{\sigma}_x\right)/2 = \frac{1}{2}\begin{pmatrix} 1 & -1 \\ -1 & 1 \end{pmatrix} \tag{B8}$$

and

$$\hat{I}^{(x)} = \frac{G}{2}\begin{pmatrix} \left(\mu^F - \mu^N\right) - pM_R - \eta_R\Sigma + i\eta_I M_I & p\left(\mu^F - \mu^N\right) - M_R + i\eta_I\Sigma - \eta_R M_I \\ p\left(\mu^F - \mu^N\right) - M_R - i\eta_I\Sigma + \eta_R M_I & \left(\mu^F - \mu^N\right) - pM_R + \eta_R\Sigma - i\eta_I M_I \end{pmatrix} \tag{B9}$$

Next, we determine the current through a symmetrical $F_1|N|F_2$ structure with parallel, antiparallel and perpendicularly oriented magnetizations, making use of the current conservation condition Eq. (90). Let us take $\mu^F = \Delta\mu$ and $\mu^F = 0$. $\hat{f}^N$ for parallel magnetizations in the $F_1^{(z)}|N|F_2^{(z)}$ configuration is determined by spin current conservation, $\hat{I}_1^{(z)} + \hat{I}_2^{(z)} = 0$.

$$\begin{pmatrix} G^{\uparrow}(\Delta\mu - f_{\uparrow\uparrow}) & -G^{\uparrow\downarrow}f_{\uparrow\downarrow} \\ -\left(G^{\uparrow\downarrow}f_{\uparrow\downarrow}\right)^* & G^{\downarrow}(\Delta\mu - f_{\downarrow\downarrow}) \end{pmatrix} = \begin{pmatrix} G^{\uparrow}f_{\uparrow\uparrow} & G^{\uparrow\downarrow}f_{\uparrow\downarrow} \\ \left(G^{\uparrow\downarrow}f_{\uparrow\downarrow}\right)^* & G^{\downarrow}f_{\downarrow\downarrow} \end{pmatrix} \tag{B10}$$

from which we conclude that $f_{\uparrow\downarrow} = 0$ and $f_{\uparrow\uparrow} = f_{\downarrow\downarrow} = \Delta\mu/2$. When antiparallel, $\hat{I}_1^{(z)} + \hat{I}_2^{(-z)} = 0$:

$$\begin{pmatrix} G^{\uparrow}(\Delta\mu - f_{\uparrow\uparrow}) & -G^{\uparrow\downarrow}f_{\uparrow\downarrow} \\ -\left(G^{\uparrow\downarrow}f_{\uparrow\downarrow}\right)^* & G^{\downarrow}(\Delta\mu - f_{\downarrow\downarrow}) \end{pmatrix} = \begin{pmatrix} G^{\downarrow}f_{\uparrow\uparrow} & G^{\uparrow\downarrow}(f_{\uparrow\downarrow})^* \\ \left(G^{\uparrow\downarrow}\right)^* f_{\uparrow\downarrow} & G^{\uparrow}f_{\downarrow\downarrow} \end{pmatrix} \tag{B11}$$

$$f_{\uparrow\uparrow} = \frac{G^{\uparrow}\Delta\mu}{G^{\uparrow}+G^{\downarrow}};\; f_{\downarrow\downarrow} = \frac{G^{\downarrow}\Delta\mu}{G^{\uparrow}+G^{\downarrow}};\; \Sigma = \frac{p\Delta\mu}{G} \tag{B12}$$

$\operatorname{Re} f_{\uparrow\downarrow} = 0$ since $-f_{\uparrow\downarrow} = (f_{\uparrow\downarrow})^*$ and $\operatorname{Im} f_{\uparrow\downarrow}$ is undetermined but irrelevant. The charge current

$$I_C = G\Delta\mu - G^{\uparrow}f_{\uparrow\uparrow} - G^{\downarrow}f_{\downarrow\downarrow} = \frac{2G^{\uparrow}G^{\downarrow}}{G^{\uparrow}+G^{\downarrow}}\Delta\mu \tag{B13}$$

vanishes for the HMF.

$\hat{f}^N$ for 90° magnetizations in $F_1^{(z)}|N|F_2^{(x)}$ structures follows again from the current conservation condition $\hat{I}_1^{(z)} + \hat{I}_2^{(x)} = 0$ and using Eq. (B9):

$$\begin{pmatrix} G^{\uparrow}(\Delta\mu - f_{\uparrow\uparrow}) & -G^{\uparrow\downarrow}f_{\uparrow\downarrow} \\ -\left(G^{\uparrow\downarrow}f_{\uparrow\downarrow}\right)^* & G^{\downarrow}(\Delta\mu - f_{\downarrow\downarrow}) \end{pmatrix} \tag{B14}$$

$$+\frac{G}{2}\begin{pmatrix} -\mu^N - pM_R - \eta_R\Sigma + i\eta_I M_I & -p\mu^N - M_R + i\eta_I\Sigma - \eta_R M_I \\ -p\mu^N - M_R - i\eta_I\Sigma + \eta_R M_I & -\mu^N - pM_R + \eta_R\Sigma - i\eta_I M_I \end{pmatrix} = 0 \tag{B15}$$

We find a spin accumulation vector

$$\vec{S} = \frac{1}{2}\frac{p\Delta\mu}{|\eta|^2 + \eta_R}\begin{pmatrix} -\eta_R \\ \eta_I \\ \eta_R \end{pmatrix} \overset{\text{HMF}}{\rightarrow} \begin{pmatrix} -1 \\ 0 \\ 1 \end{pmatrix}\frac{\Delta\mu}{6} \tag{B16}$$

and a charge current (from left to right):

$$I_C = G\left(\Delta\mu - \mu^N - \Sigma\right) = \frac{G}{2}\left(1 - \frac{p^2}{1 + \frac{|\eta|^2}{\eta_R}}\right)\Delta\mu \overset{\text{HMF}}{\rightarrow} G\frac{\Delta\mu}{3}. \tag{B17}$$

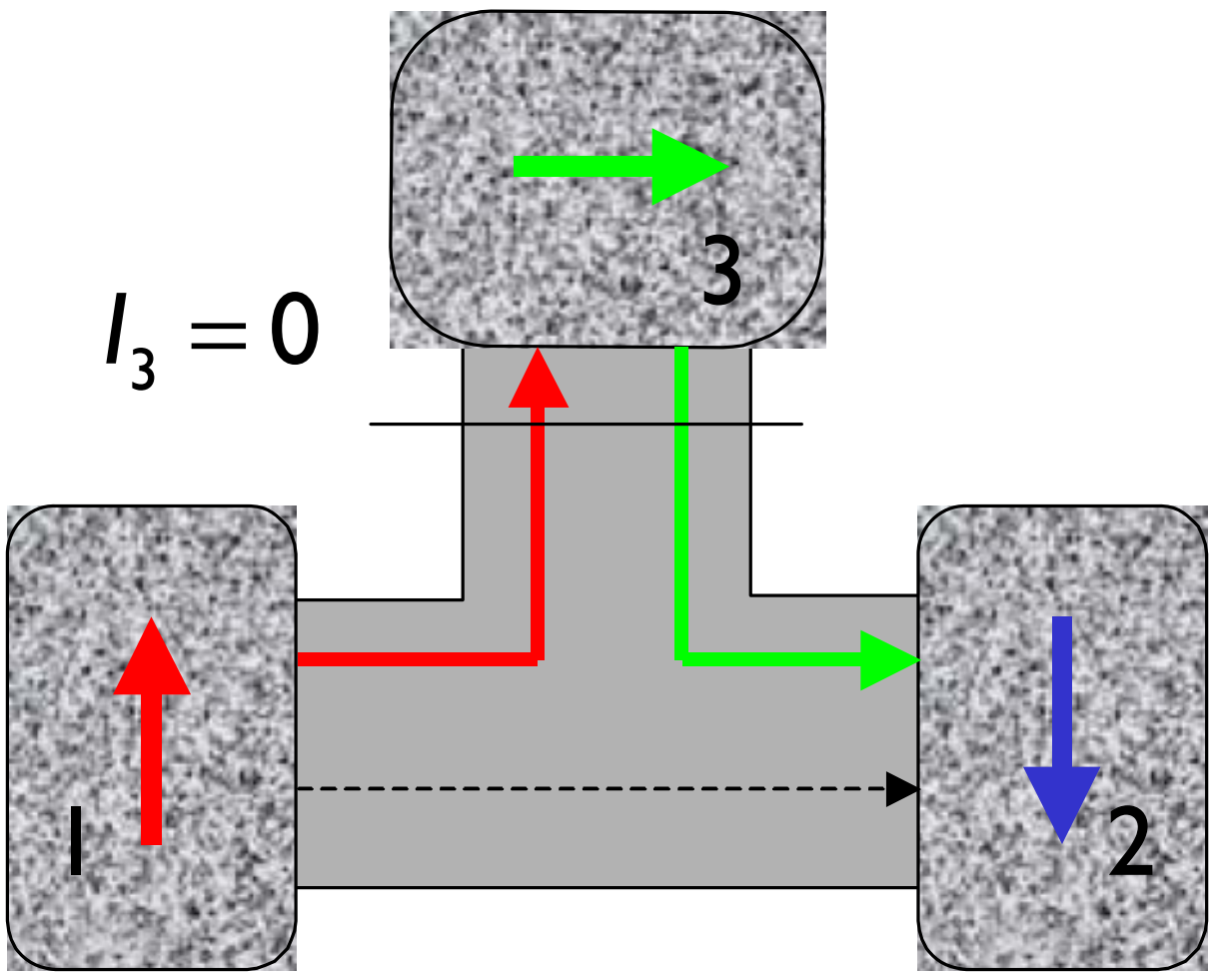

FIG. 48: The spin-flip transistor in the "on" configuration. The source, drain, and base reservoirs are numbered as 1,2 and 3, respectively. The base contact is floating such that the charge current $I_3$ vanishes.

The torque is proportional to the component normal to the magnetization that is absorbed by the ferromagnet and in the present symmetrical case equal for the right and left contact:

$$\vec{L} = \frac{\hbar}{2e^2}\frac{G}{2}\frac{|\eta|^2 p\Delta\mu}{|\eta|^2 + \eta_R}\begin{pmatrix}0\\0\\1\end{pmatrix} \overset{\text{HMF}}{\to} \frac{\hbar}{2e^2}\frac{G\Delta\mu}{3}\begin{pmatrix}0\\0\\1\end{pmatrix} = \frac{N\Delta\mu}{12\pi}\begin{pmatrix}0\\0\\1\end{pmatrix}, \tag{B18}$$

where $N$ is the number of conducting modes.

We now compute the characteristics of the device with three identical ferromagnetic terminals attached to a normal metal node. The magnetization of the third (base) terminal can be either collinear or normal to the magnetizations of source and drain. The latter are invariably taken to be antiparallel. The base contact is taken here to be metallic like source and drain contacts [12]. It is also possible to operate the spin-flip transistor with a floating contact that can read out by a tunnel junction at the back [36]. In the letter case the diagonal conductances of the base terminal may be set to zero.

For $\vec{m}_3 = (0, 0, 1)$ matrix-current conservation in the node reads:

$$\hat{I}_1^{(z)} + \hat{I}_2^{(-z)} + \hat{I}_3^{(z)} = 0, \tag{B19}$$

where

$$\hat{I}_1^{(z)} + \hat{I}_2^{(-z)} = \begin{pmatrix} G^{\uparrow}\left(\mu^F - f_{\uparrow\uparrow}\right) - G^{\downarrow} f_{\uparrow\uparrow} & 0 \\ 0 & G^{\downarrow}\left(\mu^F - f_{\downarrow\downarrow}\right) - G^{\uparrow} f_{\downarrow\downarrow} \end{pmatrix}, \tag{B20}$$

$$\hat{I}_3^{(-z)} = \begin{pmatrix} G^{\uparrow}\left(\mu_3^{(x)} - f_{\uparrow\uparrow}\right) & 0 \\ 0 & G^{\downarrow}\left(\mu_3^{(x)} - f_{\downarrow\downarrow}\right) \end{pmatrix}. \tag{B21}$$

We chose the bias of the base terminal such that its net charge current is zero

$$G^{\uparrow}\left(\mu_3^{(x)} - f_{\uparrow\uparrow}\right) + G^{\downarrow}\left(\mu_3^{(x)} - f_{\downarrow\downarrow}\right) = 0, \tag{B22}$$

$$\mu_3^{(x)} = \frac{G^{\uparrow} f_{\uparrow\uparrow} + G^{\downarrow} f_{\downarrow\downarrow}}{G^{\uparrow} + G^{\downarrow}}. \tag{B23}$$

The source-drain current then becomes

$$\frac{I_C^{(z)}}{G\Delta\mu} = \frac{3G^{\downarrow}G^{\uparrow}}{\left(G^{\downarrow}\right)^2 + 4G^{\downarrow}G^{\uparrow} + \left(G^{\uparrow}\right)^2}$$

and vanishes again for HMF terminals:

$$I_C^{(z)} = 0;\ f_{\downarrow\downarrow} = 0;\ \mu_3^{(x)} = f_{\uparrow\uparrow} = \Delta\mu;\ \Sigma = \frac{\Delta\mu}{2}. \tag{B24}$$

When the third electrode is rotated to the $x$-direction, the zero charge current condition for the base contact dictates:

$$\Delta\mu_3^{(x)} = \mu^N + \frac{pM_R}{G} \tag{B25}$$

Matrix-current conservation:

$$\hat{I}_z^{(1)} + \hat{I}_{-z}^{(2)} + \hat{I}_x^{(3)} = 0 \tag{B26}$$

leads to:

$$\Delta\mu_3^{(x)} = \mu^N = \Delta\mu/2 \tag{B27}$$

and a charge current of

$$I_C^{(x)} = \left(G - \frac{p^2}{G + \left|G^{\uparrow\downarrow}\right|^2 / \operatorname{Re} G^{\uparrow\downarrow}}\right)\frac{\Delta\mu}{2}. \tag{B28}$$

For the HMF, $\Sigma = \Delta\mu/4$ and $I_C^{(x)} = G\Delta\mu/4$. When (in this limit) the potential at the third electrode is varied, $\Sigma$ does not change, but the length of $\vec{S}$ does:

$$\left|\vec{S}\right|^2 = \Sigma^2 + M_R^2 + M_I^2 = \left(\frac{\Delta\mu}{4}\right)^2 + \left(\frac{2\Delta\mu^{(3)} - \Delta\mu}{14}\right)^2 \tag{B29}$$

The spin-accumulation is minimal for the zero charge current condition for the third terminal, which corresponds to a maximum of the spin-current through the third terminal. The minimum spin-accumulation consequently allows a maximum source-drain current without dissipation in the base. The base magnetization then experiences a maximum torque of

$$\left|\vec{L}\right|_b = \frac{\hbar}{2e^2}\frac{pG\mu^N}{2+\eta}\frac{\Delta\mu}{2\pi} \tag{B30}$$

In the HMF limit the physics of this transistor action is easily understood. In the collinear configuration the device is electrically dead: No current can flow into the drain because spin-up states can not penetrate a HMF with spin-down magnetization. When the magnetization of the base is rotated by 90° and the bias is adjusted to the zero charge current condition, the incoming spin-up current is exactly equal to the outflowing spin-down current. Although there is no direct current between source and drain, the outflowing spin-down current from the base can enter the drain. Effectively we thus have switched on a source-drain charge current by rotating the magnetization. Since the base contact operates as a perfect spin-flip, we suggest the name *spin-flip transistor* for our device. When the base contact is not connected to an external circuit, the floating potential adjusts itself automatically to fulfill the zero charge current condition.

---